\documentclass[11pt]{article}
\pdfoutput=1  
\usepackage{%
  abstract,
  amsmath,
  amssymb,
  amsthm,
  array,
  color,
  comment,
  extarrows,
  latexsym,
  mathrsfs,
  mathtools,
  multirow,
  nccmath,
  physics,
  rotating,
  setspace,
  slashed,
  textcomp,
  titlesec,
}
\usepackage[sorting=none,backend=biber,sortcites,citestyle=numeric-comp,maxbibnames=5]{biblatex}
\DeclareFieldFormat[misc]{title}{%
  {\mkbibquote{#1}}
}

\let\oldprintbibliography\printbibliography
\renewcommand{\printbibliography}{\clearpage \oldprintbibliography}

\usepackage[pdfusetitle]{hyperref}
\hypersetup{
  linktocpage=true,
  breaklinks=true,
  pdfauthor={Arif Er and Meng-Chwan Tan},
}

\makeatletter
\renewcommand{\sectionautorefname}{$\S$\@gobble}
\renewcommand{\subsectionautorefname}{$\S$\@gobble}
\renewcommand{\subsubsectionautorefname}{$\S$\@gobble}
\makeatother

\usepackage[multiple]{footmisc}
\usepackage[auth-lg,affil-it]{authblk}

\numberwithin{equation}{section}

\titleformat*{\section}{\bfseries\large}

\usepackage[a4paper,
textwidth=165mm,
tmargin=25mm,
bmargin=40mm,
]{geometry}

\allowdisplaybreaks[2]

\DeclareFontFamily{U}{MnSymbolC}{}
\DeclareSymbolFont{MnSyC}{U}{MnSymbolC}{m}{n}
\DeclareFontShape{U}{MnSymbolC}{m}{n}{
  <-6>  MnSymbolC5
  <6-7>  MnSymbolC6
  <7-8>  MnSymbolC7
  <8-9>  MnSymbolC8
  <9-10> MnSymbolC9
  <10-12> MnSymbolC10
  <12->   MnSymbolC12
}{}
\DeclareMathSymbol{\intprod}{\mathbin}{MnSyC}{'270}  

\usepackage{graphicx}
\usepackage{tikz}
\usepackage{tikz-cd}
\usepackage{subcaption}
\usepackage{adjustbox}
\usetikzlibrary{
  arrows,
  automata,
  positioning,
  calc,patterns.meta,
  arrows.meta,
  decorations.pathmorphing,
  shapes.misc
}

\newcommand{\Dv}[2]{\frac{D #1}{D #2}}
\newcommand{\longto}{\longrightarrow}

\newcommand{\subtitle}[1]{\bigskip\noindent\textit{#1}\vspace*{0.5em}}

\newcommand{\C}{\mathbb{C}}

\newcommand{\R}{\mathbb{R}}

\title{%
  Topological 8d \texorpdfstring{$\mathcal{N}=1$}{N=1} Gauge Theory: Novel Floer Homologies, %
  \texorpdfstring{\\ \bigskip}{}%
  and \texorpdfstring{$A_\infty$}{A-infty}-categories of Six, Five, and Four-Manifolds%
  \texorpdfstring{\vspace{2.5cm}}{}%
}%

\author{Arif Er\thanks{
    Email: \href{mailto:arif.er@u.nus.edu}{arif.er@u.nus.edu}
  }
}%

\author{Meng-Chwan Tan\thanks{
    Email:\href{mailto:mctan@nus.edu.sg}{mctan@nus.edu.sg}
  }
}%

\affil{%
  Department of Physics \\ \medskip%
  National University of Singapore \\ \medskip%
  2 Science Drive 3, Singapore 117551%
}%
\date{}

\begin{document}
\addtolength{\baselineskip}{1.5mm}

\maketitle
\pagenumbering{gobble} 

\begin{abstract}
  \normalsize \singlespacing \noindent%
  This work is a continuation of the program initiated in~\cite{er-2023-topol-n}.
  We show how one can define novel gauge-theoretic (holomorphic) Floer homologies of seven, six, and five-manifolds, from the physics of a topologically-twisted 8d $\mathcal{N}=1$ gauge theory on a Spin$(7)$-manifold via its supersymmetric quantum mechanics interpretation.
  They are associated with $G_2$ instanton, Donaldson-Thomas, and Haydys-Witten configurations on the seven, six, and five-manifolds, respectively.
  We also show how one can define hyperkähler Floer homologies specified by hypercontact three-manifolds, and symplectic Floer homologies of instanton moduli spaces.
  In turn, this will allow us to derive Atiyah-Floer type dualities between the various gauge-theoretic Floer homologies and symplectic intersection Floer homologies of instanton moduli spaces.
  Via a 2d gauged Landau-Ginzburg model interpretation of the 8d theory, one can derive novel Fukaya-Seidel type $A_\infty$-categories that categorify Donaldson-Thomas, Haydys-Witten, and Vafa-Witten configurations on six, five, and four-manifolds,  respectively -- thereby categorifying the aforementioned Floer homologies of six and five-manifolds, and the Floer homology of four-manifolds from \cite{er-2023-topol-n} -- where an Atiyah-Floer type correspondence for the Donaldson-Thomas case can be established.
  Last but not least, topological invariance of the theory suggests a relation amongst these Floer homologies and Fukaya-Seidel type $A_\infty$-categories for certain Spin$(7)$-manifolds.
  Our work therefore furnishes purely physical proofs and generalizations of the conjectures by Donaldson-Thomas~\cite{donaldson-1996-gauge}, Donaldson-Segal~\cite{donaldson-2009-gauge-theor-ii}, Cherkis~\cite{cherkis-2015-octon-monop-knots}, Hohloch-Noetzel-Salamon~\cite{hohloch-2009-hyper-struc}, Salamon~\cite{salamon-2013-three-dimen}, Haydys~\cite{haydys-2015-fukay-seidel}, and Bousseau~\cite{bousseau-2024-holom-floer}, and more.
\end{abstract}

\clearpage

\pagenumbering{arabic} 

\tableofcontents

\section{Introduction, Summary and Acknowledgements}
\vspace{0.4cm}
\setlength{\parskip}{5pt}

\subtitle{Introduction}

In~\cite{acharya-1997-higher-dimen}, Acharya-O'Loughlin-Spence studied an 8d $\mathcal{N} = 1$ topologically-twisted gauge theory on a certain eight-manifold where they predicted that one could physically derive from it, a Floer homology of a corresponding seven-manifold.
Incidentally, this gauge-theoretic Floer homology was also conjectured by Donaldson-Thomas in a program they initiated in~\cite{donaldson-1996-gauge}.
This program was continued by Donaldson-Segal in~\cite{donaldson-2009-gauge-theor-ii}, where they conjectured, with support from some preliminary computations, a gauge-theoretic Floer homology of a corresponding six-manifold.
Parallel to these efforts, a separate program was launched by Hohloch-Noetzel-Salamon in~\cite{hohloch-2009-hyper-struc, salamon-2013-three-dimen} to define a novel hyperkähleric Floer homology assigned to a three-four manifold pair, which they conjectured could be derived from the Donaldson-Thomas program in the adiabatic limit.

In this paper, we continue the program initiated in~\cite{er-2023-topol-n}, and aim to physically derive these mathematically conjectured Floer homologies, amongst other things.
To this end, we will study, on various decomposable eight-manifolds with Spin$(7)$ holonomy in different topological limits, the aforementioned 8d $\mathcal{N} = 1$ topologically-twisted gauge theory whose BPS equation is the Spin$(7)$ instanton equation.
As an offshoot, we would also be able to derive a web of mathematically-novel relations amongst these Floer homologies, their Atiyah-Floer dualities, and more.

The computational techniques we employ are mainly those of standard Kaluza-Klein reduction; generalizations of the topological reduction pioneered in~\cite{bershadsky-1995-topol-reduc}; recasting gauge theories as supersymmetric quantum mechanics (SQM) as pioneered in~\cite{blau-1993-topol-gauge}; the physical realization of Floer homology groups via SQM in infinite-dimensional space as elucidated in~\cite{ong-2023-vafa-witten-theor}; and the physical realization of Fukaya-Seidel type $A_{\infty}$-categories via a soliton string theory in infinite-dimensional space as elucidated in~\cite{er-2023-topol-n}.

Let us now give a brief plan and summary of the paper.

\subtitle{A Brief Plan and Summary of the Paper}

In \autoref{sec:8d theory}, we discuss general aspects of a topologically-twisted 8d $\mathcal{N}=1$ theory on an eight-manifold with Spin$(7)$ holonomy (a Spin$(7)$-manifold) with a ``trivial twist'', where the gauge group $G$ is taken to be a real, simple, compact Lie group.

In \autoref{sec:floer homology of m7}, we let the Spin$(7)$-manifold be $\text{Spin}(7) = G_2 \times \R$, where $G_2$ is a closed and compact seven-manifold with $G_2$ holonomy (a $G_2$-manifold).
We recast the aforementioned Spin$(7)$ theory as a 1d SQM in the space $\mathfrak{A}_7$ of irreducible gauge connections $A$ on the $G_2$-manifold with action~\eqref{eq:g2 x r:sqm:action}.
This will in turn allow us to express the partition function as~\eqref{eq:g2 x r:partition fn}:
\begin{equation}
  \label{summary:eq:g2 x r:partition:fn}
  \boxed{
    \mathcal{Z}_{\text{Spin}(7),G_2 \times \R}(G)
    = \sum_j \mathcal{F}^{G}_{\text{Spin}(7)}(\Psi_{G_2}^j)
    = \sum_j \text{HF}^{\text{Spin}(7)\text{-inst}}_{d_j}(G_2, G)
    = \mathcal{Z}^{\text{Floer}}_{\text{Spin}(7)\text{-inst},G_2}(G)
  }
\end{equation}
where $\text{HF}^{\text{Spin}(7)\text{-inst}}_{d_j}(G_2, G)$ is a \emph{novel} Spin$(7)$ instanton Floer homology class assigned to a $G_2$-manifold, of degree $d_j$, defined by Floer differentials described by the gradient flow equations~\eqref{%
  eq:g2 x r:sqm:flow}:
\begin{equation}
  \label{summary:eq:g2 x r:sqm:flow}
  \boxed{
    \dv{A^\alpha}{t} = -g^{\alpha\beta}_{\mathfrak{A}_7}\pdv{V_7}{A^\beta}
  }
\end{equation}
and Morse functional~\eqref{eq:g2 x r:morse fn V7}:
\begin{equation}
  \label{summary:eq:g2 x r:morse fn V7}
  \boxed{
    V_7(A, \varphi)
    = \int_{G_2} \, \Tr \, \left(
      CS(A) \wedge \star \phi_t
    \right)
  }
\end{equation}
The chains of the Spin$(7)$ instanton Floer complex are generated by fixed critical points of $V_7$, which correspond to \emph{time-invariant $G_2$ instanton configurations} on the $G_2$-manifold given by time-independent solutions to the 7d equation~\eqref{eq:g2 x r:morse fn V7:crit pts}:
\begin{equation}
  \label{summary:eq:g2 x r:morse fn V7:crit pts}
  \boxed{
    F \wedge \star \phi_t = 0
  }
\end{equation}

Note that $\text{HF}^{\text{Spin}(7)\text{-inst}}_{d_j}(G_2, G)$ was first mathematically conjectured to exist by Donaldson-Thomas~\cite[$\S3$]{donaldson-1996-gauge} as a Floer homology generated by $G_2$ instantons on a $G_2$-manifold whose flow lines are time-varying solutions to the Spin$(7)$ instanton equation on $G_2 \times \R$.
We have therefore furnished a physical proof of their mathematical conjecture.

In \autoref{sec:floer homology of m6}, we let $G_2 = CY_3 \times S^1$, where $CY_3$ is a closed and compact Calabi-Yau threefold, and perform a Kaluza-Klein (KK) dimensional reduction of Spin$(7)$ theory on $G_2 \times \R$ by shrinking $S^1$ to be infinitesimally small.
We obtain the corresponding 1d SQM theory in the space $\mathfrak{A}_6$ of irreducible $(\mathcal{A}, C)$ fields on $CY_3$ with action~\eqref{eq:cy3 x r:action:sqm}, where $\mathcal{A} \in \Omega^{(1, 0)}(CY_3, \text{ad}(G))$ and $C \in \Omega^0(CY_3, \text{ad}(G))$ are a holomorphic gauge connection and real scalar, respectively, 
that is equivalent to the resulting 7d-Spin$(7)$ theory on $CY_3 \times \R$.
As before, this will allow us to express the partition function as~\eqref{eq:cy3 x r:partition fn}:
\begin{equation}
  \label{summary:eq:cy3 x r:partition fn}
  \boxed{
    \mathcal{Z}_{\text{Spin}(7),CY_3 \times \R}(G)
    = \sum_k \mathcal{F}^{G}_{\text{7d-Spin}(7)}(\Psi_{CY_3}^k)
    = \sum_k \text{HHF}^{G_2\text{-M}}_{d_k}(CY_3, G)
    = \mathcal{Z}^{\text{Floer}}_{G_2\text{-M},CY_3}(G)
  }
\end{equation}
where $\text{HHF}^{G_2\text{-M}}_{d_k}(CY_3, G)$ is a \emph{novel} holomorphic $G_2$ monopole ($G_2$-M) Floer homology class assigned to $CY_3$, of degree $d_k$, defined by Floer differentials described by the holomorphic gradient flow equation~\eqref{eq:cy3 x r:sqm:flow}:
\begin{equation}
  \label{summary:eq:cy3 x r:sqm:flow}
  \boxed{
    \dv{\mathcal{A}^\alpha}{t}
    = - g^{\alpha\bar{\beta}}_{\mathfrak{A}_6} \left(
      \pdv{V_6}{\mathcal{A}^\beta}
    \right)^*
    \qquad
    \dv{C^\alpha}{t}
    = - g^{\alpha\bar{\beta}}_{\mathfrak{A}_6} \left(
      \pdv{V_6}{C^\beta}
    \right)^*
  }
\end{equation}
and holomorphic Morse functional~\eqref{eq:cy3 x r:morse fn V6}:
\begin{equation}
  \label{summary:eq:cy3 x r:morse fn V6}
  \boxed{
    V_6(\mathcal{A}, C)
    = \frac{1}{2} \int_{CY_3} \Tr \left(
      CS(\mathcal{A}) \wedge \bar{\star} \Lambda
      + 2 C \wedge \mathcal{F}^{(1, 1)} \wedge \bar{\star} \omega
    \right)
  }
\end{equation}
where $\Lambda$ is a holomorphic three-form and $\omega$ is the Kähler two-form of $CY_3$.
The chains of the holomorphic $G_2$-M Floer complex are generated by fixed critical points of $V_6$, which correspond to \emph{time-invariant Donaldson-Thomas (DT) configurations on $CY_3$ with the scalar being real}, given by time-independent solutions to the 6d equations~\eqref{eq:cy3 x r:morse fn V6:crit pts}:
\begin{equation}
  \label{summary:eq:cy3 x r:morse fn V6:crit pts}
  \boxed{
    \omega \wedge \bar{\star} \mathcal{F}^{(1,1)} = 0
    \qquad
    \mathcal{F}^{(2,0)} = 0
    \qquad
    \mathcal{D}_m C = 0
  }
\end{equation}

Note that when $C = 0$, $\text{HHF}^{G_2\text{-M}}_{d_k}(CY_3, G)$ becomes a holomorphic $G_2$ \emph{instanton} Floer homology, generated by holomorphic vector bundles on $CY_3$ (i.e., DT configurations on $CY_3$ with $C = 0$) whose flow lines correspond to time-varying solutions to the $G_2$ instanton equation, which was mathematically conjectured to exist by Donaldson-Segal~\cite[$\S$4]{donaldson-2009-gauge-theor-ii}.
We have therefore furnished a physical proof and generalization (when $C \neq 0$) of their mathematical conjecture.

Furthermore, when the complexification of $CY_3$ is undone and the results are instead expressed in real components, we will find that we have a \emph{real} Floer homology of $CY_3$ generated by fixed points of the \emph{real-valued} Morse functional -- the Chern-Simons-Higgs functional -- whose flow lines correspond to time-varying solutions to the $G_2$ monopole equation on $CY_3 \times \R$.
This result has previously been speculated by Cherkis~\cite[$\S$7]{cherkis-2015-octon-monop-knots}.
We have therefore furnished a physical proof of his speculation.

In \autoref{sec:floer homology of m5}, we further specialize to the case where $CY_3 = CY_2 \times S^1 \times S^1$, where $CY_2$ is a closed and compact Calabi-Yau twofold, and perform another KK dimensional reduction of 7d-Spin$(7)$ theory on $CY_3 \times \R$ by shrinking one of the $S^1$ circles to be infinitesimally small.
We obtain the corresponding 1d SQM theory in the space $\mathfrak{A}_5$ of irreducible $(\hat{\mathcal{A}}, \hat{\mathcal{B}}, \Gamma)$ fields on $CY_2 \times S^1$ that is equivalent to the resulting 6d theory on $CY_2 \times S^1 \times \R$.
Here, (i) $\hat{\mathcal{A}} \in \Omega^0(S^1, \text{ad}(G)) \otimes \Omega^{(1,0)}(CY_2, \text{ad}(G))$, (ii) $\hat{\mathcal{B}} \in \Omega^0(S^1, \text{ad}(G)) \otimes \Omega^{2, +}(CY_2, \text{ad}(G))$, and (iii) $\Gamma \in \Omega^1(S^1, \text{ad}(G)) \otimes \Omega^0(CY_2, \text{ad}(G))$ are (i) a real scalar (holomorphic gauge connection), (ii) a real scalar (real self-dual two-form), and (iii) a real gauge connection (real scalar), respectively.
Again, this will also allow us to express the partition function as~\eqref{eq:cy2 x s x r:partition fn}:
\begin{equation}
  \label{summary:eq:cy2 x s x r:partition fn}
  \boxed{
    \mathcal{Z}_{\text{Spin}(7),CY_2 \times S^1 \times \R}(G)
    = \sum_l \mathcal{F}^{G}_{\text{6d-Spin}(7)}(\Psi_{CY_2 \times S^1}^l)
    = \sum_l \text{HHF}^{\text{DT}}_{d_l}(CY_2 \times S^1, G)
    = \mathcal{Z}^{\text{Floer}}_{\text{DT},CY_2 \times S^1}(G)
  }
\end{equation}
where $\text{HHF}^{\text{DT}}_{d_l}(CY_2 \times S^1, G)$ is a \emph{novel} holomorphic DT Floer homology class assigned to $CY_2 \times S^1$, of degree $d_l$, defined by Floer differentials described by the holomorphic gradient flow equations~\eqref{eq:cy2 x s x r:sqm:flow}:
\begin{equation}
  \label{summary:eq:cy2 x s x r:sqm:flow}
  \boxed{
    \dv{\hat{\mathcal{A}}^\alpha}{t}
    = - g^{\alpha\bar{\beta}}_{\mathfrak{A}_5} \left(
      \pdv{V_5}{\hat{\mathcal{A}}^\beta}
    \right)^*
    \qquad
    \dv{\hat{\mathcal{B}}^\alpha}{t}
    = - g^{\alpha\bar{\beta}}_{\mathfrak{A}_5} \left(
      \pdv{V_5}{\hat{\mathcal{B}}^\beta}
    \right)^*
    \qquad
    \dv{\Gamma^\alpha}{t}
    = - g^{\alpha\bar{\beta}}_{\mathfrak{A}_5} \left(
      \pdv{V_5}{\Gamma^\beta}
    \right)^*
  }
\end{equation}
and holomorphic Morse functional~\eqref{eq:cy2 x s x r:morse fn V5}:
\begin{equation}
  \label{summary:eq:cy2 x s x r:morse fn V5}
  \boxed{
    V_5(\hat{\mathcal{A}}, \hat{\mathcal{B}}, \Gamma)
    = \int \frac{i}{2} d_y \left(
      \hat{\mathcal{A}} \wedge \bar{\star} \hat{\mathcal{A}}
      + \hat{\mathcal{B}} \wedge \bar{\star} \hat{\mathcal{B}}
    \right)
    - 2i \hat{\mathcal{B}} \wedge \bar{\star} \hat{\mathcal{F}}
    - \Gamma \wedge \hat{\omega} \wedge \bar{\star} \left(
      \hat{\mathcal{F}}
      - \frac{1}{4} (\hat{\mathcal{B}} \times \hat{\mathcal{B}})
    \right)
  }
\end{equation}
where $\hat{\omega}$ is the Kähler two-form of $CY_2$.
The chains of the holomorphic DT Floer complex are generated by fixed critical points of $V_5$, which correspond to \emph{time-invariant HW configurations on $CY_2 \times S^1$ with one of the linearly-independent components of the self-dual two-form field being zero}, given by time-independent solutions to the 5d equations~\eqref{eq:cy2 x s x r:morse fn V5:crit pts}:
\begin{equation}
  \label{summary:eq:cy2 x s x r:morse fn V5:crit pts}
  \boxed{
    \begin{aligned}
      d_y \hat{\mathcal{A}}
      + \bar{\star} ( \hat{\mathcal{D}} \bar{\star} \hat{\mathcal{B}} )
      &= \hat{\mathcal{D}} \Gamma
      \\
      d_y \hat{\mathcal{B}}
      + \frac{1}{2} (\hat{\mathcal{B}} \times \hat{\mathcal{B}})
      - 2 \hat{\mathcal{F}}
      &= [\hat{\mathcal{B}}, \Gamma]
    \end{aligned}
  }
\end{equation}

In~\autoref{sec:hyperkahler floer-hom}, we generalize to the case where $\text{Spin}(7) = CY_2 \times HC_3 \times \R$, with $HC_3$ being a hypercontact three-manifold.
Topologically reducing Spin$(7)$ theory along $CY_2$, we arrive at a 4d $\mathcal{N} = 2$ sigma model with target space the moduli space $\mathcal{M}^G_{\text{inst}}(CY_2)$ of instantons on $CY_2$, whose BPS equation is the Cauchy-Riemann-Fueter equation on $HC_3 \times \R$.
We then obtain the corresponding 1d SQM theory in the hypercontact three-space $\mathcal{M}(HC_3, \mathcal{M}^{G, CY_2}_{\text{inst}})$ of smooth maps from $HC_3$ to $\mathcal{M}^G_{\text{inst}}(CY_2)$.
Just as before, this will allow us to express the partition function as~\eqref{eq:hc3 x r:partition fn}
\begin{equation}
  \label{summary:eq:hc3 x r:partition fn}
  \boxed{
    \begin{aligned}
      \mathcal{Z}_{\text{Spin}(7), HC_3 \times \R}(G)
      &= \sum_s \mathcal{F}^s_{\text{4d-}\sigma, HC_3 \times \R \rightarrow \mathcal{M}^{G, CY_2}_{\text{inst}}}
      \\
      & = \sum_s \text{HHKF}_{d_s}\left(
        HC_3, \mathcal{M}^G_{\text{inst}}(CY_2)
        \right)
        = \mathcal{Z}^{\text{hyperkählerFloer}}_{HC_3, \mathcal{M}^{G, CY_2}_{\text{inst}}}
    \end{aligned}
  }
\end{equation}
where $\text{HHKF}_{d_s} (HC_3, \mathcal{M}^G_{\text{inst}}(CY_2))$ is a \emph{novel} hyperkähler Floer homology class of a hyperkähler manifold $\mathcal{M}^G_{\text{inst}}(CY_2)$ and specified by a hypercontact three-manifold $HC_3$, of degree $d_s$, defined by Floer differentials described by the gradient flow equation~\eqref{eq:hc3 x r:sqm:flow}:
\begin{equation}
  \label{summary:eq:hc3 x r:sqm:flow}
  \boxed{
    \dv{X^\alpha}{t} =
    - g^{\alpha \beta}_{\mathcal{M}(HC_3, \mathcal{M}^{G, CY_2}_{\text{inst}})}
    \pdv{V_{\sigma}}{X^\beta}
  }
\end{equation}
and Morse functional~\eqref{eq:hc3 x r:sqm:morse fn}:
\begin{equation}
  \label{summary:eq:hc3 x r:sqm:morse fn}
  \boxed{
    V_\sigma(X) = \frac{1}{2} \int_{HC_3} \dd[3]{x} \left(
       \sum_a \partial_a (X \wedge \star X) J_a
    \right)
  }
\end{equation}
where $J_a$ for $a \in \{1, 2, 3\}$ are the three complex structures of the hyperkähler $\mathcal{M}^G_{\text{inst}}(CY_2)$.
The chains of the hyperkähler Floer complex are generated by fixed critical points of $V_\sigma$, which correspond to \emph{time-invariant Fueter maps from $HC_3$ to $\mathcal{M}^G_{\text{inst}}(CY_2)$} given by time-independent solutions of the 3d equation~\eqref{eq:hc3 x r :sqm:morse fn:crit pts}:
\begin{equation}
  \label{summary:eq:hc3 x r :sqm:morse fn:crit pts}
  \boxed{
    \sum_a \partial_a X^i J_a = 0
  }
\end{equation}

Note that the existence of $\text{HHKF}_{d_s} (HC_3, \mathcal{M}^G_{\text{inst}}(CY_2))$, derived from the topological reduction of the Spin$(7)$ instanton equation on $CY_2 \times HC_3 \times \R$ along $CY_2$, was first conjectured by Hohloch-Noetzel-Salamon~\cite{hohloch-2009-hyper-struc}~\cite[$\S5$]{salamon-2013-three-dimen}.
We have therefore furnished a physical proof and realization of Hohloch-Noetzel-Salamon's mathematical conjecture.

In~\autoref{sec:symp floer-hom}, we specialize to several specific cases of $HC_3$.
First, in the case of $HC_3 = T^3$, where we can interpret the hypercontact three-space $\mathcal{M}(T^3, \mathcal{M}^{G, CY_2}_{\text{inst}})$ of maps from $T^3$ to $\mathcal{M}^G_{\text{inst}}(CY_2)$ as the triple loop space $L^3 \mathcal{M}^{G, CY_2}_{\text{inst}}$, we arrive at the following identification of a hyperkähler Floer homology of $\mathcal{M}^G_{\text{inst}}(CY_2)$ and specified by $T^3$ as a symplectic Floer homology in \eqref{eq:t3 x r:equality to hk floer-hom}:
\begin{equation}
  \label{summary:eq:hc3 x r:hk-floer-hom of t3 as sym-floer-hom of triple loop space}
  \boxed{
    \text{HHKF}_{d_s} \left(
      T^3, \mathcal{M}^G_{\text{inst}}(CY_2)
    \right)
    = \text{HSF}^{\text{Fuet}}_{d_s} \left(
      L^3 \mathcal{M}^{G, CY_2}_{\text{inst}}
    \right)
  }
\end{equation}
where $\text{HSF}^{\text{Fuet}}_{d_s} (L^3 \mathcal{M}^{G, CY_2}_{\text{inst}})$ is a \emph{novel} symplectic Floer homology class of $L^3 \mathcal{M}^{G, CY_2}_{\text{inst}}$ generated by \emph{time-invariant Fueter maps} from $T^3$ to $\mathcal{M}^G_{\text{inst}}(CY_2)$.

Second, in the case where $HC_3 = I \times T^2$, we recast the 4d sigma model as a 2d A-model on $I \times \R$ with target space being the double loop space $L^2 \mathcal{M}^{G, CY_2}_{\text{inst}}$ of maps from $T^2$ to $\mathcal{M}^G_{\text{inst}}(CY_2)$.
We then obtain the corresponding 1d SQM theory in the space $\mathcal{T}(\mathscr{L}_0, \mathscr{L}_1)_{L^2 \mathcal{M}^{G, CY_2}_{\text{inst}}}$ of smooth trajectories between isotropic-coisotropic A-branes starting at $\mathscr{L}_0$ and ending at $\mathscr{L}_1$ in $L^2 \mathcal{M}^{G, CY_2}_{\text{inst}}$ with action~\eqref{eq:t2 x i x r:a-model:sqm:action}.
This will allow us to express the partition function as~\eqref{eq:t2 x i x r:a-model:partition fn}:
\begin{equation}
  \label{summary:eq:hc3 x r:a-model:partition fn}
  \boxed{
    \begin{aligned}
      \mathcal{Z}_{\text{Spin}(7),I \times T^2 \times \R}(G)
      &= \sum_s \mathcal{F}^s_{\text{2d-}\sigma, I \times \R \rightarrow L^2 \mathcal{M}^{G, CY_2}_{\text{inst}}}
      \\
      &= \sum_s \text{HSF}^{\text{Int}}_{d_s} \left(
        L^2 \mathcal{M}^{G, CY_2}_{\text{inst}}, \mathscr{L}_0, \mathscr{L}_1
        \right)
        = \mathcal{Z}^{\text{IntSympFloer}}_{\mathscr{L}_0, \mathscr{L}_1,L^2 \mathcal{M}^{G, CY_2}_{\text{inst}}}
    \end{aligned}
  }
\end{equation}
where $\text{HSF}^{\text{Int}}_{d_s}(L^2 \mathcal{M}^{G, CY_2}_{\text{inst}}, \mathscr{L}_0, \mathscr{L}_1)$ is a \emph{novel} symplectic intersection Floer homology class generated by the intersection points of $\mathscr{L}_0$ and $\mathscr{L}_1$, of degree $d_s$, counted by the Floer differentials realized as flow lines of the SQM, whose gradient flow equations are defined by setting to zero the expression within the squared term in~\eqref{eq:t2 x i x r:a-model:action}.
Doing so, we will arrive at the following identification of a hyperkähler Floer homology of $\mathcal{M}^G_{\text{inst}}(CY_2)$ and specified by $I \times T^2$ as a symplectic intersection Floer homology of $L^2 \mathcal{M}^{G, CY_2}_{\text{inst}}$ in~\eqref{eq:t2 x i x r:equality to hk floer-hom of i x t2}:
\begin{equation}
  \label{summary:eq:hc3 x r:hk-floer-hom of t3 as sym-floer-hom of double loop space}
  \boxed{
    \text{HHKF}_{d_s} \left(
      I \times T^2, \mathcal{M}^G_{\text{inst}}(CY_2)
    \right)
    = \text{HSF}^{\text{Int}}_{d_s} \left(
      L^2 \mathcal{M}^{G, CY_2}_{\text{inst}}, \mathscr{L}_0, \mathscr{L}_1
    \right)
  }
\end{equation}

Lastly, in the case where $HC_3 = I \times S^1 \times \R$, we recast the 4d sigma model as a 2d A$_\theta$-model on $I \times \R$ with target space being the path space $\mathcal{M}(\R, L\mathcal{M}^{G, \theta, CY_2}_{\text{inst}})$ of maps from $\R$ to the loop space $L\mathcal{M}^{G, \theta, CY_2}_{\text{inst}}$, which in turn is the space of maps from $S^1$ to the $\theta$-deformed $\mathcal{M}^{G, \theta}_{\text{inst}}(CY_2)$.
We then obtain the corresponding 1d SQM theory in the space $\mathcal{T}(\mathcal{P}_0, \mathcal{P}_1)_{\mathcal{M}(\R, L\mathcal{M}^{G, \theta, CY_2}_{\text{inst}})}$ of smooth trajectories between A$_{\theta}$-branes starting at $\mathcal{P}_0(\theta)$ and ending at $\mathcal{P}_1(\theta)$ in $\mathcal{M}(\R, L\mathcal{M}^{G, \theta, CY_2}_{\text{inst}})$ with action~\eqref{eq:i x s x r2:a-model:action}.
This will allow us to express the partition function as~\eqref{eq:i x s x r2:a-model:partition fn}:
\begin{equation}
  \label{summary:eq:hc3 x r:a-model:path space:partition fn}
  \boxed{
    \begin{aligned}
      \mathcal{Z}_{\text{Spin}(7), I \times S^1 \times \R^2}(G)
      &= \sum_s \mathcal{F}^s_{\text{2d-}\sigma, I \times \R \rightarrow \mathcal{M}(\R, L\mathcal{M}^{G, \theta, CY_2}_{\text{inst}})}
      \\
      &= \sum_s \text{HSF}^{\text{Int}}_{d_s} \left(
        \mathcal{M} \left( \R, L \mathcal{M}^{G, \theta, CY_2}_{\text{inst}} \right), \mathcal{P}_0, \mathcal{P}_1
        \right)
        = \mathcal{Z}^{\text{IntSympFloer}}_{\mathcal{P}_0, \mathcal{P}_1, \mathcal{M}(\R, L \mathcal{M}^{G, CY_2}_{\text{inst}})}
    \end{aligned}
  }
\end{equation}
where $\text{HSF}^{\text{Int}}_{d_s} \big(\mathcal{M} ( \R, L \mathcal{M}^{G, \theta, CY_2}_{\text{inst}} ), \mathcal{P}_0, \mathcal{P}_1\big)$ is a \emph{novel} symplectic intersection Floer homology class generated by the intersection points of $\mathcal{P}_0(\theta)$ and $\mathcal{P}_1(\theta)$, of degree $d_s$, counted by the Floer differentials realized as flow lines of the SQM, whose gradient flow equations are defined by setting to zero the expression within the squared term in~\eqref{eq:i x s x r2:a-model:action}.
Doing so, we will arrive at the following identification of a hyperkähler Floer homology of $\mathcal{M}^{G, \theta}_{\text{inst}}(CY_2)$ and specified by $I \times S^1 \times \R$ as a symplectic intersection Floer homology of $\mathcal{M}(\R, L\mathcal{M}^{G, \theta, CY_2}_{\text{inst}})$ in~\eqref{eq:i x s x r2:equality to hk floer-hom of i x s x r}:
\begin{equation}
  \label{summary:eq:hc3 x r:hk-floer-hom of t3 as sym-floer-hom of path to loop space}
  \boxed{
    \text{HHKF}_{d_s} \left(
      I \times S^1 \times \R, \mathcal{M}^{G, \theta}_{\text{inst}}(CY_2)
    \right)
    =
    \text{HSF}^{\text{Int}}_{d_s} \left(
      \mathcal{M} \left( \R, L \mathcal{M}^{G, \theta, CY_2}_{\text{inst}} \right), \mathcal{P}_0, \mathcal{P}_1
    \right)
  }
\end{equation}

In~\autoref{sec:atiyah-floer}, we consider Spin$(7)$ theory on $CY_3 \times M_1 \times \R$, and split it into two halves by performing a Tyurin degeneration of $CY_3$ along a $CY_2$ surface.
Doing so, when $M_1 = S^1$, via the topological invariance of Spin$(7)$ theory and the results of~\autoref{sec:symp floer-hom:i x t2}, we will obtain a Spin$(7)$ Atiyah-Floer type duality of $CY_3 \times S^1$, between the gauge-theoretic Spin$(7)$ instanton Floer homology of $CY_3 \times S^1$ and the symplectic intersection Floer homology of $L^2 \mathcal{M}^{G, CY_2}_{\text{inst}}$ in~\eqref{eq:atiyah-floer:spin7}:
\begin{equation}
  \label{summary:eq:atiyah-floer:spin7}
  \boxed{
    \text{HF}^{\text{Spin}(7)\text{-inst}}_* (CY_3 \times S^1, G)
    \cong
    \text{HSF}^{\text{Int}}_* \left(
      L^2 \mathcal{M}^{G, CY_2}_{\text{inst}}, \mathscr{L}_0, \mathscr{L}_1
    \right)
  }
\end{equation}

In turn, this will lead us to a 7d-Spin$(7)$ Atiyah-Floer type duality of $CY_3$, between the gauge-theoretic holomorphic $G_2$ monopole Floer homology of $CY_3$ and the symplectic intersection Floer homology of $L \mathcal{M}^{G, CY_2}_{\text{inst}}$ in~\eqref{eq:atiyah-floer:7d-spin7}:
\begin{equation}
  \label{summary:eq:atiyah-floer:7d-spin7}
  \boxed{
    \text{HHF}^{G_2\text{-M}}_* (CY_3, G)
    \cong
    \text{HSF}^{\text{Int}}_* \left(
      L \mathcal{M}^{G, CY_2}_{\text{inst}}, \mathcal{L}_0, \mathcal{L}_1
    \right)
  }
\end{equation}
where $\mathcal{L}_*$ are isotropic-coisotropic A-branes in $L \mathcal{M}^{G, CY_2}_{\text{inst}}$.

In \autoref{sec:fs-cat of m6}, we consider the case where $\text{Spin}(7) = CY_3 \times \R^2$, and recast Spin$(7)$ theory as a 2d gauged Landau-Ginzburg (LG) model on $\R^2$ with target space $\mathfrak{A}_6$.
In turn, this 2d gauged LG model can be recast as a 1d SQM theory in the path space $\mathcal{M}(\R, \mathfrak{A}_6)$ of maps from $\R$ to $\mathfrak{A}_6$.
From the SQM and its critical points that can be interpreted as LG $\mathfrak{A}_6^\theta$-solitons in the 2d gauged LG model, we obtain~\eqref{eq:cy3 x r2:floer complex:morphism}:
\begin{equation}
  \label{summary:eq:cy3 x r2:floer complex:morphism}
  \boxed{
    \text{Hom}(\mathcal{E}^I_{\text{DT}}, \mathcal{E}^J_{\text{DT}})_\pm
    \Longleftrightarrow
    \text{HF}^{G}_{d_p}(p_{\text{DT},\pm}^{IJ})
  }
\end{equation}
Here, $\text{HF}^{G}_{d_p}(p_{\text{DT},\pm}^{IJ})$ is a Floer homology class, of degree $d_p$, generated by $p_{\text{DT},\pm}^{IJ}$, the intersection points of left and right thimbles representing LG $\mathfrak{A}_6^\theta$-solitons that can be described as morphisms $\text{Hom}(\mathcal{E}^I_{\text{DT}}, \mathcal{E}^J_{\text{DT}})_\pm$ whose endpoints $\mathcal{E}^*_{\text{DT}}$ correspond to \emph{DT configurations on $CY_3$ with $C = 0$}.
Furthermore, via this equivalent description of Spin$(7)$ theory as a 2d gauged LG model, we can interpret the normalized 8d partition function as a sum over tree-level scattering amplitudes of LG $\mathfrak{A}_6^\theta$-soliton strings given by the composition map of morphisms~\eqref{eq:cy3 x r2:mu-d maps}:
\begin{equation}
  \label{summary:eq:cy3 x r2:mu-d maps}
  \boxed{
    \mu^{n_k}_{\mathfrak{A}_6}:
    \bigotimes_{i = 1}^{n_k}
    \text{Hom}\left(
      \mathcal{E}^{I_i}_{\text{DT}}, \mathcal{E}^{I_{i + 1}}_{\text{DT}}
    \right)_-
    \longto
    \text{Hom}\left(
      \mathcal{E}^{I_1}_{\text{DT}}, \mathcal{E}^{I_{n_k+1}}_{\text{DT}}
    \right)_+
  }
\end{equation}
where $\text{Hom}(\mathcal{E}^*_{\text{DT}}, \mathcal{E}^*_{\text{DT}})_-$ and $\text{Hom}(\mathcal{E}^*_{\text{DT}}, \mathcal{E}^*_{\text{DT}})_+$ represent incoming and outgoing scattering LG $\mathfrak{A}_6^\theta$-soliton strings, as shown in \autoref{fig:cy3 x r2:mu-d maps}.

Note that \eqref{summary:eq:cy3 x r2:floer complex:morphism} and~\eqref{summary:eq:cy3 x r2:mu-d maps} underlie a \emph{novel} Fukaya-Seidel (FS) type $A_\infty$-category of DT configurations on $CY_3$ with $C = 0$ (i.e., holomorphic vector bundles on $CY_3$).
That such an FS type $A_\infty$-category of holomorphic vector bundles on $CY_3$ can be derived from Spin$(7)$ instantons on $CY_3 \times \R^2$, was conjectured by Haydys~\cite{haydys-2015-fukay-seidel}.
As such, we have furnished a purely physical proof of his mathematical conjecture.

Next, applying the results of~\autoref{sec:atiyah-floer} with $M_1 = \R$, we have the one-to-one correspondence
\begin{equation}
  \label{summary:eq:cy3 x r2:atiyah-floer:as morphism}
  \boxed{
    \text{Hom} \left(
      \mathcal{E}^I_{\text{DT}}(\theta), \mathcal{E}^J_{\text{DT}}(\theta)
    \right)_\pm
    \Longleftrightarrow
    \text{HSF}^{\text{Int}}_* \left(
      \mathcal{M} \left( \R, L \mathcal{M}^{G, \theta, CY_2}_{\text{inst}} \right), \mathcal{P}_0, \mathcal{P}_1
    \right)
  }
\end{equation}
in~\eqref{eq:cy3 x r2:atiyah-floer:as morphism}, which is a \emph{novel} Atiyah-Floer type correspondence for the FS type $A_{\infty}$-category of holomorphic vector bundles on $CY_3$!
Furthermore, we would also obtain~\eqref{eq:cy3 x r2:atiyah-floer:dt config:as morphism}:
\begin{equation}
  \label{summary:eq:cy3 x r2:atiyah-floer:dt config:as morphism}
  \boxed{
    \text{Hom}\left(
      \mathcal{E}^I_{\text{DT}}(\theta), \mathcal{E}^J_{\text{DT}}(\theta)
    \right)_\pm
    \Longleftrightarrow
    \text{Hom} \left(
      \text{Hom}\left[ \mathcal{L}^I_0(\theta), \mathcal{L}^I_1(\theta) \right],
      \text{Hom}\left[ \mathcal{L}^J_0(\theta), \mathcal{L}^J_1(\theta) \right]
    \right)_\pm
  }
\end{equation}

Lastly, via~\eqref{summary:eq:cy3 x r2:atiyah-floer:as morphism} and~\eqref{summary:eq:cy3 x r2:atiyah-floer:dt config:as morphism}, we would arrive at the one-to-one correspondence
\begin{equation}
  \label{summary:eq:cy3 x r2:atiyah-floer:intersection floer as hom-cat}
  \boxed{
    \text{HSF}^{\text{Int}}_* \left(
      \mathcal{M} \left( \R, L \mathcal{M}^{G, \theta, CY_2}_{\text{inst}} \right), \mathcal{P}_0, \mathcal{P}_1
    \right)
    \Longleftrightarrow
    \text{Hom} \left(
      \text{Hom}\left[ \mathcal{L}^I_0(\theta), \mathcal{L}^I_1(\theta) \right],
      \text{Hom}\left[ \mathcal{L}^J_0(\theta), \mathcal{L}^J_1(\theta) \right]
    \right)_\pm
  }
\end{equation}
in~\eqref{eq:cy3 x r2:atiyah-floer:intersection floer as hom-cat}, between a symplectic intersection Floer homology and a Hom-category of morphisms!

In \autoref{sec:fs-cat of m5}, we let $CY_3 = CY_2 \times S^1 \times S^1$, and perform a KK dimensional reduction of Spin$(7)$ theory on $CY_3 \times \R^2$ by shrinking one of the $S^1$ circles to be infinitesimally small.
The resulting 7d-Spin$(7)$ theory on $CY_2 \times S^1 \times \R^2$ is recast as a 2d gauged LG model on $\R^2$ with target space $\mathscr{A}_5$ of irreducible $(\mathcal{A}, \mathcal{C})$ fields on $CY_2 \times S^1$, where $\mathcal{A} \in \Omega^0(S^1, \text{ad}(G)) \otimes \Omega^{(1, 0)}(CY_2, \text{ad}(G))$ and $\mathcal{C} \in \Omega^0(S^1, \text{ad}(G_\C)) \otimes \Omega^0(CY_2, \text{ad}(G_\C))$ are a real scalar (holomorphic gauge connection) and real scalar (real scalar), respectively, with $G_\C$ being the corresponding complex Lie group.
In turn, this 2d gauged LG model can be recast as a 1d SQM theory in the path space $\mathcal{M}(\R, \mathscr{A}_5)$ of maps from $\R$ to $\mathscr{A}_5$.
From the SQM and its critical points that can be interpreted as LG $\mathscr{A}_5^\theta$-solitons in the 2d gauged LG model, we obtain~\eqref{eq:cy2 x s x r2:floer complex:morphism}:
\begin{equation}
  \label{summary:eq:cy2 x s x r2:floer complex:morphism}
  \boxed{
    \text{Hom}(\mathcal{E}^I_{\text{HW}}, \mathcal{E}^J_{\text{HW}})_\pm
    \Longleftrightarrow
    \text{HF}^{G}_{d_q} (p^{IJ}_{\text{HW},\pm})
  }
\end{equation}
Here, $\text{HF}^{G}_{d_q} (p^{IJ}_{\text{HW},\pm})$ is a Floer homology class, of degree $d_q$, generated by $p^{IJ}_{\text{HW},\pm}$, the intersection points of left and right thimbles representing LG $\mathscr{A}_5^\theta$-solitons that can be described as morphisms $\text{Hom}(\mathcal{E}^I_{\text{HW}}, \mathcal{E}^J_{\text{HW}})_\pm$ whose endpoints $\mathcal{E}^*_{\text{HW}}$ correspond to \emph{HW configurations on $CY_2 \times S^1$ with two linearly-independent components of the self-dual two-form field being zero}.
Again, via the equivalent description of 7d-Spin$(7)$ theory as a 2d gauged LG model, we can interpret the normalized 7d partition function as a sum over tree-level scattering amplitudes of LG $\mathscr{A}_5^\theta$-soliton strings given by the composition map of morphisms~\eqref{eq:cy2 x s x r2:mu-d maps}:
\begin{equation}
  \label{summary:eq:cy2 x s x r2:mu-d maps}
  \boxed{
    \mu^{n_l}_{\mathscr{A}_5}:
    \bigotimes_{i = 1}^{n_l}
    \text{Hom} \left(
      \mathcal{E}^{I_i}_{\text{HW}}, \mathcal{E}^{I_{i + 1}}_{\text{HW}}
    \right)_-
    \longto
    \text{Hom} \left(
      \mathcal{E}^{I_1}_{\text{HW}}, \mathcal{E}^{I_{n_l+1}}_{\text{HW}}
    \right)_+
  }
\end{equation}
where $\text{Hom}(\mathcal{E}^*_{\text{HW}}, \mathcal{E}^*_{\text{HW}})_-$ and $\text{Hom}(\mathcal{E}^*_{\text{HW}}, \mathcal{E}^*_{\text{HW}})_+$ represent incoming and outgoing scattering LG $\mathscr{A}_5^\theta$-soliton strings.

Together, \eqref{summary:eq:cy2 x s x r2:floer complex:morphism} and~\eqref{summary:eq:cy2 x s x r2:mu-d maps} underlie a \emph{novel} FS type $A_\infty$-category of $CY_2 \times S^1$ that categorifies HW configurations on $CY_2 \times S^1$ with two linearly-independent components of the self-dual two-form field being zero!

In \autoref{sec:fs-cat of m4}, we perform yet another KK dimensional reduction of 7d-Spin$(7)$ theory on $CY_2 \times S^1 \times \R^2$ by shrinking the remaining $S^1$ circle to be infinitesimally small.
The resulting 6d-Spin$(7)$ theory on $CY_2 \times \R^2$ is then recast as a 2d gauged LG model on $\R^2$ with target space $\mathfrak{A}_4$ of irreducible $(\mathcal{A}, \mathcal{B})$ fields on $CY_2$, where $\mathcal{A} \in \Omega^{(1,0)}(CY_2, \text{ad}(G))$ and $\mathcal{B} \in \Omega^{2, +}(CY_2, \text{ad}(G))$ are a holomorphic gauge connection and real self-dual two-form, respectively.
In turn, this 2d gauged LG model can be recast as a 1d SQM theory in the path space $\mathcal{M}(\R, \mathfrak{A}_4)$ of maps from $\R$ to $\mathfrak{A}_4$.
From the SQM and its critical points that can be interpreted as LG $\mathfrak{A}_4^\theta$-solitons in the 2d gauged LG model, we obtain~\eqref{eq:cy2 x r2:floer complex:morphism}:
\begin{equation}
  \label{summary:eq:cy2 x r2:floer complex:morphism}
  \boxed{
    \text{Hom}(\mathcal{E}^I_{\text{VW}}, \mathcal{E}^J_{\text{VW}})_\pm
    \Longleftrightarrow
    \text{HF}^{G}_{d_r}(p^{IJ}_{\text{VW}, \pm})
  }
\end{equation}
Here, $\text{HF}^{G}_{d_r}(p^{IJ}_{\text{VW},\pm})$ is a Floer homology class, of degree $d_r$, generated by $p^{IJ}_{\text{VW},\pm}$, the intersection points of left and right thimbles representing LG $\mathfrak{A}_4^\theta$-solitons that can be described as morphisms $\text{Hom}(\mathcal{E}^I_{\text{VW}}, \mathcal{E}^J_{\text{VW}})_\pm$ whose endpoints $\mathcal{E}^*_{\text{VW}}$ correspond to \emph{Vafa-Witten (VW) configurations on $CY_2$ with the scalar and one of the linearly-independent components of the self-dual two-form field being zero}.
Once again, via the equivalent description of 6d-Spin$(7)$ theory as a 2d gauged LG model, we can interpret the normalized 6d partition function as a sum over tree-level scattering amplitudes of LG $\mathfrak{A}_4^\theta$-soliton strings given by the composition map of morphisms~\eqref{eq:cy2 x r2:mu-d maps}:
\begin{equation}
  \label{summary:eq:cy2 x r2:mu-d maps}
  \boxed{
    \mu^{n_m}_{\mathfrak{A}_4}:
    \bigotimes_{i = 1}^{n_m}
    \text{Hom} \left(
      \mathcal{E}^{I_i}_{\text{VW}}, \mathcal{E}^{I_{i + 1}}_{\text{VW}}
    \right)_-
    \longto
    \text{Hom} \left(
      \mathcal{E}^{I_1}_{\text{VW}}, \mathcal{E}^{I_{n_m+1}}_{\text{VW}}
    \right)_+
  }
\end{equation}
where $\text{Hom}(\mathcal{E}^*_{\text{VW}}, \mathcal{E}^*_{\text{VW}})_-$ and $\text{Hom}(\mathcal{E}^*_{\text{VW}}, \mathcal{E}^*_{\text{VW}})_+$ represent incoming and outgoing scattering LG $\mathfrak{A}_4^\theta$-soliton strings.

Together,~\eqref{summary:eq:cy2 x r2:floer complex:morphism} and~\eqref{summary:eq:cy2 x r2:mu-d maps} underlie a \emph{novel} FS type $A_\infty$-category of $CY_2$ that categorifies VW configurations on $CY_2$ with the scalar and one of the linearly-independent components of the self-dual two-form field being zero!


In \autoref{sec:topo invariance}, we first elucidate the implications of the topological invariance of Spin$(7)$ theory on the Floer homologies obtained in \autoref{sec:floer homology of m7}--\autoref{sec:atiyah-floer}.
The results are given in~\eqref{eq:topo inv:floer partition fn:gauge-theoretic}---\eqref{eq:topo inv:fs-cat:symp int floer-hom from kk red}, and they can be summarized as follows.

(I)
\begin{equation}
  \label{summary:eq:topo inv:floer partition fn}
  \boxed{
    \begin{tikzcd}[%
      row sep=large,%
      arrows=leftrightarrow,%
      ]
      \sum_j \text{HF}^{\text{Spin}(7)\text{-inst}}_{d_j}(G_2, G)
      \arrow[d, "G_2 = CY_3 \times \widehat{S}^1"]
      \\
      \sum_k \text{HHF}^{G_2\text{-M}}_{d_k}(CY_3, G)
      \arrow[d, "CY_3 = CY_2 \times S^1 \times \widehat{S}^1"]
      \\
      \sum_l \text{HHF}^{\text{DT}}_{d_l}(CY_2 \times S^1, G)
    \end{tikzcd}
  }
\end{equation}
where $S^1$ and $\widehat{S}^1$ are circles of fixed and variable radii, respectively.

(II)
\begin{equation}
  \label{summary:eq:topo inv:floer partition fn:non-gauge-theoretic:to hk}
  \boxed{
    \begin{tikzcd}[%
      column sep=6em,%
      arrows=leftrightarrow,%
      ampersand replacement=\&,%
      ]
      \sum_j \text{HF}^{\text{Spin}(7)\text{-inst}}_{d_j}(G_2, G)
      \arrow[r, "G_2 = \widehat{CY_2} \times HC_3"]
      \&
      \sum_s \text{HHKF}_{d_s}\left(
        HC_3, \mathcal{M}^G_{\text{inst}}(CY_2)
      \right)
    \end{tikzcd}
  }
\end{equation}

\begin{equation}
  \label{summary:eq:topo inv:floer partition fn:non-gauge-theoretic:to symp}
  \boxed{
    \begin{tikzcd}[%
      row sep=large,%
      arrows=leftrightarrow,%
      ]
      \sum_j \text{HF}^{\text{Spin}(7)\text{-inst}}_{d_j}(G_2, G)
      \arrow[d, "G_2 = \widehat{CY_2} \times T^3"]
      \\
      \sum_s \text{HHKF}_{d_s}\left(
        T^3, \mathcal{M}^G_{\text{inst}}(CY_2)
      \right)
      =
      \sum_s \text{HSF}^{\text{Fuet}}_{d_s}\left(
        L^3 \mathcal{M}^{G, CY_2}_{\text{inst}}
      \right)
    \end{tikzcd}
  }
\end{equation}

\begin{equation}
  \label{summary:eq:topo inv:floer partition fn:non-gauge-theoretic:to symp-int:loop}
  \boxed{
    \begin{tikzcd}[%
      row sep=large,%
      arrows=leftrightarrow,%
      ]
      \sum_j \text{HF}^{\text{Spin}(7)\text{-inst}}_{d_j}(G_2, G)
      \arrow[d, "G_2 = \widehat{CY_2} \times I \times T^2"]
      \\
      \sum_s \text{HHKF}_{d_s}\left(
        I \times T^2, \mathcal{M}^G_{\text{inst}}(CY_2)
      \right)
      =
      \sum_s \text{HSF}^{\text{Int}}_{d_s}\left(
        L^2 \mathcal{M}^{G, CY_2}_{\text{inst}}, \mathscr{L}_0, \mathscr{L}_1
      \right)
    \end{tikzcd}
  }
\end{equation}
and
\begin{equation}
  \label{summary:eq:topo inv:floer partition fn:non-gauge-theoretic:to symp-int:path}
  \boxed{
    \begin{tikzcd}[%
      row sep=large,%
      arrows=leftrightarrow,%
      ]
      \sum_j \text{HF}^{\text{Spin}(7)\text{-inst}}_{d_j}(G_2, G)
      \arrow[d, "G_2 = \widehat{CY_2} \times I \times S^1 \times \R"]
      \\
      \sum_s \text{HHKF}_{d_s}\left(
        I \times S^1 \times \R, \mathcal{M}^{G, \theta}_{\text{inst}}(CY_2)
      \right)
      =
      \sum_s \text{HSF}^{\text{Int}}_{d_s}\left(
        \mathcal{M} \left( \R, L \mathcal{M}^{G, \theta, CY_2}_{\text{inst}} \right),
        \mathcal{P}_0, \mathcal{P}_1
      \right)
    \end{tikzcd}
  }
\end{equation}
where $\widehat{CY_2}$ is a $CY_2$ with variable size.

(III)
\begin{equation}
  \label{summary:eq:topo inv:floer partition fn:atiyah-floer}
  \boxed{
    \begin{tikzcd}[%
      column sep=3.6em,%
      arrows=leftrightarrow,%
      ampersand replacement=\&,%
      ]
      \sum_j \text{HF}^{\text{Spin}(7)\text{-inst}}_{d_j}(G_2, G)
      \arrow[rr, "G_2 = CY_3 \times S^1"]
      \arrow[rr, swap, "CY_3 = CY_3' \bigcup_{CY_2} CY_3''"]
      \&
      {}
      \arrow[d, rightarrow, "S^1 = \widehat{S}^1", shorten <= 10pt, shorten >= 10pt]
      \&
      \sum_s \text{HSF}^{\text{Int}}_{d_s} \left(
        L^2 \mathcal{M}^{G, CY_2}_{\text{inst}}, \mathscr{L}_0, \mathscr{L}_1
      \right)
      \\
      \sum_k \text{HHF}^{G_2\text{-M}}_{d_k} (CY_3, G)
      \arrow[rr, swap, "CY_3 = CY_3' \bigcup_{CY_2} CY_3''"]
      \&
      {}
      \&
      \sum_u \text{HSF}^{\text{Int}}_{d_u} \left(
        L \mathcal{M}^{G, CY_2}_{\text{inst}}, \mathcal{L}_0, \mathcal{L}_1
      \right)
    \end{tikzcd}
  }
\end{equation}
where $CY_2$ is the degeneration surface for the Tyurin degeneration of $CY_3$.

We also have
\begin{equation}
  \label{summary:eq:topo inv:floer partition fn:atiyah-floer:kk reduction}
  \boxed{
    \begin{tikzcd}[%
      column sep=3em,%
      arrows=leftrightarrow,%
      ampersand replacement=\&,%
      ]
      \sum_u \text{HSF}^{\text{Int}}_{d_u} \left(
        L \mathcal{M}^{G, CY_2}_{\text{inst}}, \mathcal{L}_0, \mathcal{L}_1
      \right)
      \arrow[r, "S^1 = \widehat{S}^1"]
      \&
      \sum_v \text{HSF}^{\text{Int}}_{d_v} \left(
        \mathcal{M}^G_{\text{inst}}(CY_2), L_0, L_1
      \right)
    \end{tikzcd}
  }
\end{equation}
where the spatial $S^1$ circle being reduced is related to the loop on the LHS, and $L_0$ and $L_1$ are isotropic-coisotropic branes in $\mathcal{M}^G_{\text{inst}}(CY_2)$, which are Lagrangian.

(IV)
\begin{equation}
  \label{summary:eq:topo inv:floer partition fn:loop spaces}
  \boxed{
    \begin{tikzcd}[%
      column sep=4.2em,%
      arrows=leftrightarrow,%
      ampersand replacement=\&,%
      ]
      \sum_s \text{HSF}^{\text{Fuet}}_{d_s}\left(
        L^3 \mathcal{M}^{G, CY_2}_{\text{inst}}
      \right)
      \arrow[r, "T^3 = T^2 \times \widehat{S}^1"]
      \&
      \sum_x \text{HSF}^{\text{hol}}_{d_x}\left(
        L^2 \mathcal{M}^{G, CY_2}_{\text{inst}}
      \right)
      \arrow[r, "T^2 = S^1 \times \widehat{S}^1"]
      \&
      \sum_y \text{HSF}^{\text{const}}_{d_y}\left(
        L \mathcal{M}^{G, CY_2}_{\text{inst}}
      \right)
    \end{tikzcd}
  }
\end{equation}
where the spatial $T^3$ is related to the triple loop on the leftmost entry;
$\text{HSF}^{\text{hol}}_{d_x} (L^2 \mathcal{M}^{G, CY_2}_{\text{inst}})$ is a symplectic Floer homology of $L^2 \mathcal{M}^{G, CY_2}_{\text{inst}}$ generated by \emph{time-invariant holomorphic maps from $T^2$ to $\mathcal{M}^G_{\text{inst}}(CY_2)$};
and $\text{HSF}^{\text{const}}_{d_y} (L \mathcal{M}^{G, CY_2}_{\text{inst}})$ is a symplectic Floer homology of $L \mathcal{M}^{G, CY_2}_{\text{inst}}$ generated by \emph{time-invariant constant maps from $S^1$ to $\mathcal{M}^G_{\text{inst}}(CY_2)$};

(V)
\begin{equation}
  \label{summary:eq:topo inv:fs-cat:symp int floer-hom from kk red}
  \boxed{
    \begin{tikzcd}[%
      column sep=large,%
      arrows=leftrightarrow,%
      ampersand replacement=\&,%
      ]
      \sum_s \text{HSF}^{\text{Int}}_{d_s} \left(
        \mathcal{M} \left( \R, L \mathcal{M}^{G, \theta, CY_2}_{\text{inst}} \right), \mathcal{P}_0, \mathcal{P}_1
      \right)
      \arrow[r, "S^1 = \widehat{S}^1"]
      \&
      \sum_r \text{HSF}^{\text{Int}}_{d_r} \left(
        \mathcal{M} \left( \R, \mathcal{M}^{G, \theta, CY_2}_{\text{inst}} \right), P_0, P_1)
      \right)
    \end{tikzcd}
  }
\end{equation}
where the spatial $S^1$ being reduced is related to the loop on the LHS, and $P_*(\theta)$ are isotropic-coisotropic A$_{\theta}$-branes in $\mathcal{M}(\R, \mathcal{M}^{G, \theta, CY_2}_{\text{inst}})$.

Note that the relation between hyperkähleric $\text{HHKF}_{d_s}(HC_3, \mathcal{M}^G_{\text{inst}}(CY_2))$ and gauge-theoretic $\text{HF}^{\text{Spin}(7)\text{-inst}}_{d_j}(HC_3 \times CY_2, G)$ in~\eqref{summary:eq:topo inv:floer partition fn:non-gauge-theoretic:to hk} was conjectured by Hohloch-Noetzel-Salamon~\cite{hohloch-2009-hyper-struc}~\cite[$\S$5]{salamon-2013-three-dimen}.
Furthermore, this relation was also conjectured by Salamon~\cite[$\S$5]{salamon-2013-three-dimen} to be analogous to an Atiyah-Floer duality.
Indeed, we do see, from~\eqref{summary:eq:topo inv:floer partition fn:non-gauge-theoretic:to symp-int:loop} and~\eqref{summary:eq:topo inv:floer partition fn:atiyah-floer}, that the two Floer homologies are related to each other by an Atiyah-Floer type duality between a gauge-theoretic and a symplectic intersection Floer homology.
We have therefore furnished a physical proof of their mathematical conjectures.

Second, we elucidate the implications of the topological invariance of Spin$(7)$ theory on the FS type $A_\infty$-categories obtained in \autoref{sec:fs-cat of m6}--\autoref{sec:fs-cat of m4}.
The results are given in~\eqref{eq:topo inv:fs-cat morphisms}--\eqref{eq:topo-inv:fs-cat:atiyah-floer}, and they can be summarized as follows.

(VI)
\begin{equation}
  \label{summary:eq:topo inv:fs-cat partition fn}
  \boxed{
    \begin{tikzcd}[%
      row sep=large,%
      arrows=rightarrow,%
      ]
      \text{Hom} \Big(
        \mathcal{E}^I_\text{DT}(\theta) , \mathcal{E}^J_\text{DT}(\theta)
      \Big)_\pm
      \arrow[d, "CY_3 = CY_2 \times S^1 \times \widehat{S}^1"]
      \\
      \text{Hom} \Big(
        \mathcal{E}^I_\text{HW}(\theta) , \mathcal{E}^J_\text{HW}(\theta)
      \Big)_\pm
      \arrow[d, "CY_2 \times S^1 = CY_2 \times \widehat{S}^1"]
      \\
      \text{Hom} \Big(
        \mathcal{E}^I_\text{VW}(\theta) , \mathcal{E}^J_\text{VW}(\theta)
      \Big)_\pm
    \end{tikzcd}
  }
\end{equation}
and
\begin{equation}
  \label{summary:eq:topo inv:mu-d maps}
  \boxed{
    \begin{tikzcd}[%
      column sep=8em,%
      arrows=Leftrightarrow,%
      ampersand replacement=\&,%
      ]
      \mu^{n_k}_{\mathfrak{A}_6}
      \arrow[r, "CY_3 = CY_2 \times S^1 \times \widehat{S}^1"]
      \&
      \mu^{n_l}_{\mathscr{A}_5}
      \arrow[r, "CY_2 \times S^1 = CY_2 \times \widehat{S}^1"]
      \&
      \mu^{n_m}_{\mathfrak{A}_4}
    \end{tikzcd}
  }
\end{equation}
where $CY_3$ is the space which the holomorphic vector bundles corresponding to the $\mathcal{E}^*_{\text{DT}}(\theta)$'s in the topmost entry of~\eqref{summary:eq:topo inv:fs-cat partition fn} are defined on.
The topmost entry of~\eqref{summary:eq:topo inv:fs-cat partition fn} in turn defines, via~\eqref{summary:eq:cy3 x r2:mu-d maps}, the leftmost entry of~\eqref{summary:eq:topo inv:mu-d maps}.

(VII)
\begin{equation}
  \label{summary:eq:topo-inv:fs-cat:atiyah-floer}
  \boxed{
    \begin{gathered}
      \text{Hom} \left(
      \mathcal{E}^I_{\text{DT}}(\theta), \mathcal{E}^J_{\text{DT}}(\theta)
      \right)_\pm
      \\
      \qquad \qquad \qquad \quad \; \,  
      \displaystyle \left\Updownarrow \vphantom{\int} \right. \scriptstyle{CY_3 = CY_3' \bigcup_{CY_2} CY_3''}
      \\
      \text{HSF}^{\text{Int}}_* \left(
      \mathcal{M} \left( \R, L \mathcal{M}^{G, \theta, CY_2}_{\text{inst}} \right), \mathcal{P}_0, \mathcal{P}_1
      \right)
      \\
      \displaystyle \left\Updownarrow \vphantom{\int} \right.
      \\
      \text{Hom} \left(
      \text{Hom}\left[ \mathcal{L}^I_0(\theta), \mathcal{L}^I_1(\theta) \right],
      \text{Hom}\left[ \mathcal{L}^J_0(\theta), \mathcal{L}^J_1(\theta) \right]
      \right)_\pm
    \end{gathered}
  }
\end{equation}

Third, from \eqref{eq:hom-cat correspondences} and \eqref{eq:cy3 x r2:atiyah-floer:dt config:as morphism}, we would obtain \eqref{eq:bousseaus conjecture}:
\begin{equation}
  \label{summary:eq:bousseaus conjecture}
  \boxed{
    \begin{gathered}
      \text{Hom} \left(
        \text{Hom}\left[ L^I_0(\theta), L^I_1(\theta) \right],
        \text{Hom}\left[ L^J_0(\theta), L^J_1(\theta) \right]
      \right)_\pm
      \\
      \displaystyle \left\Updownarrow \vphantom{\Big(} \right.
      \\
      \text{Hom} \left(
        \mathcal{E}^I_{\text{DT}}(\theta), \mathcal{E}^J_{\text{DT}}(\theta)
      \right)_\pm
      \\
      \displaystyle \left\Updownarrow \vphantom{\Big(} \right.
      \\
      \text{Hom} \left(
        \text{HHF}^{G_2\text{-inst}, \theta}(CY_3, G), \text{HHF}^{G_2\text{-inst}, \theta}(CY_3, G)
      \right)_\pm
    \end{gathered}
  }
\end{equation}
In other words, we have a correspondence amongst (i) a Hom-category of morphisms between Lagrangian submanifolds of $\mathcal{M}^{G, \theta}_{\text{inst}}(CY_2)$, (ii) an FS type $A_{\infty}$-category of $\theta$-deformed holomorphic vector bundles on $CY_3$, and (iii) a holomorphic $\theta$-generalized $G_2$ instanton Floer homology.
In particular, at $\theta = 0$, we would have a correspondence amongst (i) a Hom-category of morphisms between Lagrangian submanifolds of $\mathcal{M}^G_{\text{inst}}(CY_2)$ (which, for $CY_2$ a complex algebraic surface, span the subspace of holomorphic vector bundles on $CY_2$ that can be extended to all of $CY_3 = CY_3' \bigcup_{CY_2} CY_3''$), (ii) an FS type $A_{\infty}$-category of holomorphic vector bundles on $CY_3$, and (iii) a holomorphic $G_2$ instanton Floer homology of $CY_3$.
As such a correspondence was conjectured by Bousseau~\cite[$\S$2.8]{bousseau-2024-holom-floer}, we have therefore furnished a physical proof and generalization (for $\theta \neq 0$) of his mathematical conjecture.

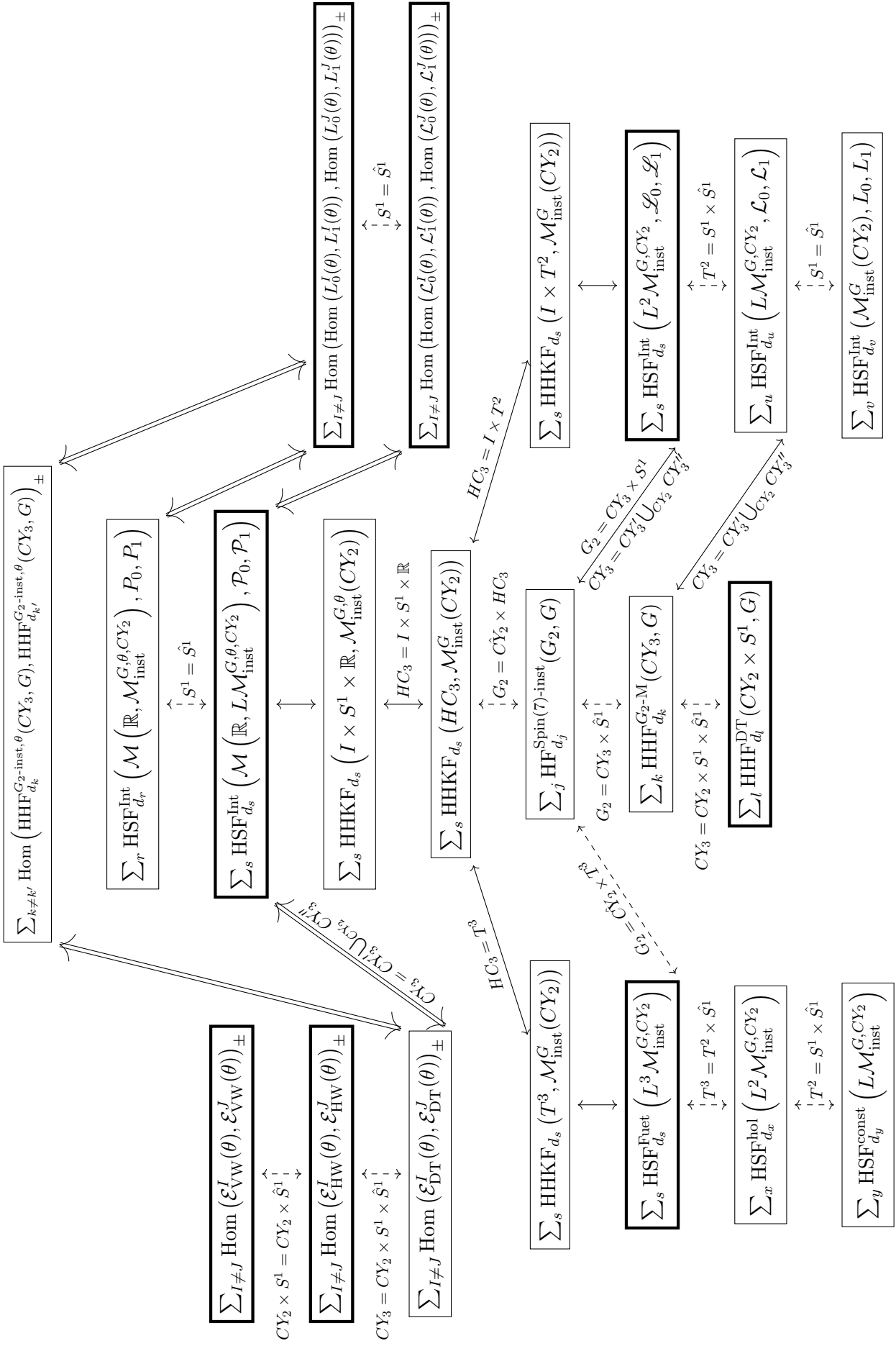
\begin{sidewaysfigure}
  \centering
  \begin{tikzpicture}[%
    auto,%
    block/.style={draw, rectangle},%
    novel/.style={draw, rectangle, ultra thick},%
    every edge/.style={draw, <->},%
    relation/.style={scale=0.8, sloped, anchor=center, align=center},%
    vertRelation/.style={scale=0.8, anchor=center, align=center},%
    horRelation/.style={scale=0.8, anchor=center, align=center},%
    shorten >=4pt,%
    shorten <=4pt,%
    ]
    \def \verRel {1} 
    \def \horRel {2.6} 
    \node[block] (8d-FT)
    {$\sum_j \text{HF}^{\text{Spin}(7)\text{-inst}}_{d_j}(G_2, G)$};
    \node[block, below={\verRel} of 8d-FT] (7d-FT)
    {$\sum_k \text{HHF}^{G_2\text{-M}}_{d_k}(CY_3, G)$};
    \node[block, novel, below={\verRel} of 7d-FT] (6d-FT)
    {$\sum_l \text{HHF}^{\text{DT}}_{d_l}(CY_2 \times S^1, G)$};
    \node[block, above={\verRel} of 8d-FT] (hK-HC3)
    {$\sum_s \text{HHKF}_{d_s}\left(
        HC_3, \mathcal{M}^G_{\text{inst}}(CY_2)
      \right)$};
    \node[block, left={\horRel} of 8d-FT] (hK-T3)
    {$\sum_s \text{HHKF}_{d_s}\left(
        T^3, \mathcal{M}^G_{\text{inst}}(CY_2)
      \right)$};
    \node[block, right={\horRel} of 8d-FT] (hK-I-T2)
    {$\sum_s \text{HHKF}_{d_s}\left(
        I \times T^2, \mathcal{M}^G_{\text{inst}}(CY_2)
      \right)$};
    \node[block, above={\verRel} of hK-HC3] (hK-ISR)
    {$\sum_s \text{HHKF}_{d_s}\left(
        I \times S^1 \times \R, \mathcal{M}^{G, \theta}_{\text{inst}}(CY_2)
      \right)$};
    \node[block, below left={0.6*\verRel} and {\horRel} of hK-ISR] (8d-FS)
    {$\sum_{I \neq J} \text{Hom} \left( \mathcal{E}^I_{\text{DT}}(\theta), \mathcal{E}^J_{\text{DT}}(\theta) \right)_\pm$};
    \node[novel, above={\verRel} of 8d-FS] (7d-FS)
    {$\sum_{I \neq J} \text{Hom} \left( \mathcal{E}^I_{\text{HW}}(\theta), \mathcal{E}^J_{\text{HW}}(\theta) \right)_\pm$};
    \node[novel, above={\verRel} of 7d-FS] (6d-FS)
    {$\sum_{I \neq J} \text{Hom} \left( \mathcal{E}^I_{\text{VW}}(\theta), \mathcal{E}^J_{\text{VW}}(\theta) \right)_\pm$};
    \node[novel, below={\verRel} of hK-T3] (HSF-T3)
    {$\sum_s \text{HSF}^{\text{Fuet}}_{d_s} \left(
        L^3 \mathcal{M}^{G, CY_2}_{\text{inst}}
      \right)$};
    \node[block, below={\verRel} of HSF-T3] (HSF-T2)
    {$\sum_x \text{HSF}^{\text{hol}}_{d_x} \left(
        L^2 \mathcal{M}^{G, CY_2}_{\text{inst}}
      \right)$};
    \node[block, below={\verRel} of HSF-T2] (HSF-S)
    {$\sum_y \text{HSF}^{\text{const}}_{d_y} \left(
        L \mathcal{M}^{G, CY_2}_{\text{inst}}
      \right)$};
    \node[novel, below={\verRel} of hK-I-T2] (HSFI-T2)
    {$\sum_s \text{HSF}^{\text{Int}}_{d_s}\left(
        L^2 \mathcal{M}^{G, CY_2}_{\text{inst}}, \mathscr{L}_0, \mathscr{L}_1
      \right)$};
    \node[block, below={\verRel} of HSFI-T2] (HSFI-S)
    {$\sum_u \text{HSF}^{\text{Int}}_{d_u}\left(
        L \mathcal{M}^{G, CY_2}_{\text{inst}}, \mathcal{L}_0, \mathcal{L}_1
      \right)$};
    \node[block, below={\verRel} of HSFI-S] (HSFI)
    {$\sum_v \text{HSF}^{\text{Int}}_{d_v}\left(
        \mathcal{M}^G_{\text{inst}}(CY_2), L_0, L_1
      \right)$};
    \node[novel, above={\verRel} of hK-ISR] (HSFI-RL)
    {$\sum_s \text{HSF}^{\text{Int}}_{d_s}\left(
        \mathcal{M} \left( \R, L \mathcal{M}^{G, \theta, CY_2}_{\text{inst}} \right), \mathcal{P}_0, \mathcal{P}_1
      \right)$};
    \node[block, above={\verRel} of HSFI-RL] (HSFI-R)
    {$\sum_r \text{HSF}^{\text{Int}}_{d_r} \left(
        \mathcal{M} \left( \R, \mathcal{M}^{G, \theta, CY_2}_{\text{inst}} \right), P_0, P_1
      \right)$};
    \node[block, novel, below right={0.6*\verRel} and {0.5*\horRel} of hK-ISR] (Hom-cat-L)
    {\footnotesize $\sum_{I \neq J} \text{Hom} \left(
      \text{Hom}\left[ \mathcal{L}^I_0(\theta), \mathcal{L}^I_1(\theta) \right],
      \text{Hom}\left[ \mathcal{L}^J_0(\theta), \mathcal{L}^J_1(\theta) \right]
    \right)_\pm$};
    \node[block, novel, above={\verRel} of Hom-cat-L] (Hom-cat)
    {\footnotesize $\sum_{I \neq J} \text{Hom} \left(
      \text{Hom}\left[ L^I_0(\theta), L^I_1(\theta) \right],
      \text{Hom}\left[ L^J_0(\theta), L^J_1(\theta) \right]
    \right)_\pm$};
    {$\text{Hom}$}
    \node[block, above={\verRel} of HSFI-R] (G2-inst)
    {\footnotesize $\sum_{k \neq k'}\text{Hom}\left(
        \text{HHF}_{d_k}^{G_2\text{-inst}, \theta}(CY_3, G),
        \text{HHF}_{d_{k'}}^{G_2\text{-inst}, \theta}(CY_3, G)
    \right)_\pm$};
    \draw
    (8d-FT) edge[dashed]
    node[vertRelation, left] {$G_2 = CY_3 \times \widehat{S}^1$}
    (7d-FT)
    (7d-FT) edge[dashed]
    node[vertRelation, left] {$CY_3 = CY_2 \times S^1 \times \widehat{S}^1$}
    (6d-FT)
    (8d-FT) edge[dashed]
    node[vertRelation, right] {$G_2 = \widehat{CY_2} \times HC_3$}
    (hK-HC3)
    (hK-HC3.south east) edge
    node[relation, above] {$HC_3 = I \times T^2$}
    (hK-I-T2)
    (hK-I-T2) edge (HSFI-T2)
    (8d-FT.south east) edge
    node[relation, above] {$G_2 = CY_3 \times S^1$}
    node[relation, below] {$CY_3 = CY_3' \bigcup_{CY_2} CY_3''$}
    (HSFI-T2.south west)
    (HSFI-T2) edge[dashed]
    node[vertRelation, right] {$T^2 = S^1 \times \widehat{S}^1$}
    (HSFI-S)
    (7d-FT.south east) edge
    node[relation, below] {$CY_3 = CY_3' \bigcup_{CY_2} CY_3''$}
    (HSFI-S.south west)
    (HSFI-S) edge [dashed]
    node[vertRelation, right] {$S^1 = \widehat{S}^1$}
    (HSFI)
    (hK-HC3.south west) edge
    node[relation, above] {$HC_3 = T^3$}
    (hK-T3)
    (hK-T3) edge (HSF-T3)
    (8d-FT.south west) edge[dashed]
    node[relation, above] {$G_2 = \widehat{CY_2} \times T^3$}
    (HSF-T3.south east)
    (HSF-T3) edge[dashed]
    node[vertRelation, right] {$T^3 = T^2 \times \widehat{S}^1$}
    (HSF-T2)
    (HSF-T2) edge[dashed]
    node[vertRelation, right] {$T^2 = S^1 \times \widehat{S}^1$}
    (HSF-S)
    (hK-HC3.north) edge
    node[vertRelation, right] {$HC_3 = I \times S^1 \times \R$}
    (hK-ISR.south)
    (hK-ISR) edge (HSFI-RL)
    (8d-FS) edge[dashed]
    node[vertRelation, left] {$CY_3 = CY_2 \times S^1 \times \widehat{S}^1$}
    (7d-FS)
    (7d-FS) edge[dashed]
    node[vertRelation, left] {$CY_2 \times S^1 = CY_2 \times \widehat{S}^1$}
    (6d-FS)
    (8d-FS.south east) edge[double equal sign distance]
    node[relation, below] {$CY_3 = CY_3' \bigcup_{CY_2} CY_3''$}
    (HSFI-RL.south west)
    (HSFI-RL) edge[dashed]
    node[vertRelation, right] {$S^1 = \widehat{S}^1$}
    (HSFI-R)
    (HSFI-RL.south east) edge[double equal sign distance]
    (Hom-cat-L.north west)
    (HSFI-R.south east) edge[double equal sign distance]
    (Hom-cat.north west)
    (Hom-cat-L) edge[dashed]
    node[vertRelation, right] {$S^1 = \widehat{S}^1$}
    (Hom-cat)
    (8d-FS.north east) edge[double equal sign distance]
    (G2-inst.south west)
    (G2-inst.south east) edge[double equal sign distance]
    ($(Hom-cat.north west)!0.20!(Hom-cat.north east)$)
    ;
  \end{tikzpicture}
  \caption{%
    A web of relations amongst the Floer homologies and FS type $A_\infty$-categories.
    \label{summary:fig:web of relations}
  }
\end{sidewaysfigure}

Fourth, we summarize the results uncovered thus far in~\autoref{summary:fig:web of relations}, where we obtain a web of relations.
Therein, the sizes of the $\widehat{S}^1$ circles and the $\widehat{CY_2}$'s are variable;
\emph{dashed lines} indicate an equivalence that is due to dimensional/topological reduction;
\emph{undashed lines} indicate an equivalence that is not due to any dimensional/topological reduction;
\emph{double lines} indicate a correspondence;
\emph{bold rectangles} indicate a result that has not been conjectured;
and \emph{regular rectangles} indicate a result that has been conjectured.

\begin{sidewaysfigure}
  \centering
  \begin{tikzpicture}[%
    auto,%
    block/.style={draw, rectangle},%
    every edge/.style={draw, ->},%
    relation/.style={scale=0.8, sloped, anchor=center, align=center},%
    vertRelation/.style={scale=0.8, anchor=center, align=center},%
    shorten >=4pt,%
    shorten <=4pt,%
    ]
    \def \verRel {2} 
    \def \horRel {3.5} 
    \node[block] (Z-Spin7)
    {$\mathcal{Z}_{\text{Spin(7)}}(G)$};
    \node[block, below={1.5*\verRel} of Z-Spin7] (8d-FS)
    {$\text{FS}^{\text{Spin}(7)\text{-inst}}(CY_3, G)$ 1-cat};
    \node[block, left={\horRel} of 8d-FS] (8d-FT)
    {$\text{HF}^{\text{Spin}(7)\text{-inst}}(G_2, G)$ 0-cat};
    \node[block, right={\horRel} of 8d-FS] (8d-Ft)
    {$\text{Fuet}^{\text{Spin}(7)\text{-inst}}(CY_2 \times S^1, G)$ 2-cat};
    \node[block, below={\verRel} of 8d-FS] (7d-FS)
    {$\text{FS}^{G_2\text{-M}}(CY_2 \times S^1, G)$ 1-cat};
    \node[block, below={\verRel} of 8d-FT] (7d-FT)
    {$\text{HHF}^{G_2\text{-M}}(CY_3, G)$ 0-cat};
    \node[block, below={\verRel} of 8d-Ft] (7d-Ft)
    {$\text{Fuet}^{G_2\text{-M}}(CY_2, G)$ 2-cat};
    \node[block, below={\verRel} of 7d-FT] (6d-FT)
    {$\text{HHF}^{\text{DT}}(CY_2 \times S^1, G)$ 0-cat};
    \node[block, below={\verRel} of 7d-FS] (6d-FS)
    {$\text{FS}^{\text{DT}}(CY_2, G)$ 1-cat};
    \node[block, below={\verRel} of 6d-FT] (5d-FT)
    {$\text{HF}^{\text{HW}}(CY_2, G)$ 0-cat};
    \draw
    (Z-Spin7.south west) edge node[relation, above]
    {$\text{Spin}(7) = G_2 \times \R$}
    (8d-FT)
    (Z-Spin7) edge node[vertRelation, left]
    {$\text{Spin}(7) = CY_3 \times \R^2$}
    (8d-FS)
    (Z-Spin7.south east) edge node[relation, above]
    {$\text{Spin}(7) = CY_2 \times S^1 \times \R^3$}
    (8d-Ft)
    (8d-FT) edge[loosely dashdotted, <->]
    node[vertRelation, left] {$G_2 = CY_3 \times \widehat{S}^1$}
    (7d-FT)
    (7d-FT) edge[loosely dashdotted, <->]
    node[vertRelation, left] {$CY_3 = CY_2 \times S^1 \times \widehat{S}^1$}
    (6d-FT)
    (6d-FT) edge[loosely dashdotted, <->]
    node[vertRelation, left] {$CY_2 \times S^1 = CY_2 \times \widehat{S}^1$}
    (5d-FT)
    (8d-FS) edge[loosely dashdotted, <->]
    node[vertRelation, left] {$CY_3 = CY_2 \times S^1 \times \widehat{S}^1$}
    (7d-FS)
    (7d-FS) edge[loosely dashdotted, <->]
    node[vertRelation, left] {$CY_2 \times S^1 = CY_2 \times \widehat{S}^1$}
    (6d-FS)
    (8d-Ft) edge[loosely dashdotted, <->]
    node[vertRelation, left] {$CY_2 \times S^1 = CY_2 \times \widehat{S}^1$}
    (7d-Ft)
    (8d-FT.east) edge
    node[relation, above] {$G_2 = CY_3 \times \R$}
    (8d-FS.west)
    (7d-FT.east) edge
    node[relation, above] {$CY_3 = CY_2 \times S^1 \times \R$}
    (7d-FS.west)
    (6d-FT.east) edge
    node[relation, above] {$CY_2 \times S^1 = CY_2 \times \R$}
    (6d-FS.west)
    (8d-FS.east) edge
    node[relation, above] {$CY_3 = CY_2 \times S^1 \times \R$}
    (8d-Ft.west)
    (7d-FS.east) edge
    node[relation, above] {$CY_2 \times S^1 = CY_2 \times \R$}
    (7d-Ft.west)
    (7d-FT.north east) edge[dotted]
    node[relation, above] {Categorification}
    (8d-FS.south west)
    (6d-FT.north east) edge[dotted]
    node[relation, above] {Categorification}
    (7d-FS.south west)
    (5d-FT.north east) edge[dotted]
    node[relation, above] {Categorification}
    (6d-FS.south west)
    (7d-FS.north east) edge[dotted]
    node[relation, above] {Categorification}
    (8d-Ft.south west)
    (6d-FS.north east) edge[dotted]
    node[relation, above] {Categorification}
    (7d-Ft.south west)
    ;
  \end{tikzpicture}
  \caption{A scheme of categorification within Spin$(7)$ theory.}
  \label{summary:fig:web of relations:categorification}
\end{sidewaysfigure}
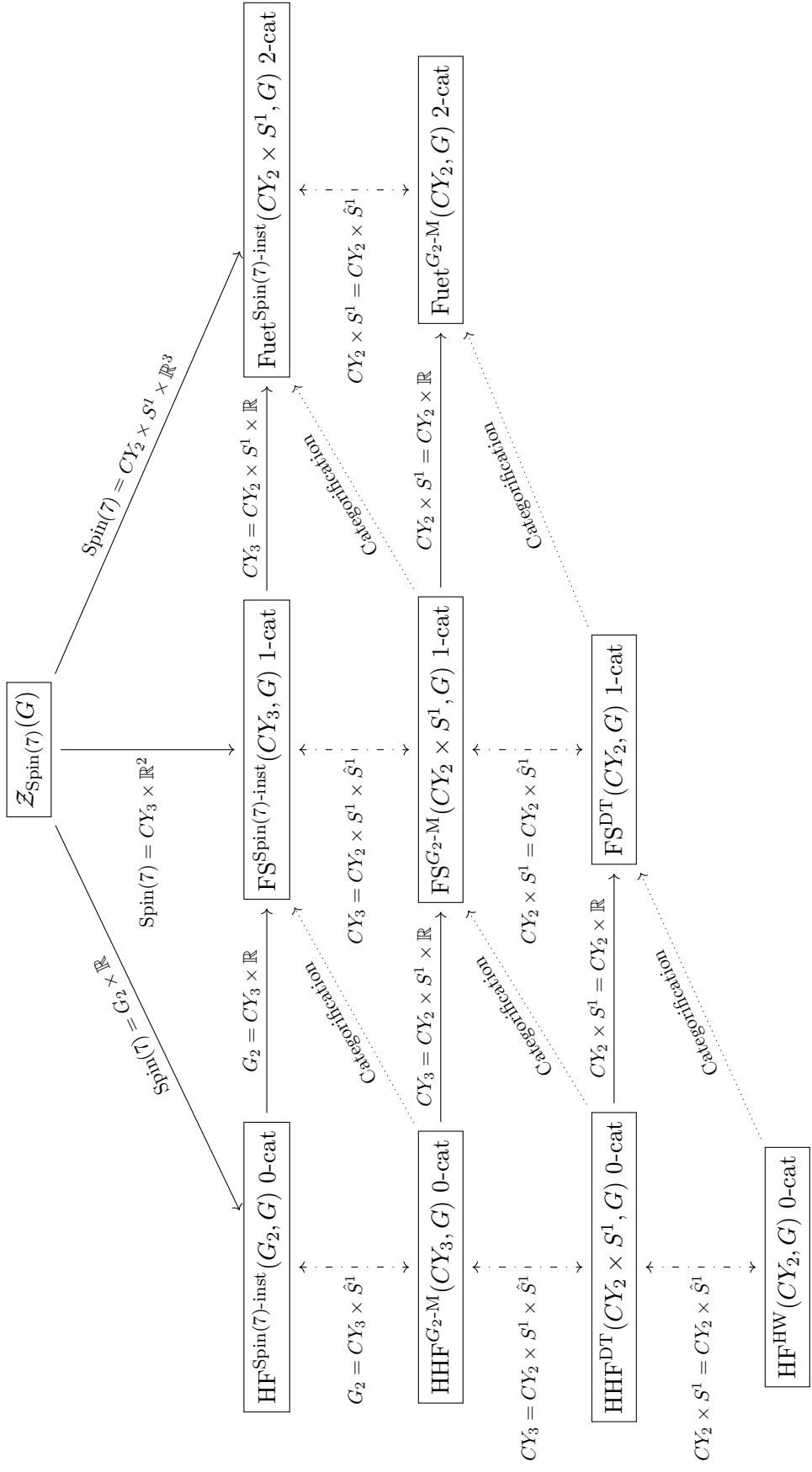

Lastly, by reviewing the various gauge-theoretic Floer homologies and FS type $A_{\infty}$-categories obtained hitherto and in \cite{er-2023-topol-n}, we observe that higher categorical structures can be introduced by taking an increasing number of spatial directions to be $\R$.
Moreover, we also find that to configurations on a $D$-manifold, $M_D$, one can associate a Floer homology of $M_D$ 0-category realized by the partition function of a gauge theory on $M_D \times \R$, which, in turn, can be categorified into an FS type $A_{\infty}$-category of $M_D$ 1-category realized by the partition function of a gauge theory on $M_D \times \R^2$.
This scheme of categorification is depicted in~\autoref{summary:fig:web of relations:categorification}.
Therein, \emph{dotted lines} are relations representing a categorification;
\emph{dash-dotted lines} are relations between categories due to dimensional reduction;
and the $\text{Fuet}^{\text{BPS-eqn}}(M_D, G)$'s are Fueter type 2-categories of $M_D$ that categorify the various FS type $A_\infty$ 1-categories.
Such 2-categories are gauge-theoretic generalizations of the Fueter 2-categories recently developed by Bousseau~\cite{bousseau-2024-holom-floer} and Doan-Rezchikov~\cite{doan-2022-holom-floer}, and will appear in a sequel paper~\cite{er-2024-topol-gauge}.

\subtitle{Acknowledgements}

We would like to thank D. Joyce, A. Haydys, and S. Hohloch for useful discussions.
A. Er would also like to thank the organizers of ``Gauge Theory and String Geometry'' for the opportunity to deliver a talk on this paper.
M.-C. Tan would also like to thank the organizers of ``String-Math 2025'' for the opportunity to deliver a plenary talk on this paper.
This work is supported in part by the MOE AcRF Tier 1 grant R-144-000-470-114.

\section{A Topological 8d \texorpdfstring{$\mathcal{N} =1$}{N=1} Gauge Theory on a \texorpdfstring{Spin$(7)$}{Spin(7)}-manifold}
\label{sec:8d theory}

In this section, we will consider a certain ``trivially-twisted'' topological 8d $\mathcal{N} = 1$ gauge theory on an eight-manifold with Spin$(7)$ holonomy, and gauge group a real, simple, compact Lie group $G$, where the BPS equation that its path integral localizes onto is the Spin$(7)$ instanton equation.
We will make use of this theory to obtain our desired results in later sections.

\subsection{The Field Composition of the Theory}
\label{sec:8d theory:fields}

The topological theory that we will consider in this paper is defined on an eight-manifold with Spin$(7)$ holonomy that is equipped with a closed Hodge self-dual structure four-form.
Such manifolds are known in the literature as Spin$(7)$-manifolds (equipped with a closed Spin$(7)$ structure) \cite[Prop. 10.5.3]{joyce-2000-compac-manif}.
The topological theory admits a ``trivial twist'', i.e., the twisted theory is equivalent to the original untwisted theory~\cite{acharya-1997-higher-dimen, elliott-2022-taxon-twist}.
The bosonic field content of the theory is a gauge connection $A_\mu \in \Omega^1(\text{Spin}(7), \text{ad}(G))$ and complex scalars $\varphi, \bar{\varphi} \in \Omega^0(\text{Spin}(7), \text{ad}(G))$.
The fermionic field content of the theory is a scalar $\eta \in \Omega^0(\text{Spin}(7), \text{ad}(G))$, a one-form $\psi \in \Omega^1(\text{Spin}(7), \text{ad}(G))$, and a self-dual two-form $\chi \in \Omega^{2, +}(\text{Spin}(7), \text{ad}(G))$.
Here, $\text{ad}(G)$ is the adjoint bundle of the underlying principal $G$-bundle.

The presence of a scalar fermion field in the theory indicates the existence of a nilpotent scalar supersymmetry generator $\mathcal{Q}$ in the theory.
The supersymmetry transformations of the topological theory under $\mathcal{Q}$ are~\cite{acharya-1997-higher-dimen}
\begin{equation}
  \label{eq:spin7 q-variations}
  \begin{aligned}
    \delta A_\mu
    &= i \psi_\mu
      \, , \\
    \delta \varphi
    &= 0
      \, , \\
    \delta \bar{\varphi}
    &= 2 i \eta
      \, , \\
    \delta \eta
    &= \frac{1}{2} [\varphi, \bar{\varphi}]
      \, , \\
    \delta \psi_\mu
    &= - D_\mu \varphi
      \, , \\
    \delta \chi_{\mu\nu}
    &= F^+_{\mu\nu}
      \, ,
  \end{aligned}
\end{equation}
where $\mu \in \{0, \dots, 7\}$ are the indices on the Spin$(7)$-manifold; $F^+_{\mu\nu} = \frac{1}{2} (F_{\mu\nu} + \frac{1}{2}\phi_{\mu\nu\rho\pi} F^{\rho\pi})$ is the self-dual part of the field strength $F_{\mu\nu}$, with $\phi$ being the closed Hodge self-dual $\text{Spin}(7)$-structure; and self-duality of $\Phi \in \{\chi, F^+\}$ means $\Phi_{\mu\nu} = \frac{1}{6} \phi_{\mu\nu\rho\pi} \Phi^{\rho\pi}$.
Our choice of $\phi$ will be that in~\cite{acharya-1997-higher-dimen}, where its non-zero components, denoted by $[\rho\pi\mu\nu] \equiv \phi_{\rho\pi\mu\nu}$, are
\begin{equation}
  \label{eq:spin7 structure}
  \begin{aligned}
    \relax  
    [0145] = [0167] = [2345] = [2367]
    &= [0246] = [1357] = [0123] = [4567] = 1
      \, , \\
    [0257] = [1346] = [0347]
    &= [0356] = [1247] = [1256] = -1
      \, .
  \end{aligned}
\end{equation}

By introducing auxiliary fields, the off-shell supersymmetry variation $\mathcal{Q}$ will be nilpotent up to gauge transformations generated by $\varphi$, i.e.,
\begin{equation}
  \label{eq:spin7 q-variations:squared}
  \begin{aligned}
    \delta^2 A
    &\propto D \varphi
      \, , \\
    \delta^2 \Psi
    &\propto [\Psi, \varphi]
      \, ,
  \end{aligned}
\end{equation}
where $\Psi$ represents all the fields that are not $A$.
As we wish to study the theory where the relevant moduli spaces are well-behaved (i.e., no reducible connections), we shall consider the case where $\varphi$ has no zero-modes.

\subsection{\texorpdfstring{Spin$(7)$}{Spin(7)} Theory and its BPS Equations}
\label{sec:8d theory:action}

The full $\mathcal{Q}$-exact topological action is~\cite{acharya-1997-higher-dimen}
\begin{equation}
  \label{eq:spin7 action}
  \begin{aligned}
    S_{\text{Spin}(7)} = \frac{1}{e^2} \int_{\text{Spin}(7)} \dd[8]{x}
    \Tr \bigg(
    & \frac{1}{2} \left| F^+_{\mu\nu} \right|^2
      - \frac{1}{2} D_\mu \varphi D^{\mu} \bar{\varphi}
      - \frac{1}{8} [\varphi, \bar{\varphi}]^2
      - i \eta D^\mu \psi_\mu
      + 2i D_\mu \psi_\nu \chi^{\mu\nu}
    \\
    & - \frac{i}{2} \varphi \{\eta, \eta\}
      - \frac{i}{4} \varphi \left\lbrace \chi_{\mu\nu}, \chi^{\mu\nu} \right\rbrace
      - \frac{i}{2} \bar{\varphi} \left\lbrace \psi_\mu, \psi^\mu \right\rbrace \bigg)
      \, .
  \end{aligned}
\end{equation}

Setting the variations of the fermions in~\eqref{eq:spin7 q-variations} to zero, we obtain the BPS equation of the 8d theory as\footnote{%
  As we are only considering the case where $\varphi$ has no zero-modes, we can take it to be zero in the variations of the fermions.
  \label{ft:ignore varphi in bps eqns}
}
\begin{equation}
  \label{eq:spin7 bps}
  \begin{aligned}
    F^+_{\mu\nu} = 0
    \, .
  \end{aligned}
\end{equation}
This is an instanton equation on the Spin$(7)$-manifold (commonly known in the literature as the Spin$(7)$ instanton equation), which is the 8d analogue of the Donaldson-Witten equation~\cite{acharya-1997-higher-dimen}.
Configurations of $A_\mu$ satisfying~\eqref{eq:spin7 bps} constitute a moduli space $\mathcal{M}_{\text{Spin}(7)}$ that the path integral of the 8d theory localizes onto, where the action in~\eqref{eq:spin7 action} is minimized.
We shall henceforth refer to this 8d theory with action $S_{\text{Spin}(7)}$ as the Spin$(7)$ theory, whose moduli space is the moduli space of irreducible Spin$(7)$ instantons.


\section{A Floer Homology of Seven-Manifolds}
\label{sec:floer homology of m7}

In this section, we shall define, purely physically, a novel Spin$(7)$ instanton Floer homology of a closed and compact $G_2$-manifold (i.e., a seven-manifold with $G_2$ holonomy) via the $\mathcal{Q}$-cohomology of Spin$(7)$ theory, through a supersymmetric quantum mechanics (SQM) interpretation of the 8d gauge theory.
In turn, this would serve as a physical proof of Donaldson-Thomas' mathematical conjecture~\cite{donaldson-1996-gauge}.

\subsection{\texorpdfstring{Spin$(7)$}{Spin(7)} Theory as a 1d SQM}
\label{sec:floer homology of m7:1d sqm}

We would like to first re-express the 8d $\mathcal{N}=1$ gauge theory on $\text{Spin}(7) = M_7 \times \mathbb{R}$ as a 1d SQM model in the space $\mathfrak{A}_7$ of irreducible gauge connections on $M_7$.
To this end, we shall employ the methods pioneered in \cite{blau-1993-topol-gauge} and further elucidated in \cite{ong-2023-vafa-witten-theor}.

We begin by noting that for $\text{Spin}(7) = M_7 \times \R$, where $M_7$ is a closed and compact seven-manifold, it necessarily has to be a $G_2$-manifold~\cite{acharya-1997-higher-dimen, cherkis-2015-octon-monop-knots, baulieu-1998-special-quant, esfahani-2022-monop-singul}.\footnote{%
  $G_2$-manifolds are equipped with a three-form structure $\tilde{\phi}_t$ that defines the Spin$(7)$-structure of a Spin$(7)$-manifold on $G_2 \times \R$ as $\phi = dt \wedge_8 \tilde{\phi_t} + \star \tilde{\phi}_t$, where $t$ is the direction along $\R$, ``$\wedge_8$'' is the exterior product on the Spin$(7)$-manifold, and ``$\star$'' is the Hodge star operator on the $G_2$-manifold.
  \label{ft:spin7 structure from g2 structure}
}
Then,~\eqref{eq:spin7 action} on $\text{Spin}(7) = G_2 \times \R$
becomes
\begin{equation}
  \label{eq:g2 x r:spin7 action}
  \begin{aligned}
    S_{\text{Spin}(7)} = \frac{1}{e^2} \int_{G_2 \times \R} \dd{t} \dd[7]{x}
    \Tr \bigg(
    & 2 \left| F^+_{ti} \right|^2
      - \frac{1}{2} D_t \varphi D^t \bar{\varphi}
      - \frac{1}{2} D_i \varphi D^i \bar{\varphi}
      - \frac{1}{8} [\varphi, \bar{\varphi}]^2
      - i \eta D_t \psi_t
      - i \eta D^i \psi_i
    \\
    & + 2i (D_t \psi_i - D_i \psi_t) \chi^{ti}
      + 2i D_i \psi_j \chi^{ij}
      - \frac{i}{2} \varphi \{\eta, \eta\}
    \\
    & - \frac{i}{2} \varphi \{\chi_{ti}, \chi^{ti}\}
      - \frac{i}{4} \varphi \{\chi_{ij}, \chi^{ij}\}
      - \frac{i}{2} \bar{\varphi} \{\psi_t, \psi_t\}
      - \frac{i}{2} \bar{\varphi} \{\psi_i, \psi^i\}
      \bigg) \, ,
  \end{aligned}
\end{equation}
where $t = x^0$ is the temporal direction along $\R$, and $x^i$ for $i \in \{1, \dots, 7\}$ are the remaining directions along the $G_2$-manifold.

Expanding out the action in~\eqref{eq:g2 x r:spin7 action} and collecting the terms without $A_t$ and $\varphi$, the action becomes\footnote{%
  We can ignore the terms with $A_t$ and $\varphi$, since they will be integrated out to furnish the terms that contribute to the Christoffel connection in the kinetic terms of the fermions as well as the four-fermi curvature term, when recast as a 1d SQM model afterwards~\cite{er-2023-topol-n, blau-1993-topol-gauge}.
  \label{ft:integrating At and varphi}
}%
\begin{equation}
  \label{eq:g2 x r:spin7 action:no A_t}
  S_{\text{Spin}(7)} = \frac{1}{2 e^2} \int_\R \dd[t] \int_{G_2} \dd[7]{x}
  \Tr \left(
    \left| \dot{A}_i + \frac{1}{2} \phi_{tijk} F^{jk} \right|^2
    + \dots
  \right) \, ,
\end{equation}
where the ``$\dots$'' contain the fermion terms in the action, and $\dot{A}_\mu = \partial_t A_{\mu}$.

After suitable rescalings, we can recast~\eqref{eq:g2 x r:spin7 action:no A_t} as a 1d SQM model, where its action will now read
\begin{equation}
  \label{eq:g2 x r:sqm:action}
  S_{\text{SQM},G_2 \text{-inst}} = \frac{1}{e^2} \int_{\R} \dd{t}
  \left(
    \left| \dot{A}^\alpha
      + g^{\alpha\beta}_{\mathfrak{A}_7} \pdv{V_7}{A^\beta}
    \right|^2
    + \dots
  \right)
  \, .
\end{equation}
Here, $A^\alpha$ and $(\alpha, \beta)$ are coordinates and indices on the space $\mathfrak{A}_7$ of irreducible $A_i$ fields on the $G_2$-manifold;\footnote{%
  Since we will ultimately consider only gauge-inequivalent configurations, $\mathfrak{A}_7$ is more precisely the space of irreducible $A_i$ fields on the $G_2$-manifold modulo gauge equivalence.
  Similar such spaces to appear in later sections should also be understood as spaces of fields modulo gauge equivalence.
  \label{ft:modulo gauge inequivalence}
}~$g_{\mathfrak{A}_7}$ is the metric on $\mathfrak{A}_7$; $V_7(A)$ is the potential function; and the `$G_2$-inst' label in the subscript will be made clear shortly.

\subsection{A \texorpdfstring{Spin$(7)$}{Spin(7)} Instanton Floer Homology of \texorpdfstring{$G_2$}{G2}-manifolds}
\label{sec:floer homology of m7:floer homology}

In a TQFT, the Hamiltonian $H$ vanishes in the $\mathcal{Q}$-cohomology, whence this means that for any state $|\mathcal{O}\rangle$ that is nonvanishing in the $\mathcal{Q}$-cohomology, we have
\begin{equation}
  \label{eq:q-cohom of operators}
  H \ket{\mathcal{O}}
  = \acomm{\mathcal{Q}}{\cdots} \ket{\mathcal{O}}
  = \mathcal{Q}(\cdots \ket{\mathcal{O}} )
  = \mathcal{Q} \ket{\mathcal{O}'}
  = \acomm{\mathcal{Q}}{\mathcal{O}'} \ket{0}
  = \ket{\acomm{\mathcal{Q}}{\mathcal{O}'}}
  \sim 0
  \, .
\end{equation}
In other words, the $|\mathcal{O}\rangle$'s which span the relevant $\mathcal{Q}$-cohomology of states in Spin$(7)$ theory are actually ground states that are therefore time-invariant.
In particular, for Spin$(7)$ theory on $\text{Spin}(7) = G_2 \times \R$ with $\mathbb R$ as the time coordinate, its relevant spectrum of states is associated only with the $G_2$-manifold.

With $\text{Spin}(7) = G_2 \times \R$, the $G_2$-manifold is the far boundary of the Spin$(7)$-manifold and one needs to specify ``boundary conditions'' on the $G_2$-manifold to compute the path integral.
We can do this by first defining a restriction of the fields to the $G_2$-manifold, which we shall denote as $\Psi_{G_2}$, and then specifying boundary values for these restrictions.
Doing this is equivalent to inserting in the path integral, an operator functional $F_7(\Psi_{G_2})$ that is nonvanishing in the ${\mathcal{Q}}$-cohomology (so that the path integral will continue to be topological).
This means that the corresponding partition function of Spin$(7)$ theory can be computed as~\cite[eqn.~(4.12)]{witten-1988-topol-quant}\footnote{%
  The partition function of Spin$(7)$ theory on $G_2 \times \mathbb{R}$ can be defined if it were a balanced TQFT like Vafa-Witten theory.
  This requires the virtual dimension of its moduli space to be zero, i.e., the difference between the number of fermion zero-modes of the theory to be zero.
  Note that fermion zero-modes are static and are thus time-invariant, i.e., they have no $t$ dependence.
  Hence, they are actually insensitive to the geometry of the $t$-direction, which is $\mathbb{R}$ in our case.
  So, if we were to replace $\mathbb{R}$ with $S^1$, the fermion zero-modes would not be affected; in particular, their numbers, and thus difference, would not change.
  On $G_2 \times S^1$, this difference between fermion zero-modes is the index of the elliptic operator associated to the (linearized) Spin$(7)$ instanton equation (modulo gauge equivalence) on $G_2 \times S^1$ \cite{lewis-1998-spin-instan}.
  Therefore, for an appropriate choice of a $G_2$-manifold and $G$ such that the index is zero, Spin$(7)$ theory on $G_2 \times \R$ is a balanced TQFT and thus, its partition function can be defined.
  We shall henceforth assume such a choice of a $G_2$-manifold and $G$; specifically, we choose a $G_2$-manifold and $G$ such that, when evaluated over the fundamental class of $G_2 \times S^1$, $p_1(G_2) \text{ch}_2(\text{ad}(G)) = 24 \text{ch}_4(\text{ad}(G))$ (where $p_1(G_2)$ is the first Pontrjagin class of the $G_2$-manifold and $\text{ch}_k(\text{ad}(G))$ is the degree-$2k$ component of the Chern character of $\text{ad}(G)$).
  \label{ft:ellipticity of spin7 instanton equation}
}
\begin{equation}
  \label{eq:partition fn:8d witten}
  \expval{1}_{F_7(\Psi_{G_2})}
  = \int_{\mathcal{M}_{\text{Spin}(7)}} F_7(\Psi_{G_2}) \, e^{-S_{\text{Spin}(7)}}
  \, .
\end{equation}

Since we have demonstrated in the previous subsection that Spin$(7)$ theory on $G_2 \times \R$ can be expressed as a 1d SQM model in $\mathfrak{A}_7$, we can thus write the partition function as
\begin{equation}
  \label{eq:g2 x r:spin7 partition fn:no homology}
  \mathcal{Z}_{\text{Spin}(7),G_2 \times \R}(G)
  = \expval{1}_{F_7(\Psi_{G_2})}
  = \sum_j \mathcal{F}^{G}_{\text{Spin}(7)}(\Psi_{G_2}^j)
  \, .
\end{equation}
Here, $\mathcal{F}^{G}_{\text{Spin}(7)}(\Psi_{G_2}^j)$, in the $\mathcal{Q}$-cohomology of Spin$(7)$ theory, is the $j^{\text{th}}$ contribution to the partition function that depends on the expression of $F_7(\Psi_{G_2})$ in the fields on the $G_2$-manifold, evaluated over the corresponding solutions to the Spin$(7)$ instanton equation in~\eqref{eq:spin7 bps} restricted to $G_2$;
and the summation in `$j$' is over all presumably isolated and non-degenerate configurations on the $G_2$-manifold in $\mathfrak{A}_7$ that the equivalent SQM localizes onto.\footnote{%
  This presumption that the configurations are isolated and non-degenerate will be justified in~\autoref{ft:g2 instanton isolation and non-degeneracy}.
  \label{ft:g2 instanton isolation and non-degeneracy presumption}
}

Let us now ascertain what the $\mathcal{F}^{G}_{\text{Spin}(7)}(\Psi_{G_2}^j)$'s correspond to.
To this end, we have to first determine the configurations that the SQM localizes onto.
These are configurations that minimize the SQM action \eqref{eq:g2 x r:sqm:action}, i.e., configurations that set the squared term therein to zero.
They are therefore given by
\begin{equation}
  \label{eq:g2 x r:sqm:flow}
  \boxed{
    \dv{A^\alpha}{t} = -g^{\alpha\beta}_{\mathfrak{A}_7}\pdv{V_7}{A^\beta}
  }
\end{equation}
where the squaring argument~\cite{witten-1988-topol-quant} means that both the LHS and RHS are \emph{simultaneously} set to zero.
In other words, the configurations that the SQM localizes onto are fixed (i.e., time-invariant) critical points of the potential $V_7$ in $\mathfrak{A}_7$.

\subtitle{$G_2$ Instanton Configurations as Critical Points of the 1d SQM}

To determine the explicit form of $V_7$, note that the squared term in~\eqref{eq:g2 x r:sqm:action} originates from the squared term in~\eqref{eq:g2 x r:spin7 action:no A_t}.
Indeed, setting the expression within the squared term in~\eqref{eq:g2 x r:spin7 action:no A_t} to zero minimizes the underlying 8d action, and this is consistent with setting the expression within the squared term in~\eqref{eq:g2 x r:sqm:action} to zero to minimize the equivalent SQM action.
Therefore, we can deduce the explicit form of $V_7$ by comparing~\eqref{eq:g2 x r:sqm:flow} with~\eqref{eq:g2 x r:spin7 action:no A_t}.
Specifically, setting to zero the expression within the squared term in~\eqref{eq:g2 x r:spin7 action:no A_t} would give us
\begin{equation}
  \label{eq:g2 x r:flow}
  \dv{A_i}{t}
  = - \frac{1}{2} \phi_{tijk} F^{jk}
  \, .
\end{equation}
Comparing~\eqref{eq:g2 x r:flow} with~\eqref{eq:g2 x r:sqm:flow}, we find that
\begin{equation}
  \label{eq:g2 x r:morse fn V7}
  \boxed{
    V_7(A, \varphi)
    = \int_{G_2} \, \Tr \, \left(
      CS(A) \wedge \star \phi_t
    \right)
  }
\end{equation}
Here, $CS(A)$ is the Chern-Simons three-form in $A$; $\phi_t$ is the three-form $G_2$-structure of the $G_2$-manifold, whose non-zero components $(\phi_t)_{ijk}$ correspond to the choice of the Spin$(7)$-structure $\phi$ in~\eqref{eq:spin7 structure} with the first index being `$0$'; and ``$\wedge$'', ``$\star$'' are the exterior product and Hodge star operator on the $G_2$-manifold, respectively.
Thus, the summation in `$j$' in~\eqref{eq:g2 x r:spin7 partition fn:no homology} is over all isolated and non-degenerate critical points of~\eqref{eq:g2 x r:morse fn V7} in $\mathfrak{A}_7$ that are also fixed.\footnote{%
  As we will explain next, the aforementioned critical points correspond to $G_2$ instanton configurations on the $G_2$-manifold.
  For them to be isolated, the actual dimension of their moduli space needs to be zero.
  This is indeed possible under appropriate transversality assumptions of the $G_2$ instantons being acyclic~\cite[$\S$3]{walpuski-2013-g2-gener}.
  We shall henceforth choose our $G_2$-manifolds to satisfy these transversality assumptions (in addition to the condition spelt out in \autoref{ft:ellipticity of spin7 instanton equation}).
  As for their non-degeneracy, a suitable perturbation of $V_7(A)$, which can be effected by introducing physically-trivial $\mathcal{Q}$-exact terms to the action, would ensure this \cite[footnote 5]{er-2023-topol-n}.
  We would like to thank D. Joyce for discussions on this point.
  \label{ft:g2 instanton isolation and non-degeneracy}
}

Critical points of $V_7(A)$ are configurations in $\mathfrak{A}_7$ that set the RHS of~\eqref{eq:g2 x r:sqm:flow} to zero, which, in turn, correspond to configurations on the $G_2$-manifold that set the RHS of~\eqref{eq:g2 x r:flow} to zero.
Such configurations span the space of solutions to the 7d $G_2$ instanton equation, for which we shall henceforth refer to as $G_2$ instanton configurations.

In summary, the partition function~\eqref{eq:g2 x r:spin7 partition fn:no homology} is an algebraic sum of \emph{fixed} $G_2$ instanton configurations on the $G_2$-manifold in $\mathfrak{A}_7$.

\subtitle{The Spin$(7)$ Instanton Floer Homology}

Notice that~\eqref{eq:g2 x r:sqm:flow} is a gradient flow equation and it governs the classical trajectory of the 1d SQM model from one time-invariant $G_2$ instanton configuration to another on the $G_2$-manifold, in $\mathfrak{A_7}$.
Hence, just as in~\cite{er-2023-topol-n, blau-1993-topol-gauge, ong-2023-vafa-witten-theor}, the equivalent 1d SQM model will physically realize a gauge-theoretic Floer homology.

Specifically, the \emph{time-invariant $G_2$ instanton configurations on the $G_2$-manifold in $\mathfrak{A}_7$}, i.e., time-independent solutions to the 7d equation
\begin{equation}
  \label{eq:g2 x r:morse fn V7:crit pts}
  \boxed{
    F \wedge \star \phi_t = 0
  }
\end{equation}
will generate the chains of a Floer complex with \emph{Morse functional} $V_7(A)$ in~\eqref{eq:g2 x r:morse fn V7}.
The Spin$(7)$ instanton flow lines, described by time-varying solutions to the \emph{gradient flow equation}~\eqref{eq:g2 x r:sqm:flow}, are the Floer differentials such that the number of outgoing flow lines at each time-invariant configuration obeying~\eqref{eq:g2 x r:morse fn V7:crit pts} is the degree $d_j$ of the corresponding chain in the Floer complex.

In other words, we can also write~\eqref{eq:g2 x r:spin7 partition fn:no homology} as
\begin{equation}
  \label{eq:g2 x r:partition fn}
  \boxed{
    \mathcal{Z}_{\text{Spin}(7),G_2 \times \R}(G)
    = \sum_j \mathcal{F}^{G}_{\text{Spin}(7)}(\Psi_{G_2}^j)
    = \sum_j \text{HF}^{\text{Spin}(7)\text{-inst}}_{d_j}(G_2, G)
    = \mathcal{Z}^{\text{Floer}}_{\text{Spin}(7)\text{-inst},G_2}(G)
  }
\end{equation}
where each $\mathcal{F}^{G}_{\text{Spin}(7)}(\Psi_{G_2}^j)$ can be identified with a \emph{novel} Floer homology class $\text{HF}^{\text{Spin}(7)\text{-inst}}_{d_j}(G_2, G)$, that we shall henceforth name a Spin$(7)$ instanton Floer homology class, assigned to the $G_2$-manifold defined by~\eqref{eq:g2 x r:sqm:flow},~\eqref{eq:g2 x r:morse fn V7}, \eqref{eq:g2 x r:morse fn V7:crit pts}, and the description above.

\subtitle{A Physical Proof of Donaldson-Thomas' Mathematical Conjecture}

Note that $\text{HF}^{\text{Spin}(7)\text{-inst}}_{d_j}(G_2, G)$ was first mathematically conjectured to exist by Donaldson-Thomas~\cite[$\S$3]{donaldson-1996-gauge} as a Floer homology generated by $G_2$ instantons on a $G_2$-manifold, whose flow lines are time-varying solutions to the Spin$(7)$ instanton equation on $G_2 \times \R$.
We have therefore furnished a physical proof of their mathematical conjecture.


\section{A Holomorphic Floer Homology of Six-Manifolds}
\label{sec:floer homology of m6}

In this section, we will specialize to $G_2 = M_6 \times S^1$, where $M_6$ is a closed and compact Calabi-Yau threefold ($CY_3$), and perform a Kaluza-Klein (KK) dimensional reduction of Spin$(7)$ theory by shrinking the $S^1$ circle to be infinitesimally small.
This will allow us to physically derive, from its topologically-invariant $\mathcal{Q}$-cohomology, a holomorphic $G_2$ monopole Floer homology of $CY_3$. In turn, this would serve as a physical proof of Donaldson-Segal's mathematical conjecture~\cite{donaldson-2009-gauge-theor-ii} and Cherkis' speculation \cite{cherkis-2015-octon-monop-knots}, and a generalization of the former.

\subsection{A KK Reduction of \texorpdfstring{Spin$(7)$}{Spin(7)} Theory along \texorpdfstring{$S^1$}{S1} and the Corresponding SQM}
\label{sec:floer homology of m6:1d sqm}

For $\text{Spin}(7) = M_6 \times S^1 \times \R$ to be a Spin$(7)$-manifold, $M_6 \times S^1$ has to be a closed and compact $G_2$-manifold.
This is possible if $M_6$ were to be a closed and compact Calabi-Yau threefold ($CY_3$)~\cite{wang-2020-modul-spaces}.\footnote{%
  $CY_3$'s are equipped with a Kähler two-form $\omega$ and holomorphic three-form $\Lambda$ that defines a $G_2$ three-form structure $\tilde{\phi}_t$ of a $G_2$-manifold on $CY_3 \times S^1$ as $\tilde{\phi}_t = ds \wedge \omega + \text{Re}(\Lambda)$, where $s$ is the direction along $S^1$ and ``$\wedge$'' is the exterior product on the $G_2$-manifold.
  \label{ft:g2 structure from cy3 structure}
}%
~We will consider this case, and start with Spin$(7)$ theory on $\text{Spin}(7) = CY_3 \times S^1 \times \R$.

Let $x^1$ be the coordinate along the $S^1$ circle.
Upon KK reduction along this $S^1$ circle, the component of the gauge field along it, i.e., $A_1$, will be interpreted as a scalar field on $CY_3 \times \R$.
The $\mathcal{Q}$-variation of the bosons are then
\begin{equation}
  \label{eq:cy3 x s x r:spin7 variations}
  \delta A_t = i \psi_t
  \, ,
  \qquad
  \delta C = i \varrho
  \, ,
  \qquad
  \delta A_a = i \psi_a
  \, ,
\end{equation}
where $x^a$ for $a \in \{2, \dots, 7\}$ are coordinates on $CY_3$, and $(A_1, \psi_1)$ have been relabeled as $(C, \varrho) \in \Omega^0(CY_3 \times \R, \text{ad}(G))$ for later convenience.

\subtitle{The BPS Equations of 7d-Spin$(7)$ Theory}

The conditions that minimize Spin$(7)$ theory when we KK reduce along $S^1$ are effectively obtained by KK reduction of the conditions that minimize the action~\eqref{eq:g2 x r:spin7 action:no A_t}~\cite[$\S$4]{er-2023-topol-n}, i.e., by the KK reduction of the BPS equation of Spin$(7)$ theory.
They are given by
\begin{equation}
  \label{eq:cy3 x r:bps}
  \dot{C} + \frac{1}{2} \phi_{t1ab} F^{ab}
  = 0
  \, ,
  \qquad
  \dot{A}_a + \phi_{t1ab} D^b C + \frac{1}{2} \phi_{tabc} F^{bc}
  = 0
  \, .
\end{equation}
These are the BPS equations of the theory that results from the KK reduction of Spin$(7)$ theory on $CY_3 \times S^1 \times \R$ along $S^1$.
Notice also that these equations can be re-expressed as a monopole equation on $G_2 = CY_3 \times \R$ (known in the literature as a $G_2$ monopole ($G_2$-M) equation~\cite{oliveira-2014-monop-higher-dimen, esfahani-2022-monop-singul}, or as an octonionic monopole equation~\cite{cherkis-2015-octon-monop-knots}) in temporal gauge, with time being along the $\R$ direction.

Let use define complex coordinates $(z^m, {\bar{z}}^{\bar{m}})$ for $CY_3$ as $z^1 = x^2 + i x^3$, $z^2 = x^4 + i x^5$, and $z^3 = x^6 + i x^7$, where their complex conjugates are $\bar{z}^{\bar 1}$, $\bar{z}^{\bar 2}$, and $\bar{z}^{\bar 3}$.
This will allow us to define $\mathcal{A} \in \Omega^{(1,0)}(CY_3, \text{ad}(G))$ and $\bar{\mathcal{A}} \in \Omega^{(0,1)}(CY_3, \text{ad}(G))$, a holomorphic and anti-holomorphic gauge connection, respectively, on $CY_3$.
The components of $\mathcal{A}$ work out to be $\mathcal{A}_1 = (A_2 - i A_3)/2$, $\mathcal{A}_2 = (A_4 - i A_5)/2$, and $\mathcal{A}_3 = (A_6 - i A_7)/2$, where their complex conjugates are $\bar{\mathcal{A}}_{\bar{1}}$, $\bar{\mathcal{A}}_{\bar{2}}$, and $\bar{\mathcal{A}}_{\bar{3}}$.
With this,~\eqref{eq:cy3 x r:bps} can be equivalently expressed as\footnote{%
  We only need to consider the holomorphic expressions, as the anti-holomorphic expressions are obtained through complex conjugations.
  \label{ft:explanation for holomorphic expressions}
}
\begin{equation}
  \label{eq:cy3 x r:bps:complexified}
  \dot{C}
  = - \omega_{m\bar{n}} \mathcal{F}^{m\bar{n}}
  \, ,
  \qquad
  \dot{\mathcal{A}}_m
  = - \omega_{m\bar{n}} \bar{\mathcal{D}}^{\bar{n}} C
  - \frac{1}{4} \varepsilon_{mpq} \mathcal{F}^{pq}
  \, ,
\end{equation}
where $(\omega_{m\bar{n}}, \varepsilon_{mpq})$ are components of the Kähler two-form and holomorphic three-form of $CY_3$, respectively; $\mathcal{F}^{pq} = \partial^p \mathcal{A}^q - \partial^q \mathcal{A}^p + [\mathcal{A}^p, \mathcal{A}^q]$ are the components of the $(2,0)$-form field strength in $\mathcal{A}$; $\mathcal{F}^{m\bar{n}} = \partial^m \bar{\mathcal{A}}^{\bar{n}} - \bar{\partial}^{\bar{m}} \mathcal{A}^n + [\mathcal{A}^m, \bar{\mathcal{A}}^{\bar{n}}]$ are the components of the $(1, 1)$-form field strength in $\mathcal{A}$ and $\bar{\mathcal{A}}$; and $\bar{\mathcal{D}} = \bar{\partial} + [\bar{\mathcal{A}}, \cdot]$ is the anti-holomorphic covariant derivative.

In other words, Spin$(7)$ theory on $CY_3 \times S^1 \times \R$, upon a KK reduction along $S^1$ to a 7d theory (henceforth referred to as 7d-Spin$(7)$ theory) on $CY_3 \times \R$, localizes onto configurations that obey~\eqref{eq:cy3 x r:bps:complexified}.

\subtitle{7d-Spin$(7)$ Theory as a 1d SQM in $\mathfrak{A}_6$}

The 7d action can thus be written as
\begin{equation}
  \label{eq:cy3 x r:action}
  \begin{aligned}
    S_{\text{7d-Spin}(7)}
    = \frac{1}{2e^2} \int_\R \dd{t} \int_{CY_3} \abs{\dd{z}}^6 \Tr \Bigg(
    & 4 \left| \dot{\mathcal{A}}_m
      + \omega_{m\bar{n}} \bar{\mathcal{D}}^{\bar{n}} C
      + \frac{1}{4} \varepsilon_{mpq} \mathcal{F}^{pq}
      \right|^2
      + \left| \dot{C}
      + \omega_{m\bar{n}} \mathcal{F}^{m\bar{n}}
      \right|^2
      + \dots
      \Bigg) \, .
  \end{aligned}
\end{equation}
After suitable rescalings, the equivalent SQM action can be obtained from~\eqref{eq:cy3 x r:action} as
\begin{equation}
  \label{eq:cy3 x r:action:sqm}
  S_{\text{SQM,DT}}= \frac{1}{e^2} \int_\R \dd{t}
  \left(
    \left| \dot{\mathcal{A}}^\alpha
      + g^{\alpha\bar{\beta}}_{\mathfrak{A}_6} \left(
        \pdv{V_6}{\mathcal{A}^\beta}
      \right)^*
    \right|^2
    + \left| \dot{C}^\alpha
      + g^{\alpha\bar{\beta}}_{\mathfrak{A}_6} \left(
        \pdv{V_6}{C^\beta}
      \right)^*
    \right|^2
    + \dots
  \right)
  \, ,
\end{equation}
where $(\mathcal{A}^\alpha, C^\alpha)$ and $(\alpha, \beta)$ are holomorphic coordinates and indices on the space $\mathfrak{A}_6$ of irreducible $(\mathcal{A}_m, C)$ fields on $CY_3$; $g_{\mathfrak{A}_6}$ is the metric on $\mathfrak{A}_6$; $V_6(\mathcal{A}, C)$ is the holomorphic potential function; and the `DT' label in the subscript will be made clear very shortly.

\subtitle{Localizing Onto Time-invariant Donaldson-Thomas Configurations on $CY_3$}

By the squaring argument~\cite{blau-1993-topol-gauge} applied to~\eqref{eq:cy3 x r:action:sqm}, the configurations that the equivalent SQM localizes onto are those that set the LHS's and RHS's of~\eqref{eq:cy3 x r:bps:complexified} \emph{simultaneously} to zero.\footnote{%
  On our choice of a closed and compact $CY_3$, setting to zero the RHS of the second equation of~\eqref{eq:cy3 x r:bps:complexified} and applying the Bianchi identity, we find that both terms within are independently zero, i.e., $\mathcal{D}_m C = 0 = \frac{1}{4} \varepsilon_{mpq} \mathcal{F}^{pq}$~\cite[Lemma 53]{esfahani-2022-monop-singul}.
  \label{ft:getting the dt eqns}
}%
~Such configurations in $\mathfrak{A}_6$ correspond to time-invariant configurations that span the space of solutions to the 6d Donaldson-Thomas (DT) equations on $CY_3$~\cite{szabo-2010-instan-topol} with the scalar being real.\footnote{%
  6d DT equations on $CY_3$ contain a holomorphic gauge connection and a \emph{complex} scalar~\cite{szabo-2010-instan-topol, oliveira-2014-monop-higher-dimen}.
  However, what we get by setting to zero the RHS's of~\eqref{eq:cy3 x r:bps:complexified} are equations involving a holomorphic gauge connection and a \emph{real} scalar (corresponding to the real gauge field $A_1$ from the KK reduction).
  Solutions to such DT equations on $CY_3$ where the complex scalar is actually real are known in the literature as Calabi-Yau monopoles on $CY_3$~\cite{esfahani-2022-monop-singul}.
  \label{ft:discrepancy of DT config in FT with full DT config}
}
We shall, in the rest of this section, refer to such configurations as DT configurations on $CY_3$.

In summary, the equivalent SQM localizes onto time-invariant DT configurations on $CY_3$ in $\mathfrak{A}_6$.

\subsection{A Holomorphic \texorpdfstring{$G_2$}{G2} Monopole Floer Homology of \texorpdfstring{$CY_3$}{CY3}}
\label{sec:floer homology of m6:floer homology}

Since the resulting 7d theory on $CY_3 \times \R$ can be interpreted as a 1d SQM in $\mathfrak{A}_6$, its partition function can, like in~\eqref{eq:g2 x r:spin7 partition fn:no homology}, be written as
\begin{equation}
  \label{eq:cy3 x r:partition fn:no homology}
  \mathcal{Z}_{\text{Spin}(7),CY_3 \times \R}(G)
  = \expval{1}_{F_6(\Psi_{CY_3})}
  = \sum_k \mathcal{F}^{G}_{\text{7d-Spin}(7)}(\Psi_{CY_3}^k)
  \, ,
\end{equation}
where $\mathcal{F}^{G}_{\text{7d-Spin}(7)}(\Psi_{CY_3}^k)$, in the $\mathcal{Q}$-cohomology of 7d-Spin$(7)$ theory, is the $k^{\text{th}}$ contribution to the partition function that depends on the expression of $F_6(\Psi_{CY_3})$ in the bosonic fields on $CY_3$, and the summation in `$k$' is over all isolated and non-degenerate DT configurations on $CY_3$ in $\mathfrak{A}_6$ that the equivalent SQM localizes onto.\footnote{%
  This presumption that the configurations will be isolated and non-degenerate is justified because the ($\mathcal{Q}$-cohomology of) Spin$(7)$ theory is topological in all directions and therefore invariant when we shrink the $S^1$.
  Thus, if $CY_3$ (where $CY_3 \times S^1 = G_2$) is chosen such as to satisfy the transversality assumptions of \autoref{ft:g2 instanton isolation and non-degeneracy}, $\mathcal{Z}_{\text{Spin}(7), CY_3 \times \R}$ will be a discrete and non-degenerate sum of contributions, just like $\mathcal{Z}_{\text{Spin}(7),G_2 \times \R}$.
  We shall henceforth assume such a choice of $CY_3$ whence the presumption would hold.
  \label{ft:dt isolation and non-degeneracy}
}

Let us now ascertain what the $\mathcal{F}^{G}_{\text{7d-Spin}(7)}(\Psi_{CY_3}^k)$'s correspond to.
Repeating here the analysis in~\autoref{sec:floer homology of m7:floer homology} with~\eqref{eq:cy3 x r:action:sqm} as the action for the equivalent SQM model, we find that we can also write~\eqref{eq:cy3 x r:partition fn:no homology} as
\begin{equation}
  \label{eq:cy3 x r:partition fn}
  \boxed{
    \mathcal{Z}_{\text{Spin}(7),CY_3 \times \R}(G)
    = \sum_k \mathcal{F}^{G}_{\text{7d-Spin}(7)}(\Psi_{CY_3}^k)
    = \sum_k \text{HHF}^{G_2\text{-M}}_{d_k}(CY_3, G)
    = \mathcal{Z}^{\text{Floer}}_{G_2\text{-M},CY_3}(C)
  }
\end{equation}
where each $\mathcal{F}^{G}_{\text{7d-Spin}(7)}(\Psi_{CY_3}^k)$ can be identified with a \emph{novel} gauge-theoretic \emph{holomorphic} Floer homology class $\text{HHF}^{G_2\text{-M}}_{d_k}(CY_3, G)$, that we shall henceforth name a holomorphic $G_2$ monopole Floer homology class, of degree $d_k$, assigned to $CY_3$.

Specifically, the \emph{time-invariant DT configurations on $CY_3$ in $\mathfrak{A}_6$} that obey the simultaneous vanishing of the LHS and RHS of the holomorphic \emph{gradient flow equations}
\begin{equation}
  \label{eq:cy3 x r:sqm:flow}
  \boxed{
    \dv{\mathcal{A}^\alpha}{t}
    = - g^{\alpha\bar{\beta}}_{\mathfrak{A}_6} \left(
      \pdv{V_6}{\mathcal{A}^\beta}
    \right)^*
    \qquad
    \dv{C^\alpha}{t}
    = - g^{\alpha\bar{\beta}}_{\mathfrak{A}_6} \left(
      \pdv{V_6}{C^\beta}
    \right)^*
  }
\end{equation}
will generate the chains of the holomorphic $G_2$ monopole Floer complex with \emph{holomorphic Morse functional}
\begin{equation}
  \label{eq:cy3 x r:morse fn V6}
  \boxed{
    V_6(\mathcal{A}, C)
    = \frac{1}{2} \int_{CY_3} \Tr \left(
      CS(\mathcal{A}) \wedge \bar{\star} \Lambda
      + 2 C \wedge \mathcal{F}^{(1,1)} \wedge \bar{\star} \omega
    \right)
  }
\end{equation}
in $\mathfrak{A}_6$, where $CS(\mathcal{A})$ is the Chern-Simons three-form in $\mathcal{A}$ and $\Lambda$ is a holomorphic $(3,0)$-form with components $\Lambda_{mpq} = \varepsilon_{mpq}$.
The $G_2$ monopole flow lines, described by time-varying solutions to~\eqref{eq:cy3 x r:sqm:flow}, are the Floer differentials such that the degree $d_k$ of the corresponding chain in the holomorphic $G_2$ monopole Floer complex is counted by the outgoing flow lines at each time-invariant DT configuration on $CY_3$ in $\mathfrak{A}_6$.
Such a configuration corresponds to a time-independent solution to the 6d equations
\begin{equation}
  \label{eq:cy3 x r:morse fn V6:crit pts}
  \boxed{
    \omega \wedge \bar{\star} \mathcal{F}^{(1,1)} = 0
    \qquad
    \mathcal{F}^{(2,0)} = 0
    \qquad
    \mathcal{D}_m C = 0
  }
\end{equation}

\subtitle{A Physical Proof of Donaldson-Segal's Mathematical Conjecture and Cherkis' Speculation, and a Generalization of the Former}

Note that if we were to restrict to the case where $C = 0$, we would instead physically realize $\text{HHF}^{G_2\text{-inst}}_{d_k}(CY_3, G)$, a holomorphic $G_2$ \emph{instanton} Floer homology of $CY_3$ generated by DT configurations on $CY_3$ with $C = 0$,\footnote{%
  On a closed and compact $CY_3$, DT configurations on $CY_3$ with $C = 0$ are also known as Donaldson-Uhlenbeck-Yau (DUY) configurations~\cite{baulieu-1998-special-quant}, Hermitian-Yang-Mills connections~\cite{wang-2020-modul-spaces, uhlenbeck-1986-exist-hermit, doan-2019-monop-fueter}, or holomorphic vector bundles on $CY_3$~\cite{donaldson-1985-anti-self, uhlenbeck-1986-exist-hermit, doan-2019-monop-fueter}.
  \label{ft:aliases for duy}
}~whose flow lines correspond to time-varying solutions to the $G_2$ instanton equation, i.e.,~\eqref{eq:cy3 x r:bps:complexified} with $C = 0$.
Such a Floer homology was first predicted to exist in the mathematical literature by Donaldson-Thomas~\cite{donaldson-1996-gauge}, and later conjectured more concretely by Donaldson-Segal~\cite[$\S$4]{donaldson-2009-gauge-theor-ii} as a Floer homology generated by holomorphic vector bundles on $CY_3$ (described in \autoref{ft:aliases for duy}), whose flow lines correspond to time-varying solutions to the $G_2$ instanton equation on $CY_3 \times \R$.
We have therefore furnished a physical proof and generalization (when $C \neq 0$) of Donaldson-Segal's mathematical conjecture.

On the other hand, in the physical literature, Cherkis~\cite[$\S$7]{cherkis-2015-octon-monop-knots} speculated that octonionic monopoles (i.e., $G_2$ monopoles) on $CY_3 \times \R$ would allow one to physically realize a \emph{real} Floer homology of $CY_3$ generated by Calabi-Yau monopoles (described in~\autoref{ft:discrepancy of DT config in FT with full DT config}), whose \emph{real} Morse functional is the Chern-Simons-Higgs functional, and flow lines are time-varying solutions to the octonionic monopole equation.
Although his results were for a \emph{real} Floer homology generated by \emph{real} configurations on $CY_3$ and a \emph{real} Morse functional, it is straightforward to verify that through a complexification of $CY_3$ in his case, our results are produced.
We have therefore furnished a physical proof of Cherkis' speculation.


\section{A Holomorphic Floer Homology of Five-Manifolds}
\label{sec:floer homology of m5}

In this section, we will further specialize to $CY_3 = M_4 \times S^1 \times S^1$, where $M_4$ is a closed and compact Calabi-Yau twofold ($CY_2$), and perform another KK dimensional reduction of Spin$(7)$ theory by shrinking one of the $S^1$ circles to be infinitesimally small.
This will allow us to physically derive, from its topologically-invariant $\mathcal{Q}$-cohomology, a novel holomorphic DT Floer homology of $CY_2 \times S^1$.

\subsection{A KK Reduction of \texorpdfstring{Spin$(7)$}{Spin(7)} Theory along \texorpdfstring{$S^1 \times S^1$}{S1 x S1} and the Corresponding SQM}
\label{sec:floer homology of m5:1d sqm}

For $\text{Spin}(7) = M_5 \times S^1 \times S^1 \times \R$ to be a Spin$(7)$-manifold, the closed and compact $M_5 \times S^1$ can be $CY_3$.
This is possible if $M_5 = CY_2 \times S^1$, where $CY_2$ is a closed and compact Calabi-Yau twofold~\cite{esfahani-2022-monop-singul}.
We will consider this case,  and start with 7d-Spin$(7)$ theory on $CY_2 \times S^1_{\mathfrak{b}} \times S^1_{\mathfrak{a}} \times \R$.
Taking $x^2$ and $x^3$ to be the coordinates along $S^1_{\mathfrak{a}}$ and $S^1_{\mathfrak{b}}$, respectively, a further KK reduction of 7d-Spin$(7)$ theory along $S^1_{\mathfrak{a}}$ amounts to setting $\partial_2 \rightarrow 0$, and relabelling $A_2 = \tilde{C} \in \Omega^0(S^1_{\mathfrak{b}}, \text{ad}(G)) \otimes \Omega^0(CY_2, \text{ad}(G))$, and $A_3 = \Gamma \in \Omega^1(S^1_{\mathfrak{b}}, \text{ad}(G)) \otimes \Omega^0(CY_2, \text{ad}(G))$.\footnote{%
  Topological invariance of 7d-Spin$(7)$ theory under yet another KK reduction along $S^1_{\mathfrak{a}}$ implies that all the fields of the theory resulting from the KK reduction, with the exception of $\Gamma$ and its fermionic supersymmetric partner, can be interpreted as scalars along $S^1_{\mathfrak{b}}$.
  On the other hand, this also means that $\Gamma$ can be interpreted as a scalar on $CY_2$.
  \label{ft:reason for fields being scalars on other cy3 circle}
}%
~Note that we can assign the scalars $(C, \tilde{C})$ to the linearly-independent components of $B \in \Omega^0(S^1_{\mathfrak{b}}, \text{ad}(G)) \otimes \Omega^{2, +}(CY_2, \text{ad}(G))$ as $B_{45} = - \tilde{C}$, $B_{46} = C$, and $B_{47} = 0$.

Using a new set of complex coordinates $(\hat{z}^p, \hat{\bar{z}}^{\bar{p}})$ for $CY_2$ defined as $\hat{z}^1 = x^4 + i x^7$, $\hat{z}^2 = x^5 + i x^6$, where their complex conjugates are $\hat{\bar{z}}^{\bar{1}}$ and $\hat{\bar{z}}^{\bar{2}}$, we are able to complexify the fields above.
In particular, we can define $\hat{\mathcal{A}} \in \Omega^0(S^1_b, \text{ad}(G)) \otimes \Omega^{(1, 0)}(CY_2, \text{ad}(G))$, where its components are $\hat{\mathcal{A}}_1 = (A_4 - i A_7)/2$, and $\hat{\mathcal{A}}_2 = (A_5 - i A_6)/2$.
We can also define $\hat{\mathcal{B}} \in \Omega^0(S^1_b, \text{ad}(G)) \otimes \Omega^{2, +}(CY_2, \text{ad}(G))$, a real scalar (\emph{real} self-dual two-form) on $S^1$ ($CY_2$), whose three linearly-independent components are $\hat{\mathcal{B}}_{12} = (\tilde{C} + i C)/2$, $\hat{\mathcal{B}}_{\bar{1}\bar{2}} = (\tilde{C} - i C)/2$, and $\hat{\mathcal{B}}_{1\bar{1}} = 0$.\footnote{%
  Note that $\hat{\mathcal{B}}$ can equally be interpreted as a one-form on an auxiliary $\R^3$ as $\mathscr{B} = \mathscr{B}_i d\mathscr{X}^i$, where each of the three components of $\mathscr{B}$ corresponds to each of the three linearly-independent components of $\hat{\mathcal{B}}$.
  We will be making use of this correspondence between $\hat{\mathcal{B}}$ and $\mathscr{B}$ shortly.
  \label{ft:introducing mathscr B}
}

Thus, upon KK reduction along $S^1_{\mathfrak{a}}$,~\eqref{eq:cy3 x r:bps} becomes
\begin{equation}
  \label{eq:cy2 x s x r:bps}
  \begin{aligned}
    \dot{\hat{\mathcal{A}}}_p
    &= - i \left(
      \partial_y \hat{\mathcal{A}}_p
      + \hat{\mathcal{D}}^q \hat{\mathcal{B}}_{qp}
      - \hat{\mathcal{D}}_p \Gamma
      \right)
      \, , \\
    \dot{\hat{\mathcal{B}}}_{pq}
    &= i \left(
      \partial_y \hat{\mathcal{B}}_{pq}
      - 2 \hat{\mathcal{F}}_{pq}
      - [\hat{\mathcal{B}}_{pq}, \Gamma]
      \right)
      \, , \\
    \dot{\Gamma}
    &= \hat{\omega}^{p\bar{q}} \left(
      \hat{\mathcal{F}}_{p\bar{q}}
      - \frac{1}{4} \left(
      \hat{\mathcal{B}} \times \hat{\mathcal{B}}
      \right)_{p\bar{q}}
      \right)
      \, ,
  \end{aligned}
\end{equation}
where $\hat{\omega}^{p\bar{q}}$ is the Kähler two-form of $CY_2$; $x^3$ is relabeled as $y$; and $\hat{\mathcal{B}} \times \hat{\mathcal{B}}$ is a shorthand defined as $\hat{\mathcal{B}} \times \hat{\mathcal{B}} \coloneq 4 \star_3 (\mathscr{B} \wedge_3 \mathscr{B}) = 2 \varepsilon_{ijk} [\mathscr{B}^i, \mathscr{B}^j] d\mathscr{X}^k$, with ``$\wedge_3$'' and ``$\star_3$'' being the exterior product and Hodge star operators, respectively, in the auxiliary $\R^3$ of $\mathscr{B}$ fields introduced in~\autoref{ft:introducing mathscr B}.\footnote{%
  The only non-zero component of $\hat{\mathcal{B}} \times \hat{\mathcal{B}}$ (since $\hat{\mathcal{B}}_{1\bar{1}} = 0$) is $(\hat{\mathcal{B}} \times \hat{\mathcal{B}})_{1\bar{1}} \propto [\hat{\mathcal{B}}_{12}, \hat{\mathcal{B}}_{\bar{1}\bar{2}}]$.
  \label{ft:components of calB x calB}
}%
~These are the BPS equations of the theory that results from the KK reduction along the $S^1_{\mathfrak{a}}$ circle.
Notice also that these equations can be expressed as DT equations on $CY_3 = CY_2 \times S^1_{\mathfrak{b}} \times \R$ in temporal gauge, with time being along the $\R$ direction.

In other words, 7d-Spin$(7)$ theory, upon a second KK reduction along $S^1_{\mathfrak{a}}$ to a 6d theory (henceforth referred to as 6d-Spin$(7)$ theory) on $CY_2 \times S^1_{\mathfrak{b}} \times \R$, localizes onto configurations that obey~\eqref{eq:cy2 x s x r:bps}.

\subtitle{6d-Spin$(7)$ Theory as a 1d SQM in $\mathfrak{A}_5$}

Just as was done in the previous sections, this means that the 6d action can be written as\footnote{%
  \label{ft:reason for omitting circle subscript}%
  The expressions and results that follow will remain the same if we had chosen to reduce along $S^1_{\mathfrak{b}}$ instead of $S^1_{\mathfrak{a}}$, up to a redefinition of the components of the fields.
  Thus, the subscripts of the $S^1_*$ circles will henceforth be omitted.
}
\begin{equation}
  \label{eq:cy2 x s x r:action}
  \begin{aligned}
    S_{\text{6d-Spin}(7)}
    =& \frac{1}{2e^2} \int_\R \dd{t} \int_{CY_2 \times S^1} \abs{\dd{\hat{z}}}^4 \dd{y} \Tr \Bigg(
       4 \left|
       \dot{\hat{\mathcal{A}}}_p
       + i \left( \partial_y \hat{\mathcal{A}}_p
       + \hat{\mathcal{D}}^{q} \hat{\mathcal{B}}_{qp}
       - \hat{\mathcal{D}}_p \Gamma
       \right)
       \right|^2
    \\
     & + 4 \left|
       \dot{\hat{\mathcal{B}}}_{pq}
       - i \left( \partial_y \hat{\mathcal{B}}_{pq}
       - 2 \hat{\mathcal{F}}_{pq}
       - [\hat{\mathcal{B}}_{pq}, \Gamma]
       \right)
       \right|^2
       + \left|
       \dot{\Gamma}
       - \hat{\omega}^{p\bar{q}} \left( \hat{\mathcal{F}}_{p\bar{q}}
       - \frac{1}{4} \left( \hat{\mathcal{B}} \times \hat{\mathcal{B}} \right)_{p\bar{q}}
       \right)
       \right|^2
       + \dots
       \Bigg) \, .
  \end{aligned}
\end{equation}
After suitable rescalings, the equivalent SQM action can be obtained from~\eqref{eq:cy2 x s x r:action} as
\begin{equation}
  \label{eq:cy2 x s x r:action:sqm}
  S_{\text{SQM,HW}} = \frac{1}{e^2} \int_\R \dd{t} \left(
    \left|
      \dot{\hat{\mathcal{A}}}^\alpha
      + g^{\alpha\bar{\beta}}_{\mathfrak{A}_5} \left(
        \pdv{V_5}{\hat{\mathcal{A}}^\beta}
      \right)^*
    \right|^2
    + \left|
      \dot{\hat{\mathcal{B}}}^\alpha
      + g^{\alpha\bar{\beta}}_{\mathfrak{A}_5} \left(
        \pdv{V_5}{\hat{\mathcal{B}}^\beta}
      \right)^*
    \right|^2
    + \left|
      \dot{\Gamma}^\alpha
      + g^{\alpha\bar{\beta}}_{\mathfrak{A}_5} \left(
        \pdv{V_5}{\Gamma^\beta}
      \right)^*
    \right|^2
    + \dots
  \right)
  \, ,
\end{equation}
where $(\hat{\mathcal{A}}^\alpha, \hat{\mathcal{B}}^\alpha, \Gamma^\alpha)$ and $(\alpha, \beta)$ are holomorphic coordinates and indices on the space $\mathfrak{A}_5$ of irreducible $(\hat{\mathcal{A}}_p, \hat{\mathcal{B}}_{pq}, \Gamma)$ fields on $CY_2 \times S^1$; $g_{\mathfrak{A}_5}$ is the metric on $\mathfrak{A}_5$; $V_5(\hat{\mathcal{A}}, \hat{\mathcal{B}}, \Gamma)$ is the holomorphic potential function; and the `HW' label in the subscript will be made clear very shortly.

\subtitle{Localizing Onto Time-invariant Haydys-Witten Configurations on $CY_2 \times S^1$}

Applying the squaring argument to~\eqref{eq:cy2 x s x r:action:sqm}, the configurations of the equivalent SQM localizes onto are those that set the LHS and RHS of~\eqref{eq:cy2 x s x r:bps} \emph{simultaneously} to zero.
In the real coordinates of $CY_2 \times S^1$, such configurations in $\mathfrak{A}_5$ are time-invariant configurations that span the space of solutions to\footnote{%
  Recall that $\Gamma$ is actually the gauge connection along the $y$-direction, i.e., $A_y$, from its definition at the start of this section.
  \label{ft:reminder that gamma is a gauge connection}
}
\begin{equation}
  \label{eq:cy2 x s x r:hw eqns}
  F_{ya} + D^b B_{ba}
  = 0
  \, ,
  \qquad
  F_{ab}^+ - \frac{1}{2} D_y B_{ab} - \frac{1}{4} [B_{ac}, B_{bd}] g^{cd}
  = 0
  \, ,
\end{equation}
where $(a, b)$ are indices on $CY_2$, with one of the three linearly-independent components of the self-dual two-form field $B$ being zero, and $[B_{ac}, B_{bd}]g^{cd} \eqcolon (B \times B)_{ab}$.
These are the 5d Haydys-Witten (HW) equations on $CY_2 \times S^1$ \cite{haydys-2015-fukay-seidel, witten-2011-fiveb-knots, er-2023-topol-n} (where the Reeb vector is along the $S^1$ direction).
We shall, in the rest of this section, refer to such configurations as HW configurations on $CY_2 \times S^1$.

In summary, the equivalent SQM localizes onto time-invariant HW configurations on $CY_2 \times S^1$ in $\mathfrak{A}_5$.

\subsection{A Holomorphic Donaldson-Thomas Floer Homology of \texorpdfstring{$CY_2 \times S^1$}{CY2 x S1}}
\label{sec:floer homology of m5:floer homology}

Since the resulting 6d theory on $CY_2 \times S^1 \times \R$ can be interpreted as a 1d SQM in $\mathfrak{A}_5$, its partition function can, like in~\eqref{eq:g2 x r:spin7 partition fn:no homology}, be written as
\begin{equation}
  \label{eq:cy2 x s x r:partition fn:no homology}
  \mathcal{Z}_{\text{Spin}(7),CY_2 \times S^1 \times \R}(G)
  = \expval{1}_{F_5(\Psi_{CY_2 \times S^1})}
  = \sum_l \mathcal{F}^{G}_{\text{6d-Spin}(7)}(\Psi_{CY_2 \times S^1}^l)
  \, ,
\end{equation}
where $\mathcal{F}^{G}_{\text{6d-Spin}(7)}(\Psi_{CY_2 \times S^1}^l)$, in the $\mathcal{Q}$-cohomology of 6d-Spin$(7)$ theory, is the $l^{\text{th}}$ contribution to the partition function that depends on the expression of $F_5(\Psi_{CY_2 \times S^1})$ in the bosonic fields of $CY_2 \times S^1$, and the summation in `$l$' is over all isolated and non-degenerate HW configurations on $CY_2 \times S^1$ in $\mathfrak{A}_5$ that the equivalent SQM localizes onto.\footnote{%
  This presumption that the configurations will be isolated and non-degenerate is justified because (the $\mathcal{Q}$-cohomology of) Spin$(7)$ theory is topological in all directions and therefore invariant when we further shrink an $S^1$ circle from 7d-Spin$(7)$ theory.
  Thus, if $CY_2$ (where $CY_2 \times T^3 = G_2$) is chosen such as to satisfy the transversality assumptions of \autoref{ft:g2 instanton isolation and non-degeneracy}, $\mathcal{Z}_{\text{Spin}(7),CY_2 \times S^1 \times \R}$ will be a discrete and non-degenerate sum of contributions, just like $\mathcal{Z}_{\text{Spin}(7),G_2 \times \R}$.
  We shall henceforth assume such a choice of $CY_2$ whence the presumption would hold.
  \label{ft:isolation and non-degeneracy of 6d-spin7}
}

Let us now ascertain what the $\mathcal{F}^{G}_{\text{6d-Spin}(7)}(\Psi_{CY_2 \times S^1}^l)$'s correspond to.
Repeating here again the analysis in \autoref{sec:floer homology of m7:floer homology} with~\eqref{eq:cy2 x s x r:action:sqm} as the action for the equivalent SQM model, we find that we can also write~\eqref{eq:cy2 x s x r:partition fn:no homology} as
\begin{equation}
  \label{eq:cy2 x s x r:partition fn}
  \boxed{
    \mathcal{Z}_{\text{Spin}(7),CY_2 \times S^1 \times \R}(G)
    = \sum_l \mathcal{F}^{G}_{\text{6d-Spin}(7)}(\Psi_{CY_2 \times S^1}^l)
    = \sum_l \text{HHF}^{\text{DT}}_{d_l}(CY_2 \times S^1, G)
    = \mathcal{Z}^{\text{Floer}}_{\text{DT},CY_2 \times S^1}(G)
  }
\end{equation}
where each $\mathcal{F}^{G}_{\text{6d-Spin}(7)}(\Psi_{CY_2 \times S^1}^l)$ can be identified with a \emph{novel} gauge-theoretic \emph{holomorphic} Floer homology class $\text{HHF}^{\text{DT}}_{d_l}(CY_2 \times S^1, G)$, that we shall henceforth name a holomorphic DT Floer homology class, of degree $d_l$, assigned to $CY_2 \times S^1$.

Specifically, the \emph{time-invariant HW configurations on $CY_2 \times S^1$ in $\mathfrak{A}_5$} that obey the simultaneous vanishing of the LHS and RHS of the holomorphic \emph{gradient flow equations}
\begin{equation}
  \label{eq:cy2 x s x r:sqm:flow}
  \boxed{
    \dv{\hat{\mathcal{A}}^\alpha}{t}
    = - g^{\alpha\bar{\beta}}_{\mathfrak{A}_5} \left(
      \pdv{V_5}{\hat{\mathcal{A}}^\beta}
    \right)^*
    \qquad
    \dv{\hat{\mathcal{B}}^\alpha}{t}
    = - g^{\alpha\bar{\beta}}_{\mathfrak{A}_5} \left(
      \pdv{V_5}{\hat{\mathcal{B}}^\beta}
    \right)^*
    \qquad
    \dv{\Gamma^\alpha}{t}
    = - g^{\alpha\bar{\beta}}_{\mathfrak{A}_5} \left(
      \pdv{V_5}{\Gamma^\beta}
    \right)^*
  }
\end{equation}
will generate the chains of the holomorphic DT Floer complex with holomorphic
\emph{Morse functional}
\begin{equation}
  \label{eq:cy2 x s x r:morse fn V5}
  \boxed{
    V_5(\hat{\mathcal{A}}, \hat{\mathcal{B}}, \Gamma)
    = \int \frac{i}{2} d_y \left(
      \hat{\mathcal{A}} \wedge \bar{\star} \hat{\mathcal{A}}
      + \hat{\mathcal{B}} \wedge \bar{\star} \hat{\mathcal{B}}
    \right)
    - 2i \hat{\mathcal{B}} \wedge \bar{\star} \hat{\mathcal{F}}
    - \Gamma \wedge \hat{\omega} \wedge \bar{\star} \left(
      \hat{\mathcal{F}}
      - \frac{1}{4} (\hat{\mathcal{B}} \times \hat{\mathcal{B}})
    \right)
  }
\end{equation}
in $\mathfrak{A}_5$, where ``$\wedge$'' is the exterior product in $CY_2 \times S^1$, and ``$\star$'' is the Hodge star operator in the complex $CY_2$.
The holomorphic DT flow lines, described by time-varying solutions to~\eqref{eq:cy2 x s x r:sqm:flow}, are the Floer differentials such that the degree $d_l$ of the corresponding chain in the holomorphic DT Floer complex is counted by the outgoing flow lines at each time-invariant HW configuration on $CY_2 \times S^1$ in $\mathfrak{A}_5$.
Such a configuration corresponds to a time-independent solution to the 5d equations
\begin{equation}
  \label{eq:cy2 x s x r:morse fn V5:crit pts}
  \boxed{
    \begin{aligned}
      d_y \hat{\mathcal{A}}
      + \bar{\star} ( \hat{\mathcal{D}} \bar{\star} \hat{\mathcal{B}} )
      &= \hat{\mathcal{D}} \Gamma
      \\
      d_y \hat{\mathcal{B}}
      + \frac{1}{2} (\hat{\mathcal{B}} \times \hat{\mathcal{B}})
      - 2 \hat{\mathcal{F}}
      &= [\hat{\mathcal{B}}, \Gamma]
    \end{aligned}
  }
\end{equation}
with a vanishing component of $\hat{\mathcal{B}}$, i.e., $\hat{\mathcal{B}}_{p\bar{p}} = 0$.

There is one remaining $S^1$ circle that we may perform KK reduction along to repeat the procedure of these past three sections.
However, we will stop here as such a case was previously studied more generally in~\cite[$\S$3]{er-2023-topol-n} as an HW Floer homology of four-manifolds.
It is indeed straightforward to show that KK reduction of 6d-Spin$(7)$ theory on $CY_2 \times S^1 \times \R$ along the remaining $S^1$ circle will physically realize an HW Floer homology of $CY_2$.


\section{A Hyperkähler Floer Homology Specified by Hypercontact Three-Manifolds}
\label{sec:hyperkahler floer-hom}

Up until now, the most specific decomposition of the Spin$(7)$-manifold considered has been $\text{Spin}(7) = CY_2 \times T^3 \times \R$.
We will relax this specification and instead, consider the case where $\text{Spin}(7) = CY_2 \times HC_3 \times \R$, with $HC_3$ being a hypercontact three-manifold.
By performing a topological reduction along $CY_2$ using a technique similar to what Bershadsky-Johansen-Sadov-Vafa (BJSV) had developed in~\cite{bershadsky-1995-topol-reduc}, we will get a 4d sigma model whose target space is the moduli space of instantons on $CY_2$.
Recasting this 4d sigma model as an SQM in the hypercontact three-space will allow us to physically derive, from its topologically-invariant $\mathcal{Q}$-cohomology, a hyperkähler Floer homology. In turn, this would serve as a physical proof of Hohloch-Noetzel-Salamon's mathematical conjecture~\cite{hohloch-2009-hyper-struc}~\cite[$\S$5]{salamon-2013-three-dimen}.

\subsection{A Topological Reduction of \texorpdfstring{Spin$(7)$}{Spin(7)} Theory along \texorpdfstring{$CY_2$}{CY2} and the Corresponding 4d Sigma Model}
\label{sec:hyperkahler floer-hom:4d sigma model}

Any closed and orientable three-manifold admits a hypercontact structure, and is thus a hypercontact three-manifold; this includes examples such as $T^3$ and $S^3$~\cite{geiges-1995-hyper-manif, hohloch-2009-hyper-struc}.
Recall however, that  $CY_2 \times HC_3 \times \R$ must admit a Spin$(7)$ structure.
This means that our choice of $HC_3$ must be such that $b_1(HC_3) > 2$.\footnote{%
  For $CY_2 \times HC_3 \times \R$ to admit a Spin$(7)$ structure, $CY_2 \times HC_3$ must admit a $G_2$ structure,
  for which it is known that $b_1(HC_3) > 2$~\cite[Remark 5.1 (ii)]{salamon-2013-three-dimen}.
  \label{ft:restriction of hc3}
}
With that said, let us begin by first performing a topological reduction along $CY_2$.\footnote{%
  Here, the compact $CY_2$'s are non-trivial, such as K3.
  \label{ft:cy2 as k3 surfaces}
}

To effect the topological reduction of Spin$(7)$ theory along $CY_2$, we have to scale the Spin$(7)$-manifold metric $g_{\text{Spin}(7)}$ appropriately.
This can be done by writing it as a block diagonal with a vanishing scale parameter $\epsilon$, i.e., $g_{\text{Spin}(7)} = \epsilon^{1/2} g_{CY_2} \oplus \epsilon^{-1/2} g_{HC_3 \times \R}$.
By doing so, in the vanishing limit of $\epsilon \rightarrow 0$, only $g_{HC_3 \times \R}$ will remain and the theory surviving the reduction process will be a 4d theory on $HC_3 \times \R$.\footnote{%
  This is similar to the BJSV topological reduction process~\cite{bershadsky-1995-topol-reduc}, where given a manifold $M = M_X \times M_Y$, instead of topologically reducing $M_X$ and leaving $M_Y$ alone, we are simultaneously topologically reducing and enlarging $M_X$ and $M_Y$, respectively.
Both processes would result in the same physical phenomenon, where the size of $M_Y$ is infinitely larger than $M_X$ in the vanishing limit of $\epsilon \rightarrow 0$.
  Indeed, we are able to reproduce the results that BJSV arrived at in~\cite{bershadsky-1995-topol-reduc} using this topological reduction process.
  \label{ft:k3 reduction as adaptation of bjsv reduction}
}

\subtitle{Topological Reduction Along $CY_2$}

Using the $\epsilon$-scaled metric in~\eqref{eq:spin7 action}, in the vanishing limit of $\epsilon \rightarrow 0$, terms with positive powers of $\epsilon$ in the action can be ignored as they will vanish, whilst terms with no power of $\epsilon$ are of interest as they will survive the reduction process.
These terms can be identified as those with a single contraction in both $CY_2$ and $HC_3 \times \R$ in~\eqref{eq:spin7 action} (and thus~\eqref{eq:g2 x r:spin7 action}), i.e.,
\begin{equation}
  \label{eq:spin7 action:cy2 reduced}
  \begin{aligned}
    S_{\text{Spin}(7), \epsilon^0}
    = \frac{1}{e^2} \int_{CY_2 \times HC_3 \times \R} \dd{t} \dd[7]{x} \Tr \bigg(
    & 2 |F_{tM}^+|^2
      + 2i (D_t \psi_M - D_M \psi_t )\chi^{tM}
      + 2i (D_a \psi_M - D_M \psi_a)  \chi^{aM}
    \\
    & - \frac{i}{2} \varphi \left\lbrace \chi_{tM}, \chi^{tM} \right\rbrace
      - \frac{i}{2} \varphi \left\lbrace \chi_{aM}, \chi^{aM} \right\rbrace
      \bigg)
      \, ,
  \end{aligned}
\end{equation}
where $M \in \{4, \dots, 7\}$ is the index on $CY_2$ and $a \in \{1, 2, 3\}$ is the index on $HC_3$.
On the other hand, terms with negative powers of $\epsilon$ will blow up in the vanishing limit of $\epsilon \rightarrow 0$, and thus need to be set to zero to ensure the finiteness of the action.
These terms are identified with those that have more than one contraction in $CY_2$ in~\eqref{eq:spin7 action}, which (for the bosonic sector of the action) work out to be
\begin{equation}
  \label{eq:finiteness cond:full}
  F_{MN}^+ = 0
  \, .
\end{equation}
Keep in mind that the self-duality of the $F^+$'s in~\eqref{eq:spin7 action:cy2 reduced} and~\eqref{eq:finiteness cond:full} is still to be understood as a self-duality on the Spin$(7)$-manifold, i.e., $F_{tM}^+ = \frac{1}{2} (F_{tM} + \phi_{tMaN} F^{aN})$ and $F_{MN}^+ = \frac{1}{2} (F_{MN} + \frac{1}{2} \phi_{MNPQ} F^{PQ} + \frac{1}{2} \phi_{MNab} F^{ab})$, with the Spin$(7)$-structure in~\eqref{eq:spin7 structure}.

When the vanishing limit of $\epsilon \rightarrow 0$ is finally taken to topologically reduce along $CY_2$, we can take $\partial_M \rightarrow 0$, and will find that the gauge connections with indices on $HC_3$ are no longer dynamical in the resulting 4d theory and thus, can be integrated out.
The finiteness condition of~\eqref{eq:finiteness cond:full} will then simply be the ASD $G$-instanton equation on $CY_{2}$, i.e.,
\begin{equation}
  \label{eq:finiteness cond}
  F_{CY_2}^+ = 0 \, .
\end{equation}

Note that the bosonic fields surviving the topological reduction process (after integrating out the gauge fields) in~\eqref{eq:spin7 action:cy2 reduced}, will consist only of connections on $CY_2$ which are also scalars on the 4d worldvolume.
Imposing the finiteness condition~\eqref{eq:finiteness cond} on the resulting 4d theory means that we can interpret the bosonic fields $\Phi$ as maps $\Phi: HC_3 \times \R \rightarrow \mathcal{M}^G_{\text{inst}}(CY_2)$ from the 4d worldvolume $HC_3 \times \R$ to the moduli space $\mathcal{M}^G_{\text{inst}}(CY_2)$ of ASD $G$-instantons on $CY_2$ (henceforth referred to as the moduli space of instantons on $CY_2$), i.e., solutions to~\eqref{eq:finiteness cond} (modulo gauge equivalence).
In other words, we will have a 4d $\mathcal{N} = 2$ topological sigma model on $HC_3 \times \R$ with target space $\mathcal{M}^G_{\text{inst}}(CY_2)$.

\subtitle{The 4d Sigma Model Action}

Let us now determine the action of the 4d sigma model.
First, in the vanishing limit of $\epsilon \rightarrow 0$,~\eqref{eq:spin7 action:cy2 reduced} becomes
\begin{equation}
  \label{eq:4d sigma model action:unconverted}
  S_{\text{4d-}\sigma} = \frac{1}{2e^2} \int_{HC_3 \times \R} \dd{t} \dd[3]{x} \bigg(
  \partial_t A_M \partial^t A^M
  + \partial_a A_M \partial^a A^M
  - \partial_t A_M \phi^{taMN} \partial_a A_N
  + \dots
  \bigg)
  \, .
\end{equation}

Next, by using the cotangent basis $\kappa_{iM}$ of $\mathcal{M}^G_{\text{inst}}(CY_2)$, we can write the boson in~\eqref{eq:4d sigma model action:unconverted} as
\begin{equation}
  \label{eq:hc3 x r:ws bosons in cotangent basis}
  A_M = X^i \kappa_{iM} \, ,
\end{equation}
where $X^i$ and $i \in 4 \mathbb{N}$ are coordinates and indices on the hyperkähler $\mathcal{M}^G_{\text{inst}}(CY_2)$.
This means that~\eqref{eq:4d sigma model action:unconverted} can therefore finally be written as
\begin{equation}
  \label{eq:4d sigma model action:converted}
    S_{\text{4d-}\sigma} = \frac{1}{2e^2} \int_{HC_3 \times \R} \dd{t} \dd[3]{x} g_{i \bar{\jmath}} \bigg(
    \partial_t X^i \partial^t X^{\bar{\jmath}}
    + \partial_a X^i \partial^a X^{\bar{\jmath}}
    + \dots
    \bigg)
    + \tau \int_{HC_3 \times \R} \sum_{a = 1}^3 \Omega_a \wedge \Phi^*(\omega_a)
    \, ,
\end{equation}
where $g_{i \bar{\jmath}}$ is the metric on $\mathcal{M}^G_{\text{inst}}(CY_2)$;
$\Omega_a$ and $\omega_a$ are the Kähler two-forms of $HC_3 \times \R$ and $\mathcal{M}^G_{\text{inst}}(CY_2)$, respectively, with respect to their three complex structures;\footnote{%
  $HC_3 \times \R$ and $\mathcal{M}^G_{\text{inst}}(CY_2)$ are both hyperkähler, and thus support three complex structures $I_1$, $I_2$, and $I_3 \equiv - I_1 I_2$ for the former, and $J_1$, $J_2$, and $J_3 = - J_1 J_2$ for the latter, satisfying the quaternionic relations $I_a^2 = J_a^2 = -1$.
  These hyperkähler manifolds also have a corresponding Kähler two-form for each of the three complex structures, $\Omega_a$ for the former, and $\omega_a$ for the latter.
  The reason for this unusual orientation of ($I_3$, $J_3$) with an additional negative sign is purely for the ease of notation.
  \label{ft:hyperkahler complex structures}
}~and the last term is a topological term containing the pullback from $\mathcal{M}^G_{\text{inst}}(CY_2)$ onto the worldvolume with some constant $\tau$.

In other words, topological reduction of Spin$(7)$ theory on $CY_2 \times HC_3 \times \R$ along $CY_2$ results in a 4d $\mathcal{N} = 2$ sigma model on $HC_3 \times \R$ with target space $\mathcal{M}^G_{\text{inst}}(CY_2)$, and action~\eqref{eq:4d sigma model action:converted}.


\subsection{A 1d SQM in the Hypercontact Three-space of Instantons on \texorpdfstring{$CY_2$}{CY2}}
\label{sec:hyperkahler floer-hom:sqm}

\subtitle{The BPS equations of the 4d Sigma Model}

First, note that the BPS equations of the 4d sigma model with action~\eqref{eq:4d sigma model action:converted}, will descend from~\eqref{eq:spin7 bps} for the fields that survive the topological reduction of Spin$(7)$ theory along $CY_2$.
The explicit components of the BPS equations are\footnote{%
  In the following expressions, there are no gauge fields in the worldvolume theory as they are integrated out. We have also lowered the indices of the fields.
  \label{ft:preamble to 4d sigma model bps equations}
}
\begin{equation}
  \label{eq:hc3 x r:bps:explicit}
  \begin{aligned}
    \partial_t A_4 - \partial_1 A_5 - \partial_2 A_6 + \partial_3 A_7
    &= 0
      \, , \\
    \partial_t A_5 + \partial_1 A_4 + \partial_2 A_7 + \partial_3 A_6
    &= 0
      \, , \\
    \partial_t A_6 - \partial_1 A_7 + \partial_2 A_4 - \partial_3 A_5
    &= 0
      \, , \\
    \partial_t A_7 + \partial_1 A_6 - \partial_2 A_5 - \partial_3 A_4
    &= 0
      \, . \\
  \end{aligned}
\end{equation}
Let us define the action of the three complex structures $J_a$ of the hyperkähler $\mathcal{M}^G_{\text{inst}}(CY_2)$ (introduced in~\autoref{ft:hyperkahler complex structures}) on the target space cotangent bases as
\begin{equation}
  \label{eq:hc3 x r:complex structure of target space}
  \begin{aligned}
    J_1 \kappa_{i4}
    &= \kappa_{i5}
      \, ,
    &\qquad
      J_1 \kappa_{i6}
    &= \kappa_{i7}
      \, ,
    \\
    J_2 \kappa_{i4}
    &= \kappa_{i6}
      \, ,
    &\qquad
      J_2 \kappa_{i5}
    &= - \kappa_{i7}
      \, ,
    \\
    J_3 \kappa_{i4}
    &= - \kappa_{i7}
      \, ,
    &\qquad
      J_3 \kappa_{i5}
    &= - \kappa_{i6}
      \, .
  \end{aligned}
\end{equation}
Then, by expressing~\eqref{eq:hc3 x r:bps:explicit} in terms of the $X^i$ coordinates via~\eqref{eq:hc3 x r:ws bosons in cotangent basis}, it becomes
\begin{equation}
  \label{eq:hc3 x r:bps}
  \partial_t X^i = - \partial_1 X^i J_1 - \partial_2 X^i J_2 - \partial_3 X^i J_3
  \, .
\end{equation}

If we also define the action of the three complex structures $I_a$ of the hyperkähler $HC_3 \times \R$ worldvolume (also introduced in~\autoref{ft:hyperkahler complex structures}) on the worldvolume cotangent bases as
\begin{equation}
  \label{eq:hc3 x r:complex structure of worldvolume}
  \begin{aligned}
    I_1 \partial_t
    &= - \partial_1
      \, ,
    &\qquad
      I_1 \partial_2
    &= \partial_3
      \, ,
    \\
    I_2 \partial_t
    &= - \partial_2
      \, ,
    &\qquad
      I_2 \partial_1
    &= - \partial_3
      \, ,
    \\
    I_3 \partial_t
    &= - \partial_3
      \, ,
    &\qquad
      I_3 \partial_1
    &= \partial_2
      \, ,
  \end{aligned}
\end{equation}
we can write \eqref{eq:hc3 x r:bps} more compactly as
\begin{equation}
  \label{eq:hc3 x r:cauchy-riemann-fueter equation}
  d X^i - \sum_{a = 1}^3 I_a d X^i J_a = 0 \, ,
\end{equation}
where `$d$' is the exterior derivative on the $HC_3 \times \R$ worldvolume.
This is the Cauchy-Riemann-Fueter equation for the aholomorphic map $X^i: HC_3 \times \R \rightarrow \mathcal{M}^G_{\text{inst}}(CY_2)$.\footnote{%
  Aholomorphic maps are also sometimes referred to in the literature as triholomorphic, quaternionic, or q-holomorphic maps~\cite{haydys-2008-nonlin-dirac}.
  \label{ft:reason for cauchy-riemann-fueter equation}
}

\subtitle{The 1d SQM Action}

Next, using the BPS equation in~\eqref{eq:hc3 x r:bps}, we can re-express the action of the 4d sigma model with target space $\mathcal{M}^G_{\text{inst}}(CY_2)$ in~\eqref{eq:4d sigma model action:converted} as
\begin{equation}
  \label{eq:hc3 x r:4d action:cauchy-riemann-fueter eq}
  S_{\text{4d-}\sigma} = \frac{1}{2e^2} \int_{\R} \dd{t} \int_{HC_3} \dd[3]{x} \left(
    \left|
      \partial_t X^i
      + \sum_{a = 1}^3 \partial_a X^i J_a
    \right|^2
  + \dots
  \right)
  \, ,
\end{equation}
where the ``$\dots$'' will now contain the fermionic \emph{and} topological terms of~\eqref{eq:4d sigma model action:converted}.
Lastly, after suitable rescalings, we obtain the equivalent SQM action from~\eqref{eq:hc3 x r:4d action:cauchy-riemann-fueter eq} as
\begin{equation}
  \label{eq:hc3 x r:sqm:action:hc3}
  S_{\text{SQM}, \text{4d-}\sigma} = \frac{1}{e^2} \int_\R \dd{t} \left(
    \left|
      \dv{X^\alpha}{t}
      + g^{\alpha \beta}_{\mathcal{M}(HC_3, \mathcal{M}^{G, CY_2}_{\text{inst}})}
      \pdv{V_{\sigma}}{X^\beta}
    \right|^2
    + \dots
  \right)
  \, ,
\end{equation}
where $X^\alpha$ and $(\alpha, \beta)$ are coordinates and indices on the hypercontact three-space $\mathcal{M}(HC_3, \mathcal{M}^{G, CY_2}_{\text{inst}})$ of instantons on $CY_2$, which is the space of smooth maps from $HC_3$ to $\mathcal{M}^G_{\text{inst}}(CY_2)$;
$g_{\mathcal{M}(HC_3, \mathcal{M}^{G, CY_2}_{\text{inst}})}$ is the metric on $\mathcal{M}(HC_3, \mathcal{M}^{G, CY_2}_{\text{inst}})$;
and $V_{\sigma}(X^\alpha)$ is the potential function.

\subtitle{Localizing onto Time-invariant Fueter Maps from $HC_3$ to $\mathcal{M}^G_{\text{inst}}(CY_2)$}

Applying again the squaring argument~\cite{blau-1993-topol-gauge} with~\eqref{eq:hc3 x r:sqm:action:hc3} as the equivalent SQM action, the configurations that the equivalent SQM localizes onto are those that set the LHS and RHS of~\eqref{eq:hc3 x r:bps} \emph{simultaneously} to zero.
In other words, the equivalent SQM localizes onto time-invariant Fueter maps from $HC_3$ to $\mathcal{M}^G_{\text{inst}}(CY_2)$.

\subsection{A Hyperkähler Floer Homology of Instantons on \texorpdfstring{$CY_2$}{CY2} and Specified by \texorpdfstring{$HC_3$}{HC3}}
\label{sec:hyperkahler floer-hom:floer-homs}

As before, since the resulting 4d theory on $HC_3 \times \R$ can be expressed as a 1d SQM in the space $\mathcal{M}(HC_3, \mathcal{M}^{G, CY_2}_{\text{inst}})$, its partition function can, like in~\eqref{eq:g2 x r:spin7 partition fn:no homology}, be written as
\begin{equation}
  \label{eq:hc3 x r:partition fn:no homology}
  \mathcal{Z}_{\text{Spin}(7), HC_3 \times \R}(G)
  = \sum_s \mathcal{F}^s_{\text{4d-}\sigma, HC_3 \times \R \rightarrow \mathcal{M}^{G, CY_2}_{\text{inst}}}
  \, ,
\end{equation}
where $\mathcal{F}^s_{\text{4d-}\sigma, HC_3 \times \R \rightarrow \mathcal{M}^{G, CY_2}_{\text{inst}}}$, in the $\mathcal{Q}$-cohomology of the 4d-sigma model, is the $s^{\text{th}}$ contribution to the partition function, and the summation in `$s$' is over all isolated and non-degenerate time-invariant Fueter maps from $HC_3$ to $\mathcal{M}^G_{\text{inst}}(CY_2)$ in $\mathcal{M}(HC_3, \mathcal{M}^{G, CY_2}_{\text{inst}})$ that the equivalent SQM localizes onto.\footnote{%
  This presumption that the configurations will be isolated and non-degenerate is justified because (the $\mathcal{Q}$-cohomology of) Spin$(7)$ theory is topological in all directions and therefore invariant when we topologically reduce along $CY_2$ from Spin$(7)$ theory.
  Thus, if $HC_3$ and $CY_2$ (where $HC_3 \times CY_2 = G_2$) are chosen such as to satisfy the transversality assumptions of~\autoref{ft:g2 instanton isolation and non-degeneracy}, $\mathcal{Z}_{\text{Spin}(7), HC_3 \times \R}$ will be a discrete and non-degenerate sum of contributions, just like $\mathcal{Z}_{\text{Spin}(7),G_2 \times \R}$.
  We shall henceforth assume such a choice of $HC_3$ and $CY_2$ whence the presumption would hold.
  \label{ft:isolation and non-degeneracy of 4d-sigma}
}

Via a similar analysis to that in \autoref{sec:floer homology of m7:floer homology} with~\eqref{eq:hc3 x r:sqm:action:hc3} as the action for the equivalent 1d SQM model, we find that we can also write~\eqref{eq:hc3 x r:partition fn:no homology} as
\begin{equation}
  \label{eq:hc3 x r:partition fn}
  \boxed{
    \begin{aligned}
      \mathcal{Z}_{\text{Spin}(7), HC_3 \times \R}(G)
      &= \sum_s \mathcal{F}^s_{\text{4d-}\sigma, HC_3 \times \R \rightarrow \mathcal{M}^{G, CY_2}_{\text{inst}}}
      \\
      & = \sum_s \text{HHKF}_{d_s}\left(
        HC_3, \mathcal{M}^G_{\text{inst}}(CY_2)
        \right)
        = \mathcal{Z}^{\text{hyperkählerFloer}}_{HC_3, \mathcal{M}^{G, CY_2}_{\text{inst}}}
    \end{aligned}
  }
\end{equation}
where each $\mathcal{F}^s_{\text{4d-}\sigma, HC_3 \times \R \rightarrow \mathcal{M}^{G, CY_2}_{\text{inst}}}$ can be identified with a \emph{novel} Floer homology class \\
$\text{HHKF}_{d_s} (HC_3, \mathcal{M}^G_{\text{inst}}(CY_2))$, of degree $d_s$, that we shall henceforth name a hyperkähler Floer homology class of the hyperkähler $\mathcal{M}^G_{\text{inst}}(CY_2)$ and specified by the hypercontact three-manifold $HC_3$.

Specifically, the \emph{time-invariant Fueter maps from $HC_3$ to $\mathcal{M}^G_{\text{inst}}(CY_2)$ in $\mathcal{M}(HC_3, \mathcal{M}^{G, CY_2}_{\text{inst}})$} that obey the simultaneous vanishing of the LHS and RHS of the \emph{gradient flow equation}
\begin{equation}
  \label{eq:hc3 x r:sqm:flow}
  \boxed{
    \dv{X^\alpha}{t} =
    - g^{\alpha \beta}_{\mathcal{M}(HC_3, \mathcal{M}^{G, CY_2}_{\text{inst}})}
    \pdv{V_{\sigma}}{X^\beta}
  }
\end{equation}
will generate the chains of the hyperkähler Floer complex with Morse functional
\begin{equation}
  \label{eq:hc3 x r:sqm:morse fn}
  \boxed{
    V_\sigma(X) = \frac{1}{2} \int_{HC_3} \dd[3]{x} \left(
       \sum_a \partial_a (X \wedge \star X) J_a
    \right)
  }
\end{equation}
in $\mathcal{M}(HC_3, \mathcal{M}^{G, CY_2}_{\text{inst}})$, where ``$\wedge$'' and ``$\star$'' are the exterior product and Hodge star operator on $\mathcal{M}^G_{\text{inst}}(CY_2)$, respectively.
The aholomorphic flow lines, described by \emph{time-varying} solutions to~\eqref{eq:hc3 x r:sqm:flow}, are the Floer differentials such that the degree $d_s$ of the corresponding chain in the hyperkähler Floer complex is counted by the outgoing flow lines at each time-invariant Fueter map in $\mathcal{M}(HC_3, \mathcal{M}^{G, CY_2}_{\text{inst}})$.
Such a Fueter map corresponds to a time-independent solution to the 3d equation
\begin{equation}
  \label{eq:hc3 x r :sqm:morse fn:crit pts}
  \boxed{
    \sum_a \partial_a X^i J_a = 0
  }
\end{equation}

\subtitle{A Physical Proof of Hohloch-Noetzel-Salamon's Mathematical Conjecture}

The existence of a Floer homology of hypercontact three-manifolds, whose chain complexes are generated by time-invariant Fueter maps from $HC_3$ to $\mathcal{M}^G_{\text{inst}}(CY_2)$, and Floer differentials that count solutions to the time-varying Cauchy-Riemann-Fueter equation for aholomorphic maps from $HC_3 \times \R$ to $\mathcal{M}^G_{\text{inst}}(CY_2)$, was conjectured by Hohloch-Noetzel-Salamon in~\cite{hohloch-2009-hyper-struc}~\cite[$\S$5]{salamon-2013-three-dimen}, where the term ``hyperkähler Floer homology'' was first coined.
We have therefore furnished a physical proof of their mathematical conjecture.


\section{Symplectic Floer Homologies of Instanton Moduli Spaces}
\label{sec:symp floer-hom}

In this section, we will specialize to several specific (not necessarily closed) $HC_3$.
In particular, we will be specializing to $HC_3$'s that contain combinations of $S^1$ circles, $I$ intervals, and $\R$ lines.
This will allow us to interpret the hyperkähler Floer homologies of $\mathcal{M}^G_{\text{inst}}(CY_2)$ and specified by these $HC_3$'s as symplectic, or symplectic intersection Floer homologies of some loop space, path space, or both, of $\mathcal{M}^G_{\text{inst}}(CY_2)$.

\subsection{A Symplectic Floer Homology in the Triple Loop Space of Instantons on \texorpdfstring{$CY_2$}{CY2}}
\label{sec:symp floer-hom:t3}

Let us look at the closed case of $HC_3 = T^3$.
First, notice that the space $\mathcal{M}(T^3, \mathcal{M}^{G, CY_2}_{\text{inst}})$ of maps from $T^3$ to $\mathcal{M}^G_{\text{inst}}(CY_2)$) can be interpreted as the triple loop space $L^3 \mathcal{M}^{G, CY_2}_{\text{inst}}$ of instantons on $CY_2$.
This means that the 1d SQM in $\mathcal{M}(T^3, \mathcal{M}^{G, CY_2}_{\text{inst}})$ with action~\eqref{eq:hc3 x r:sqm:action:hc3} can also be interpreted as a 1d SQM in $L^3 \mathcal{M}^{G, CY_2}_{\text{inst}}$. Therefore, we can interpret each $\text{HHKF}_{d_s} (T^3, \mathcal{M}^G_{\text{inst}}(CY_2))$ as a Floer homology class in $L^3 \mathcal{M}^{G, CY_2}_{\text{inst}}$, i.e.,
\begin{equation}
  \label{eq:t3 x r:equality to hk floer-hom}
  \boxed{
    \text{HHKF}_{d_s} \left(
      T^3, \mathcal{M}^G_{\text{inst}}(CY_2)
    \right)
    = \text{HSF}^{\text{Fuet}}_{d_s} \left(
      L^3 \mathcal{M}^{G, CY_2}_{\text{inst}}
    \right)
  }
\end{equation}
where each $\text{HSF}^{\text{Fuet}}_{d_s} (L^3 \mathcal{M}^{G, CY_2}_{\text{inst}})$ is a \emph{novel} symplectic Floer homology class, of degree $d_s$, in $L^3 \mathcal{M}^{G, CY_2}_{\text{inst}}$.\footnote{%
  The term ``symplectic'' is used here in the sense that $\mathcal{M}^G_{\text{inst}}(CY_2)$ is symplectic.
  \label{ft:reason for symplectic floer homology classes}
}

Specifically, the \emph{time-invariant Fueter maps from $T^3$ to $\mathcal{M}^G_{\text{inst}}(CY_2)$ in $L^3 \mathcal{M}^{G, CY_2}_{\text{inst}}$} obeying the simultaneous vanishing of LHS and RHS of~\eqref{eq:hc3 x r:sqm:flow} (with $\mathcal{M}(T^3, \mathcal{M}^{G, CY_2}_{\text{inst}}) = L^3 \mathcal{M}^{G, CY_2}_{\text{inst}}$) will generate the chains of the symplectic Floer complex with Morse functional~\eqref{eq:hc3 x r:sqm:morse fn} (with $HC_3 = T^3$).
The aholomorphic flow lines, described by \emph{time-varying} solutions to~\eqref{eq:hc3 x r:sqm:flow}, are the Floer differentials such that the degree $d_s$ of the corresponding chain in the symplectic Floer complex is counted by the outgoing flow lines at each time-invariant Fueter map in $L^3 \mathcal{M}^{G, CY_2}_{\text{inst}}$.
Such a map corresponds to a time-independent solution to~\eqref{eq:hc3 x r :sqm:morse fn:crit pts}.

In other words, a hyperkähler Floer homology of $\mathcal{M}^G_{\text{inst}}(CY_2)$ and specified by $T^3$ can be interpreted as a symplectic Floer homology of $L^3 \mathcal{M}^{G, CY_2}_{\text{inst}}$.

\subsection{A Symplectic Intersection Floer Homology in the Double Loop Space of Instantons on \texorpdfstring{$CY_2$}{CY2}}
\label{sec:symp floer-hom:i x t2}

\subtitle{The 4d Sigma Model as a 2d A-model with Branes $\mathscr{L}_0$ and $\mathscr{L}_1$}

Let us consider the case where one of the $S^1$ circles is an interval $I$ so $HC_3 = I \times T^2$.
We will take $x^3$ to be the coordinate on $I$, and relabel it as $r$ for convenience.
Doing so, the action~\eqref{eq:hc3 x r:4d action:cauchy-riemann-fueter eq} can be expressed as
\begin{equation}
  \label{eq:t2 x i x r:4d sigma:action}
  S_{\text{4d-}\sigma} = \frac{1}{2e^2} \int_{\R \times I} \dd{t} \dd{r} \int_{T^2} \dd[2]{x}
  \left(
    \left|
      \partial_t X^i + (J_3 \partial_r X^i + J_1 \partial_1 X^i + J_2 \partial_2 X^i)
    \right|^2
    + \dots
  \right)
  \, .
\end{equation}
After suitable rescalings, it can be recast as\footnote{%
  Here, we have made use of Stokes' theorem and the fact that the $S^1$ circles of $T^2$ have no boundary to note that $\partial_{\{1, 2\}} X^i$ should vanish in their integration over $T^2$.
  \label{ft:stokes thm for T2 integration}
}
\begin{equation}
  \label{eq:t2 x i x r:a-model:action}
  S_{\text{2d-}\sigma} = \frac{1}{e^2} \int_{\R} \dd{t} \int_I \dd{r}
  \left(
    \left| \partial_t X^c + J_3 \partial_r X^c \right|^2
    + \dots
  \right)
  \, ,
\end{equation}
where $X^c$ and $c$ are coordinates and indices on the double loop space $L^2 \mathcal{M}^{G, CY_2}_{\text{inst}}$ of maps from $T^2$ to $\mathcal{M}^G_{\text{inst}}(CY_2)$.
This is a 2d sigma model on $I \times \R$ with target space $L^2 \mathcal{M}^{G, CY_2}_{\text{inst}}$.

The 2d sigma model that we have obtained above describes open strings with worldsheet $I \times \R$ propagating in $L^2 \mathcal{M}^{G, CY_2}_{\text{inst}}$, starting and ending on branes $\mathscr{L}_0$ and $\mathscr{L}_1$.
To better describe this 2d sigma model, let us complexify the worldsheet, i.e., introduce complex coordinates $w = t + J_3 r$.
Then, its BPS equation, easily read off from~\eqref{eq:t2 x i x r:a-model:action}, can be expressed as
\begin{equation}
  \label{eq:t2 x i x r:a-model:bps}
  \partial_{\bar{w}} X^c = 0 \, .
\end{equation}
These are holomorphic maps from the worldsheet to the target space.
As such, our 2d sigma model is an A-model, and $\mathscr{L}_0$ and $\mathscr{L}_1$ are isotropic-coisotropic A-branes in $L^2 \mathcal{M}^{G, CY_2}_{\text{inst}}$.\footnote{%
  In the case of finite-dimensional symplectic manifolds, isotropic-coisotropic branes are Lagrangian, and thus middle-dimensional~\cite{mcduff-2017-linear-sympl-geomet}.
  \label{ft:isotropic-coistropic branes}
}

In other words, the 4d sigma model on $I \times T^2 \times \R$ with target space $\mathcal{M}^G_{\text{inst}}(CY_2)$ is equivalent to a 2d A-model on $I \times \R$ with target space $L^2 \mathcal{M}^{G, CY_2}_{\text{inst}}$ and branes $\mathscr{L}_0$ and $\mathscr{L}_1$.

\subtitle{The 2d A-model as a 1d SQM in the Interval Space of $L^2 \mathcal{M}^{G, CY_2}_{\text{inst}}$}

From the action of the A-model~\eqref{eq:t2 x i x r:a-model:action}, we can, after suitable rescalings, obtain the equivalent SQM action as
\begin{equation}
  \label{eq:t2 x i x r:a-model:sqm:action}
  S_{\text{SQM},\text{2d-}\sigma} = \frac{1}{e^2} \int_{\R} \dd{t} \left(
    \left|
      \dv{X^\alpha}{t}
      + g^{\alpha\beta}_{I \rightarrow L^2} \pdv{V_{I \rightarrow L^2}}{X^\beta}
    \right|^{2}
    + \dots
  \right)
  \, ,
\end{equation}
where $X^\alpha$ and $(\alpha, \beta)$ are coordinates and indices on the interval space $\mathcal{T}(\mathscr{L}_0, \mathscr{L}_1)_{L^2 \mathcal{M}^{G, CY_2}_{\text{inst}}}$ of smooth trajectories from $\mathscr{L}_0$ to $\mathscr{L}_1$ in $L^2 \mathcal{M}^{G, CY_2}_{\text{inst}}$;
$g_{I \rightarrow L^2}$ is the metric on $\mathcal{T}(\mathscr{L}_0, \mathscr{L}_1)_{L^2 \mathcal{M}^{G, CY_2}_{\text{inst}}}$;
and $V_{I \rightarrow L^2}(X^\alpha)$ is the potential function.
In other words, we equivalently have, from the 2d A-model, an SQM in $\mathcal{T}(\mathscr{L}_0, \mathscr{L}_1)_{L^2 \mathcal{M}^{G, CY_2}_{\text{inst}}}$.

\subtitle{Localizing Onto Constant Intervals in $\mathcal{T}(\mathscr{L}_0, \mathscr{L}_1)_{L^2 \mathcal{M}^{G, CY_2}_{\text{inst}}}$, or Intersection Points of $\mathscr{L}_0$ and $\mathscr{L}_1$}

By the squaring argument~\cite{blau-1993-topol-gauge} applied to~\eqref{eq:t2 x i x r:a-model:sqm:action}, the configurations that the equivalent SQM localizes onto are those that set to zero \emph{simultaneously} the LHS and RHS of the expression within the squared term in~\eqref{eq:t2 x i x r:a-model:action}.
In other words, the equivalent SQM localizes onto time-invariant constant intervals in $\mathcal{T}(\mathscr{L}_0, \mathscr{L}_1)_{L^2 \mathcal{M}^{G, CY_2}_{\text{inst}}}$.
In turn, these correspond to stationary trajectories between branes in $\mathcal{T}(\mathscr{L}_0, \mathscr{L}_1)_{L^2 \mathcal{M}^{G, CY_2}_{\text{inst}}}$, i.e., intersection points of $\mathscr{L}_0$ and $\mathscr{L}_1$.

\subtitle{A Symplectic Intersection Floer Homology of $L^2 \mathcal{M}^{G, CY_2}_{\text{inst}}$}

Since the resulting 2d A-model on $I \times \R$ with target space $L^2 \mathcal{M}^{G, CY_2}_{\text{inst}}$ and action~\eqref{eq:t2 x i x r:a-model:action} can be interpreted as a 1d SQM with action~\eqref{eq:t2 x i x r:a-model:sqm:action}, its partition function can, like in~\eqref{eq:g2 x r:spin7 partition fn:no homology}, be written as
\begin{equation}
  \label{eq:t2 x i x r:a-model:partition fn:no homology}
  \mathcal{Z}_{\text{Spin}(7),I \times T^2 \times \R}(G)
  =
  \sum_s \mathcal{F}^s_{\text{2d-}\sigma, I \times \R \rightarrow L^2 \mathcal{M}^{G, CY_2}_{\text{inst}}}
  \, ,
\end{equation}
where $\mathcal{F}^s_{\text{2d-}\sigma, I \times \R \rightarrow L^2 \mathcal{M}^{G, CY_2}_{\text{inst}}}$, in the $\mathcal{Q}$-cohomology, is the $s^{\text{th}}$ contribution to the partition function, and the summation in `$s$' is over all isolated and non-degenerate intersection points of $\mathscr{L}_0$ and $\mathscr{L}_1$ that the equivalent SQM localizes onto.\footnote{%
  These points can always be made isolated and non-degenerate by adding physically-inconsequential $\mathcal{Q}$-exact terms to the SQM action which will (i) correspond to a deformation to the 2d A-model worldsheet such that the $\mathscr{L}_0$ and $\mathscr{L}_1$ branes can be moved to intersect only at isolated points, and (ii) deform the SQM potential accordingly such that its critical points are non-degenerate.
  \label{ft:isolation and non-degeneracy of const. paths}
}

This then allows us, via a similar analysis to that in~\autoref{sec:floer homology of m7:floer homology} with~\eqref{eq:t2 x i x r:a-model:sqm:action} as the action for the equivalent 1d SQM, to express~\eqref{eq:t2 x i x r:a-model:partition fn:no homology} as
\begin{equation}
  \label{eq:t2 x i x r:a-model:partition fn}
  \boxed{
    \begin{aligned}
      \mathcal{Z}_{\text{Spin}(7),I \times T^2 \times \R}(G)
      &= \sum_s \mathcal{F}^s_{\text{2d-}\sigma, I \times \R \rightarrow L^2 \mathcal{M}^{G, CY_2}_{\text{inst}}}
      \\
      &= \sum_s \text{HSF}^{\text{Int}}_{d_s} \left(
        L^2 \mathcal{M}^{G, CY_2}_{\text{inst}}, \mathscr{L}_0, \mathscr{L}_1
        \right)
        = \mathcal{Z}^{\text{IntSympFloer}}_{\mathscr{L}_0, \mathscr{L}_1,L^2 \mathcal{M}^{G, CY_2}_{\text{inst}}}
    \end{aligned}
  }
\end{equation}
where each $\mathcal{F}^s_{\text{2d-}\sigma, I \times \R \rightarrow L^2 \mathcal{M}^{G, CY_2}_{\text{inst}}}$ can be identified with a \emph{novel} symplectic intersection Floer homology class $\text{HSF}^{\text{Int}}_{d_s}(L^2 \mathcal{M}^{G, CY_2}_{\text{inst}}, \mathscr{L}_0, \mathscr{L}_1)$, of degree $d_s$, in $\mathcal{T}(\mathscr{L}_0, \mathscr{L}_1)_{L^2 \mathcal{M}^{G, CY_2}_{\text{inst}}}$.

Specifically, intersections of the $\mathscr{L}_0$ and $\mathscr{L}_1$ branes in $L^2 \mathcal{M}^{G, CY_2}_{\text{inst}}$ that correspond to the simultaneous vanishing of the LHS and RHS of the squared term in~\eqref{eq:t2 x i x r:a-model:sqm:action}, will generate the chains of the symplectic intersection Floer complex in $\mathcal{T}(\mathscr{L}_0, \mathscr{L}_1)_{L^2 \mathcal{M}^{G, CY_2}_{\text{inst}}}$.
The holomorphic flow lines, described by \emph{time-varying} solutions to~\eqref{eq:t2 x i x r:a-model:bps}, are the Floer differentials such that the degree $d_s$ of the corresponding chain in the symplectic intersection Floer complex is counted by the outgoing flow lines at each intersection of $\mathscr{L}_0$ and $\mathscr{L}_1$ in $L^2 \mathcal{M}^{G, CY_2}_{\text{inst}}$.

\newpage

\subtitle{Hyperkähler Floer Homology of $\mathcal{M}^G_{\text{inst}}(CY_2)$ and Specified by $I \times T^2$ as a Symplectic Intersection Floer Homology of $L^2 \mathcal{M}^{G, CY_2}_{\text{inst}}$}

Recall that this 2d A-model is equivalent to the 4d sigma model on $I \times T^2 \times \R$ with target space $\mathcal{M}^{G, CY_2}_{\text{inst}}$.
We know from~\autoref{sec:hyperkahler floer-hom:floer-homs} that the latter will physically realize a hyperkähler Floer homology of $\mathcal{M}^G_{\text{inst}}(CY_2)$ and specified by $I \times T^2$.
This therefore allows us to interpret each $\text{HSF}^{\text{Int}}_{d_s}(L^2 \mathcal{M}^{G, CY_2}_{\text{inst}}, \mathscr{L}_0, \mathscr{L}_1)$ as a hyperkähler Floer homology class, i.e.,
\begin{equation}
  \label{eq:t2 x i x r:equality to hk floer-hom of i x t2}
  \boxed{
    \text{HHKF}_{d_s} \left(
      I \times T^2, \mathcal{M}^G_{\text{inst}}(CY_2)
    \right)
    = \text{HSF}^{\text{Int}}_{d_s} \left(
      L^2 \mathcal{M}^{G, CY_2}_{\text{inst}}, \mathscr{L}_0, \mathscr{L}_1
    \right)
  }
\end{equation}
where each $\text{HHKF}_{d_s} (I \times T^2, \mathcal{M}^G_{\text{inst}}(CY_2))$ is a hyperkähler Floer homology class, of degree $d_s$, of the hyperkähler $\mathcal{M}^G_{\text{inst}}(CY_2)$ and specified by the hypercontact three-manifold $I \times T^2$.

In other words, a hyperkähler Floer homology of $\mathcal{M}^G_{\text{inst}}(CY_2)$ and specified by $I \times T^2$ can be interpreted as a symplectic intersection Floer homology of $L^2 \mathcal{M}^{G, CY_2}_{\text{inst}}$.

\subsection{A Symplectic Intersection Floer Homology in the Path Space of Loops of Instantons on \texorpdfstring{$CY_2$}{CY2}}
\label{sec:symp floer-hom:i x s x r}

\subtitle{The 4d Sigma Model as a $\theta$-generalized 2d A-model with $\theta$-generalized Branes $\mathcal{P}_0(\theta)$ and $\mathcal{P}_1(\theta)$}

Let us consider the case where an $S^1$ circle of $T^2$ is replaced with an $\R$ line, so $HC_3 = I \times S^1 \times \R$.
We will take $x^1$ to be the coordinate on $\R$, and relabel it as $\tau$ for convenience.
Doing so, the action~\eqref{eq:t2 x i x r:4d sigma:action} can be expressed as
\begin{equation}
  \label{eq:i x s x r2:4d sigma:action:unrotated}
  S_{\text{4d-}\sigma} = \frac{1}{2e^2} \int_{\R \times I} \dd{\tau} \dd{r} \int_{\R \times S^1} \dd{t} \dd{x^2}
  \left(
    \left|
      \partial_{\tau} X^i - (J_1 \partial_t X^i - J_2 \partial_r X^i + J_3 \partial_2 X^i)
    \right|^2
    + \dots
  \right)
  \, .
\end{equation}
Notice that we can make use of the rotational symmetry of $\R^2$ to rotate the 4d sigma model about the $I$ interval.
This can be achieved by performing a quaternionic conjugation with respect to the $J_1$ complex structure by an angle $\theta$.\footnote{%
  Note that the complex structure of the worldvolume is uniquely determined by the complex structure of the target space.
  Therefore, performing a quaternionic conjugation with respect to the $J_1$ complex structure of the target space is equivalent to performing a quaternionic conjugation with respect to the $I_1$ complex structure of the $\R^2$ plane in the worldvolume.
  \label{ft:reason for quaternionic conjugation}
}
Doing so, the action~\eqref{eq:i x s x r2:4d sigma:action:unrotated} becomes
\begin{equation}
  \label{eq:i x s x r2:4d sigma:action}
  S_{\text{4d-}\sigma} = \frac{1}{2e^2} \int_{\R \times I} \dd{\tau} \dd{r} \int_{\R \times S^1} \dd{t} \dd{x^2}
  \left(
    \left|
      \partial_{\tau} X^i - (J^{\theta}_1 \partial_t X^i - J^{\theta}_2 \partial_r X^i + J^{\theta}_3 \partial_2 X^i)
    \right|^2
    + \dots
  \right)
  \, ,
\end{equation}
where $J^{\theta}_a = e^{-J_1 \theta/2} J_a e^{J_1 \theta/2}$ is a $\theta$-rotated complex structure with respect to $J_1$.
The BPS equation of the $\theta$-rotated theory, easily read off from~\eqref{eq:i x s x r2:4d sigma:action}, can be understood to be a $\theta$-deformed Cauchy-Riemann-Fueter equation of the aholomorphic maps $X^i: I \times S^1 \times \R^2 \rightarrow \mathcal{M}^{G, \theta}_{\text{inst}}(CY_2)$, where the target space also picks up the effect of the rotation.

After suitable rescalings, the $\theta$-rotated 4d sigma model can be recast as\footnote{%
  Just as in~\autoref{ft:stokes thm for T2 integration}, we make use of Stokes' theorem and the fact that $S^1 \times \R$ have no boundary to note that $\partial_{\{t, 2\}}X^i$ should vanish in their integration over $S^1 \times \R$.
  \label{ft:stokes thm for s x r integration}
}
\begin{equation}
  \label{eq:i x s x r2:a-model:action}
  S_{\text{2d-}\sigma} = \frac{1}{e^2} \int_{\R \times I} \dd{\tau} \dd{r}
  \left(
    \left|
      \partial_{\tau} X^e  + J^{\theta}_2 \partial_r X^e
    \right|^2
    + \dots
  \right)
  \, ,
\end{equation}
where $X^e$ and $e$ are coordinates and indices on the $\theta$-rotated path space $\mathcal{M}(\R, L \mathcal{M}^{G, \theta, CY_2}_{\text{inst}})$ of maps from $\R$ to the loop space $L \mathcal{M}^{G, \theta, CY_2}_{\text{inst}}$ of maps from $S^1$ to $\mathcal{M}^{G, \theta}_{\text{inst}}(CY_2)$.
This is a 2d sigma model on $I \times \R$ with target space $\mathcal{M}(\R, L \mathcal{M}^{G, \theta, CY_2}_{\text{inst}})$.

At $\theta = 0$, $\pi/2$, and $\pi$, this 2d sigma model is a 2d A-model on $I \times \R$ with target space $\mathcal{M}(\R, L \mathcal{M}^{G, CY_2}_{\text{inst}})$, describing open strings on the worldsheet propagating in $\mathcal{M}(\R, L \mathcal{M}^{G, CY_2}_{\text{inst}})$.
In general, as $\theta$ interpolates between $0$, $\pi/2$, and $\pi$, we can understand the 2d sigma model to be a $\theta$-generalized 2d A-model with $\theta$-generalized branes $\mathcal{P}_0(\theta)$ and $\mathcal{P}_1(\theta)$, which are regular isotropic-coisotropic A-branes when $\theta = 0$, $\pi/2$, and $\pi$.
Such a $\theta$-generalized A-model with $\theta$-generalized branes was also observed in a similar setting in~\cite[$\S$6.6]{er-2023-topol-n}, where it was given the name ``A$_{\theta}$-model'' with ``A$_{\theta}$-branes''.

\subtitle{A Symplectic Intersection Floer Homology of $\mathcal{M}(\R, L \mathcal{M}^{G, \theta, CY_2}_{\text{inst}})$}

We will next follow the analysis of~\autoref{sec:symp floer-hom:i x t2}, with~\eqref{eq:i x s x r2:a-model:action} as the action for the equivalent 2d sigma model.
Doing so, we will find that the 2d A$_{\theta}$-model can then be further recast as a 1d SQM in the interval space $\mathcal{T}(\mathcal{P}_0, \mathcal{P}_1)_{\mathcal{M}(\R, L \mathcal{M}^{G, \theta, CY_2}_{\text{inst}})}$ of smooth trajectories from $\mathcal{P}_0(\theta)$ to $\mathcal{P}_1(\theta)$ in $\mathcal{M}(\R, L \mathcal{M}^{G, \theta, CY_2}_{\text{inst}})$.
Thus, its partition function can therefore be expressed as\footnote{%
  In the final expression, the `$\theta$' label is omitted as the physical theory is actually equivalent for all values of $\theta$.
  \label{ft:theta-omission:symp floer hom}
}
\begin{equation}
  \label{eq:i x s x r2:a-model:partition fn}
  \boxed{
    \begin{aligned}
      \mathcal{Z}_{\text{Spin}(7), I \times S^1 \times \R^2}(G)
      &= \sum_s \mathcal{F}^s_{\text{2d-}\sigma, I \times \R \rightarrow \mathcal{M}(\R, L\mathcal{M}^{G, \theta, CY_2}_{\text{inst}})}
      \\
      &= \sum_s \text{HSF}^{\text{Int}}_{d_s} \left(
        \mathcal{M} \left( \R, L \mathcal{M}^{G, \theta, CY_2}_{\text{inst}} \right), \mathcal{P}_0, \mathcal{P}_1
        \right)
        = \mathcal{Z}^{\text{IntSympFloer}}_{\mathcal{P}_0, \mathcal{P}_1, \mathcal{M}(\R, L \mathcal{M}^{G, CY_2}_{\text{inst}})}
    \end{aligned}
  }
\end{equation}
where each $\mathcal{F}^s_{\text{2d-}\sigma, I \times \R \rightarrow \mathcal{M}(\R, L\mathcal{M}^{G, \theta, CY_2}_{\text{inst}})}$ can be identified with a \emph{novel} symplectic intersection Floer homology class $\text{HSF}^{\text{Int}}_{d_s} (\mathcal{M}(\R, L \mathcal{M}^{G, \theta, CY_2}_{\text{inst}}), \mathcal{P}_0, \mathcal{P}_1)$, of degree $d_s$, in $\mathcal{T}(\mathcal{P}_0, \mathcal{P}_1)_{\mathcal{M}(\R, L \mathcal{M}^{G, \theta, CY_2}_{\text{inst}})}$.

Specifically, intersections of the $\mathcal{P}_0(\theta)$ and $\mathcal{P}_1(\theta)$ branes in $\mathcal{M}(\R, L \mathcal{M}^{G, \theta, CY_2}_{\text{inst}})$ that correspond to the simultaneous vanishing of the LHS and RHS of the squared term in~\eqref{eq:i x s x r2:a-model:action}, will generate the chains of the symplectic intersection Floer complex in $\mathcal{T}(\mathcal{P}_0, \mathcal{P}_1)_{\mathcal{M}(\R, L \mathcal{M}^{G, \theta, CY_2}_{\text{inst}})}$.
The holomorphic flow lines, described by \emph{$\tau$-varying} solutions to the expression within the squared term in~\eqref{eq:i x s x r2:a-model:action}, are the Floer differentials such that the degree $d_s$ of the corresponding chain in the symplectic intersection Floer complex is counted by the outgoing flow lines at each intersection of $\mathcal{P}_0(\theta)$ and $\mathcal{P}_1(\theta)$ in $\mathcal{M}(\R, L \mathcal{M}^{G, \theta, CY_2}_{\text{inst}})$.

\newpage

\subtitle{Hyperkähler Floer Homology of $\mathcal{M}^{G, \theta}_{\text{inst}}(CY_2)$ and Specified by $I \times S^1 \times \R$ as a Symplectic Intersection Floer Homology of $\mathcal{M}(\R, L \mathcal{M}^{G, \theta, CY_2}_{\text{inst}})$}

Recall again that this 2d A$_{\theta}$-model is equivalent to the 4d sigma model on $I \times S^1 \times \R^2$ with target space $\mathcal{M}^{G, \theta}_{\text{inst}}(CY_2)$.
We know, again, from~\autoref{sec:hyperkahler floer-hom:floer-homs} that the latter will physically realize a hyperkähler Floer homology of $\mathcal{M}^{G, \theta}_{\text{inst}}(CY_2)$ and specified by $I \times S^1 \times \R$.
This therefore allows us to interpret each $\text{HSF}^{\text{Int}}_{d_s} (\mathcal{M}(\R, L \mathcal{M}^{G, \theta, CY_2}_{\text{inst}}), \mathcal{P}_0, \mathcal{P}_1)$ as a hyperkähler Floer homology class, i.e.,
\begin{equation}
  \label{eq:i x s x r2:equality to hk floer-hom of i x s x r}
  \boxed{
    \text{HHKF}_{d_s} \left(
      I \times S^1 \times \R, \mathcal{M}^{G, \theta}_{\text{inst}}(CY_2)
    \right)
    =
    \text{HSF}^{\text{Int}}_{d_s} \left(
      \mathcal{M} \left( \R, L \mathcal{M}^{G, \theta, CY_2}_{\text{inst}} \right), \mathcal{P}_0, \mathcal{P}_1
    \right)
  }
\end{equation}
where each $\text{HHKF}_{d_s} ( I \times S^1 \times \R, \mathcal{M}^{G, \theta}_{\text{inst}}(CY_2) )$ is a hyperkähler Floer homology class, of degree $d_s$, of the hyperkähler $\mathcal{M}^{G, \theta}_{\text{inst}}(CY_2)$ and specified by the hypercontact three-manifold $I \times S^1 \times \R$.

In other words, a hyperkähler Floer homology of $\mathcal{M}^{G, \theta}_{\text{inst}}(CY_2)$ and specified by $I \times S^1 \times \R$ can be interpreted as a symplectic intersection Floer homology of $\mathcal{M}(\R, L \mathcal{M}^{G, \theta, CY_2}_{\text{inst}})$.


\section{Atiyah-Floer Type Dualities}
\label{sec:atiyah-floer}

In this section, we will consider Spin$(7)$ theory on $CY_3 \times M_1 \times \R$, and perform a Tyurin degeneration of $CY_3$ along a $CY_2$ surface~\cite{doran-2017-mirror-symmet}.
Then, by exploiting the topological invariance of the underlying Spin$(7)$ theory under a shrinking of $CY_2$ and an $S^1$ circle, we will find that the results of the previous sections on Floer homologies naturally lead to the derivation of novel Atiyah-Floer type dualities.
In particular, we will derive a (i) Spin$(7)$ Atiyah-Floer type duality of $CY_3 \times S^1$, and (ii) 7d-Spin$(7)$ Atiyah-Floer type duality of $CY_3$.

\subsection{Splitting \texorpdfstring{Spin$(7)$}{Spin(7)} Theory on \texorpdfstring{$CY_3 \times M_1 \times \R$}{CY3 x M1 x R}}
\label{sec:atiyah-floer:tyurin-degen}


\subtitle{A Tyurin Degeneration of $CY_3$}

\begin{figure}
  \centering
  \begin{subfigure}{0.45\textwidth}
    \centering
    \begin{tikzpicture}[auto]
      \draw[blue, pattern={Lines[angle=45,distance=4pt]}, pattern color=blue, thick]
      (0,0) ellipse (0.4 and 1);
      \draw
      (0,1) arc(90:270:2 and 1);
      \draw[fill=white,draw=none]
      (0,0) circle[radius=0.3] node [scale=1.5] (sigma-left) {$\mathcal{S}$};
      \node [left of=sigma-left,scale=1.5] {$CY_3'$};
      \draw[blue, thick]
      (1,1) arc(90:270:0.4 and 1);
      \draw
      (1,-1) arc(270:450:2 and 1);%
      \node [right=0.5cm of sigma-left,scale=1.5] {$CY_3''$};
      \node [below=1.3cm of sigma-left] {};
    \end{tikzpicture}
    \caption{$CY_3$ as a connected sum of Calabi-Yau threefolds $CY_3'$ and $CY_3''$ along the surface $\mathcal{S} = CY_2$.}
    \label{fig:tyurin degenerations of cy3}
  \end{subfigure}
  \hfill
  \begin{subfigure}{0.45\textwidth}
    \centering
    \begin{tikzpicture}[auto]
      \def \eliA {2.5} 
      \def \eliB {0.4} 
      \def \eliAngle {15} 
      \def \faceLength {1.2} 
      \def \sepLength {0.8} 
      \def \nodeScale {1.3} 
      \def \leftBlockText {$\text{Spin}(7)'$} 
      \def \rightBlockText {$\text{Spin}(7)''$} 
      \def \rightBlockLift {1cm} 
      \draw ({\eliA*sin(\eliAngle) - \sepLength}, {\eliB*cos(\eliAngle) + \faceLength})
      arc({90 - \eliAngle}:{270 - \eliAngle}:{\eliA} and {\eliB});
      \draw[blue, pattern={Lines[angle=45,distance=4pt]},pattern color=blue, thick]
      ({\eliA*sin(\eliAngle) - \sepLength}, {\eliB*cos(\eliAngle) + \faceLength})
      -- ({-\eliA*sin(\eliAngle) - \sepLength}, {-\eliB*cos(\eliAngle) + \faceLength})
      -- ({-\eliA*sin(\eliAngle) - \sepLength}, {-\eliB*cos(\eliAngle) - \faceLength})
      -- ({\eliA*sin(\eliAngle) - \sepLength}, {\eliB*cos(\eliAngle) - \faceLength})
      -- ({\eliA*sin(\eliAngle) - \sepLength}, {\eliB*cos(\eliAngle) + \faceLength});
      \draw[fill=white,draw=none] ({-\sepLength},0) circle[radius=0.3] node [scale=\nodeScale] (sigma-left) {$\mathscr{B}$};
      \draw ({-\eliA - \sepLength}, {\faceLength})
      -- ({-\eliA - \sepLength}, {-\faceLength});
      \draw ({-\eliA - \sepLength}, {-\faceLength})
      arc(180:{270 - \eliAngle}:{\eliA} and {\eliB});
      \node [scale=\nodeScale] at ({-\eliA*(1 + sin(\eliAngle))/2 - \sepLength}, 0)
      (left-block) {\leftBlockText};
      \draw ({\eliA*sin(\eliAngle) + \sepLength}, {\eliB*cos(\eliAngle) + \faceLength})
      arc({360 + 90 - \eliAngle}:{270 - \eliAngle}:{\eliA} and {\eliB});
      \draw[blue, thick] ({\eliA*sin(\eliAngle) + \sepLength}, {\eliB*cos(\eliAngle) + \faceLength})
      -- ({-\eliA*sin(\eliAngle) + \sepLength}, {-\eliB*cos(\eliAngle) + \faceLength})
      -- ({-\eliA*sin(\eliAngle) + \sepLength}, {-\eliB*cos(\eliAngle) - \faceLength});
      \draw ({\eliA + \sepLength}, {\faceLength})
      -- ({\eliA + \sepLength}, {-\faceLength});
      \draw ({-\eliA*sin(\eliAngle) + \sepLength}, {-\eliB*cos(\eliAngle) - \faceLength}) arc({270 - \eliAngle}:360:{\eliA} and {\eliB});
      \node [scale=\nodeScale] at ({\eliA*(1 - sin(\eliAngle))/2 + \sepLength}, 0)
      (right-block) {\rightBlockText};
      \node [below=\rightBlockLift of right-block] {};
    \end{tikzpicture}
    \caption{Spin$(7)$-manifold splits into Spin$(7)'$ and Spin$(7)''$ along their common boundary $\mathscr{B}$.}
    \label{fig:tyurin degenerations of spin7}
  \end{subfigure}
  \caption[]{Tyurin degeneration of $CY_3$ and Spin$(7)$-manifolds.\footnotemark}
  \label{fig:tyrin degenerations}
\end{figure}
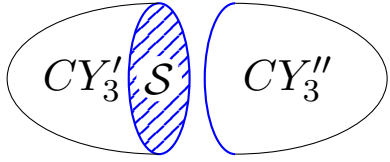
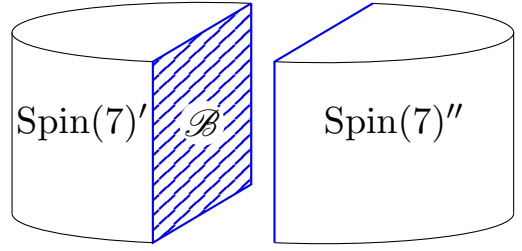

Let us begin with Spin$(7)$ theory on $CY_3 \times M_1 \times \R$, where $M_1$ is a one-manifold.
The ``splitting'' will be done on the $CY_3$ submanifold along a $CY_2$ surface.\footnote{%
  The $CY_2$ surface that we are considering in this section will exclude trivial $CY_2 = T^4$ surfaces.
  \label{ft:restriction of cy2 to k3 surfaces in atiyah-floer}
}
This is accomplished by performing a Tyurin degeneration of $CY_3$ along a $CY_2$ surface, i.e., $CY_3 = CY_3' \bigcup_{CY_2} CY_3''$ as shown in~\autoref{fig:tyurin degenerations of cy3}, whence we can view $CY_3'$ and $CY_3''$ as non-trivial fibrations of $CY_2$ over a disk $D$, where $CY_2$ goes to zero size at the boundary of the disk.
The metric on $CY_3'$ and $CY_3''$ can then be written as
\begin{equation}
  \label{eq:atiyah-floer:fibred cy3 metric}
  \dd[2]{\mathsf{s}}_{CY_3} = (\dd{r})^2 + r^2 (\dd{\vartheta})^2 + f(r, \vartheta) (g_{CY_2})_{MN} \dd{x^M} \dd{x^N}
  \, ,
\end{equation}
where $x^M$ and $(M, N)$ are coordinates and indices on $CY_2$;
$r$ and $\vartheta$ are, respectively, the radius and angle on $D$, which, in turn, can be viewed as a non-trivial $S^1$ fibration of an interval $I$ that is closed at one end;
and $f(r, \vartheta)$ is a scalar function of $r$ and $\vartheta$.
\footnotetext{%
  These figures are higher-dimensional generalizations of~\cite[Fig. 2]{er-2023-topol-n}.
  \label{ft:tyurin degen figures}
}%

\subtitle{Splitting the Spin$(7)$-manifold}

From such a splitting of $CY_3$, this means that the Spin$(7)$-manifold is split as $\text{Spin}(7) = \text{Spin}(7)' \bigcup_{\mathscr{B}} \text{Spin}(7)''$, where $\text{Spin}(7)' = CY_3' \times M_1 \times \R$, $\text{Spin}(7)'' = CY_3'' \times M_1 \times \R$, and $\mathscr{B}$ is their common boundary.
This is illustrated in~\autoref{fig:tyurin degenerations of spin7}.

We can now exploit the topological invariance of Spin$(7)$ theory to freely perform a Weyl rescaling of the corresponding Tyurin-degenerated metrics on $CY_3'$ and $CY_3''$, such that the metric on Spin$(7)'$ and Spin$(7)''$ can be expressed as
\begin{equation}
  \label{eq:atiyah-floer:fibred spin7 metric}
  \dd{\mathsf{s}}^2_{\mathrm{Spin}(7)}
  = \frac{1}{f(r, \vartheta)} \left[
    (\dd{t})^2 + (\dd{s})^2 + (\dd{r})^2 + r^2 (\dd{\vartheta})^2
  \right]
  + (g_{CY_2})_{MN} \dd{x}^M \dd{x}^N
  \, ,
\end{equation}
where $t$ and $s$ are the coordinates on $\R$ and $M_1$, respectively.
The prefactor of $f(r, \vartheta)^{-1}$ is effectively a scaling factor on $D'\times M_1 \times \R$ and $D''\times M_1 \times \R$, whence their topologies are left unchanged.
Since the theory is topological, we can replace the discs $D'$ and $D''$ with their topological equivalents $I' \times S^1$ and $I'' \times S^1$.
Thus, we can regard the Spin$(7)$-manifold as $\text{Spin}(7) = \text{Spin}(7)' \bigcup_{\mathscr{B}} \text{Spin}(7)''$, where $\text{Spin}(7)' = CY_2 \times I' \times S^1 \times M_1 \times \R$, $\text{Spin}(7)'' = CY_2 \times I'' \times S^1 \times M_1 \times \R$, and $\mathscr{B} = CY_2 \times S^1 \times M_1 \times \R$ is their common boundary.

Hence, Spin$(7)$ theory on $CY_3 \times M_1 \times \R$ can be regarded as a union of two Spin$(7)$ theories, one on $CY_2 \times I' \times S^1 \times M_1 \times \R$ and another on $CY_2 \times I'' \times S^1 \times M_1 \times \R$, along their common boundary $\mathscr{B} = CY_2 \times S^1 \times M_1 \times \R$.


\subsection{A \texorpdfstring{\texorpdfstring{Spin$(7)$}{Spin(7)}}{Spin(7)} Atiyah-Floer Type Duality of \texorpdfstring{$CY_3 \times S^1$}{CY3 x S1}}
\label{sec:atiyah-floer:spin7}

Let us now take $M_1 = S^1$.
From~\autoref{sec:symp floer-hom:i x t2}, we know that Spin$(7)$ theory on $\text{Spin}(7)' = CY_2 \times I' \times T^2 \times \R$ and $\text{Spin}(7)'' = CY_2 \times I'' \times T^2 \times \R$, when topologically reduced along $CY_2$, can be interpreted as a 4d sigma model on $I' \times T^2 \times \R$ and $I'' \times T^2 \times \R$ with target space $\mathcal{M}^G_{\text{inst}}(CY_2)$ and action~\eqref{eq:t2 x i x r:4d sigma:action}.
In turn, these 4d sigma models can be interpreted as 2d sigma models on $I' \times \R$ and $I'' \times \R$ with action~\eqref{eq:t2 x i x r:a-model:action}.
That is to say, Spin$(7)$ theory on $CY_2 \times I' \times T^2 \times \R$ and $CY_2 \times I'' \times T^2 \times \R$, when topologically reduced along $CY_2$, can also be interpreted as 2d A-models on $I' \times \R$ and $I'' \times \R$, with branes $(\mathscr{L}_0, \mathscr{L}_{1/2})$ and $(\mathscr{L}_{1/2}, \mathscr{L}_1)$ in $L^2 \mathcal{M}^{G, CY_2}_{\text{inst}}$, respectively.

\subtitle{Spin$(7)$ Theory on $CY_3 \times S^1 \times \R$ as a 2d A-model on $I \times \R$ with Branes $(\mathscr{L}_0, \mathscr{L}_1)$}

\begin{figure}
  \centering
  \begin{tikzpicture}[%
    auto,%
    shorten >=-2pt,%
    shorten <=-2pt,%
    box/.style={rectangle, text centered, minimum height=10em,text width=8mm,draw,fill=gray!40},%
    shortwave/.style={*-*,thick,decorate,decoration={snake,amplitude=5pt,segment length=1.3cm}},%
    longwave/.style={*-*,thick,decorate,decoration={snake,amplitude=5pt,segment length=1.7cm}},%
    ]
    \node [box] (ori-l0) {$\mathscr{L}_0$};
    \node [box, right=1.5cm of ori-l0] (ori-l1-prime) {$\mathscr{L}_{1/2}$};
    \node [box, right=0cm of ori-l1-prime] (ori-l0-prime) {$\mathscr{L}_{1/2}$};
    \node [box, right=1.5cm of ori-l0-prime] (ori-l1) {$\mathscr{L}_1$};
    \path (ori-l1-prime) -- (ori-l0-prime) coordinate[midway] (R-aux);
    \node [below=5em of R-aux] (R) {$\mathscr{R}$};
    \draw [shortwave] (ori-l0) -- node[below=10pt] {$\R \times I'$} (ori-l1-prime);
    \draw [shortwave] (ori-l0-prime) -- node[below=10pt] {$\R \times I''$} (ori-l1);
    \node [box, right=3cm of ori-l1] (fin-l0) {$\mathscr{L}_0$};
    \node [box, right=2cm of fin-l0] (fin-l1) {$\mathscr{L}_1$};
    \draw [longwave] (fin-l0) -- node[below=10pt] {$\R \times I$} (fin-l1);
    \draw [-{Latex[length=3mm]}, thick, shorten >=10pt, shorten <=10pt] (ori-l1) -- (fin-l0);
  \end{tikzpicture}
  \caption[]{Union of 2d A-models along their common boundary $\mathscr{R}$.}
  \label{fig:gluing 2d sigma models}
\end{figure}
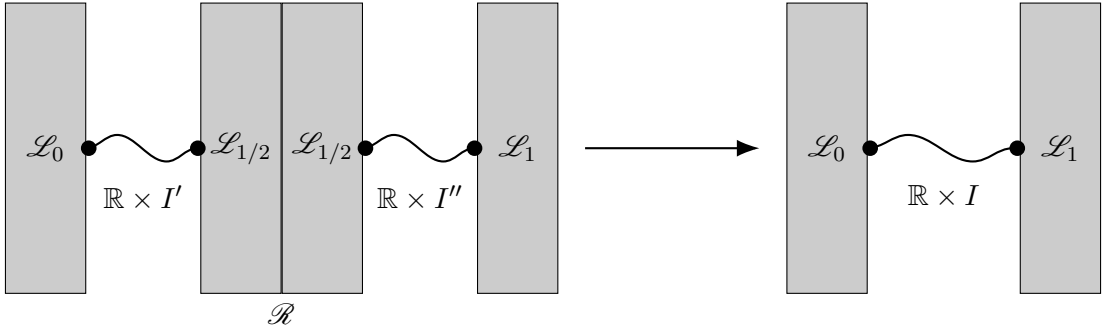

Therefore, the union of a Spin$(7)$ theory on a Spin$(7)'$-manifold and another on a Spin$(7)''$-manifold along their common boundary $\mathscr{B}$, to get a Spin$(7)$ theory on $CY_3 \times S^1 \times \R$, can be interpreted as the union of a 2d A-model on $I' \times \R$ with branes $(\mathscr{L}_0, \mathscr{L}_{1/2})$ and another on $I'' \times \R$ with branes $(\mathscr{L}_{1/2}, \mathscr{L}_1)$, to get a 2d A-model on $I \times \R$ with branes $(\mathscr{L}_0, \mathscr{L}_1)$.
This union of the 2d A-models is illustrated in~\autoref{fig:gluing 2d sigma models}.

In other words, Spin$(7)$ theory on $CY_3 \times S^1 \times \R$ can be interpreted as a 2d A-model on $I \times \R$ with branes $(\mathscr{L}_0, \mathscr{L}_1)$.
This means that we can equate their partition functions in~\eqref{eq:g2 x r:partition fn} and~\eqref{eq:t2 x i x r:a-model:partition fn}, respectively, to get
\begin{equation}
  \label{eq:atiyah-floer:spin7:equality of partition fn}
  \sum_j \text{HF}^{\text{Spin}(7)\text{-inst}}_{d_j}(CY_3 \times S^1, G)
  = \sum_s \text{HSF}^{\text{Int}}_{d_s} \left(
    L^2 \mathcal{M}^{G, CY_2}_{\text{inst}}, \mathscr{L}_0, \mathscr{L}_1
  \right)
  \, .
\end{equation}

\subtitle{A Spin$(7)$ Atiyah-Floer Type Duality of $CY_3 \times S^1$}

Let us now ascertain if there is a one-to-one correspondence between $(j, d_j)$ and $(s, d_s)$, which will in turn imply that there is a degree-to-degree isomorphism between the Spin$(7)$ instanton Floer homology of $CY_3 \times S^1$ and the symplectic intersection Floer homology of $L^2 \mathcal{M}^{G, CY_2}_{\text{inst}}$.

To ascertain if there is a one-to-one correspondence between `$j$' and `$s$', first, note that each `$j$' refers to a time-invariant critical point of $V_7$ in $\mathfrak{A}_7$, corresponding to a time-invariant solution to the 8d BPS equation on $CY_3 \times S^1 \times \R$ in~\eqref{eq:g2 x r:flow}.
Second, note that each `$s$' refers to a time-invariant critical point of $V_{I \rightarrow L^2}$ in $\mathcal{T}(\mathscr{L}_0, \mathscr{L}_1)_{L^2 \mathcal{M}^{G, CY_2}_{\text{inst}}}$, corresponding to a time-invariant solution to the 4d BPS equation on $I \times T^2 \times \R$ given by setting the LHS and RHS of the expression within the squared term in~\eqref{eq:t2 x i x r:4d sigma:action} simultaneously to zero.
Third, note that the 4d BPS equation is a direct topological reduction of the 8d BPS equation along $CY_2 \subset CY_3$, whence there is a one-to-one correspondence between the time-invariant solutions of the former and the latter.
Altogether, this means that there is a one-to-one correspondence between `$j$' and `$s$'.

To ascertain if there is a one-to-one correspondence between `$d_j$' and `$d_s$', first, note that the flow lines between time-invariant critical points of $V_7$ in $\mathfrak{A}_7$ realizing the Floer differential of $\text{HF}^{\text{Spin(7)}\text{-inst}}_{*}$, which counts `$d_j$', correspond to time-varying solutions to the gradient flow equations~\eqref{eq:g2 x r:sqm:flow}.
This, in turn, corresponds to time-varying solutions to the 8d BPS equation on $CY_3 \times S^1 \times \R$ in~\eqref{eq:g2 x r:flow}.
Second, note that the flow lines between time-invariant critical points of $V_{I \rightarrow L^2}$ in $\mathcal{T}(\mathscr{L}_0, \mathscr{L}_1)_{L^2 \mathcal{M}^{G, CY_2}_{\text{inst}}}$ realizing the Floer differentials of $\text{HSF}^{\text{Int}}_{*}$, which counts `$d_s$', correspond to time-varying solutions to the gradient flow equations defined by setting to zero the expression within the squared term in~\eqref{eq:t2 x i x r:a-model:sqm:action}.
This, in turn, corresponds to time-varying solutions to the 4d BPS equations on $I \times T^2 \times \R$ defined by setting to zero the expression within the squared term in~\eqref{eq:t2 x i x r:4d sigma:action}.
Third, note again that the 4d BPS equations are a direct topological reduction of the 8d BPS equation along $CY_2 \subset CY_3$, whence there is a one-to-one correspondence between solutions of the former and the latter.
Altogether, this means that there is a one-to-one correspondence between `$d_j$' and `$d_s$'.

In other words, we do indeed have a one-to-one correspondence between $(j, d_j)$ and $(s, d_s)$ in~\eqref{eq:atiyah-floer:spin7:equality of partition fn}, whence we would have the following degree-to-degree isomorphism between the Spin$(7)$ instanton Floer homology classes of $CY_3 \times S^1$ and the symplectic intersection Floer homology classes of $L^2 \mathcal{M}^{G, CY_2}_{\text{inst}}$
\begin{equation}
  \label{eq:atiyah-floer:spin7}
  \boxed{
    \text{HF}^{\text{Spin}(7)\text{-inst}}_* (CY_3 \times S^1, G)
    \cong
    \text{HSF}^{\text{Int}}_* \left(
      L^2 \mathcal{M}^{G, CY_2}_{\text{inst}}, \mathscr{L}_0, \mathscr{L}_1
    \right)
  }
\end{equation}

We thus have an Atiyah-Floer type duality, that we shall henceforth name a Spin$(7)$ Atiyah-Floer type duality of $CY_3 \times S^1$!

\subsection{A \texorpdfstring{7d-Spin$(7)$}{7d-Spin(7)} Atiyah-Floer Type Duality of \texorpdfstring{$CY_3$}{CY3}}
\label{sec:atiyah-floer:7d-spin7}

Let us now perform a KK reduction of Spin$(7)$ theory on $CY_3 \times S^1 \times \R$ by shrinking the $S^1$ circle to be infinitesimally small.
We know that, according to~\autoref{sec:floer homology of m6}, the LHS of~\eqref{eq:atiyah-floer:spin7:equality of partition fn} becomes~\eqref{eq:cy3 x r:partition fn}, i.e.,
\begin{equation}
  \label{eq:atiyah-floer:7d-spin7:kk reduction:gauge}
  \sum_j \text{HF}^{\text{Spin}(7)\text{-inst}}_{d_j} (CY_3 \times S^1, G)
  \xrightarrow{S^1 \rightarrow 0}
  \sum_k \text{HHF}^{G_2\text{-M}}_{d_k} (CY_3, G)
  \, ,
\end{equation}
whilst the RHS of~\eqref{eq:atiyah-floer:spin7:equality of partition fn} will simply become\footnote{%
  This can be seen by performing a straightforward KK reduction along an $S^1$ circle of the 4d sigma model on $I \times T^2 \times \R$ from~\autoref{sec:symp floer-hom:i x t2}.
  The resulting 3d sigma model on $I \times S^1 \times \R$ can be recast as a 2d A-model on $I \times \R$, this time with target space $L \mathcal{M}^{G, CY_2}_{\text{inst}}$.
  Following the analysis of~\autoref{sec:symp floer-hom:i x t2}, we will physically realize a symplectic intersection Floer homology of $L \mathcal{M}^{G, CY_2}_{\text{inst}}$ generated by intersections of isotropic-coisotropic A-branes $\mathcal{L}_*$ in $L \mathcal{M}^{G, CY_2}_{\text{inst}}$.
  \label{ft:symp-int floer-hom of loop space from kk reduction}
}
\begin{equation}
  \label{eq:atiyah-floer:7d-spin7:kk reduction:symplectic}
  \sum_s \text{HSF}^{\text{Int}}_{d_s} \left(
    L^2 \mathcal{M}^{G, CY_2}_{\text{inst}}, \mathscr{L}_0, \mathscr{L}_1
  \right)
  \xrightarrow{S^1 \rightarrow 0}
  \sum_u \text{HSF}^{\text{Int}}_{d_u} \left(
    L \mathcal{M}^{G, CY_2}_{\text{inst}}, \mathcal{L}_0, \mathcal{L}_1
  \right)
  \, ,
\end{equation}
where the branes $\mathcal{L}_0$ and $\mathcal{L}_1$ are isotropic-coisotropic A-branes in $L \mathcal{M}^{G, CY_2}_{\text{inst}}$.

In other words, we will have, from the KK reduction of Spin$(7)$ theory on $CY_3 \times S^1 \times \R$,
\begin{equation}
  \label{eq:atiyah-floer:7d-spin7:equality of partition fn}
  \sum_k \text{HHF}^{G_2\text{-M}}_{d_k} (CY_3, G)
  =
  \sum_u \text{HSF}^{\text{Int}}_{d_u} \left(
    L \mathcal{M}^{G, CY_2}_{\text{inst}}, \mathcal{L}_0, \mathcal{L}_1
  \right)
  \, .
\end{equation}
By application of the same arguments that led us from~\eqref{eq:atiyah-floer:spin7:equality of partition fn} to~\eqref{eq:atiyah-floer:spin7}, we will find that~\eqref{eq:atiyah-floer:7d-spin7:equality of partition fn} will mean that
\begin{equation}
  \label{eq:atiyah-floer:7d-spin7}
  \boxed{
    \text{HHF}^{G_2\text{-M}}_* (CY_3, G)
    \cong
    \text{HSF}^{\text{Int}}_* \left(
      L \mathcal{M}^{G, CY_2}_{\text{inst}}, \mathcal{L}_0, \mathcal{L}_1
    \right)
  }
\end{equation}

In other words, we have an Atiyah-Floer type duality, that we shall henceforth name a 7d-Spin$(7)$ Atiyah-Floer type duality of $CY_3$!


\section{A Fukaya-Seidel Type \texorpdfstring{$A_\infty$}{A-infty}-category of Six-Manifolds}
\label{sec:fs-cat of m6}

In this section, we will consider the case where $\text{Spin}(7) = CY_3 \times \R^2$, and recast Spin$(7)$ theory as either a 2d gauged Landau-Ginzburg (LG) model on $\R^2$ or a 1d LG SQM in path space.
Following the program in~\cite[$\S$9]{er-2023-topol-n}, we will, via the $\text{Spin}(7)$ partition function, be able to physically realize a novel Fukaya-Seidel (FS) type  $A_\infty$-category of $CY_3$ whose objects correspond to DT configurations on $CY_3$.
In doing so, we will furnish a physical proof of Haydys' mathematical conjecture~\cite{haydys-2015-fukay-seidel}.
Furthermore, by exploiting one of the Atiyah-Floer type dualities in~\autoref{sec:atiyah-floer}, we will find that this FS type $A_{\infty}$-category physically manifests a Hom-category.

\subsection{\texorpdfstring{Spin$(7)$}{Spin(7)} Theory on \texorpdfstring{$CY_3 \times \R^2$}{CY3 x R2} as a 2d Model on \texorpdfstring{$\R^2$}{R2} or SQM in Path Space}
\label{sec:fs-cat of m6:2d model or 1d sqm}

For $\text{Spin}(7) = M_6 \times \R^2$ to be a Spin$(7)$-manifold, $M_6 \times \R$ has to be a $G_2$-manifold.
This is possible if $M_6$ is a closed and compact $CY_3$~\cite{cherkis-2015-octon-monop-knots, acharya-1997-higher-dimen, esfahani-2022-monop-singul}.
We will consider this case, and study Spin$(7)$ theory on $\text{Spin}(7) = CY_3 \times \R^2$.

By choosing $x^0 = t$ and $x^1 = \tau$ as the directions of $\R^2$, and collecting the terms without $\varphi$,~\eqref{eq:spin7 action} becomes
\begin{equation}
  \label{eq:cy3 x r2:action}
  S_{\text{Spin}(7), CY_3 \times \R^2}
  = \frac{1}{e^2} \int_{CY_3 \times \R^2} \dd{t} \dd{\tau} \dd[6]{x} \Tr \Big(
    |F^+_{t \tau}|^2
    + |F^+_{t i}|^2
    + |F^+_{\tau i}|^2
    + \dots
  \Big) \, ,
\end{equation}
where ``$\dots$'' contain the fermion terms in~\eqref{eq:spin7 action}, and $x^i$ for $i \in \{2, \dots, 7\}$ are coordinates on $CY_3$.
We will now like to recast Spin$(7)$ theory on $CY_3 \times \R^2$ with action~\eqref{eq:cy3 x r2:action} as a 2d model on $\R^2$.

\subtitle{Spin$(7)$ Theory on $CY_3 \times \R^2$ as a 2d Model}

To this end, first note that by expanding the self-dual $F^+$'s in~\eqref{eq:cy3 x r2:action}, we are able to re-express the action as
\begin{equation}
  \label{eq:cy3 x r2:action:expanded}
  \begin{aligned}
    S_{\text{Spin}(7), CY_3 \times \R^2}
    = \frac{1}{4e^2} \int_{\R^2} \dd{t} \dd{\tau} \int_{CY_3} \dd[6]{x} \Tr \bigg(
    & \left|F_{t\tau} + \frac{1}{2} \phi_{t\tau ij} F^{ij} \right|^2
      + \left|F_{\tau i} + \phi_{\tau i t j} F^{t j} \right|^2
    \\
    & + \left|F_{t i}
      + \phi_{t i \tau j} F^{\tau j}
      + \frac{1}{2} \phi_{t i j k} F^{jk}
      \right|^2
      + \dots
      \bigg) \, ,
  \end{aligned}
\end{equation}
where the conditions that minimize the action~\eqref{eq:cy3 x r2:action:expanded} (and thus \eqref{eq:cy3 x r2:action}) are easily identified to be
\begin{equation}
  \label{eq:cy3 x r2:bps}
  \begin{aligned}
    F_{t\tau}
    &= - \frac{1}{2} \phi_{t\tau ij} F^{ij}
      \, , \\
    F_{t i} + \phi_{t i \tau j} F^{\tau j}
    &= - \frac{1}{2} \phi_{t i j k} F^{jk}
      \, , \\
    F_{\tau i} + \phi_{\tau i t j} F^{t j}
    &= - \frac{1}{2} \phi_{\tau i j k} F^{jk}
      \, .
  \end{aligned}
\end{equation}

Second, using the $(z^1, z^2, z^3)$ holomorphic coordinates on $CY_3$ as defined in~\autoref{sec:floer homology of m6},~\eqref{eq:cy3 x r2:bps} becomes
\begin{equation}
  \label{eq:cy3 x r2:bps:complex}
  \begin{aligned}
    F_{\tau t}
    &= \omega_{m \bar{n}} \mathcal{F}^{m \bar{n}}
      \, , \\
    2 (D_t \mathcal{A}_m - \partial_m A_t) 
    - 2i (D_\tau \mathcal{A}_m - \partial_m A_\tau) 
    &= - \frac{1}{2} \varepsilon_{mpq}\mathcal{F}^{pq}
      \, , \\
    2 (D_\tau \mathcal{A}_m - \partial_m A_\tau) 
    + 2i (D_t \mathcal{A}_m - \partial_m A_t) 
    &= - \frac{i}{2} \varepsilon_{mpq}\mathcal{F}^{pq}
      \, .
  \end{aligned}
\end{equation}

Third, noticing that the last two equations are actually identical up to an overall factor of $i$, and that we are physically free to rotate $\R^2$ about the origin, the linearly-independent equations of~\eqref{eq:cy3 x r2:bps:complex} become
\begin{equation}
  \label{bps eqns on cy3 x r2 complex rotated}
  \begin{aligned}
    F_{\tau t}
    &= \omega_{m \bar{n}} \mathcal{F}^{m \bar{n}}
      \, , \\
    2 (D_\tau \mathcal{A}_m - \partial_m A_\tau) 
    + 2i (D_t \mathcal{A}_m - \partial_m A_t) 
    &= - \frac{ie^{i\theta}}{2} \varepsilon_{mpq}\mathcal{F}^{pq}
      \, ,
  \end{aligned}
\end{equation}
where $\theta$ is the angle of rotation.
This allows us to write~\eqref{eq:cy3 x r2:action:expanded} as
\begin{equation}
  \label{eq:cy3 x r2:action:compact}
  S_{\text{Spin}(7), CY_3 \times \R^2}
  = \frac{1}{4e^2} \int_{\R^2} \dd{t} \dd{\tau} \int_{CY_3} \abs{\dd{z}}^6 \Tr \Big(
  |F_{\tau t} + \kappa |^2
  + 8 |D_\tau \mathcal{A}_m + i D_t \mathcal{A}_m + u_m|^2
  + \dots
  \Big) \, ,
\end{equation}
where
\begin{equation}
  \label{eq:cy3 x r2:action:compact:components}
  \kappa = - \omega_{m\bar{n}} \mathcal{F}^{m \bar{n}}
  \, ,
  \qquad
  u_m
  = - \partial_m A_\tau
  - i \partial_m A_t
  + \frac{ie^{i\theta}}{4} \varepsilon_{mpq} \mathcal{F}^{pq}
  \, .
\end{equation}

Lastly, after suitable rescalings, we can recast~\eqref{eq:cy3 x r2:action:compact} as a 2d model, where the action is\footnote{%
  To arrive at the following expression for the action, we have (i) employed Stokes' theorem and the fact that $CY_3$ has no boundary to omit terms with $\partial_m A_{\{t, \tau\}}$ as they will vanish when integrated over $CY_3$, and (ii) integrated out an auxiliary scalar field $\mathfrak{H}_6(\kappa) = \omega_{a\bar{b}} \mathcal{F}^{a\bar{b}}$ corresponding to the scalar $\kappa$, whose contribution to the action is $|\mathfrak{H}_6(\kappa)|^2$.
  \label{ft:stokes theorem on cy3 x r2}
}
\begin{equation}
  \label{eq:cy3 x r2:2d action}
  S_{\text{2d}, \mathfrak{A}_6}
  = \frac{1}{e^2} \int_{\R^2} \dd{t} \dd{\tau} \Big(
  |F_{\tau t}|^2
  + |D_\tau \mathcal{A}^a + i D_t \mathcal{A}^a + u^a|^2
  + \dots
  \Big) \, .
\end{equation}
Here, $\mathcal{A}^a$ and $a$ are holomorphic coordinates and indices on the space $\mathfrak{A}_6$ of irreducible $\mathcal{A}_m$ fields on $CY_3$; and
\begin{equation}
  \label{eq:cy3 x r2:2d action:components}
  u^a
  = \frac{ie^{i\theta}}{4} \varepsilon^{abc} \mathcal{F}_{bc}
\end{equation}
will correspond to $u_m$ in~\eqref{eq:cy3 x r2:action:compact:components}.

In other words, Spin$(7)$ theory on $CY_3 \times \R^2$ can be regarded as a 2d gauged sigma model along the $(t, \tau)$-directions with target space $\mathfrak{A}_6$ and action \eqref{eq:cy3 x r2:2d action}.
We will now further recast this 2d gauged sigma model as a 1d SQM.

\subtitle{The 2d Model on $\R^2$ with Target Space $\mathfrak{A}_6$ as a 1d SQM}

Singling out $\tau$ as the direction in ``time'', the equivalent SQM action can be obtained from~\eqref{eq:cy3 x r2:2d action} after suitable rescalings as\footnote{%
  In the resulting SQM, as $A_\tau$ has no field strength and is thus non-dynamical, it will be integrated out to furnish the Christoffel connection for the fermions in the SQM~\cite{er-2023-topol-n}, leaving us with an SQM without $A_{\tau}$.
  We have also omitted a term $|\partial_{\tau} A_t|^2$ in the following expression as it will just lead to the trivial condition $\partial_{\tau} A_t = 0$.
  \label{ft:aux fields of cy3 x r2 sqm}
}
\begin{equation}
  \label{eq:cy3 x r2:sqm action}
  S_{\text{SQM}, \mathcal{M}(\R, \mathfrak{A}_6)}
  = \frac{1}{e^2} \int \dd{\tau} \left(
    \left| \partial_\tau \mathcal{A}^\alpha
      + g^{\alpha \beta}_{\mathcal{M}(\R, \mathfrak{A}_6)}
      \pdv{h_6}{\mathcal{A}^\beta} \right|^2
    + \dots
  \right) \, ,
\end{equation}
where $\mathcal{A}^\alpha$ and $(\alpha, \beta)$ are holomorphic coordinates and indices on the path space $\mathcal{M}(\R, \mathfrak{A}_6)$ of maps from $\R$ to $\mathfrak{A}_6$; $g_{\mathcal{M}(\R, \mathfrak{A}_6)}$ is the metric on $\mathcal{M}({\R,\mathfrak{A}_6})$; and $h_6(\mathcal{A})$ is the potential function.

In other words, Spin$(7)$ theory on $CY_3 \times \R^2$ can also be regarded as a 1d SQM along $\tau$ in $\mathcal{M}(\R, \mathfrak{A}_6)$ whose action is \eqref{eq:cy3 x r2:sqm action}.


\subsection{Non-constant Paths, Solitons, and DT Configurations}
\label{sec:fs-cat of m6:solitons and dt configs}

\subtitle{$\theta$-deformed, Non-constant Paths in the SQM}

The squaring argument \cite{blau-1993-topol-gauge} applied to \eqref{eq:cy3 x r2:sqm action} tells us that the equivalent SQM localizes onto configurations that set both the LHS and RHS of the expression within the squared term \emph{simultaneously} to zero, i.e., the SQM localizes onto $\tau$-invariant critical points of $h_6(\mathcal{A})$ that obey
\begin{equation}
  \label{eq:cy3 x r2:non-constant paths}
  \partial_t \mathcal{A}^\alpha
  = - \frac{e^{i\theta}}{4} \varepsilon^{\alpha \beta \gamma} \mathcal{F}_{\beta \gamma}
  \, .
\end{equation}
These are \emph{$\tau$-invariant, $\theta$-deformed}, non-constant paths in $\mathcal{M}(\R, \mathfrak{A}_6)$.

\subtitle{$\mathfrak{A}_6^\theta$-solitons in the 2d Gauged Model}

By comparing \eqref{eq:cy3 x r2:sqm action} with \eqref{eq:cy3 x r2:2d action}, we find that such $\tau$-invariant, $\theta$-deformed, non-constant paths in the SQM defined by \eqref{eq:cy3 x r2:non-constant paths}, will correspond, in the 2d gauged sigma model with target space $\mathfrak{A}_6$, to configurations defined by
\begin{equation}
  \label{eq:cy3 x r2:soliton:eqn:components}
  \relax
  [A_\tau, \mathcal{A}^a] + i D_t \mathcal{A}^a + u^a
  = 0
  \, .
\end{equation}
Via \eqref{eq:cy3 x r2:2d action:components}, we can write this as
\begin{equation}
  \label{eq:cy3 x r2:soliton:eqn}
  \partial_t \mathcal{A}^a
  = - [A_t, \mathcal{A}^a]
  + i [A_\tau, \mathcal{A}^a]
  - \frac{e^{i\theta}}{4} \varepsilon^{abc} \mathcal{F}_{bc}
  \, .
\end{equation}
These are $\tau$-invariant, $\theta$-deformed solitons along the $t$-direction in the 2d gauged sigma model, which also satisfy the condition
\begin{equation}
  \label{eq:cy3 x r2:soliton:aux eqn}
  F_{\tau t}
  = 0
  = \omega^{a\bar{b}} \mathcal{F}_{a\bar{b}}
  \, ,
\end{equation}
where $\omega^{a\bar{b}} \mathcal{F}_{a\bar{b}} = \mathfrak{H}_6(\kappa)$ is the auxiliary scalar field in~\autoref{ft:stokes theorem on cy3 x r2}.

We shall henceforth refer to such $\tau$-invariant, $\theta$-deformed solitons in the 2d gauged sigma model with target space $\mathfrak{A}_6$, defined by~\eqref{eq:cy3 x r2:soliton:eqn} and~\eqref{eq:cy3 x r2:soliton:aux eqn}, as $\mathfrak{A}_6^\theta$-solitons.

\subtitle{$\tau$-independent, $\theta$-deformed Spin$(7)$ Configurations in Spin$(7)$ Theory}

In turn, by comparing \eqref{eq:cy3 x r2:2d action} with \eqref{eq:cy3 x r2:action:compact}, we find that the 2d configurations defined by~\eqref{eq:cy3 x r2:soliton:eqn:components}, will correspond, in Spin$(7)$ theory, to 8d configurations defined by
\begin{equation}
  \label{eq:cy3 x r2:soliton:config:components}
  \relax
  [A_\tau, A_m] + i D_t \mathcal{A}_m + u_m
  = 0
  \, .
\end{equation}
Via~\eqref{eq:cy3 x r2:action:compact:components}, we can write this as
\begin{equation}
  \label{eq:cy3 x r2:soliton:config}
  \partial_t \mathcal{A}_m
  = \mathcal{D}_m \mathcal{A}_t
  - i \mathcal{D}_m A_\tau
  - \frac{e^{i\theta}}{4} \varepsilon_{mpq} \mathcal{F}^{pq}
  \, .
\end{equation}
These are $\tau$-independent, $\theta$-deformed Spin$(7)$ configurations on $CY_3 \times \R^2$ which also satisfy the conditions
\begin{equation}
  \label{eq:cy3 x r2:soliton:aux config}
  \partial_t A_\tau
  = [A_\tau, A_t]
  \, ,
  \qquad
  \omega_{m\bar{n}} \mathcal{F}^{m\bar{n}}
  = 0
  \, .
\end{equation}

\subtitle{Spin$(7)$ Configurations, $\mathfrak{A}_6^\theta$-solitons, and Non-constant Paths}

In short, these \emph{$\tau$-independent, $\theta$-deformed} Spin$(7)$ configurations on $CY_3 \times \R^2$ that are defined by~\eqref{eq:cy3 x r2:soliton:config} and~\eqref{eq:cy3 x r2:soliton:aux config}, will correspond to the $\mathfrak{A}_6^\theta$-solitons defined by~\eqref{eq:cy3 x r2:soliton:eqn} and~\eqref{eq:cy3 x r2:soliton:aux eqn}, which, in turn, will correspond to the $\tau$-invariant, $\theta$-deformed, non-constant paths in $\mathcal{M}(\R, \mathfrak{A}_6)$ defined by~\eqref{eq:cy3 x r2:non-constant paths}.

\subtitle{$\mathfrak{A}_6^\theta$-soliton Endpoints Corresponding to $\theta$-deformed DT Configurations on $CY_3$}

Consider now the fixed endpoints of the $\mathfrak{A}_6^\theta$-solitons at $t = \pm \infty$, where we also expect the finite-energy 2d gauge fields $A_t, A_\tau$ to decay to zero.
They are given by~\eqref{eq:cy3 x r2:soliton:eqn} and~\eqref{eq:cy3 x r2:soliton:aux eqn} with $\partial_t \mathcal{A}_a = 0$ and $A_t, A_\tau \rightarrow 0$, i.e.,
\begin{equation}
  \label{eq:cy3 x r2:soliton:endpts}
  \begin{aligned}
    e^{i\theta} \varepsilon^{abc}\mathcal{F}_{bc}
    &= 0
      \, , \\
    \omega^{a\bar{b}} \mathcal{F}_{a\bar{b}}
    &= 0
      \, .
  \end{aligned}
\end{equation}
In turn, they will correspond, in Spin$(7)$ theory, to $(t, \tau)$-independent,
$\theta$-deformed configurations that obey
\begin{equation}
  \label{eq:cy3 x r2:soliton:endpts:config}
  \begin{aligned}
    e^{i\theta} \varepsilon_{mpq} \mathcal{F}^{pq}
    &= 0
      \, , \\
    \omega_{m\bar{n}}\mathcal{F}^{m\bar{n}}
    &= 0
      \, .
  \end{aligned}
\end{equation}
Notice that~\eqref{eq:cy3 x r2:soliton:endpts:config} can also be obtained from~\eqref{eq:cy3 x r2:soliton:config} and~\eqref{eq:cy3 x r2:soliton:aux config} with $\partial_t \mathcal{A}_m = 0$ and $A_t, A_\tau \rightarrow 0$.

The equations in~\eqref{eq:cy3 x r2:soliton:endpts:config} are a $\theta$-deformed version of Donaldson-Uhlenbeck-Yau (DUY) equations on $CY_3$.
At $\theta = 0, \pi$, they become the regular DUY equations on $CY_3$, which are simply DT equations on $CY_3$ with the scalar being zero.
Configurations spanning the space of solutions to these equations shall, in the rest of this section, be referred to as DT configuration on $CY_3$.

In other words, the $(t, \tau)$-independent, $\theta$-deformed Spin$(7)$ configurations corresponding to the endpoints of the $\mathfrak{A}_6^\theta$-solitons, are $\theta$-deformed DT configurations on $CY_3$.
We will also assume choices of $CY_3$ satisfying~\autoref{ft:dt isolation and non-degeneracy} whereby such configurations are isolated and non-degenerate.\footnote{%
  At $\theta = 0$, the moduli space of such configurations is the moduli space of undeformed DT configurations on $CY_3$.
  For such a choice of $CY_3$, this moduli space will be made of isolated and non-degenerate points.
  Therefore, at $\theta = 0$, the endpoints of the $\mathfrak{A}_6^{\theta}$-solitons will be isolated and non-degenerate.
  As the physical theory is symmetric under a variation of $\theta$, this observation about the endpoints of the $\mathfrak{A}_6^\theta$-solitons will continue to hold true for any value of $\theta$. Hence, the presumption that the moduli space of $\theta$-deformed DT configurations on $CY_3$ will be made of isolated and non-degenerate points, is justified.
  \label{ft:duy isolation and non-degeneracy}
}

In short, from the equivalent 1d SQM of Spin$(7)$ theory on $CY_3 \times \R^2$, the theory localizes onto $\tau$-invariant, $\theta$-deformed, non-constant paths in $\mathcal{M}(\R, \mathfrak{A}_6)$, which, in turn, will correspond to $\mathfrak{A}_6^\theta$-solitons in the 2d gauged sigma model whose endpoints correspond to $\theta$-deformed DT configurations on $CY_3$.


\subsection{The 2d Model on \texorpdfstring{$\R^2$}{R2} and an Open String Theory in \texorpdfstring{$\mathfrak{A}_6$}{A6}}
\label{sec:fs-cat of m6:open string theory}

\subtitle{Flow Lines of the SQM as BPS Worldsheets of the 2d Model}

The classical trajectories or flow lines of the equivalent SQM are governed by the gradient flow equation (defined by setting to zero the expression within the squared term in \eqref{eq:cy3 x r2:sqm action}), i.e.,
\begin{equation}
  \label{eq:cy3 x r2:sqm flow}
  \dv{\mathcal{A}^\alpha}{\tau}
  = - g^{\alpha \beta}_{\mathcal{M}(\R, \mathfrak{A}_6)} \pdv{h_6}{\mathcal{A}^\beta}
  \, ,
\end{equation}
and they go from one $\tau$-invariant critical point of $h_6$ to another in $\mathcal{M}(\R, \mathfrak{A}_6)$.
In the 2d gauged sigma model with target space $\mathfrak{A}_6$, these flow lines will correspond to worldsheets that have, at $\tau = \pm \infty$, $\mathfrak{A}_6^\theta$-solitons.\footnote{%
  The $\mathfrak{A}_6^\theta$-soliton can translate in the $\tau$-direction due to its ``center of mass'' motion, and because it is $\tau$-invariant, it is effectively degenerate. This reflects the fact that generically, each critical point of $h_6$ is degenerate and does not correspond to a point but a real line $\R$ in $\mathcal{M} (\R, {\mathfrak A}_6)$. Nonetheless, one can perturb $h_6$ via the addition of physically-inconsequential $\mathcal{Q}$-exact terms to the SQM action, and collapse the degeneracy such that the critical points really correspond to points in $\mathcal{M} (\R, {\mathfrak A}_6)$. This is equivalent to factoring out the center of mass degree of freedom of the $\mathfrak{A}_6^{\theta}$-soliton, and fixing it at $\tau = \pm \infty$.\label{ft:fixing frakA6-soliton centre of mass dof}
}
These solitons shall be denoted as $\gamma_\pm(t, \theta, \mathfrak{A}_6)$, and are defined by~\eqref{eq:cy3 x r2:soliton:eqn} with finite-energy gauge fields $A_t, A_\tau \rightarrow 0$, i.e.,
\begin{equation}
  \label{eq:cy3 x r2:soliton:eqns no gauge}
  \dv{\mathcal{A}^a}{t}
  = - \frac{e^{i\theta}}{4} \varepsilon^{abc} \mathcal{F}_{bc}
  \, .
\end{equation}
Their endpoints $\gamma(\pm\infty, \theta, \mathfrak{A}_6)$ at $t = \pm\infty$ are defined by
\begin{equation}
  \label{eq:cy3 x r2:soliton:endpts no gauge}
  e^{i\theta} \varepsilon^{abc} \mathcal{F}_{bc} = 0
  \, ,
\end{equation}
which is simply~\eqref{eq:cy3 x r2:soliton:eqns no gauge} with $d_t \mathcal{A}^a = 0$.

Note that the flow lines are governed by the gradient flow equations, which are actually the BPS equations of the 1d SQM.
This means that the worldsheets that they will correspond to are governed by the BPS equations of the equivalent 2d gauged sigma model with target space $\mathfrak{A}_6$ (defined by setting to zero the expression within the squared terms in \eqref{eq:cy3 x r2:2d action}), i.e.,
\begin{equation}
  \label{eq:cy3 x r2:worldsheet:eqn:components}
  \begin{aligned}
    F_{\tau t}
    &= 0
      \, , \\
    D_\tau \mathcal{A}^a + i D_t \mathcal{A}^a + u^a
    &= 0
      \, , \\
    \mathfrak{H}_6(\kappa)
    &= 0
      \, ,
  \end{aligned}
\end{equation}
or more explicitly,
\begin{equation}
  \label{eq:cy3 x r2:worldsheet:eqn}
  \begin{aligned}
    \dv{A_\tau}{t} - \dv{A_t}{\tau} + [A_t, A_\tau]
    &= 0
      \, , \\
    \Dv{\mathcal{A}^a}{\tau} + i \Dv{\mathcal{A}^a}{t}
    &= - \frac{ie^{i\theta}}{4} \varepsilon^{abc} \mathcal{F}_{bc}
      \, , \\
    0
    &= \omega^{a\bar{b}}\mathcal{F}_{a\bar{b}}
      \, .
  \end{aligned}
\end{equation}
In~\cite[$\S$9.3]{er-2023-topol-n}, we coined such worldsheets corresponding to the classical trajectories of 2d gauged sigma models, as BPS worldsheets. We shall do the same here.

\subtitle{BPS Worldsheets with Boundaries Corresponding to $\theta$-deformed DT Configurations on $CY_3$}

The boundaries of the BPS worldsheets are traced out by the endpoints of the $\mathfrak{A}_6^\theta$-solitons as they propagate in $\tau$.
As we have seen at the end of \autoref{sec:fs-cat of m6:solitons and dt configs}, at $\theta = 0$, these endpoints correspond to DT configurations on $CY_3$.
If there are `$k$' such configurations $\{ \mathcal{E}^1_{\text{DT}}(0), \mathcal{E}^2_{\text{DT}}(0), \dots, \mathcal{E}^k_{\text{DT}}(0) \}$, we can further specify the undeformed $\mathfrak{A}_6^0$-solitons at $\tau = \pm \infty$ as $\gamma^{IJ}_\pm(t, 0, \mathfrak{A}_6)$, where $I, J \in \{1, \dots, k\}$ indicates that its left and right endpoints, given by $\gamma^I(-\infty, 0, \mathfrak{A}_6)$ and $\gamma^J(+\infty, 0, \mathfrak{A}_6)$, would correspond to the configurations $\mathcal{E}^I_{\text{DT}}(0)$ and $\mathcal{E}^J_{\text{DT}}(0)$, respectively.
As the physical theory is symmetric under a variation of $\theta$, this would be true at any value of $\theta$.
In other words, we can also further specify any $\mathfrak{A}_6^\theta$-soliton at $\tau = \pm \infty$ as $\gamma^{IJ}_\pm(t, \theta, \mathfrak{A}_6)$, where its left and right endpoints, given by $\gamma^I(-\infty, \theta, \mathfrak{A}_6)$ and $\gamma^J(+\infty, \theta, \mathfrak{A}_6)$, would correspond to $\mathcal{E}^I_{\text{DT}}(\theta)$ and $\mathcal{E}^J_{\text{DT}}(\theta)$, respectively, with the $\mathcal{E}^*_{\text{DT}}(\theta)$'s being $k$ number of $\theta$-deformed DT configurations on $CY_3$.

\begin{figure}
  \centering
  \begin{tikzpicture}
    \coordinate (lt) at (0,4) {};
    \coordinate (rt) at (4,4) {}
    edge node[pos=0.5, above] {$\gamma^{IJ}_+(t, \theta, \mathfrak{A}_6)$}
    (lt) {};
    \coordinate (lb) at (0,0) {}
    edge node[pos=0.5, left] {$\mathcal{E}^I_{\text{DT}}(\theta)$}
    (lt) {};
    \coordinate (rb) at (4,0) {}
    edge node[pos=0.5, below] {$\gamma^{IJ}_-(t, \theta, \mathfrak{A}_6)$}
    (lb) {}
    edge node[pos=0.5, right] {$\mathcal{E}^J_{\text{DT}}(\theta)$}
    (rt) {};
    \draw (lb) -- (lt);
    \draw (rb) -- (rt);
    \coordinate (co) at (5,0);
    \coordinate (cx) at (5.5,0);
    \node at (cx) [right=2pt of cx] {$t$};
    \coordinate (cy) at (5,0.5);
    \node at (cy) [above=2pt of cy] {$\tau$};
    \draw[->] (co) -- (cx);
    \draw[->] (co) -- (cy);
  \end{tikzpicture}
  \caption[]{BPS worldsheet with solitons $\gamma^{IJ}_\pm(t, \theta, \mathfrak{A}_6)$ and boundaries corresponding to $\mathcal{E}^I_{\text{DT}}(\theta)$ and $\mathcal{E}^J_{\text{DT}}(\theta)$.
  }
  \label{fig:cy3 x r2:mu-1 map}
\end{figure}
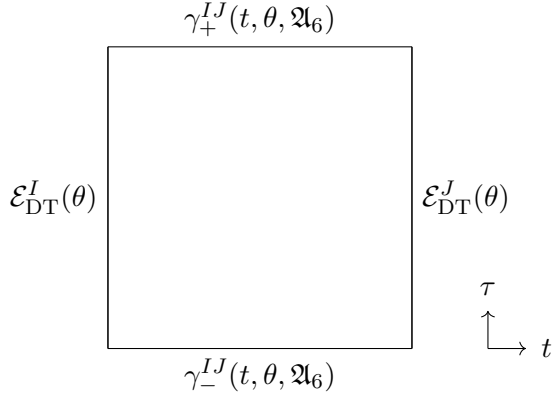

Since the $\mathcal{E}^*_{\text{DT}}(\theta)$'s are $\tau$-independent and therefore, have the same values for all $\tau$, we have BPS worldsheets of the kind shown in~\autoref{fig:cy3 x r2:mu-1 map}.

\subtitle{The 2d Model on $\R^2$ and an Open String Theory in $\mathfrak{A}_6$}

Hence, one can understand the 2d gauged sigma model on $\R^2$ with target space $\mathfrak{A}_6$ to define an open string theory in $\mathfrak{A}_6$, with \emph{effective} worldsheet and boundaries shown in~\autoref{fig:cy3 x r2:mu-1 map}, where $\tau$ and $t$ are the temporal and spatial directions, respectively.


\subsection{Soliton String Theory, the \texorpdfstring{$\text{Spin}(7)$}{Spin(7)} Partition Function, and an FS Type \texorpdfstring{$A_\infty$}{A-infty}-category of DT Configurations on \texorpdfstring{$CY_3$}{CY3}}
\label{sec:fs-cat of m6:fs cat of dt configs}

\subtitle{The 2d Model as a Gauged LG Model}

Notice that we can also express the action of the 2d gauged sigma model with target space $\mathfrak{A}_6$ in~\eqref{eq:cy3 x r2:2d action} as
\begin{equation}
  \label{eq:cy3 x r2:lg:2d action}
  \begin{aligned}
    S_{\text{LG},\mathfrak{A}_6}
    =& \frac{1}{e^2} \int \dd{t} \dd{\tau}
       \left(
       \left| D_\tau \mathcal{A}^a
       + i D_t \mathcal{A}^a
       + \frac{i e^{i\theta}}{4} \varepsilon^{abc} \mathcal{F}_{bc}
       \right|^2
       + \left| F_{\tau t} \right|^2
       + \dots
       \right)
    \\
    =& \frac{1}{e^2} \int \dd{t} \dd{\tau}
       \left(
       \left| D_\tau \mathcal{A}^a
       + i D_t \mathcal{A}^a
       - g^{a\bar{b}}_{\mathfrak{A}_6} \left(
       \frac{i \zeta}{2} \pdv{W_6}{\mathcal{A}^b}
       \right)^*
       \right|^2
       + \left| F_{\tau t} \right|^2
       + \dots
       \right)
    \\
    =& \frac{1}{e^2} \int \dd{t} \dd{\tau}
       \left(
       \left| D_\sigma \mathcal{A}^a \right|^2
       + \left| \pdv{W_6}{\mathcal{A}^a} \right|^2
       + \left| F_{\tau t} \right|^2
       + \dots
       \right)
       \, ,
  \end{aligned}
\end{equation}
where $\sigma$ is the index on the worldsheet, and $\zeta = e^{-i\theta}$.
In other words, the 2d gauged sigma model with target space $\mathfrak{A}_6$ can also be interpreted as a 2d gauged LG model in $\mathfrak{A}_6$ with holomorphic superpotential $W_6(\mathcal{A})$.
Noting that the gradient vector field of $W_6(\mathcal{A})$ is the $\theta$-independent part of $u^a$ from~\eqref{eq:cy3 x r2:2d action:components}, i.e., $\mathcal{F}$, we find that the holomorphic superpotential $W_6(\mathcal{A})$ must therefore be $CS(\mathcal{A})$, a \emph{Chern-Simons function of $\mathcal{A}$}.

By setting $d_\tau \mathcal{A}^a = 0$ and $A_t, A_\tau \rightarrow 0$ in the expression within the squared terms in~\eqref{eq:cy3 x r2:lg:2d action}, we can read off the LG $\mathfrak{A}_6^\theta$-soliton equations corresponding to $\gamma^{IJ}_\pm(t, \theta, \mathfrak{A}_6)$ (that re-expresses~\eqref{eq:cy3 x r2:soliton:eqns no gauge}) as
\begin{equation}
  \label{eq:cy3 x r2:lg soliton:eqn}
  \dv{\mathcal{A}^a}{t}
  = - i g^{a\bar{b}}_{\mathfrak{A}_6} \left(
    \frac{i \zeta}{2} \pdv{W_6}{\mathcal{A}^{b}}
  \right)^* \, .
\end{equation}
By setting $d_t \mathcal{A}^a = 0$ in \eqref{eq:cy3 x r2:lg soliton:eqn}, we get the LG $\mathfrak{A}_6^\theta$-soliton endpoint equations corresponding to $\gamma^{IJ}(\pm\infty, \theta, \mathfrak{A}_6)$ (that re-expresses \eqref{eq:cy3 x r2:soliton:endpts no gauge}) as
\begin{equation}
  \label{eq:cy3 x r2:lg soliton:endpts}
  g^{a\bar{b}}_{\mathfrak{A}_6} \left(
    \frac{i \zeta}{2} \pdv{W_6}{\mathcal{A}^{b}}
  \right)^*
  = 0
  \, .
\end{equation}

Recall from the end of \autoref{sec:fs-cat of m6:solitons and dt configs} that we are only considering certain $CY_3$ such that the endpoints $\gamma^{IJ}(\pm\infty, \theta, \mathfrak{A}_6)$ are isolated and non-degenerate.
Therefore, from their definition in~\eqref{eq:cy3 x r2:lg soliton:endpts} which tells us that they are critical points of $W_6(\mathcal{A})$, we conclude that $W_6(\mathcal{A})$ can be regarded as a holomorphic Morse function in $\mathfrak{A}_6$.

A consequence of being able to write the 2d model as a 2d gauged LG model with holomorphic superpotential $W_6$, is that it is known that such LG solitons map to straight line segments in the complex $W_6$-plane.
Specifically, an LG $\mathfrak{A}_6^\theta$-soliton defined in \eqref{eq:cy3 x r2:lg soliton:eqn} maps to a straight line segment $[W_6^I(\theta), W_6^J(\theta)]$ in the complex $W_6$-plane that starts and ends at the critical values $W_6^I(\theta) \coloneq W_6(\gamma^I(-\infty, \theta, \mathfrak{A}_6))$ and $W_6^J(\theta) \coloneq W_6(\gamma^J(+\infty, \theta, \mathfrak{A}_6))$, respectively, where its slope depends on $\theta$ (via $\zeta$).
This fact will be useful shortly.
We shall also assume that $\text{Re}(W_6^I(\theta)) < \text{Re}(W_6^J(\theta))$.

\subtitle{The Gauged LG Model as an LG SQM}

Last but not least, after suitable rescalings, we can recast \eqref{eq:cy3 x r2:lg:2d action} as a 1d LG SQM (that re-expresses \eqref{eq:cy3 x r2:sqm action}), where its action will be given by\footnote{%
  Just as in~\autoref{ft:aux fields of cy3 x r2 sqm}, we have integrated out $A_{\tau}$ and omitted the term containing $A_t$ in the resulting SQM.
  \label{ft:auxiliary fields of cy3 x r2 lg sqm}
}
\begin{equation}
  \label{eq:cy3 x r2:lg sqm:action}
  S_{\text{LG SQM}, \mathcal{M}(\R, \mathfrak{A}_6)} = \frac{1}{e^2} \int \dd{\tau}
  \left(
    \left| \partial_\tau \mathcal{A}^\alpha
      + g^{\alpha\beta}_{\mathcal{M}(\R, \mathfrak{A}_6)} \pdv{H_6}{\mathcal{A}^\beta}
    \right|^2
    + \dots
  \right) \, ,
\end{equation}
where $H_6(\mathcal{A})$ is the \emph{real-valued} potential in $\mathcal{M}(\R, \mathfrak{A}_6)$.

The LG SQM will localize onto configurations that \emph{simultaneously} set to zero the LHS and RHS of the expression within the squared term in~\eqref{eq:cy3 x r2:lg sqm:action}.
In other words, it will localize onto $\tau$-invariant critical points of $H_6(\mathcal{A})$ that will correspond to the LG $\mathfrak{A}_6^\theta$-solitons defined by~\eqref{eq:cy3 x r2:lg soliton:eqn}.
For our choice of $CY_3$, the LG $\mathfrak{A}_6^\theta$-solitons, just like their endpoints, will be isolated and non-degenerate.
Thus, $H_6(\mathcal{A})$ can be regarded as a real-valued Morse functional in $\mathcal{M}(\R, \mathfrak{A}_6)$.

\subtitle{Morphisms from $\mathcal{E}^I_{\text{DT}}(\theta)$ to $\mathcal{E}^J_{\text{DT}}(\theta)$ as Floer Homology Classes of Intersecting Thimbles}

Note that we can also describe an LG $\mathfrak{A}_6^\theta$-soliton in terms of the intersection of thimbles, as was done in~\cite[$\S$9.4]{er-2023-topol-n}.
One can understand such thimbles as submanifolds of a certain fiber space over the complex $W_6$-plane.
Solutions satisfying
\begin{equation}
  \label{eq:cy3 x r2:lg soliton:left endpts}
  \lim_{t \rightarrow -\infty} \gamma_{\pm}(t, \theta, \mathfrak{A}_6)
  = \gamma^I(-\infty, \theta, \mathfrak{A}_6)
\end{equation}
are known as left thimbles, and those satisfying
\begin{equation}
  \label{eq:cy3 x r2:lg soliton:right endpst}
  \lim_{t \rightarrow +\infty} \gamma_{\pm}(t, \theta, \mathfrak{A}_6)
  = \gamma^J(+\infty, \theta, \mathfrak{A}_6)
\end{equation}
are known as right thimbles.
Such a description makes it clear that they correspond, respectively, to the left and right endpoints of an LG $\mathfrak{A}_6^\theta$-soliton solution $\gamma^{IJ}_\pm(t, \theta, \mathfrak{A}_6)$.

Clearly, an LG $\mathfrak{A}_6^\theta$-soliton solution, which would correspond to $\gamma^{IJ}_\pm(t, \theta, \mathfrak{A}_6)$, must simultaneously be in a left and right thimble.
Thus, it can be represented as a transversal intersection of the left and right thimble in the fiber space over the line segment $[W^I_6(\theta), W^J_6(\theta)]$.\footnote{%
  This intersection is guaranteed at some $\theta$, for which we can freely tune as the physical theory is symmetric under its variation.
  \label{ft:intersection in W-plane guaranteed by theta}
}%
~Denoting such intersections as $S^{IJ}_{\text{DT}}$, each LG $\mathfrak{A}_6^\theta$-soliton pair $\gamma^{IJ}_\pm(t, \theta, \mathfrak{A}_6)$, whose left and right endpoints correspond to $\mathcal{E}^I_{\text{DT}}(\theta)$ and $\mathcal{E}^J_{\text{DT}}(\theta)$ on a BPS worldsheet as shown in~\autoref{fig:cy3 x r2:mu-1 map}, will correspond to a pair of intersection points $p^{IJ}_{\text{DT}, \pm}(\theta) \in S^{IJ}_{\text{DT}}$.

Hence, as in earlier sections, the LG SQM in $\mathcal{M}(\R, \mathfrak{A}_6)$ with action~\eqref{eq:cy3 x r2:lg sqm:action} will physically realize a Floer homology that we shall name an $\mathfrak{A}_6$-LG Floer homology.
The chains of the $\mathfrak{A}_6$-LG Floer complex will be generated by LG $\mathfrak{A}_6^\theta$-solitons which we can identify with $p_{\text{DT},\pm}^{**}(\theta)$, and the $\mathfrak{A}_6$-LG Floer differential will be realized by the flow lines governed by the gradient flow equation satisfied by $\tau$-varying configurations which set the expression within the squared term in~\eqref{eq:cy3 x r2:lg sqm:action} to zero.
In particular, the SQM partition function of the LG SQM in $\mathcal{M}(\R, \mathfrak{A}_6)$ will be given by\footnote{%
  The `$\theta$' label is omitted in the LHS of the following expression, as the physical theory is actually equivalent for all values of $\theta$.
  \label{ft:theta omission in lg sqm 8d partition fn}
}
\begin{equation}
  \label{eq:cy3 x r2:lg sqm:partition fn}
  \mathcal{Z}_{\text{LG SQM}, \mathcal{M}(\R, \mathfrak{A}_6)}(G)
  = \sum_{I \neq J = 1}^k \sum_{p^{IJ}_{\text{DT}, \pm} \in S^{IJ}_{\text{DT}}}
  \text{HF}^{G}_{d_p}(p_{\text{DT},\pm}^{IJ}(\theta))
  \, ,
\end{equation}
where the contribution $\text{HF}^{G}_{d_q}(p_{\text{DT},\pm}^{IJ}(\theta))$ can be identified with a homology class in an $\mathfrak{A}_6$-LG Floer homology generated by intersection points of thimbles.
These intersection points represent LG $\mathfrak{A}_6^\theta$-solitons whose endpoints correspond to $\theta$-deformed DT configurations on $CY_3$.
The degree of each chain in the complex is $d_p$, and is counted by the number of outgoing flow lines from the fixed critical points of $H_6(\mathcal{A})$ in $\mathcal{M}(\R, \mathfrak{A}_6)$ which can also be identified as $p_{\text{DT},\pm}^{IJ}(\theta)$.

Therefore, $\mathcal{Z}_{\text{LG SQM}, \mathcal{M}(\R, \mathfrak{A}_6)}(G)$ in~\eqref{eq:cy3 x r2:lg sqm:partition fn} is a sum of LG $\mathfrak{A}_6^\theta$-solitons defined by~\eqref{eq:cy3 x r2:lg soliton:eqn} with endpoints~\eqref{eq:cy3 x r2:lg soliton:endpts}, or equivalently, $\gamma^{IJ}_\pm(t, \theta, \mathfrak{A}_6)$-solitons defined by~\eqref{eq:cy3 x r2:soliton:eqns no gauge} with endpoints~\eqref{eq:cy3 x r2:soliton:endpts no gauge}, whose start and end correspond to $\mathcal{E}^I_{\text{DT}}(\theta)$ and $\mathcal{E}^J_{\text{DT}}(\theta)$, respectively.
In other words, we can write
\begin{equation}
  \label{eq:cy3 x r2:floer complex:vector}
  \text{CF}_{\mathfrak{A}_6}\left(
    \mathcal{E}^I_{\text{DT}}(\theta), \mathcal{E}^J_{\text{DT}}(\theta)
  \right)_\pm
  = \text{HF}^{G}_{d_p}(p_{\text{DT},\pm}^{IJ}(\theta))
  \, ,
\end{equation}
where $\text{CF}_{\mathfrak{A}_6}( \mathcal{E}^I_{\text{DT}}(\theta), \mathcal{E}^J_{\text{DT}}(\theta) )_\pm$ is a vector representing a $\gamma^{IJ}_\pm(t, \theta, \mathfrak{A}_6)$-soliton, such that $\text{Re}(W^I_6(\theta)) < \text{Re}(W^J_6(\theta))$.

Recall that a soliton can be regarded as a morphism between its endpoints.
Specifically, the pair of $\gamma^{IJ}_\pm(t, \theta, \mathfrak{A}_6)$-solitons can be regarded as a pair of morphisms $\text{Hom}(\mathcal{E}^I_{\text{DT}}(\theta), \mathcal{E}^J_{\text{DT}}(\theta))_\pm$ from $\mathcal{E}^I_{\text{DT}}(\theta)$ to $\mathcal{E}^J_{\text{DT}}(\theta)$.
Thus, we have the following one-to-one identification\footnote{%
  The `$\theta$' label is once again omitted in the following expression, as the physical theory is actually equivalent for all values of $\theta$.
  \label{ft:theta omission in 6d fs-cat morphism}
}
\begin{equation}
  \label{eq:cy3 x r2:floer complex:morphism}
  \boxed{
    \text{Hom}(\mathcal{E}^I_{\text{DT}}, \mathcal{E}^J_{\text{DT}})_\pm
    \Longleftrightarrow
    \text{HF}^{G}_{d_p}(p_{\text{DT},\pm}^{IJ})
  }
\end{equation}
where the RHS is proportional to the identity class when $I = J$, and zero when $I \leftrightarrow J$ (since the $\gamma^{IJ}_\pm(t, \theta, \mathfrak{A}_6)$-soliton only moves in one direction from $\mathcal{E}^I_{\text{DT}}(\theta)$ to $\mathcal{E}^J_{\text{DT}}(\theta)$ as depicted in~\autoref{fig:cy3 x r2:mu-1 map}).

\subtitle{Soliton String Theory from the 2d LG Model}

Just like the 2d gauged sigma model, the equivalent 2d gauged LG model will define an open string theory in $\mathfrak{A}_6$ with effective worldsheets and boundaries shown in~\autoref{fig:cy3 x r2:mu-1 map}, where $\tau$ and $t$ are the temporal and spatial directions, respectively.

The dynamics of this open string theory in $\mathfrak{A}_6$ will be governed by the BPS worldsheet equations determined by setting to zero the expression within the squared term in \eqref{eq:cy3 x r2:lg:2d action}, where $\mathcal{A}^a$ are scalars on the worldsheet corresponding to the holomorphic coordinates on $\mathfrak{A}_6$.
At an arbitrary instant in time whence $d_\tau \mathcal{A}^a = 0$ therein, the dynamics of $\mathcal{A}^a$ along $t$ will be governed by the soliton equation
\begin{equation}
  \label{eq:cy3 x r2:string soliton:eqn}
  \dv{\mathcal{A}^a}{t}
  = - [A_t - i A_\tau, \mathcal{A}^a]
  - i g^{a\bar{b}}_{\mathfrak{A}_6} \left(
    \frac{i \zeta}{2} \pdv{W_6}{\mathcal{A}^{b}}
  \right)^* \, .
\end{equation}

Hence, just as a topological A-model can be interpreted as an instanton string theory whose corresponding dynamics of the ${\mathcal A}^a$ fields along the spatial $t$-direction  will be governed by the instanton equation $d{\mathcal A^a} / dt = 0$, our LG model can be interpreted as a \textit{soliton} string theory.

\subtitle{The Normalized Spin$(7)$ Partition Function, LG $\mathfrak{A}_6^\theta$-soliton String Scattering, and Maps of an $A_\infty$-structure}

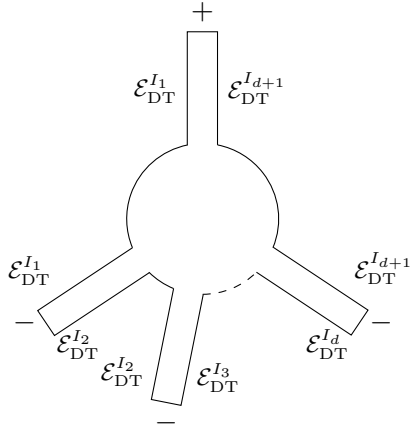
\begin{figure}
  \centering
  \begin{tikzpicture}[declare function={
      lenX(\legLength,\leftAngle,\rightAngle)
      = \legLength * cos((\leftAngle + \rightAngle)/2);
      lenY(\legLength,\leftAngle,\rightAngle)
      = \legLength * sin((\leftAngle + \rightAngle)/2);
      lenLX(\segAngle,\leftAngle,\rightAngle)
      = -2 * tan(\segAngle/2) * sin((\leftAngle + \rightAngle)/2);
      lenLY(\segAngle,\leftAngle,\rightAngle)
      = 2 * tan(\segAngle/2) * cos((\leftAngle + \rightAngle)/2);
    }]
    \def \NumSeg {8}                                
    \def \Rad {1}                                   
    \def \Leg {1.5}                                 
    \def \SegAngle {180/\NumSeg}                    
    \def \TpRtAngle {{(\NumSeg - 1)*\SegAngle/2}}   
    \def \TpLtAngle {{(\NumSeg + 1)*\SegAngle/2}}   
    \def \BaLtAngle {{(\NumSeg + 1)*\SegAngle}}     
    \def \BaRtAngle {{(\NumSeg + 2)*\SegAngle}}     
    \def \BbLtAngle {{(\NumSeg + 3)*\SegAngle}}     
    \def \BbRtAngle {{(\NumSeg + 4)*\SegAngle}}     
    \def \BcLtAngle {{(2 * \NumSeg - 2)*\SegAngle}} 
    \def \BcRtAngle {(2 * \NumSeg - 1)*\SegAngle}   
    \draw ([shift=({\BcRtAngle-360}:\Rad)]3,3) arc ({\BcRtAngle-360}:\TpRtAngle:\Rad);
    \draw ([shift=(\TpLtAngle:\Rad)]3,3) arc (\TpLtAngle:\BaLtAngle:\Rad);
    \draw ([shift=(\BaRtAngle:\Rad)]3,3) arc (\BaRtAngle:\BbLtAngle:\Rad);
    \draw[dashed] ([shift=(\BbRtAngle:\Rad)]3,3) arc (\BbRtAngle:\BcLtAngle:\Rad);
    \draw ([shift=(\TpRtAngle:\Rad)]3cm,3cm)
    -- node[right] {\footnotesize $\mathcal{E}^{I_{n_k + 1}}_{\text{DT}}$}
    ++(
    {lenX(\Leg,\TpLtAngle,\TpRtAngle)},
    {lenY(\Leg,\TpLtAngle,\TpRtAngle)}
    )
    -- node[above] {$+$}
    ++(
    {lenLX(\SegAngle,\TpLtAngle,\TpRtAngle)},
    {lenLY(\SegAngle,\TpLtAngle,\TpRtAngle)}
    )
    -- node[left] {\footnotesize $\mathcal{E}^{I_1}_{\text{DT}}$}
    ++(
    -{lenX(\Leg,\TpLtAngle,\TpRtAngle)},
    -{lenY(\Leg,\TpLtAngle,\TpRtAngle)}
    );
    \draw ([shift=(\BaLtAngle:\Rad)]3,3)
    -- node[near end, above left] {\footnotesize $\mathcal{E}^{I_1}_{\text{DT}}$}
    ++(
    {lenX(\Leg,\BaLtAngle,\BaRtAngle)},
    {lenY(\Leg,\BaLtAngle,\BaRtAngle)}
    )
    -- node[left] {$-$}
    ++(
    {lenLX(\SegAngle,\BaLtAngle,\BaRtAngle)},
    {lenLY(\SegAngle,\BaLtAngle,\BaRtAngle)}
    )
    -- node[near start, below] {\footnotesize $\mathcal{E}^{I_2}_{\text{DT}}$}
    ++(
    -{lenX(\Leg,\BaLtAngle,\BaRtAngle)},
    -{lenY(\Leg,\BaLtAngle,\BaRtAngle)}
    );
    \draw ([shift=(\BbLtAngle:\Rad)]3,3)
    -- node[near end, left=1pt] {\footnotesize $\mathcal{E}^{I_2}_{\text{DT}}$}
    ++(
    {lenX(\Leg,\BbLtAngle,\BbRtAngle)},
    {lenY(\Leg,\BbLtAngle,\BbRtAngle)}
    )
    -- node[below] {$-$}
    ++(
    {lenLX(\SegAngle,\BbLtAngle,\BbRtAngle)},
    {lenLY(\SegAngle,\BbLtAngle,\BbRtAngle)}
    )
    -- node[near start, right] {\footnotesize $\mathcal{E}^{I_3}_{\text{DT}}$}
    ++(
    -{lenX(\Leg,\BbLtAngle,\BbRtAngle)},
    -{lenY(\Leg,\BbLtAngle,\BbRtAngle)}
    );
    \draw ([shift=(\BcLtAngle:\Rad)]3,3)
    -- node[near end, below] {\footnotesize $\mathcal{E}^{I_{n_k}}_{\text{DT}}$}
    ++(
    {lenX(\Leg,\BcLtAngle,\BcRtAngle)},
    {lenY(\Leg,\BcLtAngle,\BcRtAngle)}
    )
    -- node[right] {$-$}
    ++(
    {lenLX(\SegAngle,\BcLtAngle,\BcRtAngle)},
    {lenLY(\SegAngle,\BcLtAngle,\BcRtAngle)}
    )
    -- node[near start, above right] {\footnotesize $\mathcal{E}^{I_{n_k + 1}}_{\text{DT}}$}
    ++(
    -{lenX(\Leg,\BcLtAngle,\BcRtAngle)},
    -{lenY(\Leg,\BcLtAngle,\BcRtAngle)}
    );
  \end{tikzpicture}
  \caption[]{Tree-level scattering BPS worldsheet of incoming ($-$) and outgoing ($+$) LG $\mathfrak{A}_6^\theta$-soliton strings.}
  \label{fig:cy3 x r2:mu-d maps}
\end{figure}

The spectrum of Spin$(7)$ theory is given by the $\mathcal{Q}$-cohomology of operators.
In particular, its normalized 8d partition function will be a sum over the free-field correlation functions of these operators.\footnote{%
  Recall from \autoref{ft:ellipticity of spin7 instanton equation} that Spin$(7)$ theory is a balanced TQFT, whence the (normalized) 8d Spin$(7)$ partition function can be computed by bringing down interaction terms to absorb fermion pair zero-modes in the path integral measure.
  These interaction terms can be regarded as operators of the free-field theory that are necessarily in the $\mathcal{Q}$-cohomology (since the non-vanishing partition function ought to remain $\mathcal{Q}$-invariant), where their contribution to the partition function can be understood as free-field correlation functions.
  \label{ft:8d part-fn as sum of free-field correlators}
}%
~As our Spin$(7)$ theory is semi-classical, these correlation functions will correspond to tree-level scattering only.
From the equivalent LG SQM and 2d gauged LG model perspective, the $\mathcal{Q}$-cohomology will be spanned by the LG $\mathfrak{A}_6^\theta$-soliton strings defined by~\eqref{eq:cy3 x r2:lg soliton:eqn}.
In turn, this means that the normalized Spin$(7)$ partition function can also be regarded as a sum over tree-level scattering amplitudes of these LG $\mathfrak{A}_6^{\theta}$-soliton strings.
The BPS worldsheet underlying such a tree-level scattering amplitude is shown in~\autoref{fig:cy3 x r2:mu-d maps}.\footnote{%
  Here, we have exploited the topological and hence conformal invariance of the soliton string theory to replace the outgoing LG $\mathfrak{A}_6^\theta$-soliton  strings with their vertex operators on the disc, then used their coordinate-independent operator products to reduce them to a single vertex operator, before finally translating it back as a single outgoing LG $\mathfrak{A}_6$-soliton string.
  \label{ft:reason for single outgoing lg soliton in mud map}
}

In other words, we can express the normalized Spin$(7)$ partition function as
\begin{equation}
  \label{eq:cy3 x r2:partition fn}
  \widetilde{\mathcal{Z}}_{\text{Spin}(7), CY_3 \times \R^2}(G)
  = \sum_{n_k} \mu^{n_k}_{\mathfrak{A}_6},
  \qquad n_k \in \{1, 2, \dots, k-1\}
\end{equation}
where each
\begin{equation}
  \label{eq:cy3 x r2:mu-d maps}
  \boxed{
    \mu^{n_k}_{\mathfrak{A}_6}:
    \bigotimes_{i = 1}^{n_k}
    \text{Hom}\left(
      \mathcal{E}^{I_i}_{\text{DT}}, \mathcal{E}^{I_{i + 1}}_{\text{DT}}
    \right)_-
    \longto
    \text{Hom}\left(
      \mathcal{E}^{I_1}_{\text{DT}}, \mathcal{E}^{I_{n_k+1}}_{\text{DT}}
    \right)_+
  }
\end{equation}
is a scattering amplitude of $n_k$ incoming LG $\mathfrak{A}_6^\theta$-soliton strings $\text{Hom} ( \mathcal{E}^{I_1}_{\text{DT}}, \mathcal{E}^{I_{2}}_{\text{DT}})_-, \dots, \text{Hom} (\mathcal{E}^{I_{n_k}}_{\text{DT}}, \mathcal{E}^{I_{n_k+1}}_{\text{DT}})_-$ and a single outgoing LG $\mathfrak{A}^\theta_6$-soliton string $\text{Hom}(\mathcal{E}^{I_1}_{\text{DT}}, \mathcal{E}^{I_{n_k+1}}_{\text{DT}})_+$ with left and right boundaries as labeled, whose underlying worldsheet shown in~\autoref{fig:cy3 x r2:mu-d maps} can be regarded as a disc with $n_k+1$ vertex operators at the boundary.
That is, $\mu^{n_k}_{\mathfrak{A}_6}$ counts pseudoholomorphic discs with $n_k + 1$ punctures at the boundary that are mapped to $\mathfrak{A}_6$ according to the BPS worldsheet equations~\eqref{eq:cy3 x r2:worldsheet:eqn}.

In turn, this means that $\mu^{n_k}_{\mathfrak{A}_6}$ counts the moduli of solutions to~\eqref{eq:cy3 x r2:bps:complex} (or equivalently~\eqref{eq:cy3 x r2:bps}) with $n_k + 1$ boundary conditions that can be described as follows.
First, note that we can regard $\R^2$ as the effective worldsheet in~\autoref{fig:cy3 x r2:mu-d maps} that we shall denote as $\Omega$, so the Spin$(7)$-manifold can be interpreted as a trivial $CY_3$ fibration over $\Omega$.
Then, at the $n_k + 1$ $\mathfrak{A}_6^\theta$-soliton strings on $\Omega$ where $\tau = \pm \infty$,~\eqref{eq:cy3 x r2:bps:complex} will become~\eqref{eq:cy3 x r2:soliton:config} and~\eqref{eq:cy3 x r2:soliton:aux config} with $A_t, A_\tau \rightarrow 0$, and over the $\mathfrak{A}^\theta_6$-soliton string boundaries on $\Omega$ where $t = \pm \infty$,~\eqref{eq:cy3 x r2:bps:complex} will become~\eqref{eq:cy3 x r2:soliton:endpts:config} which defines $\theta$-deformed DT configurations on $CY_3$.

Note at this point that  the collection of $\mu^{n_k}_{\mathfrak{A}_6}$ maps in~\eqref{eq:cy3 x r2:mu-d maps} can be regarded as composition maps defining an $A_\infty$-structure.

\subtitle{An FS Type $A_\infty$-category of DT Configurations on $CY_3$}

Altogether, this means that the normalized partition function of Spin$(7)$ theory on $CY_3 \times \R^2$ as expressed in~\eqref{eq:cy3 x r2:partition fn}, manifests a \emph{novel} FS type $A_\infty$-category defined by the $\mu^{n_k}_{\mathfrak{A}_6}$ maps~\eqref{eq:cy3 x r2:mu-d maps} and the one-to-one identification~\eqref{eq:cy3 x r2:floer complex:morphism}, where the $k$ objects $\{\mathcal{E}^1_{\text{DT}}, \mathcal{E}^2_{\text{DT}}, \dots, \mathcal{E}^k_{\text{DT}}\}$ correspond to ($\theta$-deformed) DT configurations on $CY_3$ (with the scalar being zero)!

\subtitle{A Physical Proof and Generalization of Haydys' Mathematical Conjecture}

Note that the existence of an FS type $A_\infty$-category of holomorphic vector bundles on $CY_3$ (i.e., DT configurations on $CY_3$ with the scalar being zero) was conjectured by Haydys as an extension of his program of constructing an FS type $A_\infty$-category of three-manifolds in~\cite{haydys-2015-fukay-seidel} to higher dimensions.
Thus, we have furnished a purely physical proof and generalization (when $\theta \neq 0$) of Haydys' mathematical conjecture.


\subsection{An Atiyah-Floer Type Correspondence for the FS Type \texorpdfstring{$A_\infty$}{A-infty}-category of DT Configurations on \texorpdfstring{$CY_3$}{CY3}, and a Hom-category}
\label{sec:fs-cat of m6:atiyah-floer}

\subtitle{Intersecting Thimbles as Intersecting Branes}

Notice that the setting of this section, i.e., Spin$(7)$ theory on $CY_3 \times \R^2$, is the same as that in~\autoref{sec:atiyah-floer:tyurin-degen} with $M_1 = \R$, where we performed  a Tyurin degeneration of $CY_3$ along a $CY_2$ surface.
By the same arguments that made use of~\eqref{eq:t2 x i x r:a-model:partition fn} to lead us to~\eqref{eq:atiyah-floer:spin7:equality of partition fn}, we can make use of~\eqref{eq:i x s x r2:a-model:partition fn} to get
\begin{equation}
  \label{eq:cy3 x r2:atiyah-floer:equality of partition fn}
  \sum_j \text{HF}^{\text{Spin}(7)\text{-inst}, \theta}_{d_j} (CY_3 \times \R, G)
  =
  \sum_s \text{HSF}^{\text{Int}}_{d_s} \left(
    \mathcal{M} \left( \R, L \mathcal{M}^{G, \theta, CY_2}_{\text{inst}} \right), \mathcal{P}_0, \mathcal{P}_1
  \right)
  \, ,
\end{equation}
where ``$\text{Spin}(7)\text{-inst}, \theta$'' in the superscript of the LHS refers to the fact that it is a $\theta$-generalized Spin$(7)$ instanton Floer homology class ``assigned to'' $CY_3 \times \R$, derived from Spin$(7)$ theory on $CY_3 \times \R^2$ with the $\R^2$ plane rotated by an angle $\theta$, as per our formulation in this section thus far.
This is simply a $\theta$-generalization of~\eqref{eq:g2 x r:partition fn}, i.e., we can write~\eqref{eq:cy3 x r2:atiyah-floer:equality of partition fn} as
\begin{equation}
  \label{eq:cy3 x r2:atiyah-floer:partition fn}
  \mathcal{Z}_{\text{Spin}(7)^{\theta}, CY_3 \times \R^2}(G)
  =
  \sum_s \text{HSF}^{\text{Int}}_{d_s} \left(
    \mathcal{M} \left( \R, L \mathcal{M}^{G, \theta, CY_2}_{\text{inst}} \right), \mathcal{P}_0, \mathcal{P}_1
  \right)
  \, .
\end{equation}
In turn, from~\eqref{eq:cy3 x r2:lg sqm:partition fn}, this can be written as
\begin{equation}
  \label{eq:cy3 x r2:atiyah-floer:partition fn:with solitons}
  \sum_{I \neq J = 1}^k \sum_{p^{IJ}_{\text{DT}, \pm} \in S^{IJ}_{\text{DT}}}
  \text{HF}^{G}_{d_p}(p_{\text{DT},\pm}^{IJ}(\theta))
  =
  \sum_s \text{HSF}^{\text{Int}}_{d_s} \left(
    \mathcal{M} \left( \R, L \mathcal{M}^{G, \theta, CY_2}_{\text{inst}} \right), \mathcal{P}_0, \mathcal{P}_1
  \right)
  \, ,
\end{equation}
which implies that
\begin{equation}
  \label{eq:cy3 x r2:atiyah-floer:correspondence}
  \text{HF}^{G}_* (p_{\text{DT},\pm}^{IJ}(\theta))
  \cong
  \text{HSF}^{\text{Int}}_* \left(
    \mathcal{M} \left( \R, L \mathcal{M}^{G, \theta, CY_2}_{\text{inst}} \right), \mathcal{P}_0, \mathcal{P}_1
  \right)
  \, .
\end{equation}
Thus, we have a correspondence between a gauge-theoretic Floer homology generated by intersecting \emph{thimbles} and a symplectic intersection Floer homology generated by intersecting \emph{branes}!

\subtitle{An Atiyah-Floer Type Correspondence for the FS Type $A_{\infty}$-category of DT Configurations on $CY_3$}

Moreover, via~\eqref{eq:cy3 x r2:floer complex:morphism}, we would have the following one-to-one identification\footnote{%
  We have restored the `$\theta$' label in the following expression, as both sides of the correspondence depend on the choice of $\theta$ used to rotate the $\R^2$ plane.
  \label{ft:restoration of theta label in fs-cat of cy3}
}
\begin{equation}
  \label{eq:cy3 x r2:atiyah-floer:as morphism}
  \boxed{
    \text{Hom} \left(
      \mathcal{E}^I_{\text{DT}}(\theta), \mathcal{E}^J_{\text{DT}}(\theta)
    \right)_\pm
    \Longleftrightarrow
    \text{HSF}^{\text{Int}}_* \left(
      \mathcal{M} \left( \R, L \mathcal{M}^{G, \theta, CY_2}_{\text{inst}} \right), \mathcal{P}_0, \mathcal{P}_1
    \right)
  }
\end{equation}
This means that we now have an FS type $A_{\infty}$-category defined by the $\mu^{n_k}_{\mathfrak{A}_6}$ composition maps~\eqref{eq:cy3 x r2:mu-d maps} and the one-to-one identification~\eqref{eq:cy3 x r2:atiyah-floer:as morphism}.
The $k$ objects $\{ \mathcal{E}^1_{\text{DT}}(\theta), \mathcal{E}^2_{\text{DT}}(\theta), \dots, \mathcal{E}^k_{\text{DT}}(\theta) \}$ corresponding to $\theta$-deformed DT configurations on $CY_3$, are now related to intersecting A$_{\theta}$-branes $\mathcal{P}_{*}(\theta)$ in $\mathcal{M}(\R, L \mathcal{M}^{G, \theta, CY_2}_{\text{inst}})$.

In other words, we have a \emph{novel} Atiyah-Floer type correspondence for the FS type $A_{\infty}$-category of DT configurations on $CY_3$!

\subtitle{The Soliton as a Hom-category}

At $\theta = 0, \pi$, $\mathcal{E}^*_{\text{DT}}(\theta)$, corresponding to $\theta$-deformed DT configurations on $CY_3$, are regular DT configurations on $CY_3$.
Such configurations, according to~\autoref{sec:floer homology of m6}, will generate a holomorphic $G_2$ \emph{instanton} Floer homology of $CY_3$.
Therefore, via the 7d-Spin$(7)$ Atiyah-Floer duality of $CY_3$ in~\eqref{eq:atiyah-floer:7d-spin7}, we can identify $\mathcal{E}^I_{\text{DT}}(0)$ and $\mathcal{E}^I_{\text{DT}}(\pi)$ with a class in a symplectic intersection Floer homology generated by intersecting isotropic-coisotropic branes $\mathcal{L}^I_0$ and $\mathcal{L}^I_1$ in $L \mathcal{M}^{G, CY_2}_{\text{inst}}$, i.e.,
\begin{equation}
  \label{eq:cy3 x r2:atiyah-floer:dt config:symp floer-hom}
  \mathcal{E}^I_{\text{DT}}(0)
  \Longleftrightarrow
  \text{HSF}^{\text{Int}}_* \left(
    L \mathcal{M}^{G, CY_2}_{\text{inst}}, \mathcal{L}^I_0, \mathcal{L}^I_1
  \right)
  \Longleftrightarrow
  \mathcal{E}^I_{\text{DT}}(\pi)
  \, .
\end{equation}
Then, this means that for general $\theta$, $\mathcal{E}^*_{\text{DT}}(\theta)$ can be identified with a class in a symplectic intersection Floer homology generated by $\theta$-deformed isotropic-coisotropic branes $\mathcal{L}^I_0(\theta)$ and $\mathcal{L}^I_1(\theta)$ in $L \mathcal{M}^{G, CY_2, \theta}_{\text{inst}}$, i.e.,
\begin{equation}
  \label{eq:cy3 x r2:atiyah-floer:dt config:symp floer-hom:theta}
  \mathcal{E}^I_{\text{DT}}(\theta)
  \Longleftrightarrow
  \text{HSF}^{\text{Int}}_* \left(
    L \mathcal{M}^{G, \theta, CY_2}_{\text{inst}}, \mathcal{L}^I_0(\theta), \mathcal{L}^I_1(\theta)
  \right)
  \, .
\end{equation}

However, notice that the classes on the RHS of~\eqref{eq:cy3 x r2:atiyah-floer:dt config:symp floer-hom:theta} correspond to open string states of the 2d A$_\theta$-model with branes $\mathcal{L}^I_0(\theta)$ and $\mathcal{L}^I_1(\theta)$, whence we can interpret them as $\text{Hom}[\mathcal{L}^I_0(\theta), \mathcal{L}^I_1(\theta)]$, i.e.,
\begin{equation}
  \label{eq:cy3 x r2:atiyah-floer:2d model as morphism}
  \text{HSF}^{\text{Int}}_* \left(
    L \mathcal{M}^{G, \theta, CY_2}_{\text{inst}}, \mathcal{L}^I_0(\theta), \mathcal{L}^I_1(\theta)
  \right)
  \Longleftrightarrow
  \text{Hom}\left[ \mathcal{L}^I_0(\theta), \mathcal{L}^I_1(\theta) \right]
  \, .
\end{equation}
This must mean that the LHS of~\eqref{eq:cy3 x r2:atiyah-floer:as morphism} can be identified as
\begin{equation}
  \label{eq:cy3 x r2:atiyah-floer:dt config:as morphism}
  \boxed{
    \text{Hom}\left(
      \mathcal{E}^I_{\text{DT}}(\theta), \mathcal{E}^J_{\text{DT}}(\theta)
    \right)_\pm
    \Longleftrightarrow
    \text{Hom} \left(
      \text{Hom}\left[ \mathcal{L}^I_0(\theta), \mathcal{L}^I_1(\theta) \right],
      \text{Hom}\left[ \mathcal{L}^J_0(\theta), \mathcal{L}^J_1(\theta) \right]
    \right)_\pm
  }
\end{equation}

In other words, the morphisms defining an FS type $A_{\infty}$-category of $\theta$-deformed DT configurations on $CY_3$ can be identified as a Hom-category with objects themselves being morphisms between isotropic-coisotropic branes in $L \mathcal{M}^{G, \theta, CY_2}_{\text{inst}}$.

\subtitle{Intersecting Branes as a Hom-category}

Finally, by applying~\eqref{eq:cy3 x r2:atiyah-floer:dt config:as morphism} to~\eqref{eq:cy3 x r2:atiyah-floer:as morphism}, we would have the one-to-one identification
\begin{equation}
  \label{eq:cy3 x r2:atiyah-floer:intersection floer as hom-cat}
  \boxed{
    \text{HSF}^{\text{Int}}_* \left(
      \mathcal{M} \left( \R, L \mathcal{M}^{G, \theta, CY_2}_{\text{inst}} \right), \mathcal{P}_0, \mathcal{P}_1
    \right)
    \Longleftrightarrow
    \text{Hom} \left(
      \text{Hom}\left[ \mathcal{L}^I_0(\theta), \mathcal{L}^I_1(\theta) \right],
      \text{Hom}\left[ \mathcal{L}^J_0(\theta), \mathcal{L}^J_1(\theta) \right]
    \right)_\pm
  }
\end{equation}
between a symplectic intersection Floer homology of intersecting branes and a Hom-category!

This identification is indeed a consistent one as follows.
Recall that the LHS of~\eqref{eq:cy3 x r2:atiyah-floer:intersection floer as hom-cat} actually corresponds to open three-brane states of a 4d sigma model on $I \times S^1 \times \R^2$ with target space $\mathcal{M}^{G, \theta}_{\text{inst}}(CY_2)$.
In turn, these open three-brane states can be understood as morphisms between the open string states of two 2d sigma models on $I \times \R$ with target space $L \mathcal{M}^{G, \theta, CY_2}_{\text{inst}}$ and branes $\mathcal{L}^*_0(\theta)$ and $\mathcal{L}^*_1(\theta)$.
From~\autoref{sec:atiyah-floer:7d-spin7}  and the generalization to general $\theta$ above, such open string states of 2d sigma models are given by symplectic intersection Floer homology classes generated by intersections of isotropic-coisotropic branes of $L \mathcal{M}^{G, \theta, CY_2}_{\text{inst}}$.

In other words, we would have
\begin{equation}
  \label{eq:cy3 x r2:atiyah-floer:intersection floer as hom-cat:verification}
  \begin{gathered}
    \text{HSF}^{\text{Int}}_* \left(
      \mathcal{M} \left( \R, L \mathcal{M}^{G, \theta, CY_2}_{\text{inst}} \right), \mathcal{P}_0, \mathcal{P}_1
    \right)
    \\
    \cong
    \text{Hom} \left(
      \text{HSF}^{\text{Int}}_* \left(
        L \mathcal{M}^{G, \theta, CY_2}_{\text{inst}}, \mathcal{L}^M_0(\theta), \mathcal{L}^M_1(\theta)
      \right),
      \text{HSF}^{\text{Int}}_* \left(
        L \mathcal{M}^{G, \theta, CY_2}_{\text{inst}}, \mathcal{L}^N_0(\theta), \mathcal{L}^N_1(\theta)
      \right)
    \right)
    \, .
  \end{gathered}
\end{equation}
Then, via~\eqref{eq:cy3 x r2:atiyah-floer:2d model as morphism}, the bottom line of~\eqref{eq:cy3 x r2:atiyah-floer:intersection floer as hom-cat:verification} will become the RHS of~\eqref{eq:cy3 x r2:atiyah-floer:intersection floer as hom-cat}, thus concluding the consistency check.


\section{A Fukaya-Seidel Type \texorpdfstring{$A_\infty$}{A-infty}-category of Five-Manifolds}
\label{sec:fs-cat of m5}

In this section, we will specialize to the case where $CY_3 = CY_2 \times S^1 \times S^1$, and perform a KK dimensional reduction of Spin$(7)$ theory by shrinking one of the $S^1$ circles to be infinitesimally small.
Recasting the resulting 7d-Spin$(7)$ theory as either a 2d gauged LG model on $\R^2$ or a 1d LG SQM in path space, we will, via the 7d-Spin$(7)$ partition function, physically realize a novel FS type $A_\infty$-category of $CY_2 \times S^1$ whose objects correspond to HW configurations on $CY_2 \times S^1$.

\subsection{7d-\texorpdfstring{Spin$(7)$}{Spin(7)} Theory on \texorpdfstring{$CY_2 \times S^1 \times \R^2$}{CY2 x S1 x R2} as a 2d Model on \texorpdfstring{$\R^2$}{R2} or SQM in Path Space}
\label{sec:fs-cat of m5:2d model or 1d sqm}


\subtitle{7d-Spin$(7)$ Theory on $CY_2 \times S^1 \times \R^2$}

Let us take $x^2$ and $x^3$ as the coordinates on $S^1 \times S^1$, and relabel $x^3$ as $y$.
We first perform a KK reduction of Spin$(7)$ theory on $CY_2 \times S^1 \times S^1 \times \R^2$ along the circle in the direction of $x^2$, i.e., set $\partial_2 \rightarrow 0$, and relabel $A_2 = C \in \Omega^0(S^1, \text{ad}(G)) \otimes \Omega^0(CY_2, \text{ad}(G))$ and $A_y = \Gamma \in \Omega^1(S^1, \text{ad}(G)) \otimes \Omega^0(CY_2, \text{ad}(G))$.\footnote{%
  That $C$ is a scalar (scalar) and $\Gamma$ is a one-form (scalar) on $S^1$ ($CY_2$) is explained in~\autoref{ft:reason for fields being scalars on other cy3 circle}.
  \label{ft:quoting reason for fields being scalars on other cy3 cicle.}
}%
~Doing so, we will get 7d-Spin$(7)$ theory on $CY_2 \times S^1 \times \R^2$.

Second, using the $(z^2, z^3)$ coordinates defined in~\autoref{sec:floer homology of m6} as the holomorphic coordinates on $CY_2$, we will have fields $(\mathcal{A}, \mathcal{B}, \Gamma)$ similar to what we had in~\autoref{sec:floer homology of m5}.
Here, the components of $\mathcal{A} \in \Omega^0(S^1, \text{ad}(G)) \otimes \Omega^{(1, 0)}(CY_2, \text{ad}(G))$ are as defined in~\autoref{sec:floer homology of m6}, and the linearly-independent components of $\mathcal{B} \in \Omega^0(S^1, \text{ad}(G)) \otimes \Omega^{2, +}(CY_2, \text{ad}(G))$ are $\mathcal{B}_{2\bar{2}} = \frac{i}{2} C$ and $\mathcal{B}_{23} = 0 = \mathcal{B}_{2\bar{3}}$.
The conditions that minimize the 7d-Spin$(7)$ action on $CY_2 \times S^1 \times \R$ are obtained by performing a KK reduction along an $S^1$ circle of~\eqref{eq:cy3 x r2:bps:complex} when $CY_3 = CY_2 \times S^1 \times S^1$, i.e.,
\begin{equation}
  \label{eq:cy2 x s x r2:bps}
  \begin{aligned}
    F_{\tau t}
    &= - \frac{1}{2} \omega^{p\bar{q}} \left(
      \partial_y \mathcal{B}_{p\bar{q}}
      - 2 \mathcal{F}_{p\bar{q}}
      + [\Gamma, \mathcal{B}_{p\bar{q}}]
      \right)
      \, , \\
    (D_\tau \mathcal{A}_p - \partial_p A_\tau)
    + i (D_t \mathcal{A}_p - \partial_p A_t)
    &= \frac{1}{2} \varepsilon_{pq} \left(
      \mathcal{D}^q \Gamma
      - \mathcal{D}_{\bar{s}} \mathcal{B}^{\bar{s}q}
      - \partial_y \mathcal{A}^q
    \right)
      \, , \\
    \frac{1}{2} \omega^{p\bar{q}} D_\tau \mathcal{B}_{p\bar{q}}
    + D_t \Gamma - \partial_y A_t
    &= - \frac{1}{2} \text{Im}(\varepsilon_{pq} \mathcal{F}^{pq})
      \, , \\
    D_\tau \Gamma - \partial_y A_\tau
    - \frac{1}{2} \omega^{p\bar{q}} D_t \mathcal{B}_{p\bar{q}}
    &= \frac{1}{2} \text{Re}(\varepsilon_{pq} \mathcal{F}^{pq})
      \, .
  \end{aligned}
\end{equation}

Third, noting that we are physically free to rotate $\R^2$ about the origin by an angle $\theta$,~\eqref{eq:cy2 x s x r2:bps} becomes
\begin{equation}
  \label{eq:cy2 x s x r2:bps:rotated}
  \begin{aligned}
    F_{\tau t}
    &= - \frac{1}{2} \omega^{p\bar{q}} \left(
      \partial_y \mathcal{B}_{p\bar{q}}
      - 2 \mathcal{F}_{p\bar{q}}
      + [\Gamma, \mathcal{B}_{p\bar{q}}]
      \right)
      \, , \\
    (D_\tau \mathcal{A}_p - \partial_p A_\tau)
    + i (D_t \mathcal{A}_p - \partial_p A_t)
    &= \frac{1}{2} e^{i\theta} \varepsilon_{pq} \left(
      \mathcal{D}^q \Gamma
      - \mathcal{D}_{\bar{s}} \mathcal{B}^{\bar{s}q}
      - \partial_y \mathcal{A}^q
      \right)
      \, , \\
    \frac{1}{2} \omega^{p\bar{q}} D_\tau \mathcal{B}_{p\bar{q}}
    + D_t \Gamma
    - \partial_y A_t
    &= - \frac{1}{2} \text{Im}(e^{i\theta} \varepsilon_{pq} \mathcal{F}^{pq})
      \, , \\
    D_\tau \Gamma
    - \frac{1}{2} \omega^{p\bar{q}} D_t \mathcal{B}_{p\bar{q}}
    - \partial_y A_\tau
    &= \frac{1}{2} \text{Re}(e^{i\theta} \varepsilon_{pq} \mathcal{F}^{pq})
      \, .
  \end{aligned}
\end{equation}
This allows us to write the action of 7d-Spin$(7)$ theory on $CY_2 \times S^1 \times \R^2$ as
\begin{equation}
  \label{eq:cy2 x s x r2:action}
  \begin{aligned}
    S_{\text{7d-Spin}(7), CY_2 \times S^1 \times \R^2}
    &= \frac{1}{4e^2} \int_{\R^2} \dd{t} \dd{\tau} \int_{CY_2 \times S^1} \dd{y} \abs{\dd{z}}^4
    \Tr \Bigg(
    \left|F_{\tau t} + \tilde{\kappa} \right|^2
    + 8 \left|D_\tau \mathcal{A}_p
      + i D_t \mathcal{A}_p
      + v_p
    \right|^2
    \\
    & \qquad \qquad
    + \left|\frac{1}{2} \omega^{p\bar{q}} D_\tau \mathcal{B}_{p\bar{q}}
      + D_t \Gamma
      + r
    \right|^2
    + \left|D_\tau \Gamma
      - \frac{1}{2} \omega^{p\bar{q}} D_t \mathcal{B}_{p\bar{q}}
      + \tilde{r}
    \right|^2
    + \dots
    \Bigg) \, ,
  \end{aligned}
\end{equation}
where
\begin{equation}
  \label{eq:cy2 x s x r2:action:components}
  \begin{aligned}
    \tilde{\kappa}
    &= \frac{1}{2} \omega^{p\bar{q}} \left(
      \partial_y \mathcal{B}_{p\bar{q}}
      - 2 \mathcal{F}_{p\bar{q}}
      + [\Gamma, \mathcal{B}_{p\bar{q}}]
      \right) \, ,
    &\qquad
      v_p
    &= - \partial_p \mathcal{A}_\tau
      - i \partial_p A_t
      - \frac{1}{2} e^{i\theta} \varepsilon_{pq} \left(
      \mathcal{D}^q \Gamma
      - \mathcal{D}_{\bar{s}} \mathcal{B}^{\bar{s}q}
      - \partial_y \mathcal{A}^q
      \right) \, , \\
    r
    &= - \partial_y A_t
      + \frac{1}{2} \mathrm{Im} \left(
      e^{i\theta} \varepsilon_{pq} \mathcal{F}^{pq}
      \right) \, ,
    &\qquad
      \tilde{r}
    &= - \partial_y A_\tau
      - \frac{1}{2} \mathrm{Re} \left(
      e^{i\theta} \varepsilon_{pq} \mathcal{F}^{pq}
      \right) \, .
  \end{aligned}
\end{equation}

\subtitle{7d-Spin(7) Theory on $CY_2 \times S^1 \times \R^2$ as a 2d Model}

After suitable rescalings, we can recast~\eqref{eq:cy2 x s x r2:action} as a 2d model on $\R^2$, where its action now reads\footnote{%
  Just as in~\autoref{ft:stokes theorem on cy3 x r2}, we have (i) omitted terms with $\partial_{\{y, p, \bar{p}\}}A_{\{t, \tau\}}$, as these boundary terms will vanish when integrated over $CY_2 \times S^1$, and (ii) integrated out an auxiliary scalar field $\mathfrak{H}_5(\tilde{\kappa})$ corresponding to the scalar $\tilde{\kappa}$ of~\eqref{eq:cy2 x s x r2:action:components} in $\mathfrak{A}_5$, whose contribution to the action is $|\mathfrak{H}_5(\tilde{\kappa})|^2$.
  \label{ft:stokes theorem on cy2 x s1 x r2}
}
\begin{equation}
  \label{eq:cy2 x s x r2:action:2d}
  \begin{aligned}
    S_{\text{2d}, \mathfrak{A}_5}
    = \frac{1}{e^2} \int_{\R^2} \dd{t} \dd{\tau}
    \Bigg(
    & |F_{\tau t}|^2
      + |D_\tau \mathcal{A}^P + i D_t \mathcal{A}^P + v^P|^2
      + |D_\tau \mathcal{B}^P + D_t \Gamma^P + r^P|^2
    \\
    & + |D_\tau \Gamma^P - D_t \mathcal{B}^P + \tilde{r}^P|^2
      + \dots
      \Bigg) \, ,
  \end{aligned}
\end{equation}
where $(\mathcal{A}^P, \mathcal{B}^P, \Gamma^P)$ and $P$ are holomorphic coordinates and indices on the space $\mathfrak{A}_5$ of irreducible $(\mathcal{A}_p, \mathcal{B}_{p\bar{q}}, \Gamma)$ fields on $CY_2 \times S^1$, and $(v^P, r^P, \tilde{r}^P)$ will correspond to $(v_p, r, \tilde{r})$ in~\eqref{eq:cy2 x s x r2:action:components}.

In other words, 7d-Spin$(7)$ theory on $CY_2 \times S^1 \times \R^2$ can be regarded as a 2d gauged sigma model along the $(t, \tau)$-directions with target space $\mathfrak{A}_5$ and action~\eqref{eq:cy2 x s x r2:action:2d}.
We will now further recast this 2d gauged sigma model as a 1d SQM.

\subtitle{The 2d Model on $\R^2$ with Target Space $\mathfrak{A}_5$ as a 1d SQM}

Singling out $\tau$ as the direction in ``time'', the equivalent SQM action can be obtained from~\eqref{eq:cy2 x s x r2:action:2d} after suitable rescalings as\footnote{%
  Just as in~\autoref{ft:aux fields of cy3 x r2 sqm}, we have integrated out $A_{\tau}$ and omitted the term containing $A_t$ in the resulting SQM.
}
\begin{equation}
  \label{eq:cy2 x s x r2:sqm:action}
  \begin{aligned}
    S_{\text{SQM}, \mathcal{M}(\R, \mathfrak{A}_5)}
    = \frac{1}{e^2} \int \dd{\tau}
    \Bigg(
    & \left| \partial_\tau \mathcal{A}^\alpha
      + g^{\alpha\beta}_{\mathcal{M}(\R, \mathfrak{A}_5)} \pdv{h_5}{\mathcal{A}^\beta}
      \right|^2
      + \left| \partial_\tau \mathcal{B}^\alpha
      + g^{\alpha\beta}_{\mathcal{M}(\R, \mathfrak{A}_5)} \pdv{h_5}{\mathcal{B}^\beta}
      \right|^2
    \\
    & + \left| \partial_\tau \Gamma^\alpha
      + g^{\alpha\beta}_{\mathcal{M}(\R, \mathfrak{A}_5)} \pdv{h_5}{\Gamma^\beta}
      \right|^2
      + \dots
      \Bigg) \, ,
  \end{aligned}
\end{equation}
where $(\mathcal{A}^\alpha, \mathcal{B}^\alpha, \Gamma^\alpha)$ and $(\alpha, \beta)$ are holomorphic coordinates and indices on the path space $\mathcal{M}(\R, \mathfrak{A}_5)$ of maps from $\R$ to $\mathfrak{A}_5$; $g_{\mathcal{M}(\R, \mathfrak{A}_5)}$ is the metric on $\mathcal{M}(\R, \mathfrak{A}_5)$; and $h_5(\mathcal{A}, \mathcal{B}, \Gamma)$ is the potential function.

In other words, 7d-Spin$(7)$ theory on $CY_2 \times S^1 \times \R^2$ can also be regarded as a 1d SQM along $\tau$ in $\mathcal{M}(\R, \mathfrak{A}_5)$ whose action is~\eqref{eq:cy2 x s x r2:sqm:action}.

\subsection{Non-constant Paths, Solitons, and HW Configurations}
\label{sec:fs-cat of m5:solitons and hw configs}

\subtitle{$\theta$-deformed, Non-constant Paths in the SQM}

By following the same analysis in~\autoref{sec:fs-cat of m6:solitons and dt configs}, we find that the equivalent 1d SQM of 7d-Spin$(7)$ theory on $CY_2 \times S^1 \times \R^2$ will localize onto \emph{$\tau$-invariant, $\theta$-deformed}, non-constant paths in $\mathcal{M}(\R, \mathfrak{A}_5)$ which will correspond, in the 2d gauged sigma model with target space $\mathfrak{A}_5$, to $\tau$-invariant, $\theta$-deformed solitons along the $t$-direction. We shall refer to these solitons as $\mathfrak{A}_5^\theta$-solitons.

\subtitle{$\mathfrak{A}_5^\theta$-solitons in the 2d Gauged Model}

Specifically, such $\mathfrak{A}_5^\theta$-solitons are defined by
\begin{equation}
  \label{eq:cy2 x s x r2:soliton:eqn}
  \begin{aligned}
    \relax
    [A_\tau, \mathcal{A}^P] + i D_t \mathcal{A}^P + v^P
    &= 0
      \, , \\
    [A_\tau, \mathcal{B}^P] + D_t \Gamma^P + r^P
    &= 0
      \, , \\
    [A_\tau, \Gamma^P] - D_t \mathcal{B}^P + \tilde{r}^P
    &= 0
      \, , \\
  \end{aligned}
\end{equation}
and the condition
\begin{equation}
  \label{eq:cy2 x s x r2:soliton:aux}
  F_{\tau t} = 0 = \mathfrak{H}_5(\tilde{\kappa})
  \, ,
\end{equation}
where $\mathfrak{H}_5(\tilde{\kappa})$ is the auxiliary scalar field defined in~\autoref{ft:stokes theorem on cy2 x s1 x r2}.

\subtitle{$\tau$-independent, $\theta$-deformed 7d-Spin$(7)$ Configurations in 7d-Spin$(7)$ Theory}

In turn, they will correspond, in 7d-Spin$(7)$ theory, to \emph{$\tau$-independent, $\theta$-deformed} 7d-Spin$(7)$ configurations on $CY_2 \times S^1 \times \R^2$ that are defined by
\begin{equation}
  \label{eq:cy2 x s x r2:soliton:config}
  \begin{aligned}
    \partial_t \mathcal{A}_p
    &= \mathcal{D}_p A_t
      - i \mathcal{D}_p \mathcal{A}_\tau
      - \frac{i}{2} e^{i\theta} \varepsilon_{pq} \left(
      \mathcal{D}^q \Gamma
      - \mathcal{D}_{\bar{s}} \mathcal{B}^{\bar{s}q}
      - \partial_y \mathcal{A}^q
      \right) \, , \\
    \partial_t \Gamma
    &= - [A_t, \Gamma]
      - \frac{1}{2} \omega^{p\bar{q}} [A_\tau, \mathcal{B}_{p\bar{q}}]
      + \partial_y A_t
      - \frac{1}{2} \text{Im} \left(
      e^{i\theta} \varepsilon_{pq} \mathcal{F}^{pq}
      \right) \, , \\
    \frac{1}{2} \omega^{p\bar{q}} \partial_t \mathcal{B}_{p\bar{q}}
    &= - \frac{1}{2} \omega^{p\bar{q}} [A_t, \mathcal{B}_{p\bar{q}}]
      + [A_\tau, \Gamma]
      - \partial_y A_\tau
      - \frac{1}{2} \text{Re} \left(
      e^{i\theta} \varepsilon_{pq} \mathcal{F}^{pq}
      \right) \, ,
  \end{aligned}
\end{equation}
 and the conditions
\begin{equation}
  \label{eq:cy2 x s x r2:soliton:aux:config}
  \begin{aligned}
    \partial_t A_\tau
    &= [A_\tau, A_t]
      \, , \\
    0
    &= \omega^{p\bar{q}} \left(
      \partial_y \mathcal{B}_{p\bar{q}}
      - 2 \mathcal{F}_{p\bar{q}}
      + [\Gamma, \mathcal{B}_{p\bar{q}}]
      \right) \, .
  \end{aligned}
\end{equation}

\subtitle{7d-Spin$(7)$ Configurations, $\mathfrak{A}_5^\theta$-solitons, and Non-constant Paths}

In short, these \emph{$\tau$-independent, $\theta$-deformed} 7d-Spin$(7)$ configurations on $CY_2 \times S^1 \times \R^2$ that are defined by~\eqref{eq:cy2 x s x r2:soliton:config} and~\eqref{eq:cy2 x s x r2:soliton:aux:config}, will correspond to the $\mathfrak{A}_5^\theta$-solitons defined by~\eqref{eq:cy2 x s x r2:soliton:eqn} and~\eqref{eq:cy2 x s x r2:soliton:aux}, which, in turn, will correspond to the $\tau$-invariant, $\theta$-deformed, non-constant paths in $\mathcal{M}(\R, \mathfrak{A}_5)$ defined by setting both the LHS and RHS of the expression within the squared terms in~\eqref{eq:cy2 x s x r2:sqm:action} \emph{simultaneously} to zero.

\subtitle{$\mathfrak{A}^\theta_5$-soliton Endpoints Corresponding to $\theta$-deformed HW Configurations on $CY_2 \times S^1$}

Consider now the fixed endpoints of the $\mathfrak{A}_5^\theta$-solitons at $t = \pm \infty$, where we also expect the finite-energy 2d gauge fields $A_t, A_\tau$ to decay to zero.
They are given by~\eqref{eq:cy2 x s x r2:soliton:eqn} and~\eqref{eq:cy2 x s x r2:soliton:aux} with $\partial_t \mathcal{A}^P = 0 = \partial_t \mathcal{B}^P = \partial_t \Gamma^P$ and $A_t, A_\tau \rightarrow 0$.
In turn, they will correspond, in 7d-Spin$(7)$ theory, to $(t, \tau)$-independent, $\theta$-deformed configurations that obey~\eqref{eq:cy2 x s x r2:soliton:config} and~\eqref{eq:cy2 x s x r2:soliton:aux:config} with $\partial_t \mathcal{A}_p = 0 = \partial_t \mathcal{B}_{p\bar{q}} = \partial_t \Gamma$ and $A_t, A_\tau \rightarrow 0$, i.e.,
\begin{equation}
  \label{eq:cy2 x s x r2:soliton:endpts:config}
  \begin{aligned}
    i e^{i\theta} \varepsilon^{pq} \left(
    \mathcal{D}_q \Gamma
    - \mathcal{D}^{\bar{s}} \mathcal{B}_{\bar{s}q}
    - \partial_y \mathcal{A}_q
    \right)
    &= 0
      \, ,
    \\
    e^{i\theta} \varepsilon^{pq} \mathcal{F}_{pq}
    &= 0
      \, ,
    \\
    \omega^{p\bar{q}} \left( \partial_y \mathcal{B}_{p\bar{q}}
    - 2 \mathcal{F}_{p\bar{q}}
    + [\Gamma, \mathcal{B}_{p\bar{q}}]
    \right)
    &= 0
      \, .
  \end{aligned}
\end{equation}

At $\theta = 0, \pi$, \eqref{eq:cy2 x s x r2:soliton:endpts:config} can be written, in the real coordinates of $CY_2 \times S^1$, as\footnote{%
  Recall that $\Gamma$ is actually the gauge connection along the $y$-direction, as explained in \autoref{ft:reminder that gamma is a gauge connection}.
  \label{ft:reminder that gamma is gauge connection in fs cat}
}
\begin{equation}
  \label{eq:cy2 x s x r2:hw eqns}
  F_{ya} + D^b B_{ba} = 0
  \, ,
  \qquad
  F^+_{ab} - \frac{1}{2} D_y B_{ab} = 0
  \, .
\end{equation}
These are the HW equations on $CY_2 \times S^1$ with two of the three linearly-independent components of the self-dual two-form field $B$ being zero.
Configurations spanning the space of solutions to \eqref{eq:cy2 x s x r2:hw eqns} shall, in the rest of this section, be referred to as HW configurations on $CY_2 \times S^1$.

In other words, the $(t, \tau)$-independent, $\theta$-deformed 7d-Spin$(7)$ configurations corresponding to the endpoints of the $\mathfrak{A}_5^\theta$-solitons, are $\theta$-deformed HW configurations on $CY_2 \times S^1$.
We will also assume choices of $CY_2$ satisfying~\autoref{ft:isolation and non-degeneracy of 6d-spin7} whereby such configurations are isolated and non-degenerate.\footnote{%
  At $\theta = 0$, the moduli space of such configurations is the moduli space of undeformed HW configurations on $CY_2 \times S^1$.
  Hence, we can apply the same reasoning as in~\autoref{ft:duy isolation and non-degeneracy} to see that this presumption that the moduli space of $\theta$-deformed HW configurations on $CY_2 \times S^1$ will be made of isolated and non-degenerate points, is justified.
  \label{ft:theta-hw isolation and non-degeneracy}
}

In short, from the equivalent 1d SQM of 7d-Spin$(7)$ theory on $CY_2 \times S^1 \times \R^2$, the theory localizes onto $\tau$-invariant, $\theta$-deformed, non-constant paths in $\mathcal{M}(\R, \mathfrak{A}_5)$, which, in turn, will correspond to $\mathfrak{A}_5^\theta$-solitons in the 2d gauged sigma model whose endpoints correspond to $\theta$-deformed HW configurations on $CY_2 \times S^1$.


\subsection{The 2d Model on \texorpdfstring{$\R^2$}{R2} and an Open String Theory in \texorpdfstring{$\mathfrak{A}_5$}{A5}}
\label{sec:fs-cat of m5:open string theory}


By following the same analysis in~\autoref{sec:fs-cat of m6:open string theory} with~\eqref{eq:cy2 x s x r2:action:2d} as the action for the 2d gauged sigma model on $\R^2$ with target space $\mathfrak{A}_5$, we find that it will define an open string theory in $\mathfrak{A}_5$.
We will now work out the details pertaining to the BPS worldsheets and their boundaries that are necessary to define this open string theory.

\subtitle{BPS Worldsheets of the 2d Model}

The BPS worldsheets of the 2d gauged sigma model with target space $\mathfrak{A}_5$ correspond to its classical trajectories.
Specifically, these are defined by setting to zero the expression within the squared terms in~\eqref{eq:cy2 x s x r2:action:2d}, i.e.,
\begin{equation}
  \label{eq:cy2 x s x r2:worldsheet:bps}
  \begin{gathered}
    F_{\tau t}
    = 0
    \, ,
    \qquad
    \mathfrak{H}_5(\tilde{\kappa})
    = 0
    \, , \\
    \Dv{\mathcal{A}^P}{\tau} + i \Dv{\mathcal{A}^P}{t}
    = - v^P
    \, ,
    \qquad
    \Dv{\mathcal{B}^P}{\tau} + \Dv{\Gamma^P}{t}
    = - r^P
    \, ,
    \qquad
    \Dv{\Gamma^P}{\tau} - \Dv{\mathcal{B}^P}{t}
    = - \tilde{r}^P
    \, .
  \end{gathered}
\end{equation}

\subtitle{BPS Worldsheets with Boundaries Corresponding to $\theta$-deformed HW Configurations on $CY_2 \times S^1$}

The boundaries of the BPS worldsheets are traced out by the endpoints of the $\mathfrak{A}_5^\theta$-solitons as they propagate in $\tau$.
As we have seen at the end of~\autoref{sec:fs-cat of m5:solitons and hw configs}, these endpoints correspond to $\theta$-deformed HW configurations on $CY_2 \times S^1$.
If there are `$l$' such configurations $\{\mathcal{E}^1_{\text{HW}}(\theta), \mathcal{E}^2_{\text{HW}}(\theta), \dots, \mathcal{E}^l_{\text{HW}}(\theta)\}$, just as in \autoref{sec:fs-cat of m6:open string theory}, we can further specify any $\mathfrak{A}_5^\theta$-soliton at $\tau = \pm \infty$ as $\gamma^{IJ}_\pm(t, \theta, \mathfrak{A}_5)$,\footnote{%
  Just as in~\autoref{ft:fixing frakA6-soliton centre of mass dof}, the $\tau$-invariant $\mathfrak{A}_5^\theta$-solitons can be fixed at $\tau = \pm \infty$ by adding physically inconsequential $\mathcal{Q}$-exact terms to the SQM action.
  \label{ft:fixing frakA5-soliton centre of mass dof}
}%
~where its left and right endpoints would correspond to $\mathcal{E}^I_{\text{HW}}(\theta)$ and $\mathcal{E}^J_{\text{HW}}(\theta)$, respectively.

Since the $\mathcal{E}^*_{\text{HW}}(\theta)$'s are $\tau$-independent and therefore, have the same values for all $\tau$, we will have BPS worldsheets of the kind similar to~\autoref{fig:cy3 x r2:mu-1 map}.
This time, however, instead of the boundaries being $\mathcal{E}^*_{\text{DT}}(\theta)$, we will have $\mathcal{E}^*_{\text{HW}}(\theta)$.
And, instead of the solitons at $\tau = \pm \infty$ being $\gamma^{**}_\pm(t, \theta, \mathfrak{A}_6)$, we will have $\gamma^{**}_\pm(t, \theta, \mathfrak{A}_5)$.

\subtitle{The 2d Model on $\R^2$ and an Open String Theory in $\mathfrak{A}_5$}

Thus, like  in \autoref{sec:fs-cat of m6:open string theory}, one can understand the 2d gauged sigma model on $\R^2$ with target space $\mathfrak{A}_5$ to define an open string theory in $\mathfrak{A}_5$ as described above, whose \emph{effective} worldsheet and boundaries are similar to~\autoref{fig:cy3 x r2:mu-1 map}, where $\tau$ and $t$ are the temporal and spatial directions, respectively.


\subsection{Soliton String Theory, the 7d-\texorpdfstring{Spin$(7)$}{Spin(7)} Partition Function, and an FS Type \texorpdfstring{$A_\infty$}{A-infty}-category of HW Configurations on \texorpdfstring{$CY_2 \times S^1$}{CY2 x S1}}
\label{sec:fs-cat of m5:fs cat of hw configs}

\subtitle{The 2d Model as a Gauged LG Model}

Notice that we can also express~\eqref{eq:cy2 x s x r2:worldsheet:bps} as
\begin{equation}
  \label{eq:cy2 x s x r2:bps:complex}
  \begin{aligned}
    F_{\tau t}
    &= 0
      \, ,
    &\qquad
      \mathfrak{H}_5(\tilde{\kappa})
    &= 0
      \, , \\
    \Dv{\mathcal{A}^P}{\tau} + i \Dv{\mathcal{A}^P}{t}
    &= - v^P
      \, ,
    &\qquad
      \Dv{\mathcal{C}^P}{\tau}
      + i \Dv{\mathcal{C}^P}{t}
    &= - (\tilde{r}^P + i r^P)
      \, .
  \end{aligned}
\end{equation}
Here, $\mathcal{A}^P$ and $\mathcal{C}^P = \Gamma^P + i \mathcal{B}^P$ can be interpreted as holomorphic coordinates on the space $\mathscr{A}_5$ of irreducible $(\mathcal{A}_p, \mathcal{C})$ fields on $CY_2 \times S^1$, where $\mathcal{C} = \Gamma + \frac{i}{2} \omega^{p\bar{q}} \mathcal{B}_{p\bar{q}} \in \Omega^0(S^1, \text{ad}(G_\C)) \otimes \Omega^0(CY_2, \text{ad}(G_\C))$ is a scalar on $CY_2 \times S^1$ valued $\text{ad}(G_{\C})$, where $G_{\C}$ is the complexification of $G$.
In turn, this means that we can express the action of the 2d gauged sigma model with target space $\mathfrak{A}_5$ in~\eqref{eq:cy2 x s x r2:action:2d} as
\begin{equation}
  \label{eq:cy2 x s x r2:lg:2d action}
  \begin{aligned}
    S_{\text{LG}, \mathfrak{A}_5}
    = \frac{1}{e^2} \int \dd{t} \dd{\tau} \Bigg(
    & \left| D_\tau \mathcal{A}^P
      + i D_t \mathcal{A}^P
      + i g^{P\bar{Q}}_{\mathscr{A}_5} \left(
      \frac{i \zeta}{2} \pdv{W_5}{\mathcal{A}^Q}
      \right)^*
      \right|^2 \\
    & + \left| D_\tau \mathcal{C}^P
      + i D_t \mathcal{C}^P
      + i g^{P\bar{Q}}_{\mathscr{A}_5} \left(
      \frac{i \zeta}{2} \pdv{W_5}{\mathcal{C}^Q}
      \right)^*
      \right|^2
      + \left| F_{\tau t} \right|^2
      + \dots
      \Bigg) \\
    = \frac{1}{e^2} \int \dd{t} \dd{\tau} \Bigg(
    & \left| D_\sigma \mathcal{A}^P  \right|^2
      + \left| D_\sigma \mathcal{C}^P  \right|^2
      + \left| \pdv{W_5}{\mathcal{A}^P} \right|^2
      + \left| \pdv{W_5}{\mathcal{C}^P} \right|^2
      + \left| F_{\tau t} \right|^2
      + \dots
      \Bigg) \, ,
  \end{aligned}
\end{equation}
where $g_{\mathscr{A}_5}$ is the metric on $\mathscr{A}_5$.
In other words, the 2d gauged sigma model with target space $\mathfrak{A}_5$ can also be interpreted as a 2d gauged LG model in $\mathscr{A}_5$ with holomorphic superpotential $W_5(\mathcal{A}, \mathcal{C})$.

By setting $d_\tau \mathcal{A}^P = 0 = d_\tau \mathcal{C}^P$ and $A_t, A_\tau \rightarrow 0$ in the expression within the squared terms in~\eqref{eq:cy2 x s x r2:lg:2d action}, we can read off the LG $\mathscr{A}_5^\theta$-soliton equations corresponding to $\gamma^{IJ}_\pm(t, \theta, \mathfrak{A}_5)$ (that re-expresses~\eqref{eq:cy2 x s x r2:soliton:eqn} with $A_t, A_\tau \rightarrow 0$) as
\begin{equation}
  \label{eq:cy2 x s x r2:lg soliton:eqn}
  \dv{\mathcal{A}^P}{t}
  = - g^{P\bar{Q}}_{\mathscr{A}_5} \left(
    \frac{i\zeta}{2} \pdv{W_5}{\mathcal{A}^Q}
  \right)^* \, ,
  \qquad
  \dv{\mathcal{C}^P}{t}
  = - g^{P\bar{Q}}_{\mathscr{A}_5} \left(
    \frac{i\zeta}{2} \pdv{W_5}{\mathcal{C}^Q}
  \right)^* \, .
\end{equation}
By setting $d_t \mathcal{A}^P = 0 = d_t \mathcal{C}^P$ in~\eqref{eq:cy2 x s x r2:lg soliton:eqn}, we get the LG $\mathscr{A}_5^\theta$-soliton endpoint equations corresponding to $\gamma^{IJ}(\pm\infty, \theta, \mathfrak{A}_5)$ as
\begin{equation}
  \label{eq:cy2 x s x r2:lg soliton:endpts}
  g^{P\bar{Q}}_{\mathscr{A}_5} \left(
    \frac{i\zeta}{2} \pdv{W_5}{\mathcal{A}^Q}
  \right)^*
  = 0
  \, ,
  \qquad
  g^{P\bar{Q}}_{\mathscr{A}_5} \left(
    \frac{i\zeta}{2} \pdv{W_5}{\mathcal{C}^Q}
  \right)^*
  = 0
  \, .
\end{equation}

Recall from the end of \autoref{sec:fs-cat of m5:solitons and hw configs} that we are only considering certain $CY_2$ such that the endpoints $\gamma^{IJ}(\pm \infty, \theta, \mathfrak{A}_5)$ are isolated and non-degenerate.
Therefore, from their definitions in~\eqref{eq:cy2 x s x r2:lg soliton:endpts} which tell us that they are critical points of $W_5(\mathcal{A},\mathcal{C})$, we conclude that $W_5(\mathcal{A},\mathcal{C})$ can be regarded as a holomorphic Morse function in $\mathscr{A}_5$.

Just as in \autoref{sec:fs-cat of m6:fs cat of dt configs}, this means that an LG $\mathscr{A}_5^\theta$-soliton defined in~\eqref{eq:cy2 x s x r2:lg soliton:eqn} maps to a straight line segment $[W^I_5(\theta), W^J_5(\theta)]$ in the complex $W_5$-plane that starts and ends at the critical values $W^I_5(\theta) \coloneq W_5\left(\gamma^I(-\infty, \theta, \mathfrak{A}_5)\right)$ and $W^J_5(\theta) \coloneq W_5\left(\gamma^J(+\infty, \theta, \mathfrak{A}_5)\right)$, respectively, where its slope depends on $\theta$ (via $\zeta$).
We shall also assume that $\text{Re}\left(W^I_5(\theta)\right) < \text{Re}\left(W^J_5(\theta)\right)$.

\subtitle{The Gauged LG Model as an LG SQM}

With suitable rescalings, we can recast~\eqref{eq:cy2 x s x r2:lg:2d action} as a 1d LG SQM (that re-expresses~\eqref{eq:cy2 x s x r2:sqm:action}), where its action will be given by\footnote{%
  Just as in~\autoref{ft:aux fields of cy3 x r2 sqm}, we have integrated out $A_{\tau}$ and omitted the term containing $A_t$ in the resulting SQM.
  \label{ft:aux fields of cy2 x s1 x r2 sqm}
}
\begin{equation}
  \label{eq:cy2 x s x r2:lg:sqm action}
  S_{\text{LG SQM}, \mathcal{M}(\R, \mathscr{A}_5)} = \frac{1}{e^2} \int \dd{\tau}
  \left(
    \left| \partial_\tau \mathcal{A}^\alpha
      + g^{\alpha \beta}_{\mathcal{M}(\R, \mathscr{A}_5)} \pdv{H_5}{\mathcal{A}^\beta}
    \right|^2
    + \left| \partial_\tau \mathcal{C}^\alpha
      + g^{\alpha \beta}_{\mathcal{M}(\R, \mathscr{A}_5)} \pdv{H_5}{\mathcal{C}^\beta}
    \right|^2
    + \dots
  \right) \, ,
\end{equation}
where $g_{\mathcal{M}(\R, \mathscr{A}_5)}$ is the metric on the path space $\mathcal{M}(\R, \mathscr{A}_5)$ of maps from $\R$ to $\mathscr{A}_5$, and $H_5(\mathcal{A}, \mathcal{C})$ is the \emph{real-valued} potential in $\mathcal{M}(\R, \mathscr{A}_5)$.

The LG SQM will localize onto configurations that \emph{simultaneously} set to zero the LHS and RHS of the expression within the squared terms in~\eqref{eq:cy2 x s x r2:lg:sqm action}.
In other words, it will localize onto $\tau$-invariant critical points of $H_5(\mathcal{A}, \mathcal{C})$ that correspond to the LG $\mathscr{A}_5^\theta$-solitons defined by~\eqref{eq:cy2 x s x r2:lg soliton:eqn}.
For our choice of $CY_2$, the LG $\mathscr{A}_5^\theta$-solitons, just like their endpoints, will be isolated and non-degenerate.
Thus, $H_5(\mathcal{A}, \mathcal{C})$ can be regarded as a real-valued Morse functional in $\mathcal{M}(\R, \mathscr{A}_5)$.

\subtitle{Morphisms from $\mathcal{E}^I_{\text{HW}}(\theta)$ to $\mathcal{E}^J_{\text{HW}}(\theta)$ as Floer Homology Classes of Intersecting Thimbles}

Repeating here the analysis in \autoref{sec:fs-cat of m6:fs cat of dt configs} with~\eqref{eq:cy2 x s x r2:lg:sqm action} as the action for the LG SQM, we find that we can interpret the LG $\mathscr{A}_5^\theta$-soliton solutions as intersections of thimbles.
Specifically, an LG $\mathscr{A}_5^\theta$-soliton pair (corresponding to an $\mathfrak{A}_{5}$-soliton pair $\gamma^{IJ}_\pm(t, \theta, \mathfrak{A}_5)$), whose left and right endpoints correspond to $\mathcal{E}^I_{\text{HW}}(\theta)$ and $\mathcal{E}^J_{\text{HW}}(\theta)$, respectively, can be identified as a pair of transversal intersection points $p^{IJ}_{\text{HW},\pm}(\theta) \in S^{IJ}_{\text{HW}}$ of a left and right thimble in the fiber space over the line segment $[W^I_5(\theta), W^J_5(\theta)]$.

This means that the LG SQM in $\mathcal{M}(\R, \mathscr{A}_5)$ with action~\eqref{eq:cy2 x s x r2:lg:sqm action} will physically realize a Floer homology that we shall name an $\mathscr{A}_5$-LG Floer homology.
The chains of the $\mathscr{A}_5$-LG Floer complex will be generated by LG $\mathscr{A}_5^\theta$-solitons which we can identify with $p^{**}_{\text{HW},\pm}(\theta)$, and the $\mathscr{A}_5$-LG Floer differential will be realized by the flow lines governed by the gradient flow equations satisfied by $\tau$-varying configurations which set the expression within the squared terms in~\eqref{eq:cy2 x s x r2:lg:sqm action} to zero.
In particular, the SQM partition function of the LG SQM in $\mathcal{M}(\R, \mathscr{A}_5)$ will be given by\footnote{%
  Just as in~\autoref{ft:theta omission in lg sqm 8d partition fn}, the `$\theta$' label is omitted in the LHS of the following expression.
  \label{ft:theta omission in lg sqm 7d partition fn}
}
\begin{equation}
  \label{eq:cy2 x s x r2:lg sqm:partition fn}
  \mathcal{Z}_{\text{LG SQM}, \mathcal{M}(\R, \mathscr{A}_5)}(G)
  = \sum_{I \neq J = 1}^l \sum_{p^{IJ}_{\text{HW}, \pm} \in S^{IJ}_{\text{HW}}}
  \text{HF}^{G}_{d_q} (p^{IJ}_{\text{HW},\pm}(\theta))
  \, ,
\end{equation}
where the contribution $\text{HF}^{G}_{d_q} (p^{IJ}_{\text{HW},\pm}(\theta))$ can be identified with a homology class in an $\mathscr{A}_5$-LG Floer homology generated by intersection points of thimbles.
These intersection points represent LG $\mathscr{A}_5^\theta$-solitons whose endpoints correspond to $\theta$-deformed HW configurations on $CY_2 \times S^1$.
The degree of each chain in the complex is $d_q$, and is counted by the number of outgoing flow lines from the fixed critical points of $H_5(\mathcal{A}, \mathcal{C})$ in $\mathcal{M}(\R, \mathscr{A}_5)$ which can also be identified as $p^{IJ}_{\text{HW},\pm}(\theta)$.

Therefore, $\mathcal{Z}_{\text{LG SQM}, \mathcal{M}(\R, \mathscr{A}_5)}(G)$ in~\eqref{eq:cy2 x s x r2:lg sqm:partition fn} is a sum of LG $\mathscr{A}_5^\theta$-solitons defined by~\eqref{eq:cy2 x s x r2:lg soliton:eqn} with endpoints~\eqref{eq:cy2 x s x r2:lg soliton:endpts}, or equivalently, $\gamma^{IJ}_\pm(t, \theta, \mathfrak{A}_5)$-solitons defined by~\eqref{eq:cy2 x s x r2:soliton:config} and~\eqref{eq:cy2 x s x r2:soliton:aux:config} with endpoints~\eqref{eq:cy2 x s x r2:soliton:endpts:config}, whose start and end correspond to $\mathcal{E}^I_{\text{HW}}(\theta)$ and $\mathcal{E}^J_{\text{HW}}(\theta)$, respectively.
In other words, we can write
\begin{equation}
  \label{eq:cy2 x s x r2:floer complex:vector}
  \text{CF}_{\mathfrak{A}_5} \left(
    \mathcal{E}^I_{\text{HW}}(\theta), \mathcal{E}^J_{\text{HW}}(\theta)
  \right)_\pm
  = \text{HF}^{G}_{d_q} (p^{IJ}_{\text{HW},\pm}(\theta))
  \, ,
\end{equation}
where $\text{CF}_{\mathfrak{A}_5}(\mathcal{E}^I_{\text{HW}}(\theta), \mathcal{E}^J_{\text{HW}}(\theta))_\pm$ is a vector representing a $\gamma^{IJ}_\pm(t, \theta, \mathfrak{A}_5)$-soliton, such that $\text{Re}\left(W^I_5(\theta)\right) < \text{Re}\left(W^J_5(\theta)\right)$.
This will lead us to the one-to-one identification\footnote{%
  Just as in \autoref{ft:theta omission in 6d fs-cat morphism}, the `$\theta$' label is omitted in the following expression.
  \label{ft:theta omission in 5d fs-cat morphism}
}
\begin{equation}
  \label{eq:cy2 x s x r2:floer complex:morphism}
  \boxed{
    \text{Hom}(\mathcal{E}^I_{\text{HW}}, \mathcal{E}^J_{\text{HW}})_\pm
    \Longleftrightarrow
    \text{HF}^{G}_{d_q} (p^{IJ}_{\text{HW},\pm})
  }
\end{equation}
where the RHS is proportional to the identity class when $I = J$, and zero when $I \leftrightarrow J$ since the $\mathfrak{A}_5^\theta$-soliton only moves in one direction from $\mathcal{E}^I_{\text{HW}}(\theta)$ to $\mathcal{E}^J_{\text{HW}}(\theta)$.

\subtitle{Soliton String Theory from the 2d LG Model}

Just as in \autoref{sec:fs-cat of m6:fs cat of dt configs}, the 2d gauged LG model in $\mathscr{A}_5$ with action~\eqref{eq:cy2 x s x r2:lg:2d action} can be interpreted as a soliton string theory in $\mathscr{A}_5$.
The dynamics of this soliton string theory in $\mathscr{A}_5$ will be governed by the BPS worldsheet equations determined by setting to zero the expression within the squared terms in \eqref{eq:cy2 x s x r2:lg:2d action}, where $(\mathcal{A}^P, \mathcal{C}^P)$ are scalars on the worldsheet corresponding to the holomorphic coordinates on $\mathscr{A}_5$.
At an arbitrary instant in time whence $d_\tau \mathcal{A}^P = 0 = d_\tau \mathcal{C}^P$ therein, the dynamics of $(\mathcal{A}^P, \mathcal{C}^P)$ along $t$ will be governed by the soliton equations
\begin{equation}
  \label{eq:cy2 x s x r2:string soliton:eqn}
  \dv{\mathcal{A}^P}{t}
  = - [A_t - i A_\tau, \mathcal{A}^P]
  - g^{P\bar{Q}}_{\mathscr{A}_5} \left(
    \frac{i \zeta}{2} \pdv{W_5}{\mathcal{A}^Q}
  \right)^* \, ,
  \quad
  \dv{\mathcal{C}^P}{t}
  = - [A_t - i A_\tau, \mathcal{C}^P]
  - g^{P\bar{Q}}_{\mathscr{A}_5} \left(
    \frac{i \zeta}{2} \pdv{W_5}{\mathcal{C}^Q}
  \right)^* \, .
\end{equation}

\newpage

\subtitle{The Normalized 7d-Spin$(7)$ Partition Function, LG $\mathscr{A}_5^\theta$-soliton String Scattering, and Maps of an $A_\infty$-structure}

Since our 7d-Spin$(7)$ theory on $CY_2 \times S^1 \times \R^2$ is semi-classical, its normalized partition function can be regarded as a sum over tree-level scattering amplitudes of the LG $\mathscr{A}_5^\theta$-soliton strings defined by~\eqref{eq:cy2 x s x r2:lg soliton:eqn}.
The BPS worldsheet underlying such a tree-level scattering amplitude is similar to~\autoref{fig:cy3 x r2:mu-d maps}, where instead of the endpoints of each string being $\mathcal{E}^*_{\text{DT}}$, it is now understood to be $\mathcal{E}^*_{\text{HW}}$.

In other words, we can, like in~\eqref{eq:cy3 x r2:partition fn}, express the normalized 7d-Spin$(7)$ partition function as
\begin{equation}
  \label{eq:cy2 x s x r2:partition fn}
  \widetilde{\mathcal{Z}}_{\text{7d-Spin}(7), CY_2 \times S^1 \times \R^2}(G)
  = \sum_{n_l} \mu^{n_l}_{\mathscr{A}_5},
  \qquad
  n_l \in \{1, 2, \dots, l-1\}
\end{equation}
where each
\begin{equation}
  \label{eq:cy2 x s x r2:mu-d maps}
  \boxed{
    \mu^{n_l}_{\mathscr{A}_5}:
    \bigotimes_{i = 1}^{n_l}
    \text{Hom} \left(
      \mathcal{E}^{I_i}_{\text{HW}}, \mathcal{E}^{I_{i + 1}}_{\text{HW}}
    \right)_-
    \longto
    \text{Hom} \left(
      \mathcal{E}^{I_1}_{\text{HW}}, \mathcal{E}^{I_{n_l+1}}_{\text{HW}}
    \right)_+
  }
\end{equation}
is a scattering amplitude of $n_l$ incoming LG $\mathscr{A}_5^\theta$-soliton strings $\text{Hom}(\mathcal{E}^{I_1}_{\text{HW}}, \mathcal{E}^{I_2}_{\text{HW}})_-, \dots, \text{Hom}(\mathcal{E}^{I_{n_l}}_{\text{HW}}, \mathcal{E}^{I_{n_l+1}}_{\text{HW}})_-$ and a single outgoing LG $\mathscr{A}_5^\theta$-soliton string $\text{Hom}(\mathcal{E}^{I_1}_{\text{HW}}, \mathcal{E}^{I_{n_l+1}}_{\text{HW}})_+$ with left and right boundaries as labeled, whose underlying worldsheet can be regarded as a disc with $n_l+1$ vertex operators at the boundary.
That is, $\mu^{n_l}_{\mathscr{A}_5}$ counts pseudoholomorphic discs with $n_l+1$ punctures at the boundary that are mapped to $\mathscr{A}_5$ according to the BPS worldsheet equations~\eqref{eq:cy2 x s x r2:bps:complex}.

Just as in \autoref{sec:fs-cat of m6:fs cat of dt configs}, the collection of $\mu^{n_l}_{\mathscr{A}_5}$ maps in~\eqref{eq:cy2 x s x r2:mu-d maps} can be regarded as composition maps defining an $A_\infty$-structure.

\subtitle{An FS Type $A_\infty$-category of HW Configurations on $CY_2 \times S^1$}

Altogether, this means that the normalized partition function of 7d-Spin$(7)$ theory on $CY_2 \times S^1 \times \R^2$ as expressed in~\eqref{eq:cy2 x s x r2:partition fn}, manifests a \emph{novel} FS type $A_\infty$-category defined by the $\mu^{n_l}_{\mathscr{A}_5}$ maps~\eqref{eq:cy2 x s x r2:mu-d maps} and the one-to-one identification~\eqref{eq:cy2 x s x r2:floer complex:morphism}, where the $l$ objects $\{\mathcal{E}^1_{\text{HW}}, \mathcal{E}^2_{\text{HW}}, \dots, \mathcal{E}^l_{\text{HW}}\}$ correspond to ($\theta$-deformed) HW configurations on $CY_2 \times S^1$ (with two of the three linearly-independent components of the self-dual two-form field being zero)!

\section{A Fukaya-Seidel Type \texorpdfstring{$A_\infty$}{A-infty}-category of Four-Manifolds}
\label{sec:fs-cat of m4}

In this section, we will perform yet another KK dimensional reduction of 7d-Spin$(7)$ theory by shrinking the remaining $S^1$ circle to be infinitesimally small.
Recasting the resulting 6d-Spin$(7)$ theory as either a 2d gauged LG model on $\R^2$ or a 1d LG SQM in path space, we will, via the  6d-Spin$(7)$ partition function, physically realize a novel FS type $A_\infty$-category whose objects correspond to Vafa-Witten (VW) configurations on $CY_2$.

\subsection{6d-\texorpdfstring{Spin$(7)$}{Spin(7)} Theory on \texorpdfstring{$CY_2 \times \R^2$}{CY2 x R2} as a 2d Model on \texorpdfstring{$\R^2$}{R2} or SQM in Path Space}
\label{sec:fs-cat of m4:2d model or sqm}

\subtitle{6d-Spin$(7)$ Theory on $CY_2 \times \R^2$}

First, note that KK reduction of 7d-Spin$(7)$ theory along the $y$-direction means we have to set $\partial_y \rightarrow 0$.
The two scalar bosons of the resulting 6d-Spin$(7)$ theory on $CY_2 \times \R^2$, i.e., $(C, \Gamma) \in \Omega^0(CY_2, \text{ad}(G))$, just like the two scalar bosons in \autoref{sec:floer homology of m5}, can be assigned to the linearly-independent components of a self-dual two-form.

Second, using the $(z^2, z^3)$ coordinates as the holomorphic coordinates on $CY_2$, we will have fields $(\mathcal{A}, \mathcal{B})$.
Here, the components of $\mathcal{A} \in \Omega^{(1, 0)}(CY_2, \text{ad}(G))$ are as defined in \autoref{sec:floer homology of m6}, and the linearly-independent components of $\mathcal{B} \in \Omega^{2, +}(CY_2, \text{ad}(G))$ are $\mathcal{B}_{2{\bar{2}}} = 0$, $\mathcal{B}_{23} = \frac{1}{2} (\Gamma - i C)$ and $\mathcal{B}_{\bar{2}\bar{3}} = \frac{1}{2} (\Gamma + i C)$.
The conditions that minimize the 6d-Spin$(7)$ theory on $CY_2 \times \R^2$ are obtained by performing a KK reduction along the remaining $S^1$ circle of~\eqref{eq:cy2 x s x r2:bps}, i.e.,
\begin{equation}
  \label{eq:cy2 x r2:bps}
  \begin{aligned}
    F_{\tau t}
    &= \omega^{p\bar{q}} \left(
      \mathcal{F}_{p\bar{q}}
      - \frac{1}{4} (\mathcal{B} \times \mathcal{B})_{p\bar{q}}
    \right)
    \, , \\
    (D_\tau \mathcal{A}_p - \partial_p A_\tau)
    + i (D_t \mathcal{A}_p - \partial_p A_t)
    &= - \mathcal{D}^q \mathcal{B}_{qp}
    \, , \\
    D_\tau \mathcal{B}_{pq} - i D_t \mathcal{B}_{pq}
    &= 2 \mathcal{F}_{pq}
      \, .
  \end{aligned}
\end{equation}

Third, noting that we are physically free to rotate $\R^2$ about the origin by an angle $\theta$,~\eqref{eq:cy2 x r2:bps} becomes
\begin{equation}
  \label{eq:cy2 x r2:bps:rotated}
  \begin{aligned}
    F_{\tau t}
    &= \omega^{p\bar{q}} \left(
      \mathcal{F}_{p\bar{q}}
      - \frac{1}{4} (\mathcal{B} \times \mathcal{B})_{p\bar{q}}
    \right)
    \, , \\
    (D_\tau \mathcal{A}_p - \partial_p A_\tau)
    + i (D_t \mathcal{A}_p - \partial_p A_t)
    &= - e^{i\theta} \mathcal{D}^q \mathcal{B}_{qp}
    \, , \\
    D_\tau \mathcal{B}_{pq} - i D_t \mathcal{B}_{pq}
    &= 2 e^{i\theta} \mathcal{F}_{pq}
      \, .
  \end{aligned}
\end{equation}
This allows us to write the action of 6d-Spin$(7)$ theory on $CY_2 \times \R^2$ as
\begin{equation}
  \label{eq:cy2 x r2:action}
  \begin{aligned}
    S_{\text{6d-Spin}(7),CY_2 \times \R^2}
    = \frac{1}{4e^2} \int_{\R^2} \dd{t} \dd{\tau} \int_{CY_2} \abs{\dd{z}}^4 \Tr \Big(
    &\left| F_{\tau t} + \hat{\kappa} \right|^2
      + 8 \left| D_\tau \mathcal{A}_p
      + i D_t \mathcal{A}_p + w_p
      \right|^2 \\
    & + 4 \left| D_\tau \mathcal{B}_{pq}
      - i D_t \mathcal{B}_{pq}
      + s
      \right|^2
      + \dots
      \Big) \, ,
  \end{aligned}
\end{equation}
where
\begin{equation}
  \label{eq:cy2 x r2:action:components}
  \begin{aligned}
    \hat{\kappa}
    &= - \omega^{p\bar{q}} \left(
      \mathcal{F}_{p\bar{q}}
      - \frac{1}{4} (\mathcal{B} \times \mathcal{B})_{p\bar{q}}
      \right)
      \, , \\
    w_p
    &= - \partial_p A_\tau
      - i \partial_p A_t
      + e^{i\theta} \mathcal{D}^q \mathcal{B}_{qp}
      \, , \\
    s
    &= - 2 e^{i\theta} \mathcal{F}_{pq}
      \, .
  \end{aligned}
\end{equation}

\subtitle{6d-Spin$(7)$ Theory on $CY_2 \times \R^2$ as a 2d Model}

After suitable rescalings, we can recast~\eqref{eq:cy2 x r2:action} as a 2d model on $\R^2$, where its action now reads\footnote{%
  Just as in~\autoref{ft:stokes theorem on cy3 x r2}, we have (i) omitted terms with $\partial_p A_{\{t, \tau\}}$ as these boundary terms will vanish when integrated over $CY_2$, and (ii) integrated out an auxiliary scalar field $\mathfrak{H}_4(\hat{\kappa})$ corresponding to the scalar $\hat{\kappa}$ of~\eqref{eq:cy2 x r2:action:components}, whose contribution to the action is $|\mathfrak{H}_4(\hat{\kappa})|^2$.
  \label{ft:stokes theorem on cy2 x r2}
}
\begin{equation}
  \label{eq:cy2 x r2:2d:action}
  S_{\text{2d},\mathfrak{A}_4}
  = \frac{1}{e^2} \int_{\R^2} \dd{t} \dd{\tau} \Big(
    \left| F_{\tau t} \right|^2
    + \left| D_\tau \mathcal{A}^M
      + i D_t \mathcal{A}^M
      + w^M
    \right|^2
    + \left| D_\tau \mathcal{B}^M
      - i D_t \mathcal{B}^M
      + s^M
    \right|^2
    + \dots
  \Big) \, ,
\end{equation}
where $(\mathcal{A}^M, \mathcal{B}^M)$ and $M$ are holomorphic coordinates and indices on the space $\mathfrak{A}_4$ of irreducible $(\mathcal{A}_p, \mathcal{B}_{pq})$ fields on $CY_2$,
and $(w^M, s^M)$ will correspond to $(w_p, s)$ in~\eqref{eq:cy2 x r2:action:components}.

In other words, 6d-Spin$(7)$ theory on $CY_2 \times \R^2$ can be regarded as a 2d gauged sigma model along the $(t, \tau)$-directions with target space $\mathfrak{A}_4$ and action~\eqref{eq:cy2 x r2:2d:action}.
We will now further recast this 2d gauged sigma model as a 1d SQM.

\subtitle{The 2d Model on $\R^2$ with Target Space $\mathfrak{A}_4$ as a 1d SQM}

Singling out $\tau$ as the direction in ``time'', the equivalent SQM action can be obtained from~\eqref{eq:cy2 x r2:2d:action} after suitable rescalings as\footnote{%
  Just as in~\autoref{ft:aux fields of cy3 x r2 sqm}, we have integrated out $A_{\tau}$ and omitted the term containing $A_t$ in the resulting SQM.
  \label{ft:aux fields of cy2 x r2 sqm}
}
\begin{equation}
  \label{eq:cy2 x r2:sqm:action}
  S_{\text{SQM},\mathcal{M}(\R, \mathfrak{A}_4)}
  = \frac{1}{e^2} \int \dd{\tau} \left(
    \left| \partial_\tau \mathcal{A}^\alpha
      + g^{\alpha\beta}_{\mathcal{M}(\R, \mathfrak{A}_4)} \pdv{h_4}{\mathcal{A}^\beta}
    \right|^2
    + \left| \partial_\tau \mathcal{B}^\alpha
      + g^{\alpha\beta}_{\mathcal{M}(\R, \mathfrak{A}_4)} \pdv{h_4}{\mathcal{B}^\beta}
    \right|^2
    + \dots
  \right) \, ,
\end{equation}
where $(\mathcal{A}^\alpha, \mathcal{B}^\alpha)$ and $(\alpha, \beta)$ are holomorphic coordinates and indices on the path space $\mathcal{M}(\R, \mathfrak{A}_4)$ of maps from $\R$ to $\mathfrak{A}_4$; $g_{\mathcal{M}(\R, \mathfrak{A}_4)}$ is the metric on $\mathcal{M}(\R, \mathfrak{A}_4)$; and $h_4(\mathcal{A}, \mathcal{B})$ is the potential function.

In other words, 6d-Spin$(7)$ theory on $CY_2 \times \R^2$ can also be regarded as a 1d SQM along $\tau$ in $\mathcal{M}(\R, \mathfrak{A}_4)$ whose action is~\eqref{eq:cy2 x r2:sqm:action}.

\subsection{Non-constant Paths, Solitons, and VW Configurations}
\label{sec:fs-cat of m4:solitons and kw configs}

\subtitle{$\theta$-deformed, Non-constant Paths in the SQM}

By following the analysis in \autoref{sec:fs-cat of m6:solitons and dt configs}, we find that the equivalent 1d SQM of 6d-Spin$(7)$ theory on $CY_2 \times \R^2$ will localize onto \emph{$\tau$-invariant, $\theta$-deformed}, non-constant paths in $\mathcal{M}(\R, \mathfrak{A}_4)$ which will correspond, in the 2d gauged sigma model with target space $\mathfrak{A}_4$, to $\tau$-invariant, $\theta$-deformed solitons along the $t$-direction.
We shall refer to these solitons as $\mathfrak{A}_4^\theta$-solitons.

\subtitle{$\mathfrak{A}_4^\theta$-solitons in the 2d Gauged Model}

Specifically, such $\mathfrak{A}_4^\theta$-solitons are defined by
\begin{equation}
  \label{eq:cy2 x r2:soliton:eqn}
  \relax
  [A_\tau, \mathcal{A}^M] + i D_t \mathcal{A}^M + w^M
  = 0
  \, ,
  \qquad
  [A_\tau, \mathcal{B}^M] - i D_t \mathcal{B}^M + s^M
  = 0
  \, ,
\end{equation}
and the condition
\begin{equation}
  \label{eq:cy2 x r2:soliton:aux}
  F_{\tau t} = 0 = \mathfrak{H}_4(\hat{\kappa})
  \, ,
\end{equation}
where $\mathfrak{H}_4(\hat{\kappa})$ is the auxiliary scalar field defined in~\autoref{ft:stokes theorem on cy2 x r2}.

\subtitle{$\tau$-independent, $\theta$-deformed 6d-Spin$(7)$ Configurations in 6d-Spin$(7)$ Theory}

In turn, they will correspond, in 6d-Spin$(7)$ theory, to \emph{$\tau$-independent, $\theta$-deformed} 6d-Spin$(7)$ configurations on $CY_2 \times \R^2$ that are defined by
\begin{equation}
  \label{eq:cy2 x r2:soliton:config}
  \begin{aligned}
    \partial_t \mathcal{A}_p
    &= \mathcal{D}_p A_t
      - i \mathcal{D}_p A_\tau
      + i e^{i\theta} \mathcal{D}^q \mathcal{B}_{pq}
      \, , \\
    \partial_t \mathcal{B}_{pq}
    &= - [A_t + i A_\tau, \mathcal{B}_{pq}]
      + 2 i e^{i\theta} \mathcal{F}_{pq}
      \, ,
  \end{aligned}
\end{equation}
and the conditions
\begin{equation}
  \label{eq:cy2 x r2:soliton:aux:config}
  \begin{aligned}
    \partial_\tau A_t
    &= [A_t, A_\tau]
      \, , \\
    0
    &= \omega^{p\bar{q}} \left(
      \mathcal{F}_{p\bar{q}}
      - \frac{1}{4} (\mathcal{B} \times \mathcal{B})_{p\bar{q}}
      \right)
      \, .
  \end{aligned}
\end{equation}

\subtitle{6d-Spin$(7)$ Configurations, $\mathfrak{A}_4^\theta$-solitons, and Non-constant Paths}

In short, these \emph{$\tau$-independent, $\theta$-deformed} 6d-Spin$(7)$ configurations on $CY_2 \times \R^2$ that are defined by~\eqref{eq:cy2 x r2:soliton:config} and~\eqref{eq:cy2 x r2:soliton:aux:config}, will correspond to the $\mathfrak{A}_4^\theta$-solitons defined by~\eqref{eq:cy2 x r2:soliton:eqn} and~\eqref{eq:cy2 x r2:soliton:aux}, which, in turn, will correspond to the $\tau$-invariant, $\theta$-deformed, non-constant paths in $\mathcal{M}(\R, \mathfrak{A}_4)$ defined by setting both the LHS and RHS of the expression within the squared terms in~\eqref{eq:cy2 x r2:sqm:action} \emph{simultaneously} to zero.

\subtitle{$\mathfrak{A}^\theta_4$-soliton Endpoints Corresponding to $\theta$-deformed VW Configurations on $CY_2$}

Consider now the fixed endpoints of the $\mathfrak{A}_4^\theta$-solitons at $t = \pm \infty$, where we also expect the finite-energy 2d gauge fields $A_t, A_\tau$ to decay to zero.
They are given by~\eqref{eq:cy2 x r2:soliton:eqn} and~\eqref{eq:cy2 x r2:soliton:aux} with $\partial_t \mathcal{A}^M = 0 = \partial_t \mathcal{B}^M$ and $A_t, A_\tau \rightarrow 0$.
In turn, they will correspond, in 6d-Spin$(7)$ theory, to $(t, \tau)$-independent, $\theta$-deformed configurations that obey~\eqref{eq:cy2 x r2:soliton:config} and~\eqref{eq:cy2 x r2:soliton:aux:config} with $\partial_t \mathcal{A}_p = 0 = \partial_t \mathcal{B}_{pq}$ and $A_t, A_\tau \rightarrow 0$, i.e.,
\begin{equation}
  \label{eq:cy2 x r2:soliton:endpts:config}
  \begin{aligned}
    i e^{i\theta} \mathcal{D}^q \mathcal{B}_{qp}
    &= 0
      \, ,
    \\
    i e^{i\theta} \mathcal{F}_{pq}
    &= 0
      \, ,
    \\
    \omega^{p\bar{q}} \left(
    \mathcal{F}_{p\bar{q}} - \frac{1}{4} (\mathcal{B} \times \mathcal{B})_{p\bar{q}}
    \right)
    &= 0
      \, .
  \end{aligned}
\end{equation}

At $\theta = 0, \pi$, \eqref{eq:cy2 x r2:soliton:endpts:config} can be written, in the real coordinates of $CY_2$, as
\begin{equation}
  \label{eq:cy2 x r2:vw eqns}
  D^b B_{ba} = 0
  \, ,
  \qquad
  F^+_{ab} - \frac{1}{4} [B_{ac}, B_{bd}] g^{cd} = 0
  \, .
\end{equation}
These are the 4d VW equations on $CY_2$ \cite{vafa-1994-stron-coupl, ong-2023-vafa-witten-theor, er-2023-topol-n, tanaka-2019-vafa-witten} with the scalar and one of the linearly-independent components of the self-dual two-form field being zero.
Configurations spanning the space of solutions to these equations shall, in the rest of this section, be referred to as VW configurations on $CY_2$.

In other words, the $(t, \tau)$-independent, $\theta$-deformed 6d-Spin$(7)$ configurations corresponding to the endpoints of the $\mathfrak{A}_4^\theta$-solitons, are $\theta$-deformed VW configurations on $CY_2$.
For our choice of $CY_2$, such configurations are isolated and non-degenerate.\footnote{%
  At $\theta = 0$, such a moduli space is the undeformed moduli space of VW configurations on $CY_2$.
  Note that such configurations are obtained by a KK reduction along $S^1$ of the undeformed HW configurations on $CY_2 \times S^1$ from~\autoref{sec:fs-cat of m5}.
  Since our choice of $CY_2$ therein satisfies~\autoref{ft:theta-hw isolation and non-degeneracy} such that these undeformed HW configurations on $CY_2 \times S^1$ are isolated, it would mean that the undeformed VW configurations on $CY_2$ must also be isolated.
  We can then apply the same reasoning as~\autoref{ft:duy isolation and non-degeneracy} again to see that this presumption that the moduli space of $\theta$-deformed VW configurations on $CY_2$ will be made of isolated and non-degenerate points, is justified.
  \label{ft:isolation and non-degeneracy of vw configs}
}

In short, from the equivalent 1d SQM of 6d-Spin$(7)$ theory on $CY_2 \times \R^2$, the theory localizes onto $\tau$-invariant, $\theta$-deformed, non-constant paths in $\mathcal{M}(\R, \mathfrak{A}_4)$, which, in turn, will correspond to $\mathfrak{A}_4^\theta$-solitons in the 2d gauged sigma model whose endpoints correspond to $\theta$-deformed VW configurations on $CY_2$.


\subsection{The 2d Model on \texorpdfstring{$\R^2$}{R2} and an Open String Theory in \texorpdfstring{$\mathfrak{A}_4$}{A4}}
\label{sec:fs-cat of m4:open string theory}

By following the same analysis in \autoref{sec:fs-cat of m6:open string theory} with~\eqref{eq:cy2 x r2:2d:action} as the action for the 2d gauged sigma model on $\R^2$ with target space $\mathfrak{A}_4$, we find that it will define an open string theory in $\mathfrak{A}_4$.
We will now work out the details pertaining to the BPS worldsheets and their boundaries that are necessary to define this open string theory.

\subtitle{BPS Worldsheets of the 2d Model}

The BPS worldsheets of the 2d gauged sigma model with target space $\mathfrak{A}_4$ correspond to its classical trajectories.
Specifically, these are defined by setting to zero the expression within the squared terms in~\eqref{eq:cy2 x r2:2d:action}, i.e.,
\begin{equation}
  \label{eq:cy2 x r2:worldsheet:eqn}
  \begin{aligned}
    F_{\tau t}
    &= 0
      \, ,
    &\qquad
      \mathfrak{H}_4(\hat{u})
    &= 0
      \, ,
    \\
    \Dv{\mathcal{A}^M}{\tau} + i \Dv{\mathcal{A}^M}{t}
    &= - w^M
      \, ,
    &\qquad
      \Dv{\mathcal{B}^M}{\tau} - i \Dv{\mathcal{B}^M}{t}
    &= - s^M
      \, .
  \end{aligned}
\end{equation}

\subtitle{BPS Worldsheets with Boundaries Corresponding to $\theta$-deformed VW Configurations on $CY_2$}

The boundaries of the BPS worldsheets are traced out by the endpoints of the $\mathfrak{A}_4^\theta$-solitons as they propagate in $\tau$.
As we have seen at the end of~\autoref{sec:fs-cat of m4:solitons and kw configs}, these endpoints correspond to $\theta$-deformed VW configurations on $CY_2$.
If there are `$m$' such configurations $\{\mathcal{E}^1_{\text{VW}}(\theta), \mathcal{E}^2_{\text{VW}}(\theta), \dots, \mathcal{E}^m_{\text{VW}}(\theta)\}$, just as in \autoref{sec:fs-cat of m6:open string theory}, we can further specify any $\mathfrak{A}_4^\theta$-soliton at $\tau = \pm \infty$ as $\gamma^{IJ}_\pm(t, \theta, \mathfrak{A}_4)$,\footnote{%
  Just as in~\autoref{ft:fixing frakA6-soliton centre of mass dof}, the $\tau$-invariant $\mathfrak{A}_4^\theta$-solitons can be fixed at $\tau = \pm \infty$ by adding physically inconsequential $\mathcal{Q}$-exact terms to the SQM action.
  \label{ft:fixing frakA4-soliton centre of mass dof}
}%
~where its left and right endpoints would correspond to $\mathcal{E}_{\text{VW}}^I(\theta)$ and $\mathcal{E}_{\text{VW}}^J(\theta)$, respectively.

Since the $\mathcal{E}^*_{\text{VW}}(\theta)$'s are $\tau$-independent and therefore, have the same values for all $\tau$, we will have BPS worldsheets of the kind similar to~\autoref{fig:cy3 x r2:mu-1 map}.
This time, however, instead of the boundaries being $\mathcal{E}^*_{\text{DT}}(\theta)$, we will have $\mathcal{E}^*_{\text{VW}}(\theta)$.
And, instead of the solitons at $\tau = \pm \infty$ being $\gamma^{**}_\pm(t, \theta, \mathfrak{A}_6)$, we will have $\gamma^{**}_\pm(t, \theta, \mathfrak{A}_4)$.

\subtitle{The 2d Model on $\R^2$ and an Open String Theory in $\mathfrak{A}_4$}

Thus, like in~\autoref{sec:fs-cat of m6:open string theory}, one can understand the 2d gauged sigma model on $\R^2$ with target space $\mathfrak{A}_4$ to define an open string theory in $\mathfrak{A}_4$ as described above, whose \emph{effective} worldsheet and boundaries are similar to~\autoref{fig:cy3 x r2:mu-1 map}, where $\tau$ and $t$ are the temporal and spatial directions, respectively.

\subsection{Soliton String Theory, the 6d-\texorpdfstring{Spin$(7)$}{Spin(7)} Partition Function, and an FS Type \texorpdfstring{$A_\infty$}{A-infty}-category of VW Configurations on \texorpdfstring{$CY_2$}{CY2}}
\label{sec:fs-cat of m4:fs cat of kw configs}

\subtitle{The 2d Model as a Gauged LG Model}

Notice that we can also express~\eqref{eq:cy2 x r2:2d:action} as
\begin{equation}
  \label{eq:cy2 x r2:lg:2d action}
  \begin{aligned}
    S_{\text{LG},\mathfrak{A}_4}
    = \frac{1}{e^2} \int \dd{t} \dd{\tau} \Bigg(
    & \left| D_\tau \mathcal{A}^M
      + i D_t \mathcal{A}^M
      + i g^{M\bar{N}}_{\mathfrak{A}_4} \left(
      \frac{i \zeta}{2} \pdv{W_4}{\mathcal{A}^N}
      \right)^*
      \right|^2 \\
    & + \left| D_\tau \mathcal{B}^M
      - i D_t \mathcal{B}^M
      + i g^{M\bar{N}}_{\mathfrak{A}_4} \left(
      \frac{i \zeta}{2} \pdv{W_4}{\mathcal{B}^N}
      \right)^*
      \right|^2
      + \left| F_{\tau t} \right|^2
      + \dots
      \Bigg) \\
    = \frac{1}{e^2} \int \dd{t} \dd{\tau}
    &  \left(
      \left| D_\sigma \mathcal{A}^M \right|^2
      + \left| D_\sigma \mathcal{B}^M \right|^2
      + \left| \pdv{W_4}{\mathcal{A}^M} \right|^2
      + \left| \pdv{W_4}{\mathcal{B}^M} \right|^2
      + \left| F_{\tau t} \right|^2
      + \dots
      \right) \, .
  \end{aligned}
\end{equation}
In other words, the 2d gauged sigma model with target space $\mathfrak{A}_4$ can also be interpreted as a 2d gauged LG model in $\mathfrak{A}_4$ with holomorphic superpotential $W_4(\mathcal{A}, \mathcal{B})$.

By setting $d_\tau \mathcal{A}^M = 0 = d_\tau \mathcal{B}^M$ and $A_t, A_\tau \rightarrow 0$ in the expression within the squared terms in~\eqref{eq:cy2 x r2:lg:2d action}, we can read off the LG $\mathfrak{A}_4^\theta$-soliton equations corresponding to $\gamma^{IJ}_\pm(t, \theta, \mathfrak{A}_4)$ (that re-expresses~\eqref{eq:cy2 x r2:soliton:eqn} with $A_t, A_\tau \rightarrow 0$) as
\begin{equation}
  \label{eq:cy2 x r2:lg soliton:eqn}
  \dv{\mathcal{A}^M}{t}
  = - g^{M\bar{N}}_{\mathfrak{A}_4} \left(
    \frac{i\zeta}{2} \pdv{W_4}{\mathcal{A}^N}
  \right)^* \, ,
  \qquad
  \dv{\mathcal{B}^M}{t}
  = g^{M\bar{N}}_{\mathfrak{A}_4} \left(
    \frac{i\zeta}{2} \pdv{W_4}{\mathcal{B}^N}
  \right)^* \, .
\end{equation}
By setting $d_t \mathcal{A}^M = 0 = d_t \mathcal{B}^M$ in~\eqref{eq:cy2 x r2:lg soliton:eqn}, we get the LG $\mathfrak{A}_4^\theta$-soliton endpoint equations corresponding to $\gamma^{IJ}(\pm \infty, \theta, \mathfrak{A}_4)$ as
\begin{equation}
  \label{eq:cy2 x r2:lg soliton:endpts}
  g^{M\bar{N}}_{\mathfrak{A}_4} \left(
    \frac{i\zeta}{2} \pdv{W_4}{\mathcal{A}^N}
  \right)^*
  = 0
  \, ,
  \qquad
  g^{M\bar{N}}_{\mathfrak{A}_4} \left(
    \frac{i\zeta}{2} \pdv{W_4}{\mathcal{B}^N}
  \right)^*
  = 0
  \, .
\end{equation}

Recall from the end of~\autoref{sec:fs-cat of m4:solitons and kw configs} that we are only considering certain $CY_2$ such that the endpoints $\gamma^{IJ}(\pm \infty, \theta, \mathfrak{A}_4)$ are isolated and non-degenerate.
Therefore, from their definitions in~\eqref{eq:cy2 x r2:lg soliton:endpts} which tell us that they are critical points of $W_4(\mathcal{A}, \mathcal{B})$, we conclude that $W_4(\mathcal{A}, \mathcal{B})$ can be regarded as a holomorphic Morse function in $\mathfrak{A}_4$.

Just as in \autoref{sec:fs-cat of m6:fs cat of dt configs}, this means that an LG $\mathfrak{A}_4^\theta$-soliton defined in~\eqref{eq:cy2 x r2:lg soliton:eqn} maps to a straight line segment $[W^I_4(\theta), W^J_4(\theta)]$ in the complex $W_4$-plane that starts and ends at the critical values $W^I_4(\theta) \coloneq W_4(\gamma^I(-\infty, \theta, \mathfrak{A}_4))$ and $W^J_4(\theta) \coloneq W_4(\gamma^J(+\infty, \theta, \mathfrak{A}_4))$, respectively, where its slope depends on $\theta$ (via $\zeta$).
We shall also assume that $\text{Re} \left( W^I_4(\theta) \right) < \text{Re} \left(W^J_4(\theta) \right)$.

\subtitle{The Gauged LG Model as an LG SQM}

With suitable rescalings, we can recast~\eqref{eq:cy2 x r2:lg:2d action} as a 1d LG SQM (that re-expresses~\eqref{eq:cy2 x r2:sqm:action}), where its action will be given by\footnote{%
  Just as in~\autoref{ft:aux fields of cy3 x r2 sqm}, we have integrated out $A_{\tau}$ and omitted the term containing $A_t$ in the resulting SQM.
  \label{ft:auxiliary fields of cy2 x r2 lg sqm}
}
\begin{equation}
  \label{eq:cy2 x r2:lg:sqm:action}
  S_{\text{LG SQM}, \mathcal{M}(\R, \mathfrak{A}_4)}
  = \frac{1}{e^2} \int \dd{\tau} \left(
    \left| \partial_\tau \mathcal{A}^\alpha
      + g^{\alpha\beta}_{\mathcal{M}(\R, \mathfrak{A}_4)} \pdv{H_4}{\mathcal{A}^\beta}
    \right|^2
    + \left| \partial_\tau \mathcal{B}^\alpha
      + g^{\alpha\beta}_{\mathcal{M}(\R, \mathfrak{A}_4)} \pdv{H_4}{\mathcal{B}^\beta}
    \right|^2
    + \dots
  \right) \, ,
\end{equation}
where $H_4(\mathcal{A}, \mathcal{B})$ is the \emph{real-valued} potential in $\mathcal{M}(\R, \mathfrak{A}_4)$.

The LG SQM will localize onto configurations that \emph{simultaneously} set to zero the LHS and RHS of the expression within the squared terms in~\eqref{eq:cy2 x r2:lg:sqm:action}.
In other words, it will localize onto $\tau$-invariant critical points of $H_4(\mathcal{A}, \mathcal{B})$ that correspond to the LG $\mathfrak{A}_4^\theta$-solitons defined by~\eqref{eq:cy2 x r2:lg soliton:eqn}.
For our choice of $CY_2$, the LG $\mathfrak{A}_4^\theta$-solitons, just like their endpoints, will be isolated and non-degenerate.
Thus, $H_4(\mathcal{A}, \mathcal{B})$ can be regarded as a real-valued Morse functional in $\mathcal{M}(\R, \mathfrak{A}_4)$.

\subtitle{Morphisms from $\mathcal{E}^I_{\text{VW}}(\theta)$ to $\mathcal{E}^J_{\text{VW}}(\theta)$ as Floer Homology Classes of Intersecting Thimbles}

Repeating here the analysis in \autoref{sec:fs-cat of m6:fs cat of dt configs} with~\eqref{eq:cy2 x r2:lg:sqm:action} as the action for the LG SQM, we find that we can interpret the LG $\mathfrak{A}_4^\theta$-soliton solutions as intersections of thimbles.
Specifically, an LG $\mathfrak{A}_4^\theta$-soliton pair $\gamma^{IJ}_\pm(t, \theta, \mathfrak{A}_4)$, whose left and right endpoints correspond to $\mathcal{E}^I_{\text{VW}}(\theta)$ and $\mathcal{E}^J_{\text{VW}}(\theta)$, respectively, can be identified as a pair of transversal intersection points $p^{IJ}_{\text{VW},\pm}(\theta) \in S^{IJ}_{\text{VW}}$ of a left and right thimble in the fiber space over the line segment $[W^I_4(\theta), W^J_4(\theta)]$.

This means that the LG SQM in $\mathcal{M}(\R, \mathfrak{A}_4)$ with action~\eqref{eq:cy2 x r2:lg:sqm:action} will physically realize a Floer homology that we shall name an $\mathfrak{A}_4$-LG Floer homology.
The chains of the $\mathfrak{A}_4$-LG Floer complex will be generated by LG $\mathfrak{A}_4^\theta$-solitons which we can identify with $p^{**}_{\text{VW},\pm}(\theta)$, and the $\mathfrak{A}_4$-LG Floer differential will be realized by the flow lines governed by the gradient flow equations satisfied by $\tau$-varying configurations which set the expression within the squared terms in~\eqref{eq:cy2 x r2:lg:sqm:action} to zero.
In particular, the SQM partition function of the LG SQM in $\mathcal{M}(\R, \mathfrak{A}_4)$ will be given by\footnote{%
  Just as in~\autoref{ft:theta omission in lg sqm 8d partition fn}, the `$\theta$' label is omitted in the LHS of the following expression.
  \label{ft:theta omission in lg sqm 6d partition fn}
}
\begin{equation}
  \label{eq:cy2 x r2:lg:sqm:partition fn}
  \mathcal{Z}_{\text{LG SQM}, \mathcal{M}(\R, \mathfrak{A}_4)}(G)
  = \sum_{I \neq J = 1}^m \sum_{p^{IJ}_{VW, \pm} \in S^{IJ}_{\text{VW}}}
  \text{HF}^{G}_{d_r}(p^{IJ}_{\text{VW}, \pm}(\theta))
  \, ,
\end{equation}
where the contribution $\text{HF}^{G}_{d_r} (p^{IJ}_{\text{VW},\pm}(\theta))$ can be identified with a homology class in an $\mathfrak{A}_4$-LG Floer homology generated by intersection points of thimbles.
These intersection points represent LG $\mathfrak{A}_4^\theta$-solitons whose endpoints correspond to $\theta$-deformed VW configurations on $CY_2$.
The degree of each chain in the complex is $d_r$, and is counted by the number of outgoing flow lines from the fixed critical points of $H_4(\mathcal{A}, \mathcal{B})$ in $\mathcal{M}(\R, \mathfrak{A}_4)$ which can also be identified as $p^{IJ}_{\text{VW},\pm}(\theta)$.

Therefore, $\mathcal{Z}_{\text{LG SQM}, \mathcal{M}(\R, \mathfrak{A}_4)}(G)$ in~\eqref{eq:cy2 x r2:lg:sqm:partition fn} is a sum of LG $\mathfrak{A}_4^\theta$-solitons defined by~\eqref{eq:cy2 x r2:lg soliton:eqn} with endpoints~\eqref{eq:cy2 x r2:lg soliton:endpts}, or equivalently, $\gamma^{IJ}_\pm(t, \theta, \mathfrak{A}_4)$-solitons defined by~\eqref{eq:cy2 x r2:soliton:config} and~\eqref{eq:cy2 x r2:soliton:aux:config} with endpoints~\eqref{eq:cy2 x r2:soliton:endpts:config}, whose start and end correspond to $\mathcal{E}^I_{\text{VW}}(\theta)$ and $\mathcal{E}^J_{\text{VW}}(\theta)$, respectively.
In other words, we can write
\begin{equation}
  \label{eq:cy2 x r2:floer complex:vector}
  \text{CF}_{\mathfrak{A}_4} \left(
    \mathcal{E}^I_{\text{VW}}(\theta), \mathcal{E}^J_{\text{VW}}(\theta)
  \right)_\pm
  = \text{HF}^{G}_{d_r}(p^{IJ}_{\text{VW}, \pm}(\theta))
  \, ,
\end{equation}
where $\text{CF}_{\mathfrak{A}_4}(\mathcal{E}^I_{\text{VW}}(\theta), \mathcal{E}^J_{\text{VW}}(\theta))_\pm$ is a vector representing a $\gamma^{IJ}_\pm(t, \theta, \mathfrak{A}_4)$-soliton, such that $\text{Re} \left( W^I_4(\theta) \right) < \text{Re} \left( W^J_4(\theta) \right)$.
This will lead us to the one-to-one identification\footnote{%
  Once again, just as in~\autoref{ft:theta omission in 6d fs-cat morphism}, the `$\theta$' label is omitted in the following expression.
  \label{ft:theta omission in 4d fs-cat morphism}
}
\begin{equation}
  \label{eq:cy2 x r2:floer complex:morphism}
  \boxed{
    \text{Hom}(\mathcal{E}^I_{\text{VW}}, \mathcal{E}^J_{\text{VW}})_\pm
    \Longleftrightarrow
    \text{HF}^{G}_{d_r}(p^{IJ}_{\text{VW}, \pm})
  }
\end{equation}
where the RHS is proportional to the identity class when $I = J$, and zero when $I \leftrightarrow J$ since the $\mathfrak{A}_4^\theta$-soliton only moves in one direction from $\mathcal{E}^I_{\text{VW}}(\theta)$ to $\mathcal{E}^J_{\text{VW}}(\theta)$.

\subtitle{Soliton String Theory from the 2d LG Model}

Just as in \autoref{sec:fs-cat of m6:fs cat of dt configs}, the 2d gauged LG model in $\mathfrak{A}_4$ with action~\eqref{eq:cy2 x r2:lg:2d action} can be interpreted as a soliton string theory in $\mathfrak{A}_4$.
The dynamics of this soliton string theory in $\mathfrak{A}_4$ will be governed by the BPS worldsheet equations determined by setting to zero the expression within the squared terms of \eqref{eq:cy2 x r2:lg:2d action}, where $(\mathcal{A}^M, \mathcal{B}^M)$ are scalars on the worldsheet corresponding to the holomorphic coordinates on $\mathfrak{A}_4$.
At an arbitrary instant in time whence $d_\tau\mathcal{A}^M = 0 = d_\tau\mathcal{B}^M$ therein, the dynamics of $(\mathcal{A}^M, \mathcal{B}^M)$ along $t$ will be governed by the soliton equations
\begin{equation}
  \label{eq:cy2 x r2:string soliton:eqn}
  \dv{\mathcal{A}^M}{t}
  = - [A_t - i A_\tau, \mathcal{A}^M]
  - g^{M\bar{N}}_{\mathfrak{A}_4} \left(
    \frac{i\zeta}{2} \pdv{W_4}{\mathcal{A}^N}
  \right)^* \, ,
  \qquad
  \dv{\mathcal{B}^M}{t}
  = - [A_t + i A_\tau, \mathcal{B}^M]
  + g^{M\bar{N}}_{\mathfrak{A}_4} \left(
    \frac{i\zeta}{2} \pdv{W_4}{\mathcal{B}^N}
  \right)^* \, .
\end{equation}

\subtitle{The Normalized 6d-Spin$(7)$ Partition Function, LG $\mathfrak{A}_4^\theta$-soliton String Scattering, and Maps of an $A_\infty$-structure}

Since our 6d-Spin$(7)$ theory on $CY_2 \times \R^2$ is semi-classical, its normalized partition function can be regarded as a sum over tree-level scattering amplitudes of the LG $\mathfrak{A}_4^\theta$-soliton strings defined by~\eqref{eq:cy2 x r2:lg soliton:eqn}.
The BPS worldsheet underlying such a tree-level scattering amplitude is similar to~\autoref{fig:cy3 x r2:mu-d maps}, where instead the endpoints of each string being $\mathcal{E}^*_{\text{DT}}$, it is now understood to be $\mathcal{E}^*_{\text{VW}}$.

In other words, we can, like in~\eqref{eq:cy3 x r2:partition fn}, express the normalized 6d-Spin$(7)$ partition function as
\begin{equation}
  \label{eq:cy2 x r2:partition fn}
  \widetilde{\mathcal{Z}}_{\text{6d-Spin}(7),CY_2 \times \R^2}(G) =
  \sum_{n_m} \mu^{n_m}_{\mathfrak{A}_4}
  \, ,
  \qquad
  n_m \in \{1, 2, \dots, m-1\}
\end{equation}
where each
\begin{equation}
  \label{eq:cy2 x r2:mu-d maps}
  \boxed{
    \mu^{n_m}_{\mathfrak{A}_4}:
    \bigotimes_{i = 1}^{n_m}
    \text{Hom} \left(
      \mathcal{E}^{I_i}_{\text{VW}}, \mathcal{E}^{I_{i + 1}}_{\text{VW}}
    \right)_-
    \longto
    \text{Hom} \left(
      \mathcal{E}^{I_1}_{\text{VW}}, \mathcal{E}^{I_{n_m+1}}_{\text{VW}}
    \right)_+
  }
\end{equation}
is a scattering amplitude of $n_m$ incoming LG $\mathfrak{A}_4^\theta$-soliton strings $\text{Hom}( \mathcal{E}^{I_1}_{\text{VW}}, \mathcal{E}^{I_2}_{\text{VW}} )_-, \dots, \text{Hom}(\mathcal{E}^{I_{n_m}}_{\text{VW}}, \mathcal{E}^{I_{n_m+1}}_{\text{VW}} )_-$ and a single outgoing LG $\mathfrak{A}_4^\theta$-soliton string $\text{Hom}(\mathcal{E}^{I_1}_{\text{VW}}, \mathcal{E}^{I_{n_m+1}}_{\text{VW}})_+$ with left and right boundaries as labeled, whose underlying worldsheet can be regarded as a disc with $n_m + 1$ vertex operators at the boundary.
That is, $\mu^{n_m}_{\mathfrak{A}_4}$ counts pseudoholomorphic discs with $n_m + 1$ punctures at the boundary that are mapped to $\mathfrak{A}_4$ according to the BPS worldsheet equations~\eqref{eq:cy2 x r2:worldsheet:eqn}.

Just as in \autoref{sec:fs-cat of m6:fs cat of dt configs}, the collection of $\mu^{n_m}_{\mathfrak{A}_4}$ maps in~\eqref{eq:cy2 x r2:mu-d maps} can be regarded as composition maps defining an $A_\infty$-structure.

\subtitle{An FS Type $A_\infty$-category of VW Configurations on $CY_2$}

Altogether, this means that the normalized partition function of 6d-Spin$(7)$ theory on $CY_2 \times \R^2$ as expressed in~\eqref{eq:cy2 x r2:partition fn}, manifests a \emph{novel} FS type $A_\infty$-category defined by the $\mu^{n_m}_{\mathfrak{A}_4}$ maps~\eqref{eq:cy2 x r2:mu-d maps} and the one-to-one identification~\eqref{eq:cy2 x r2:floer complex:morphism}, where the $m$ objects $\{\mathcal{E}^1_{\text{VW}}, \mathcal{E}^2_{\text{VW}}, \dots, \mathcal{E}^m_{\text{VW}}\}$ correspond to ($\theta$-deformed) VW configurations on $CY_2$ (with the scalar and one of the linearly-independent components of the self-dual two-form field being zero)!

We may continue to repeat the procedure of these past three sections by performing a KK reduction of $CY_2$ along an $S^1$ circle within, to physically derive an FS type $A_{\infty}$-category of a three-manifold.
However, for a compact $CY_2$, that would mean we would need to specialize to the trivial case of $CY_2 = T^4$ to derive an FS type $A_{\infty}$-category of objects on $T^3$.
We will choose not to consider this here, as we have previously studied such a case more generally as an FS type $A_\infty$-category of Hitchin configurations on non-trivial three-manifolds in~\cite[$\S$9]{er-2023-topol-n}.


\section{Topological Invariance and a Relation Amongst the Floer Homologies and \texorpdfstring{$A_\infty$}{A-infty}-categories}
\label{sec:topo invariance}

In this section, we will first exploit the topological invariance of Spin$(7)$ theory to relate the Floer homologies derived in \autoref{sec:floer homology of m7}--\autoref{sec:atiyah-floer} to one another.
Then, we will do the same for the FS type $A_\infty$-categories derived in \autoref{sec:fs-cat of m6}--\autoref{sec:fs-cat of m4}.
After which, we will explain how these FS type $A_\infty$-categories would categorify the Floer homologies.
Finally, we will summarize all our results and obtain a web of relations and correspondences amongst the Floer homologies and FS type $A_\infty$-categories.
In turn, these results would serve as physical proofs and generalizations of the mathematical conjectures by Hohloch-Noetzel-Salamon~\cite{hohloch-2009-hyper-struc}, Salamon~\cite{salamon-2013-three-dimen}, and Bousseau~\cite{bousseau-2024-holom-floer}.

\subsection{Topological Invariance of \texorpdfstring{Spin$(7)$}{Spin(7)} Theory and the Floer Homologies}
\label{sec:topo invariance:floer homology}

\subtitle{Relating the Gauge-theoretic Floer Homologies of \autoref{sec:floer homology of m7}--\autoref{sec:floer homology of m5}}

Recall that the topological invariance of (the $\mathcal{Q}$-cohomology of) Spin$(7)$ theory in all directions means that we can relate the partition functions~\eqref{eq:g2 x r:partition fn},~\eqref{eq:cy3 x r:partition fn}, and~\eqref{eq:cy2 x s x r:partition fn} as
\begin{equation}
  \label{eq:topo inv:floer partition fn:gauge-theoretic}
  \boxed{
    \begin{tikzcd}[%
      row sep=large,%
      arrows=leftrightarrow,%
      ]
      \sum_j \text{HF}^{\text{Spin}(7)\text{-inst}}_{d_j}(G_2, G)
      \arrow[d, "G_2 = CY_3 \times \widehat{S}^1"]
      \\
      \sum_k \text{HHF}^{G_2\text{-M}}_{d_k}(CY_3, G)
      \arrow[d, "CY_3 = CY_2 \times S^1 \times \widehat{S}^1"]
      \\
      \sum_l \text{HHF}^{\text{DT}}_{d_l}(CY_2 \times S^1, G)
    \end{tikzcd}
  }
\end{equation}
where $S^1$ and $\widehat{S}^1$ are circles of fixed and variable radii, respectively.

The relations in~\eqref{eq:topo inv:floer partition fn:gauge-theoretic} are consistent, in that they have a one-to-one correspondence in their summations over `$j$', `$k$', and `$l$'.
Specifically, each `$j$', `$k$', and `$l$' corresponds to a solution of (the simultaneous vanishing of the LHS and RHS of)~\eqref{eq:g2 x r:flow},~\eqref{eq:cy3 x r:bps:complexified}, and~\eqref{eq:cy2 x s x r:bps}, respectively, where~\eqref{eq:cy2 x s x r:bps} is obtained via a KK reduction of~\eqref{eq:cy3 x r:bps:complexified}, which in turn is obtained via a KK reduction of~\eqref{eq:g2 x r:flow}.

In short, we have a \emph{novel} equivalence amongst gauge-theoretic Floer homologies of seven, six, and five-manifolds.

\subtitle{Relating the Floer Homologies of \autoref{sec:hyperkahler floer-hom} and \autoref{sec:symp floer-hom}, and a Physical Proof of Hohloch-Noetzel-Salamon's Mathematical Conjecture}

Topological invariance (of the $\mathcal{Q}$-cohomology of) Spin$(7)$ theory in all directions also means that we can relate the partition functions~\eqref{eq:g2 x r:partition fn} and~\eqref{eq:hc3 x r:partition fn} as
\begin{equation}
  \label{eq:topo inv:floer partition fn:non-gauge-theoretic:to hk}
  \boxed{
    \begin{tikzcd}[%
      column sep=6em,%
      arrows=leftrightarrow,%
      ampersand replacement=\&,%
      ]
      \sum_j \text{HF}^{\text{Spin}(7)\text{-inst}}_{d_j}(G_2, G)
      \arrow[r, "G_2 = \widehat{CY_2} \times HC_3"]
      \&
      \sum_s \text{HHKF}_{d_s}\left(
        HC_3, \mathcal{M}^G_{\text{inst}}(CY_2)
      \right)
    \end{tikzcd}
  }
\end{equation}
Then, via~\eqref{eq:t3 x r:equality to hk floer-hom}, we will have
\begin{equation}
  \label{eq:topo inv:floer partition fn:non-gauge-theoretic:to symp}
  \boxed{
    \begin{tikzcd}[%
      row sep=large,%
      arrows=leftrightarrow,%
      ]
      \sum_j \text{HF}^{\text{Spin}(7)\text{-inst}}_{d_j}(G_2, G)
      \arrow[d, "G_2 = \widehat{CY_2} \times T^3"]
      \\
      \sum_s \text{HHKF}_{d_s}\left(
        T^3, \mathcal{M}^G_{\text{inst}}(CY_2)
      \right)
      =
      \sum_s \text{HSF}^{\text{Fuet}}_{d_s}\left(
        L^3 \mathcal{M}^{G, CY_2}_{\text{inst}}
      \right)
    \end{tikzcd}
  }
\end{equation}
Via~\eqref{eq:t2 x i x r:equality to hk floer-hom of i x t2}, we will have \begin{equation}
  \label{eq:topo inv:floer partition fn:non-gauge-theoretic:to symp-int:loop}
  \boxed{
    \begin{tikzcd}[%
      row sep=large,%
      arrows=leftrightarrow,%
      ]
      \sum_j \text{HF}^{\text{Spin}(7)\text{-inst}}_{d_j}(G_2, G)
      \arrow[d, "G_2 = \widehat{CY_2} \times I \times T^2"]
      \\
      \sum_s \text{HHKF}_{d_s}\left(
        I \times T^2, \mathcal{M}^G_{\text{inst}}(CY_2)
      \right)
      =
      \sum_s \text{HSF}^{\text{Int}}_{d_s}\left(
        L^2 \mathcal{M}^{G, CY_2}_{\text{inst}}, \mathscr{L}_0, \mathscr{L}_1
      \right)
    \end{tikzcd}
  }
\end{equation}
Via~\eqref{eq:i x s x r2:equality to hk floer-hom of i x s x r}, we will have
\begin{equation}
  \label{eq:topo inv:floer partition fn:non-gauge-theoretic:to symp-int:path}
  \boxed{
    \begin{tikzcd}[%
      row sep=large,%
      arrows=leftrightarrow,%
      ]
      \sum_j \text{HF}^{\text{Spin}(7)\text{-inst}}_{d_j}(G_2, G)
      \arrow[d, "G_2 = \widehat{CY_2} \times I \times S^1 \times \R"]
      \\
      \sum_s \text{HHKF}_{d_s}\left(
        I \times S^1 \times \R, \mathcal{M}^{G, \theta}_{\text{inst}}(CY_2)
      \right)
      =
      \sum_s \text{HSF}^{\text{Int}}_{d_s}\left(
        \mathcal{M} \left( \R, L \mathcal{M}^{G, \theta, CY_2}_{\text{inst}} \right),
        \mathcal{P}_0, \mathcal{P}_1
      \right)
    \end{tikzcd}
  }
\end{equation}
where $\widehat{CY_2}$ is a $CY_2$ with variable size.

The relations in~\eqref{eq:topo inv:floer partition fn:non-gauge-theoretic:to hk} to~\eqref{eq:topo inv:floer partition fn:non-gauge-theoretic:to symp-int:path} are consistent as well, in that they have a one-to-one correspondence in their summations over `$j$' and `$s$'.
Specifically, each `$j$' and `$s$' corresponds to a solution of (the simultaneous vanishing of the LHS and RHS of)~\eqref{eq:g2 x r:flow} and~\eqref{eq:hc3 x r:cauchy-riemann-fueter equation} (on the specific $HC_3$'s), respectively, where~\eqref{eq:hc3 x r:cauchy-riemann-fueter equation} is obtained via a topological reduction of~\eqref{eq:g2 x r:flow} along $CY_2$.

In short, we have a \emph{novel} equivalence amongst (i) gauge-theoretic Floer homologies of various seven-manifolds, (ii) hyperkähler Floer homologies of instanton moduli spaces specified by hypercontact three-manifolds, and (iii) symplectic and symplectic intersection Floer homologies of certain spaces of instantons.

Note that the relation between hyperkähleric $\text{HHKF}_{d_s}(HC_3, \mathcal{M}^G_{\text{inst}}(CY_2))$ and gauge-theoretic $\text{HF}^{\text{Spin}(7)\text{-inst}}_{d_j}(HC_3 \times CY_2, G)$ was conjectured by Hohloch-Noetzel-Salamon~\cite{hohloch-2009-hyper-struc}~\cite[$\S$5]{salamon-2013-three-dimen}.
Therefore, in arriving at~\eqref{eq:topo inv:floer partition fn:non-gauge-theoretic:to hk}, we have furnished a physical proof of their mathematical conjecture.

\subtitle{The Floer Homologies of \autoref{sec:atiyah-floer}, and a Physical Proof of Salamon's Mathematical Conjecture}

From~\eqref{eq:atiyah-floer:spin7:equality of partition fn} and~\eqref{eq:atiyah-floer:7d-spin7:equality of partition fn}, we have
\begin{equation}
  \label{eq:topo inv:floer partition fn:atiyah-floer}
  \boxed{
    \begin{tikzcd}[%
      column sep=3.6em,%
      arrows=leftrightarrow,%
      ampersand replacement=\&,%
      ]
      \sum_j \text{HF}^{\text{Spin}(7)\text{-inst}}_{d_j}(G_2, G)
      \arrow[rr, "G_2 = CY_3 \times S^1"]
      \arrow[rr, swap, "CY_3 = CY_3' \bigcup_{CY_2} CY_3''"]
      \&
      {}
      \arrow[d, rightarrow, "S^1 = \widehat{S}^1", shorten <= 10pt, shorten >= 10pt]
      \&
      \sum_s \text{HSF}^{\text{Int}}_{d_s} \left(
        L^2 \mathcal{M}^{G, CY_2}_{\text{inst}}, \mathscr{L}_0, \mathscr{L}_1
      \right)
      \\
      \sum_k \text{HHF}^{G_2\text{-M}}_{d_k} (CY_3, G)
      \arrow[rr, swap, "CY_3 = CY_3' \bigcup_{CY_2} CY_3''"]
      \&
      {}
      \&
      \sum_u \text{HSF}^{\text{Int}}_{d_u} \left(
        L \mathcal{M}^{G, CY_2}_{\text{inst}}, \mathcal{L}_0, \mathcal{L}_1
      \right)
    \end{tikzcd}
  }
\end{equation}
where $CY_2$ is the degeneration surface of the Tyurin degeneration of $CY_3$.

Also, notice that we can perform yet another KK dimensional reduction along an $S^1$ circle on the RHS of~\eqref{eq:atiyah-floer:7d-spin7:equality of partition fn} to get\footnote{%
  This can be seen by performing a straightforward KK dimensional reduction of the remaining $S^1$ circle from the 3d sigma model on $I \times S^1 \times \R$ in~\autoref{ft:symp-int floer-hom of loop space from kk reduction}.
  The resulting 2d sigma model on $I \times \R$ is a 2d A-model, this time with target space $\mathcal{M}^G_{\text{inst}}(CY_2)$.
  Following again the analysis of~\autoref{sec:symp floer-hom:i x t2}, we will physically realize a symplectic intersection Floer homology of $\mathcal{M}^G_{\text{inst}}(CY_2)$ generated by intersections of isotropic-coisotropic A-branes $L_*$ in $\mathcal{M}^G_{\text{inst}}(CY_2)$, which are Lagrangian.
  \label{ft:symp-int floer-hom from kk reduction}
}
\begin{equation}
  \label{eq:topo inv:floer partition fn:atiyah-floer:kk reduction}
  \boxed{
    \begin{tikzcd}[%
      column sep=3em,%
      arrows=leftrightarrow,%
      ampersand replacement=\&,%
      ]
      \sum_u \text{HSF}^{\text{Int}}_{d_u} \left(
        L \mathcal{M}^{G, CY_2}_{\text{inst}}, \mathcal{L}_0, \mathcal{L}_1
      \right)
      \arrow[r, "S^1 = \widehat{S}^1"]
      \&
      \sum_v \text{HSF}^{\text{Int}}_{d_v} \left(
        \mathcal{M}^G_{\text{inst}}(CY_2), L_0, L_1
      \right)
    \end{tikzcd}
  }
\end{equation}
where $L_0$ and $L_1$ are Lagrangian branes in $\mathcal{M}^G_{\text{inst}}(CY_2)$.

In the above, just as there is a one-to-one correspondence in the summations over `$j$' and `$s$', for similar reasons, there would be a one-to-one correspondence in the summations over `$k$' and `$u$', as well as over `$u$' and `$v$'.

In short, we have (i) a \emph{novel} equivalence of Atiyah-Floer dualities between gauge-theoretic Floer homologies and symplectic intersection Floer homologies, and (ii) a \emph{novel} equivalence of symplectic intersection Floer homologies of certain spaces of instantons.

Note that the relation between the hyperkähleric $\text{HHKF}_{d_s}(HC_3, \mathcal{M}^G_{\text{inst}}(CY_2))$ and gauge-theoretic $\text{HF}^{\text{Spin}(7)\text{-inst}}(CY_2 \times HC_3, G)$ was conjectured by Salamon as being analogous to an Atiyah-Floer duality~\cite[$\S$5]{salamon-2013-three-dimen}.
Indeed, we do find, from~\eqref{eq:topo inv:floer partition fn:non-gauge-theoretic:to symp-int:loop} and~\eqref{eq:topo inv:floer partition fn:atiyah-floer}, that the two Floer homologies (with $HC_3 = I \times T^2$) are related to each other by an Atiyah-Floer type duality between a symplectic intersection and a gauge-theoretic Floer homology.
Therefore, we have furnished a physical proof of his mathematical conjecture.

\subtitle{Relating Symplectic Floer Homologies}

Because of the topological invariance of Spin$(7)$ theory in all directions, we would also have a series of relations involving the symplectic Floer homologies starting with the one on the RHS of~\eqref{eq:t3 x r:equality to hk floer-hom}.
Specifically, by performing KK dimensional reductions along the $S^1$ circles of the 4d sigma model on $T^3 \times \R$ with target space $\mathcal{M}^G_{\text{inst}}(CY_2)$ and action~\eqref{eq:hc3 x r:4d action:cauchy-riemann-fueter eq} (with $HC_3 = T^3$), we will get the following series of relations involving symplectic Floer homologies of certain spaces of instantons on $CY_2$:
\begin{equation}
  \label{eq:topo inv:floer partition fn:loop spaces}
  \boxed{
    \begin{tikzcd}[%
      column sep=4.2em,%
      arrows=leftrightarrow,%
      ampersand replacement=\&,%
      ]
      \sum_s \text{HSF}^{\text{Fuet}}_{d_s}\left(
        L^3 \mathcal{M}^{G, CY_2}_{\text{inst}}
      \right)
      \arrow[r, "T^3 = T^2 \times \widehat{S}^1"]
      \&
      \sum_x \text{HSF}^{\text{hol}}_{d_x}\left(
        L^2 \mathcal{M}^{G, CY_2}_{\text{inst}}
      \right)
      \arrow[r, "T^2 = S^1 \times \widehat{S}^1"]
      \&
      \sum_y \text{HSF}^{\text{const}}_{d_y}\left(
        L \mathcal{M}^{G, CY_2}_{\text{inst}}
      \right)
    \end{tikzcd}
  }
\end{equation}
where $\text{HSF}^{\text{hol}}_{d_x} (L^2 \mathcal{M}^{G, CY_2}_{\text{inst}})$ and $\text{HSF}^{\text{const}}_{d_y} (L \mathcal{M}^{G, CY_2}_{\text{inst}})$ are symplectic Floer homologies of $L^2 \mathcal{M}^{G, CY_2}_{\text{inst}}$ and $L \mathcal{M}^{G, CY_2}_{\text{inst}}$, generated by time-invariant holomorphic and constant maps from $T^2$ and $S^1$ to $\mathcal{M}^G_{\text{inst}}(CY_2)$, respectively.

The relations in~\eqref{eq:topo inv:floer partition fn:loop spaces} are consistent, in that they have a one-to-one correspondence in their summations over `$s$', `$x$', and `$y$', for reasons similar to that given earlier.

In short, we have a \emph{novel} equivalence amongst symplectic Floer homologies of certain spaces of instantons.

\subtitle{Relating Symplectic Intersection Floer Homologies}

Once again, because of the topological invariance of (the $\mathcal{Q}$-cohomology of) Spin$(7)$ theory in all directions, we can derive yet another relation starting with the bottom RHS of~\eqref{eq:topo inv:floer partition fn:non-gauge-theoretic:to symp-int:path}.
Specifically, by KK reducing the spatial $S^1$ circle of the 4d sigma model on $I \times S^1 \times \R^2$ in~\autoref{sec:symp floer-hom:i x s x r}, we will obtain a 3d sigma model on $I \times \R^2$ with target space $L \mathcal{M}^{G, \theta}_{\text{inst}}(CY_2)$.
In turn, this 3d sigma model can be recast as a 2d A-model on $I \times \R$ with target $L \mathcal{M}(\R, \mathcal{M}^{G, \theta, CY_2}_{\text{inst}})$, which will enable us to physically realize an intersection Floer homology of intersecting A$_{\theta}$-branes $\mathcal{P}_*(\theta) \subset \mathcal{M}(\R, L \mathcal{M}^{G, \theta, CY_2}_{\text{inst}})$, whence we will have
\begin{equation}
  \label{eq:topo inv:fs-cat:symp int floer-hom from kk red}
  \boxed{
    \begin{tikzcd}[%
      column sep=large,%
      arrows=leftrightarrow,%
      ampersand replacement=\&,%
      ]
      \sum_s \text{HSF}^{\text{Int}}_{d_s} \left(
        \mathcal{M} \left( \R, L \mathcal{M}^{G, \theta, CY_2}_{\text{inst}} \right), \mathcal{P}_0, \mathcal{P}_1
      \right)
      \arrow[r, "S^1 = \widehat{S}^1"]
      \&
      \sum_r \text{HSF}^{\text{Int}}_{d_r} \left(
        \mathcal{M} \left( \R, \mathcal{M}^{G, \theta, CY_2}_{\text{inst}} \right), P_0, P_1)
      \right)
    \end{tikzcd}
  }
\end{equation}
where $\text{HSF}^{\text{Int}}_{d_r} (\mathcal{M} ( \R, \mathcal{M}^{G, \theta, CY_2}_{\text{inst}} ), P_0, P_1)$ is a symplectic instersection Floer homology of intersecting A$_{\theta}$-branes $P_{*}(\theta) \subset \mathcal{M} ( \R, \mathcal{M}^{G, \theta, CY_2}_{\text{inst}} )$.

The relation in~\eqref{eq:topo inv:fs-cat:symp int floer-hom from kk red} is consistent, in that the summations over `$s$' and `$r$' have a one-to-one correspondence, for reasons similar to that given earlier.

In short, we have a \emph{novel} equivalence between symplectic intersection Floer homologies of certain spaces of instantons.

\subtitle{Summarizing the Relations Amongst the Floer Homologies}

\begin{sidewaysfigure}
  \centering
  \begin{tikzpicture}[%
    auto,%
    block/.style={draw, rectangle},%
    novel/.style={draw, rectangle, ultra thick},%
    every edge/.style={draw, <->},%
    relation/.style={scale=0.8, sloped, anchor=center, align=center},%
    vertRelation/.style={scale=0.8, anchor=center, align=center},%
    horRelation/.style={scale=0.8, anchor=center, align=center},%
    shorten >=4pt,%
    shorten <=4pt,%
    ]
    \def \verRel {1} 
    \def \horRel {2.6} 
    \node[block] (8d-FT)
    {$\sum_j \text{HF}^{\text{Spin}(7)\text{-inst}}_{d_j}(G_2, G)$};
    \node[block, below={\verRel} of 8d-FT] (7d-FT)
    {$\sum_k \text{HHF}^{G_2\text{-M}}_{d_k}(CY_3, G)$};
    \node[novel, below={\verRel} of 7d-FT] (6d-FT)
    {$\sum_l \text{HHF}^{\text{DT}}_{d_l}(CY_2 \times S^1, G)$};
    \node[block, above={\verRel} of 8d-FT] (hK-HC3)
    {$\sum_s \text{HHKF}_{d_s}\left(
        HC_3, \mathcal{M}^G_{\text{inst}}(CY_2)
      \right)$};
    \node[block, left={\horRel} of 8d-FT] (hK-T3)
    {$\sum_s \text{HHKF}_{d_s}\left(
        T^3, \mathcal{M}^G_{\text{inst}}(CY_2)
      \right)$};
    \node[block, right={\horRel} of 8d-FT] (hK-I-T2)
    {$\sum_s \text{HHKF}_{d_s}\left(
        I \times T^2, \mathcal{M}^G_{\text{inst}}(CY_2)
      \right)$};
    \node[block, above={\verRel} of hK-HC3] (hK-ISR)
    {$\sum_s \text{HHKF}_{d_s}\left(
        I \times S^1 \times \R, \mathcal{M}^{G, \theta}_{\text{inst}}(CY_2)
      \right)$};
    \node[novel, below={\verRel} of hK-T3] (HSF-T3)
    {$\sum_s \text{HSF}^{\text{Fuet}}_{d_s} \left(
        L^3 \mathcal{M}^{G, CY_2}_{\text{inst}}
      \right)$};
    \node[block, below={\verRel} of HSF-T3] (HSF-T2)
    {$\sum_x \text{HSF}^{\text{hol}}_{d_x} \left(
        L^2 \mathcal{M}^{G, CY_2}_{\text{inst}}
      \right)$};
    \node[block, below={\verRel} of HSF-T2] (HSF-S)
    {$\sum_y \text{HSF}^{\text{const}}_{d_y} \left(
        L \mathcal{M}^{G, CY_2}_{\text{inst}}
      \right)$};
    \node[novel, below={\verRel} of hK-I-T2] (HSFI-T2)
    {$\sum_s \text{HSF}^{\text{Int}}_{d_s}\left(
        L^2 \mathcal{M}^{G, CY_2}_{\text{inst}}, \mathscr{L}_0, \mathscr{L}_1
      \right)$};
    \node[block, below={\verRel} of HSFI-T2] (HSFI-S)
    {$\sum_u \text{HSF}^{\text{Int}}_{d_u}\left(
        L \mathcal{M}^{G, CY_2}_{\text{inst}}, \mathcal{L}_0, \mathcal{L}_1
      \right)$};
    \node[block, below={\verRel} of HSFI-S] (HSFI)
    {$\sum_v \text{HSF}^{\text{Int}}_{d_v}\left(
        \mathcal{M}^G_{\text{inst}}(CY_2), L_0, L_1
      \right)$};
    \node[novel, above={\verRel} of hK-ISR] (HSFI-RL)
    {$\sum_s \text{HSF}^{\text{Int}}_{d_s}\left(
        \mathcal{M} \left( \R, L \mathcal{M}^{G, \theta, CY_2}_{\text{inst}} \right), \mathcal{P}_0, \mathcal{P}_1
      \right)$};
    \node[block, above={\verRel} of HSFI-RL] (HSFI-R)
    {$\sum_r \text{HSF}^{\text{Int}}_{d_r} \left(
      \mathcal{M} \left( \R, \mathcal{M}^{G, \theta, CY_2}_{\text{inst}} \right), P_0, P_1)
    \right)$};
    \draw
    (8d-FT) edge[dashed]
    node[vertRelation, left] {$G_2 = CY_3 \times \widehat{S}^1$}
    (7d-FT)
    (7d-FT) edge[dashed]
    node[vertRelation, left] {$CY_3 = CY_2 \times S^1 \times \widehat{S}^1$}
    (6d-FT)
    (8d-FT) edge[dashed]
    node[vertRelation, right] {$G_2 = \widehat{CY_2} \times HC_3$}
    (hK-HC3)
    (hK-HC3.south east) edge
    node[relation, above] {$HC_3 = I \times T^2$}
    (hK-I-T2)
    (hK-I-T2) edge (HSFI-T2)
    (8d-FT.south east) edge
    node[relation, above] {$G_2 = CY_3 \times S^1$}
    node[relation, below] {$CY_3 = CY_3' \bigcup_{CY_2} CY_3''$}
    (HSFI-T2.south west)
    (HSFI-T2) edge[dashed]
    node[vertRelation, right] {$T^2 = S^1 \times \widehat{S}^1$}
    (HSFI-S)
    (7d-FT.south east) edge
    node[relation, below] {$CY_3 = CY_3' \bigcup_{CY_2} CY_3''$}
    (HSFI-S.south west)
    (HSFI-S) edge [dashed]
    node[vertRelation, right] {$S^1 = \widehat{S}^1$}
    (HSFI)
    (hK-HC3.south west) edge
    node[relation, above] {$HC_3 = T^3$}
    (hK-T3)
    (hK-T3) edge (HSF-T3)
    (8d-FT.south west) edge[dashed]
    node[relation, above] {$G_2 = \widehat{CY_2} \times T^3$}
    (HSF-T3.south east)
    (HSF-T3) edge[dashed]
    node[vertRelation, right] {$T^3 = T^2 \times \widehat{S}^1$}
    (HSF-T2)
    (HSF-T2) edge[dashed]
    node[vertRelation, right] {$T^2 = S^1 \times \widehat{S}^1$}
    (HSF-S)
    (hK-HC3) edge
    node[vertRelation, right] {$HC_3 = I \times S^1 \times \R$}
    (hK-ISR)
    (hK-ISR) edge (HSFI-RL)
    (HSFI-RL) edge[dashed]
    node[vertRelation, right] {$S^1 = \widehat{S}^1$}
    (HSFI-R)
    ;
  \end{tikzpicture}
  \caption{%
    Relations amongst Floer homologies via the topological invariance of Spin$(7)$ theory.
    \label{fig:web of relations:floer}
  }
\end{sidewaysfigure}
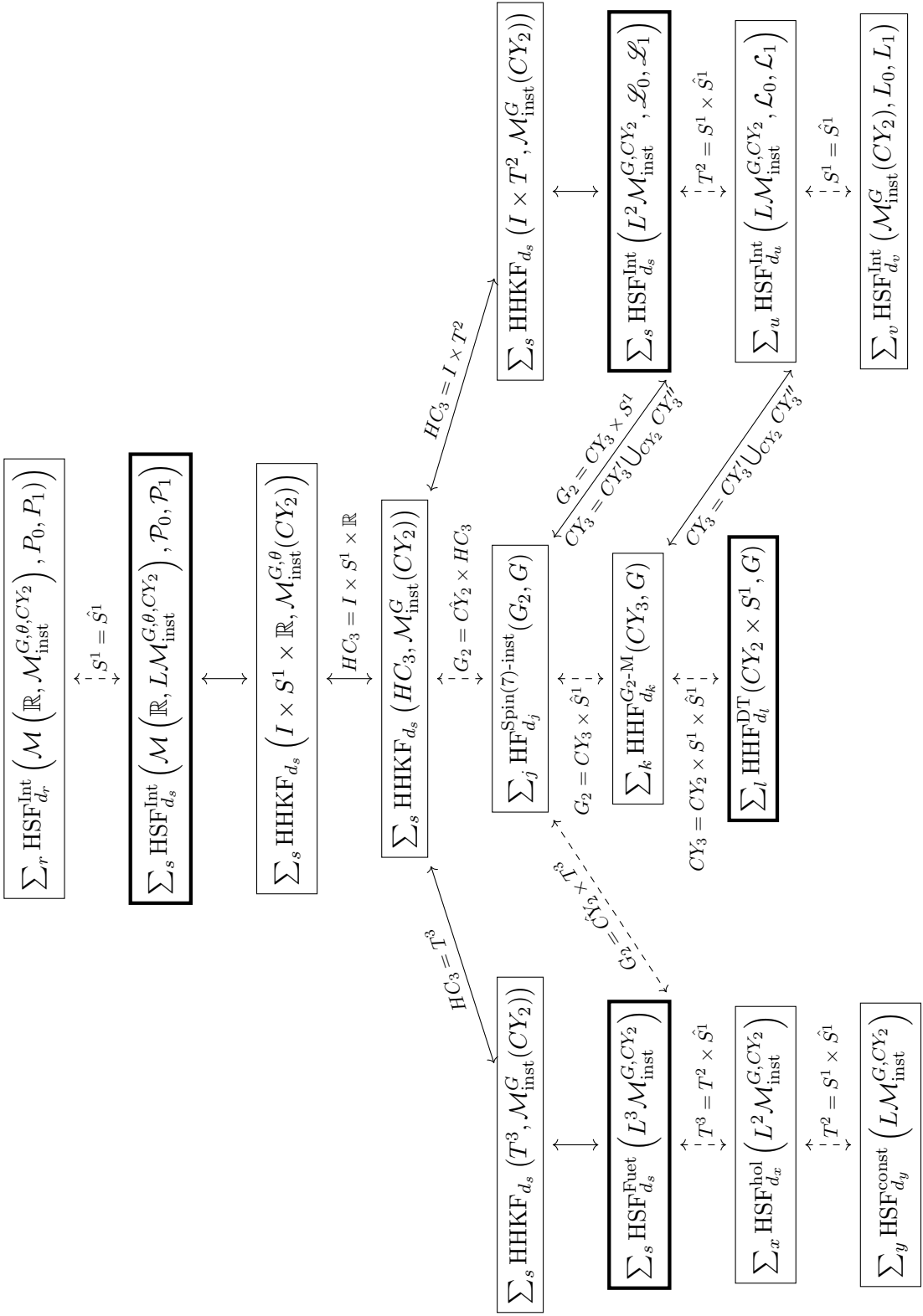

We can summarize the various relations amongst the Floer homologies obtained hitherto in~\autoref{fig:web of relations:floer}, where the radii of the $\widehat{S}^1$ circles and sizes of the $\widehat{CY_2}$'s can be varied;
\emph{dashed lines} indicate an equivalence that is due to dimensional/topological reduction;
\emph{undashed lines} indicate an equivalence that  is not due to any dimensional/topological reduction;
\emph{bold rectangles} indicate a novel Floer homology that has not been conjectured before;
and \emph{regular rectangles} indicate a Floer homology that has been conjectured.

\subsection{Topological Invariance of \texorpdfstring{Spin$(7)$}{Spin(7)} Theory and the FS Type \texorpdfstring{$A_\infty$}{A-infty}-categories}
\label{sec:topo invariance:fs-cat}

\subtitle{Relating the FS Type $A_\infty$-categories of \autoref{sec:fs-cat of m6}--\autoref{sec:fs-cat of m4}}

The topological invariance of (the $\mathcal{Q}$-cohomology of) Spin$(7)$ theory means that we can relate the partition functions \eqref{eq:cy3 x r2:lg sqm:partition fn}, \eqref{eq:cy2 x s x r2:lg sqm:partition fn}, and \eqref{eq:cy2 x r2:lg:sqm:partition fn}. In turn, via \eqref{eq:cy3 x r2:floer complex:morphism}, \eqref{eq:cy2 x s x r2:floer complex:morphism}, and \eqref{eq:cy2 x r2:floer complex:morphism}, it would mean that
\begin{equation}
  \label{eq:topo inv:fs-cat morphisms}
  \boxed{
    \begin{tikzcd}[%
      row sep=large,%
      arrows=rightarrow,%
      ]
      \text{Hom} \Big(
        \mathcal{E}^I_\text{DT}(\theta) , \mathcal{E}^J_\text{DT}(\theta)
      \Big)_\pm
      \arrow[d, "CY_3 = CY_2 \times S^1 \times \widehat{S}^1"]
      \\
      \text{Hom} \Big(
        \mathcal{E}^I_\text{HW}(\theta) , \mathcal{E}^J_\text{HW}(\theta)
      \Big)_\pm
      \arrow[d, "CY_2 \times S^1 = CY_2 \times \widehat{S}^1"]
      \\
      \text{Hom} \Big(
        \mathcal{E}^I_\text{VW}(\theta) , \mathcal{E}^J_\text{VW}(\theta)
      \Big)_\pm
    \end{tikzcd}
  }
\end{equation}
Here, $CY_3$ is the space over which the DT configurations (with scalar being zero) that correspond to the $\mathcal{E}^*_{\text{DT}}(\theta)$'s are defined on, while $CY_2 \times S^1$ is the space over which the HW configurations (with two of the three linearly-independent components of the self-dual two-form field being zero) that correspond to the $\mathcal{E}^*_{\text{HW}}(\theta)$'s are defined on.

Notice that \eqref{eq:topo inv:fs-cat morphisms} also allows us to identify the composition maps~\eqref{eq:cy3 x r2:mu-d maps},~\eqref{eq:cy2 x s x r2:mu-d maps}, and~\eqref{eq:cy2 x r2:mu-d maps}, as
\begin{equation}
  \label{eq:topo inv:mu-d maps}
  \boxed{
    \begin{tikzcd}[%
      column sep=8em,%
      arrows=Leftrightarrow,%
      ampersand replacement=\&,%
      ]
      \mu^{n_k}_{\mathfrak{A}_6}
      \arrow[r, "CY_3 = CY_2 \times S^1 \times \widehat{S}^1"]
      \&
      \mu^{n_l}_{\mathscr{A}_5}
      \arrow[r, "CY_2 \times S^1 = CY_2 \times \widehat{S}^1"]
      \&
      \mu^{n_m}_{\mathfrak{A}_4}
    \end{tikzcd}
  }
\end{equation}

The relations in \eqref{eq:topo inv:fs-cat morphisms} are consistent, in that there is a one-to-one correspondence amongst the various $\mathcal{E}^*_{\text{XX}}(\theta)$'s.
Specifically, each $\mathcal{E}^*_{\text{DT}}(\theta)$, $\mathcal{E}^*_{\text{HW}}(\theta)$, and $\mathcal{E}^*_{\text{VW}}(\theta)$ corresponds to a solution of \eqref{eq:cy3 x r2:soliton:endpts:config}, \eqref{eq:cy2 x s x r2:soliton:endpts:config}, and \eqref{eq:cy2 x r2:soliton:endpts:config}, respectively, where \eqref{eq:cy2 x r2:soliton:endpts:config} is obtained via a KK reduction of \eqref{eq:cy2 x s x r2:soliton:endpts:config}, which in turn is obtained via a KK reduction of \eqref{eq:cy3 x r2:soliton:endpts:config}.

The identifications in \eqref{eq:topo inv:mu-d maps} are also consistent, in that there is also a one-to-one correspondence amongst the various $\mu^{n_*}_{\mathfrak{A}_*}$'s.
Specifically, each `$n_k$', `$n_l$', and `$n_m$', corresponds to a pseudoholomorphic disc with $n_* + 1$ punctures at the boundary whose definition is rooted in~\eqref{eq:cy3 x r2:bps:complex},~\eqref{eq:cy2 x s x r2:bps}, and~\eqref{eq:cy2 x r2:bps}, respectively, where \eqref{eq:cy2 x r2:bps} is obtained via KK dimensional reduction of~\eqref{eq:cy2 x s x r2:bps}, which in turn is obtained via a KK reduction of~\eqref{eq:cy3 x r2:bps:complex}.

In short, we have (i) a \emph{novel} equivalence amongst the morphisms of FS type $A_{\infty}$-categories of six, five, and four-manifolds, and (ii) a \emph{novel} identification amongst their corresponding composition maps.

Thus, we have a \emph{novel} correspondence amongst FS type $A_{\infty}$-categories of six, five, and four-manifolds. In particular, one can expect the existence of a mathematical functor between these FS type $A_{\infty}$-categories (which maps their objects, morphisms, and $A_{\infty}$ composition maps), that has a physical interpretation as an invariance under KK dimensional reduction.

\subtitle{An Atiyah-Floer Type Correspondence of the FS Type $A_{\infty}$-category of  \autoref{sec:fs-cat of m6}}

From~\eqref{eq:cy3 x r2:atiyah-floer:as morphism} and~\eqref{eq:cy3 x r2:atiyah-floer:intersection floer as hom-cat}, we have
\begin{equation}
  \label{eq:topo-inv:fs-cat:atiyah-floer}
  \boxed{
    \begin{gathered}
      \text{Hom} \left(
      \mathcal{E}^I_{\text{DT}}(\theta), \mathcal{E}^J_{\text{DT}}(\theta)
      \right)_\pm
      \\
      \qquad \qquad \qquad \quad \; \,  
      \displaystyle \left\Updownarrow \vphantom{\Big(} \right. \scriptstyle{CY_3 = CY_3' \bigcup_{CY_2} CY_3''}
      \\
      \text{HSF}^{\text{Int}}_* \left(
      \mathcal{M} \left( \R, L \mathcal{M}^{G, \theta, CY_2}_{\text{inst}} \right), \mathcal{P}_0, \mathcal{P}_1
      \right)
      \\
      \displaystyle \left\Updownarrow \vphantom{\Big(} \right.
      \\
      \text{Hom} \left(
        \text{Hom}\left[ \mathcal{L}^I_0(\theta), \mathcal{L}^I_1(\theta) \right],
        \text{Hom}\left[ \mathcal{L}^J_0(\theta), \mathcal{L}^J_1(\theta) \right]
      \right)_\pm
    \end{gathered}
  }
\end{equation}

In short, we have a \emph{novel} Atiyah-Floer type correspondence amongst (i) the morphisms of an FS type $A_{\infty}$-category of a six-manifold, (ii) a symplectic intersection Floer homology of the loop space of instantons, and (iii) a Hom-category of morphisms between isotropic-coisotropic branes.

\subtitle{Another Hom-category and a Physical Proof of Bousseau's Mathematical Conjecture}

Applying the arguments used to establish the one-to-one identification of~\eqref{eq:cy3 x r2:atiyah-floer:intersection floer as hom-cat} to the RHS of~\eqref{eq:topo inv:fs-cat:symp int floer-hom from kk red}, we will have yet another one-to-one identification
\begin{equation}
  \label{eq:cy2 x s x r2:atiyah-floer:intersection floer as hom-cat}
    \text{HSF}^{\text{Int}}_* \left(
    \mathcal{M} \left( \R, \mathcal{M}^{G, \theta, CY_2}_{\text{inst}} \right), P_0, P_1
  \right)
  \Longleftrightarrow
  \text{Hom} \left(
    \text{Hom}\left[ L^I_0(\theta), L^I_1(\theta) \right],
    \text{Hom}\left[ L^J_0(\theta), L^J_1(\theta) \right]
  \right)
  \, .
\end{equation}
This, \eqref{eq:cy3 x r2:atiyah-floer:intersection floer as hom-cat}, and \eqref{eq:topo inv:fs-cat:symp int floer-hom from kk red}, would then mean that we have the following relation between the Hom-category of morphisms between isotropic-coisotropic branes of $L\mathcal{M}^{G, \theta, CY_2}_{\text{inst}}$ and $\mathcal{M}^{G, \theta}_{\text{inst}}(CY_2)$:
\begin{equation}
  \label{eq:hom-cat correspondences}
    \begin{gathered}
    \text{Hom} \left(
      \text{Hom}\left[ \mathcal{L}^I_0(\theta), \mathcal{L}^I_1(\theta) \right],
      \text{Hom}\left[ \mathcal{L}^J_0(\theta), \mathcal{L}^J_1(\theta) \right]
    \right)_\pm
    \\
    \qquad \qquad \displaystyle\left\downarrow \vphantom{\Big(} \right. S^1 = \widehat{S}^1
    \\
    \text{Hom} \left(
      \text{Hom}\left[ L^I_0(\theta), L^I_1(\theta) \right],
      \text{Hom}\left[ L^J_0(\theta), L^J_1(\theta) \right]
    \right)_\pm
    \, ,
  \end{gathered}
\end{equation}
where the spatial $S^1$ is related to the loop in $L\mathcal{M}^{G, \theta, CY_2}_{\text{inst}}$.

Therefore, from~\eqref{eq:cy3 x r2:atiyah-floer:dt config:as morphism} and the identification of $\mathcal{E}_{\text{DT}}^*(\theta)$ as generators of the $\theta$-generalized holomorphic $G_2$ instanton Floer homology of $CY_3$,\footnote{Recall from \autoref{sec:floer homology of m6:floer homology} that the $\theta$-generalized holomorphic $G_2$ instanton Floer homology of $CY_3$ is generated by $\theta$-deformed DT configurations on $CY_3$ with the scalar being zero. \label{ft:theta-deformed g2-instanton floer-hom}}
we have
\begin{equation}
  \label{eq:bousseaus conjecture}
  \boxed{
    \begin{gathered}
      \text{Hom} \left(
        \text{Hom}\left[ L^I_0(\theta), L^I_1(\theta) \right],
        \text{Hom}\left[ L^J_0(\theta), L^J_1(\theta) \right]
      \right)_\pm
      \\
      \displaystyle \left\Updownarrow \vphantom{\Big(} \right.
      \\
      \text{Hom} \left(
        \mathcal{E}^I_{\text{DT}}(\theta), \mathcal{E}^J_{\text{DT}}(\theta)
      \right)_\pm
      \\
      \displaystyle \left\Updownarrow \vphantom{\Big(} \right.
      \\
      \text{Hom} \left(
        \text{HHF}^{G_2\text{-inst}, \theta}(CY_3, G), \text{HHF}^{G_2\text{-inst}, \theta}(CY_3, G)
      \right)_\pm
    \end{gathered}
  }
\end{equation}

This is a \emph{novel} correspondence amongst (i) a Hom-category of morphisms between Lagrangian submanifolds of $\mathcal{M}^{G, \theta}_{\text{inst}}(CY_2)$, (ii) an FS type $A_{\infty}$-category whose objects are $\theta$-deformed holomorphic vector bundles on $CY_3$ (i.e., $\theta$-deformed DT configurations on $CY_3$ with the scalar being zero), and (iii) a $\theta$-generalized holomorphic  $G_2$ instanton Floer homology of $CY_3$.

When $\theta = 0$, we would have a correspondence amongst (i) a Hom-category of morphisms between Lagrangian submanifolds of $\mathcal{M}^G_{\text{inst}}(CY_2)$ (which, for $CY_2$ a complex algebraic surface, span the space of holomorphic vector bundles~\cite{donaldson-1985-anti-self} that can be extended to all of $CY_3 = CY_3' \bigcup_{CY_2} CY_3''$), (ii) an FS type $A_{\infty}$-category whose objects are holomorphic vector bundles on $CY_3$, and (iii) a holomorphic $G_2$ instanton Floer homology of $CY_3$, as conjectured by Bousseau~\cite[$\S$2.8]{bousseau-2024-holom-floer}. Thus, we have furnished a physical proof and generalization (for general $\theta \neq 0$) of Bousseau's mathematical conjecture.

\subtitle{Summarizing the Relations Amongst the FS Type $A_\infty$-categories}

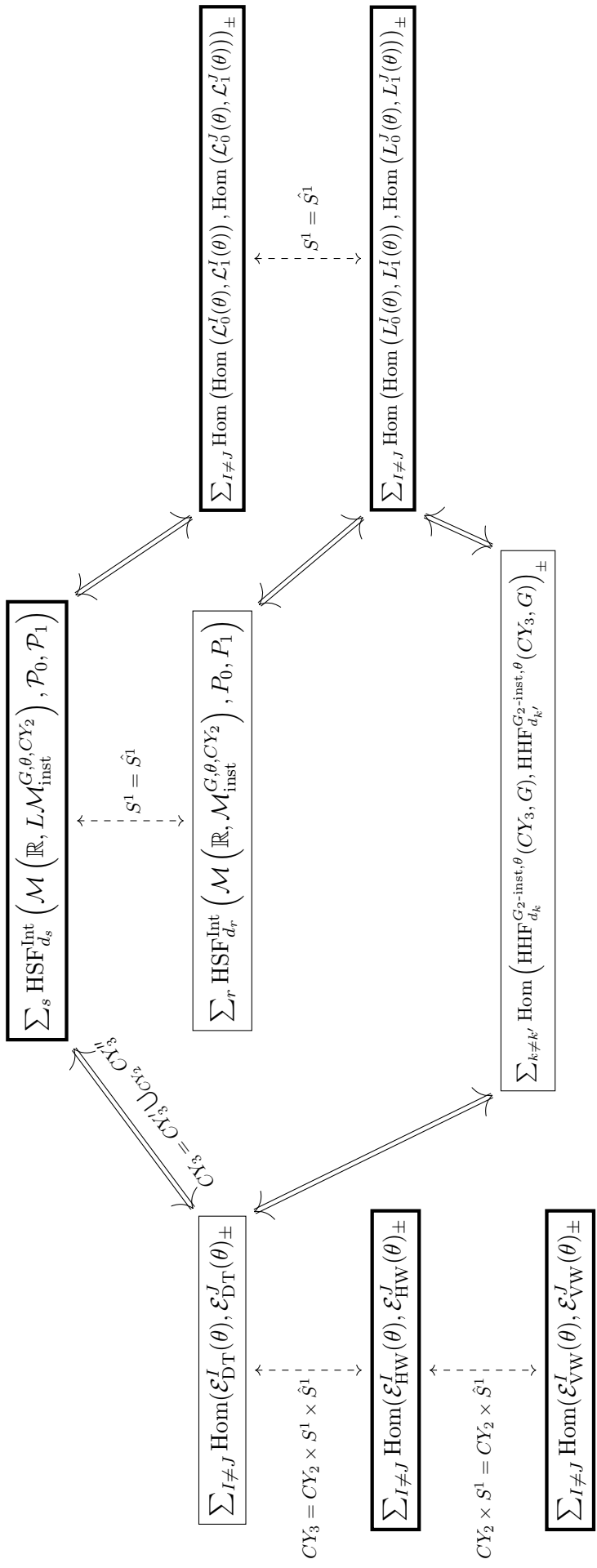
\begin{sidewaysfigure}
  \centering
  \begin{tikzpicture}[%
    auto,%
    block/.style={draw, rectangle},%
    novel/.style={draw, rectangle, ultra thick},%
    every edge/.style={draw, <->},%
    relation/.style={scale=0.8, sloped, anchor=center, align=center},%
    vertRelation/.style={scale=0.8, anchor=center, align=center},%
    horRelation/.style={scale=0.8, anchor=center, align=center},%
    shorten >=4pt,%
    shorten <=4pt,%
    ]
    \def \verRel {2} 
    \def \horRel {2} 
    \node[novel] (HSFI-RL)
    {$\sum_s \text{HSF}^{\text{Int}}_{d_s} \left(
        \mathcal{M} \left( \R, L \mathcal{M}^{G, \theta, CY_2}_{\text{inst}} \right), \mathcal{P}_0, \mathcal{P}_1
      \right)$};
    \node[block, below={\verRel} of HSFI-RL] (HSFI-R)
    {$\sum_r \text{HSF}^{\text{Int}}_{d_r} \left(
      \mathcal{M} \left( \R, \mathcal{M}^{G, \theta, CY_2}_{\text{inst}} \right), P_0, P_1
      \right)$};
    \node[block, left={1.5*\horRel} of HSFI-R] (8d-FS)
    {$\sum_{I \neq J} \text{Hom}\left(\mathcal{E}^I_{\text{DT}}(\theta), \mathcal{E}^J_{\text{DT}}(\theta)\right)_\pm$};
    \node[block, novel, below={\verRel} of 8d-FS] (7d-FS)
    {$\sum_{I \neq J} \text{Hom}\left(\mathcal{E}^I_{\text{HW}}(\theta), \mathcal{E}^J_{\text{HW}}(\theta)\right)_\pm$};
    \node[block, novel, below={\verRel} of 7d-FS] (6d-FS)
    {$\sum_{I \neq J} \text{Hom}\left(\mathcal{E}^I_{\text{VW}}(\theta), \mathcal{E}^J_{\text{VW}}(\theta)\right)_\pm$};
    \node[block, novel, right={0.8*\horRel} of HSFI-R] (Hom-cat-L)
    {\footnotesize $\sum_{I \neq J} \text{Hom} \left(
      \text{Hom}\left[ \mathcal{L}^I_0(\theta), \mathcal{L}^I_1(\theta) \right],
      \text{Hom}\left[ \mathcal{L}^J_0(\theta), \mathcal{L}^J_1(\theta) \right]
    \right)_\pm$};
    \node[novel, below={\verRel} of Hom-cat-L] (Hom-cat)
    {\footnotesize $\sum_{I \neq J} \text{Hom} \left(
      \text{Hom}\left[ L^I_0(\theta), L^I_1(\theta) \right],
      \text{Hom}\left[ L^J_0(\theta), L^J_1(\theta) \right]
    \right)_\pm$};
    \node[block, below={2*\verRel} of HSFI-R] (G2-inst)
    {\footnotesize $\sum_{k \neq k'}\text{Hom}\left(
        \text{HHF}_{d_k}^{G_2\text{-inst}, \theta}(CY_3, G),
        \text{HHF}_{d_{k'}}^{G_2\text{-inst}, \theta}(CY_3, G)
    \right)_\pm$};
    \draw
    (8d-FS) edge[dashed]
    node[vertRelation, left] {$CY_3 = CY_2 \times S^1 \times \widehat{S}^1$}
    (7d-FS)
    (7d-FS) edge[dashed]
    node[vertRelation, left] {$CY_2 \times S^1 = CY_2 \times \widehat{S}^1$}
    (6d-FS)
    (8d-FS.north east) edge[double equal sign distance]
    node[relation, below=2mm] {$CY_3 = CY_3' \bigcup_{CY_2} CY_3''$}
    (HSFI-RL.south west)
    (HSFI-RL) edge[dashed]
    node[vertRelation, right] {$S^1 = \widehat{S}^1$}
    (HSFI-R)
    (HSFI-RL.south east) edge[double equal sign distance]
    (Hom-cat-L.north west)
    (HSFI-R.south east) edge[double equal sign distance]
    (Hom-cat.north west)
    (Hom-cat-L) edge[dashed]
    node[vertRelation, right] {$S^1 = \widehat{S}^1$}
    (Hom-cat)
    (G2-inst.north east) edge[double equal sign distance]
    (Hom-cat.south west)
    (8d-FS.south east) edge[double equal sign distance]
    (G2-inst.north west)
    ;
  \end{tikzpicture}
  \caption{%
    Relations amongst FS type $A_\infty$-categories from the topological invariance of Spin$(7)$ theory.
    \label{fig:web of relations:fs-cat}
  }
\end{sidewaysfigure}

We can summarize the various relations amongst the FS type $A_\infty$-categories obtained hitherto in~\autoref{fig:web of relations:fs-cat}, where the radii of the $\widehat{S}^1$ circles and sizes of the $\widehat{CY_2}$'s can be varied;
\emph{dashed lines} indicate an equivalence that is due to dimensional/topological reduction;
\emph{undashed lines} indicate an equivalence that is not due to any dimensional/topological reduction;
\emph{double lines} indicate a correspondence;
\emph{bold rectangles} indicate a result that has not been conjectured;
and \emph{regular rectangles} indicate a result that has been conjectured.

\subsection{FS Type \texorpdfstring{$A_\infty$}{A-infty}-categories Categorifying Floer Homologies}
\label{sec:topo invariance:floer homologies and fs-cat}

\subtitle{Introducing Higher Categorical Structures in Spin$(7)$ Theory}

Referring to \autoref{sec:floer homology of m7}, we see that when we replace one of the spatial directions of the general Spin$(7)$-manifold with an $\R$ line to consider $\text{Spin}(7) = G_2 \times \mathbb{R}$, the Spin$(7)$ partition function will be associated with a Spin$(7)$ instanton Floer homology 0-category (see~\eqref{eq:g2 x r:partition fn}).

Referring to \autoref{sec:fs-cat of m6}, we see that if we further replace one of the spatial directions of the general $G_2$-manifold with an $\R$ line to consider $\text{Spin}(7) = CY_3 \times \mathbb{R}^2$, the Spin$(7)$ partition function will be associated with an FS type $A_\infty$-category of $CY_3$ 1-category (see~\eqref{eq:cy3 x r2:partition fn},~\eqref{eq:cy3 x r2:mu-d maps}, and~\eqref{eq:cy3 x r2:floer complex:morphism}).

Clearly, one can introduce higher categorical structures in Spin$(7)$ theory by replacing spatial directions of the Spin$(7)$-manifold with $\R$ lines. The more $\R$ lines there are, the higher the categorical structure. In fact, this scheme works for any gauge theory, not just Spin$(7)$ theory (see~\cite{er-2023-topol-n}).

\subtitle{The FS Type $A_\infty$-category of \autoref{sec:fs-cat of m6} Categorifying the Gauge-theoretic Floer Homology of \autoref{sec:floer homology of m6}}

Recall that the objects of the FS type $A_\infty$-category of $CY_3$ 1-category in~\autoref{sec:fs-cat of m6} are the endpoints of the $\mathfrak{A}_6^0$-solitons corresponding to DT configurations on $CY_3$ with the scalar being zero. Recall also that these configurations generate the holomorphic $G_2$ instanton Floer homology of $CY_3$ 0-category in~\autoref{sec:floer homology of m6}. In other words, the FS type $A_\infty$-category of $CY_3$ categorifies the holomorphic $G_2$ instanton Floer homology of $CY_3$.

\subtitle{The FS Type $A_\infty$-category of \autoref{sec:fs-cat of m5} Categorifying the Gauge-theoretic Floer Homology of \autoref{sec:floer homology of m5}}

Recall that the objects of the FS type $A_\infty$-category of $CY_2 \times S^1$ 1-category in~\autoref{sec:fs-cat of m5} are the endpoints of the $\mathfrak{A}_5^0$-solitons corresponding to HW configurations on $CY_2 \times S^1$ with two of the three linearly-independent components of the self-dual two-form field being zero. Recall also that these configurations generate the holomorphic DT Floer homology of $CY_2 \times S^1$ 0-category in~\autoref{sec:floer homology of m5}.\footnote{%
  The configurations which generate  the holomorphic DT Floer homology of $CY_2 \times S^1$ are HW configurations  with one of the three linearly-independent components of the self-dual two-form field being zero, which thus include the aforementioned configurations associated with the objects of the  FS type $A_\infty$-category of $CY_2 \times S^1$.
  \label{ft:hw configs condition for categorification}
}
In other words, the FS type $A_\infty$-category of $CY_2 \times S^1$ categorifies the holomorphic DT Floer homology of $CY_2 \times S^1$.

\subtitle{The FS Type $A_\infty$-category of \autoref{sec:fs-cat of m4} Categorifying the Gauge-theoretic HW Floer Homology of~\cite{er-2023-topol-n} }

Recall that the objects of the FS type $A_\infty$-category of $CY_2$ 1-category in~\autoref{sec:fs-cat of m4} are the endpoints of the $\mathfrak{A}_4^0$-solitons corresponding to VW configurations on $CY_2$ with the scalar and one of the linearly-independent components of the self-dual two-form field being zero. Recall also that these configurations generate the HW Floer homology of $CY_2$ 0-category in~\cite{er-2023-topol-n}.\footnote{%
  The configurations which generate  the HW Floer homology of $CY_2$ in~\cite{er-2023-topol-n} are VW configurations with the scalar being zero and no restrictions on the components of the self-dual two-form field, which thus include the aforementioned configurations associated with the objects of the  FS type $A_\infty$-category of $CY_2$.
  \label{ft:vw configs condition for categorification}
}
In other words, the FS type $A_\infty$-category of $CY_2$ categorifies the HW Floer homology of $CY_2$ in~\cite{er-2023-topol-n}.

\subtitle{Summarizing the Scheme of Categorification, and a 2-category}

In short, we see that to configurations on a $D$-manifold, $M_D$, one can associate a Floer homology of $M_D$ 0-category realized by the partition function of a gauge theory on $M_D \times \R$, which, in turn, can be categorified into an FS type $A_\infty$-category of $M_D$ 1-category realized by the partition function of a gauge theory on $M_D \times \R^2$.

By continuing this scheme of categorification, the FS type $A_\infty$-category of $M_D$ 1-category can be further  categorified into a 2-category of $M_D$ that is realized by the partition function of a gauge theory on $M_D \times \R^3$.
Indeed, as we will show in a sequel paper~\cite{er-2024-topol-gauge}, such a 2-category would be one that is a gauge-theoretic generalization of the Fueter 2-category recently developed by Bousseau~\cite{bousseau-2024-holom-floer} and Doan-Rezchikov~\cite{doan-2022-holom-floer}.

\subsection{A Web of Relations}

By concatenating \autoref{fig:web of relations:floer} and \autoref{fig:web of relations:fs-cat} along the common node, we get a richer web of relations depicted in \autoref{fig:web of relations}, while the results of~\autoref{sec:topo invariance:floer homologies and fs-cat} fit into a broader scheme of categorification depicted in \autoref{fig:web of relations:categorification}.
In the latter figure, \emph{dotted lines} are relations representing a categorification;
\emph{dash-dotted lines} are relations between categories due to dimensional reduction;
and the $\text{Fuet}^{\text{BPS-eqn}}(M_D, G)$'s are the expected Fueter type 2-categories of $M_D$.

\begin{sidewaysfigure}
  \centering
  \begin{tikzpicture}[%
    auto,%
    block/.style={draw, rectangle},%
    novel/.style={draw, rectangle, ultra thick},%
    every edge/.style={draw, <->},%
    relation/.style={scale=0.8, sloped, anchor=center, align=center},%
    vertRelation/.style={scale=0.8, anchor=center, align=center},%
    horRelation/.style={scale=0.8, anchor=center, align=center},%
    shorten >=4pt,%
    shorten <=4pt,%
    ]
    \def \verRel {1} 
    \def \horRel {2.6} 
    \node[block] (8d-FT)
    {$\sum_j \text{HF}^{\text{Spin}(7)\text{-inst}}_{d_j}(G_2, G)$};
    \node[block, below={\verRel} of 8d-FT] (7d-FT)
    {$\sum_k \text{HHF}^{G_2\text{-M}}_{d_k}(CY_3, G)$};
    \node[block, novel, below={\verRel} of 7d-FT] (6d-FT)
    {$\sum_l \text{HHF}^{\text{DT}}_{d_l}(CY_2 \times S^1, G)$};
    \node[block, above={\verRel} of 8d-FT] (hK-HC3)
    {$\sum_s \text{HHKF}_{d_s}\left(
        HC_3, \mathcal{M}^G_{\text{inst}}(CY_2)
      \right)$};
    \node[block, left={\horRel} of 8d-FT] (hK-T3)
    {$\sum_s \text{HHKF}_{d_s}\left(
        T^3, \mathcal{M}^G_{\text{inst}}(CY_2)
      \right)$};
    \node[block, right={\horRel} of 8d-FT] (hK-I-T2)
    {$\sum_s \text{HHKF}_{d_s}\left(
        I \times T^2, \mathcal{M}^G_{\text{inst}}(CY_2)
      \right)$};
    \node[block, above={\verRel} of hK-HC3] (hK-ISR)
    {$\sum_s \text{HHKF}_{d_s}\left(
        I \times S^1 \times \R, \mathcal{M}^{G, \theta}_{\text{inst}}(CY_2)
      \right)$};
    \node[block, below left={0.6*\verRel} and {\horRel} of hK-ISR] (8d-FS)
    {$\sum_{I \neq J} \text{Hom} \left( \mathcal{E}^I_{\text{DT}}(\theta), \mathcal{E}^J_{\text{DT}}(\theta) \right)_\pm$};
    \node[novel, above={\verRel} of 8d-FS] (7d-FS)
    {$\sum_{I \neq J} \text{Hom} \left( \mathcal{E}^I_{\text{HW}}(\theta), \mathcal{E}^J_{\text{HW}}(\theta) \right)_\pm$};
    \node[novel, above={\verRel} of 7d-FS] (6d-FS)
    {$\sum_{I \neq J} \text{Hom} \left( \mathcal{E}^I_{\text{VW}}(\theta), \mathcal{E}^J_{\text{VW}}(\theta) \right)_\pm$};
    \node[novel, below={\verRel} of hK-T3] (HSF-T3)
    {$\sum_s \text{HSF}^{\text{Fuet}}_{d_s} \left(
        L^3 \mathcal{M}^{G, CY_2}_{\text{inst}}
      \right)$};
    \node[block, below={\verRel} of HSF-T3] (HSF-T2)
    {$\sum_x \text{HSF}^{\text{hol}}_{d_x} \left(
        L^2 \mathcal{M}^{G, CY_2}_{\text{inst}}
      \right)$};
    \node[block, below={\verRel} of HSF-T2] (HSF-S)
    {$\sum_y \text{HSF}^{\text{const}}_{d_y} \left(
        L \mathcal{M}^{G, CY_2}_{\text{inst}}
      \right)$};
    \node[novel, below={\verRel} of hK-I-T2] (HSFI-T2)
    {$\sum_s \text{HSF}^{\text{Int}}_{d_s}\left(
        L^2 \mathcal{M}^{G, CY_2}_{\text{inst}}, \mathscr{L}_0, \mathscr{L}_1
      \right)$};
    \node[block, below={\verRel} of HSFI-T2] (HSFI-S)
    {$\sum_u \text{HSF}^{\text{Int}}_{d_u}\left(
        L \mathcal{M}^{G, CY_2}_{\text{inst}}, \mathcal{L}_0, \mathcal{L}_1
      \right)$};
    \node[block, below={\verRel} of HSFI-S] (HSFI)
    {$\sum_v \text{HSF}^{\text{Int}}_{d_v}\left(
        \mathcal{M}^G_{\text{inst}}(CY_2), L_0, L_1
      \right)$};
    \node[novel, above={\verRel} of hK-ISR] (HSFI-RL)
    {$\sum_s \text{HSF}^{\text{Int}}_{d_s}\left(
        \mathcal{M} \left( \R, L \mathcal{M}^{G, \theta, CY_2}_{\text{inst}} \right), \mathcal{P}_0, \mathcal{P}_1
      \right)$};
    \node[block, above={\verRel} of HSFI-RL] (HSFI-R)
    {$\sum_r \text{HSF}^{\text{Int}}_{d_r} \left(
        \mathcal{M} \left( \R, \mathcal{M}^{G, \theta, CY_2}_{\text{inst}} \right), P_0, P_1
      \right)$};
    \node[block, novel, below right={0.6*\verRel} and {0.5*\horRel} of hK-ISR] (Hom-cat-L)
    {\footnotesize $\sum_{I \neq J} \text{Hom} \left(
      \text{Hom}\left[ \mathcal{L}^I_0(\theta), \mathcal{L}^I_1(\theta) \right],
      \text{Hom}\left[ \mathcal{L}^J_0(\theta), \mathcal{L}^J_1(\theta) \right]
    \right)_\pm$};
    \node[block, novel, above={\verRel} of Hom-cat-L] (Hom-cat)
    {\footnotesize $\sum_{I \neq J} \text{Hom} \left(
      \text{Hom}\left[ L^I_0(\theta), L^I_1(\theta) \right],
      \text{Hom}\left[ L^J_0(\theta), L^J_1(\theta) \right]
    \right)_\pm$};
    {$\text{Hom}$}
    \node[block, above={\verRel} of HSFI-R] (G2-inst)
    {\footnotesize $\sum_{k \neq k'}\text{Hom}\left(
        \text{HHF}_{d_k}^{G_2\text{-inst}, \theta}(CY_3, G),
        \text{HHF}_{d_{k'}}^{G_2\text{-inst}, \theta}(CY_3, G)
    \right)_\pm$};
    \draw
    (8d-FT) edge[dashed]
    node[vertRelation, left] {$G_2 = CY_3 \times \widehat{S}^1$}
    (7d-FT)
    (7d-FT) edge[dashed]
    node[vertRelation, left] {$CY_3 = CY_2 \times S^1 \times \widehat{S}^1$}
    (6d-FT)
    (8d-FT) edge[dashed]
    node[vertRelation, right] {$G_2 = \widehat{CY_2} \times HC_3$}
    (hK-HC3)
    (hK-HC3.south east) edge
    node[relation, above] {$HC_3 = I \times T^2$}
    (hK-I-T2)
    (hK-I-T2) edge (HSFI-T2)
    (8d-FT.south east) edge
    node[relation, above] {$G_2 = CY_3 \times S^1$}
    node[relation, below] {$CY_3 = CY_3' \bigcup_{CY_2} CY_3''$}
    (HSFI-T2.south west)
    (HSFI-T2) edge[dashed]
    node[vertRelation, right] {$T^2 = S^1 \times \widehat{S}^1$}
    (HSFI-S)
    (7d-FT.south east) edge
    node[relation, below] {$CY_3 = CY_3' \bigcup_{CY_2} CY_3''$}
    (HSFI-S.south west)
    (HSFI-S) edge [dashed]
    node[vertRelation, right] {$S^1 = \widehat{S}^1$}
    (HSFI)
    (hK-HC3.south west) edge
    node[relation, above] {$HC_3 = T^3$}
    (hK-T3)
    (hK-T3) edge (HSF-T3)
    (8d-FT.south west) edge[dashed]
    node[relation, above] {$G_2 = \widehat{CY_2} \times T^3$}
    (HSF-T3.south east)
    (HSF-T3) edge[dashed]
    node[vertRelation, right] {$T^3 = T^2 \times \widehat{S}^1$}
    (HSF-T2)
    (HSF-T2) edge[dashed]
    node[vertRelation, right] {$T^2 = S^1 \times \widehat{S}^1$}
    (HSF-S)
    (hK-HC3.north) edge
    node[vertRelation, right] {$HC_3 = I \times S^1 \times \R$}
    (hK-ISR.south)
    (hK-ISR) edge (HSFI-RL)
    (8d-FS) edge[dashed]
    node[vertRelation, left] {$CY_3 = CY_2 \times S^1 \times \widehat{S}^1$}
    (7d-FS)
    (7d-FS) edge[dashed]
    node[vertRelation, left] {$CY_2 \times S^1 = CY_2 \times \widehat{S}^1$}
    (6d-FS)
    (8d-FS.south east) edge[double equal sign distance]
    node[relation, below] {$CY_3 = CY_3' \bigcup_{CY_2} CY_3''$}
    (HSFI-RL.south west)
    (HSFI-RL) edge[dashed]
    node[vertRelation, right] {$S^1 = \widehat{S}^1$}
    (HSFI-R)
    (HSFI-RL.south east) edge[double equal sign distance]
    (Hom-cat-L.north west)
    (HSFI-R.south east) edge[double equal sign distance]
    (Hom-cat.north west)
    (Hom-cat-L) edge[dashed]
    node[vertRelation, right] {$S^1 = \widehat{S}^1$}
    (Hom-cat)
    (8d-FS.north east) edge[double equal sign distance]
    (G2-inst.south west)
    (G2-inst.south east) edge[double equal sign distance]
    ($(Hom-cat.north west)!0.20!(Hom-cat.north east)$)
    ;
  \end{tikzpicture}
  \caption{%
    A web of relations amongst the Floer homologies and FS type $A_\infty$-categories.
    \label{fig:web of relations}
  }
\end{sidewaysfigure}

\begin{sidewaysfigure}
  \centering
  \begin{tikzpicture}[%
    auto,%
    block/.style={draw, rectangle},%
    every edge/.style={draw, ->},%
    relation/.style={scale=0.8, sloped, anchor=center, align=center},%
    vertRelation/.style={scale=0.8, anchor=center, align=center},%
    shorten >=4pt,%
    shorten <=4pt,%
    ]
    \def \verRel {2} 
    \def \horRel {3.5} 
    \node[block] (Z-Spin7)
    {$\mathcal{Z}_{\text{Spin(7)}}(G)$};
    \node[block, below={1.5*\verRel} of Z-Spin7] (8d-FS)
    {$\text{FS}^{\text{Spin}(7)\text{-inst}}(CY_3, G)$ 1-cat};
    \node[block, left={\horRel} of 8d-FS] (8d-FT)
    {$\text{HF}^{\text{Spin}(7)\text{-inst}}(G_2, G)$ 0-cat};
    \node[block, right={\horRel} of 8d-FS] (8d-Ft)
    {$\text{Fuet}^{\text{Spin}(7)\text{-inst}}(CY_2 \times S^1, G)$ 2-cat};
    \node[block, below={\verRel} of 8d-FS] (7d-FS)
    {$\text{FS}^{G_2\text{-M}}(CY_2 \times S^1, G)$ 1-cat};
    \node[block, below={\verRel} of 8d-FT] (7d-FT)
    {$\text{HHF}^{G_2\text{-M}}(CY_3, G)$ 0-cat};
    \node[block, below={\verRel} of 8d-Ft] (7d-Ft)
    {$\text{Fuet}^{G_2\text{-M}}(CY_2, G)$ 2-cat};
    \node[block, below={\verRel} of 7d-FT] (6d-FT)
    {$\text{HHF}^{\text{DT}}(CY_2 \times S^1, G)$ 0-cat};
    \node[block, below={\verRel} of 7d-FS] (6d-FS)
    {$\text{FS}^{\text{DT}}(CY_2, G)$ 1-cat};
    \node[block, below={\verRel} of 6d-FT] (5d-FT)
    {$\text{HF}^{\text{HW}}(CY_2, G)$ 0-cat};
    \draw
    (Z-Spin7.south west) edge node[relation, above]
    {$\text{Spin}(7) = G_2 \times \R$}
    (8d-FT)
    (Z-Spin7) edge node[vertRelation, left]
    {$\text{Spin}(7) = CY_3 \times \R^2$}
    (8d-FS)
    (Z-Spin7.south east) edge node[relation, above]
    {$\text{Spin}(7) = CY_2 \times S^1 \times \R^3$}
    (8d-Ft)
    (8d-FT) edge[loosely dashdotted, <->]
    node[vertRelation, left] {$G_2 = CY_3 \times \widehat{S}^1$}
    (7d-FT)
    (7d-FT) edge[loosely dashdotted, <->]
    node[vertRelation, left] {$CY_3 = CY_2 \times S^1 \times \widehat{S}^1$}
    (6d-FT)
    (6d-FT) edge[loosely dashdotted, <->]
    node[vertRelation, left] {$CY_2 \times S^1 = CY_2 \times \widehat{S}^1$}
    (5d-FT)
    (8d-FS) edge[loosely dashdotted, <->]
    node[vertRelation, left] {$CY_3 = CY_2 \times S^1 \times \widehat{S}^1$}
    (7d-FS)
    (7d-FS) edge[loosely dashdotted, <->]
    node[vertRelation, left] {$CY_2 \times S^1 = CY_2 \times \widehat{S}^1$}
    (6d-FS)
    (8d-Ft) edge[loosely dashdotted, <->]
    node[vertRelation, left] {$CY_2 \times S^1 = CY_2 \times \widehat{S}^1$}
    (7d-Ft)
    (8d-FT.east) edge
    node[relation, above] {$G_2 = CY_3 \times \R$}
    (8d-FS.west)
    (7d-FT.east) edge
    node[relation, above] {$CY_3 = CY_2 \times S^1 \times \R$}
    (7d-FS.west)
    (6d-FT.east) edge
    node[relation, above] {$CY_2 \times S^1 = CY_2 \times \R$}
    (6d-FS.west)
    (8d-FS.east) edge
    node[relation, above] {$CY_3 = CY_2 \times S^1 \times \R$}
    (8d-Ft.west)
    (7d-FS.east) edge
    node[relation, above] {$CY_2 \times S^1 = CY_2 \times \R$}
    (7d-Ft.west)
    (7d-FT.north east) edge[dotted]
    node[relation, above] {Categorification}
    (8d-FS.south west)
    (6d-FT.north east) edge[dotted]
    node[relation, above] {Categorification}
    (7d-FS.south west)
    (5d-FT.north east) edge[dotted]
    node[relation, above] {Categorification}
    (6d-FS.south west)
    (7d-FS.north east) edge[dotted]
    node[relation, above] {Categorification}
    (8d-Ft.south west)
    (6d-FS.north east) edge[dotted]
    node[relation, above] {Categorification}
    (7d-Ft.south west)
    ;
  \end{tikzpicture}
  \caption{A scheme of categorification within Spin$(7)$ theory.}
  \label{fig:web of relations:categorification}
\end{sidewaysfigure}

\printbibliography

\end{document}